\documentclass[rmp,aps,twocolumn]{revtex4-1}

\pdfoutput=1
\usepackage{amsmath}
\usepackage{graphicx}
\usepackage{color}
\usepackage{soul}
\usepackage{units}
\usepackage[colorlinks, linkcolor=blue, citecolor=blue, urlcolor=blue,breaklinks=true]{hyperref}

\newcommand{\ket}[1]{\left\vert #1\right\rangle}
\newcommand{\bra}[1]{\left\langle #1\right\vert}

\newcommand{\abs}[1]{\left\vert #1 \right\vert}

\newcommand{\lp}{\left(}
\newcommand{\rp}{\right)}
\newcommand{\sub}[1]{{\rm #1}}
\newcommand{\re}{\mathrm{Re}}

\renewcommand{\vec}[1]{{\bf {#1}}}
\newcommand{\vecr}{\vec{r}}
\newcommand{\vecp}{\vec{p}}

\newcommand{\mat}[1]{\mathbf{#1}}
\newcommand{\rev}[1]{\vec{r}^{(#1)}}
\newcommand{\lev}[1]{\vec{l}^{(#1)}}

\newcommand{\be}{\begin{equation}}
\newcommand{\ee}{\end{equation}}
\newcommand{\bea}{\begin{eqnarray}}
\newcommand{\eea}{\end{eqnarray}}

\newcommand\f{{\mathbf f}}

\newcommand\kB{k_{\mathrm B}}

\newcommand\XXi{{\boldsymbol\Xi}}

\newcommand\DA{\Delta_{A}}

\newcommand\DC{\Delta_{C}}
\newcommand{\etaeff}{\eta_\sub{eff}}

\def\eref#1{Eq.~(\ref{#1})}

\newcommand\vect[1]{\ensuremath{\mathbf{#1}}}
\newcommand\nab{\ensuremath{\vect{\nabla}}}

\newcommand{\ie}{i.e.\ }

\newcommand{\eq}[1]{\begin{equation}{#1}\end{equation}}

\newcommand{\bd}{\begin{equation*}}
\newcommand{\ed}{\end{equation*}}

\newcommand{\vp}{\operatorname{vp}}

\newcommand{\ew}[1]{\left\langle{#1}\right\rangle}

\newcommand{\omrec}{\omega_R}

\newcommand{\vT}{v_\mathrm{T}}

\newcommand{\Tkin}{T_\mathrm{kin}}

\graphicspath{{./figures/}}

\begin{document}

\title{Cold atoms in cavity-generated dynamical optical potentials}

\author{Helmut Ritsch}
\email{helmut.ritsch@uibk.ac.at}
\affiliation{Institut f\"ur theoretische Physik, Universit\"at Innsbruck, Technikerstr.~25, A-6020 Innsbruck, Austria}
\author{Peter Domokos}
\email{domokos.peter@wigner.mta.hu}
\affiliation{Wigner Research Centre for Physics, Hungarian Academy of Sciences, H-1525 Budapest P.O. Box 49, Hungary}
\author{Ferdinand Brennecke}
\email{brennecke@phys.ethz.ch}
\author{Tilman Esslinger}
\email{esslinger@phys.ethz.ch}
\affiliation{Institute for Quantum Electronics, ETH Z\"urich, CH-8093 Z\"urich, Switzerland}

\begin{abstract}
We review state-of-the-art theory and experiment of the motion of
cold and ultracold atoms coupled to the radiation field within a
high-finesse optical resonator in the dispersive regime of the
atom-field interaction with small internal excitation.  The
optical dipole force on the atoms together with the back-action of
atomic motion onto the light field gives rise to a complex
nonlinear coupled dynamics. As the resonator constitutes an open
driven and damped system, the dynamics is non-conservative and in
general enables cooling and confining the motion of polarizable
particles. In addition the emitted cavity field allows for
real-time monitoring of the particle's position with minimal
perturbation up to sub-wavelength accuracy. For many-body systems,
the resonator field mediates controllable long-range atom-atom
interactions, which set the stage for collective phenomena.
Besides correlated motion of distant particles, one finds critical
behavior and non-equilibrium phase transitions between states of
different atomic order in conjunction with superradiant light
scattering. Quantum degenerate gases inside optical resonators can
be used to emulate opto-mechanics as well as novel quantum phases
like supersolids and spin glasses. Non-equilibrium quantum phase
transitions as predicted by e.g.~the Dicke Hamiltonian can be
controlled and explored in real-time via monitoring the cavity
field. In combination with optical lattices, the cavity field can
be utilized for non-destructive probing Hubbard physics and
tailoring long-range interactions for ultracold quantum systems.
\end{abstract}

\date{28. September 2012}

\maketitle

\tableofcontents

\section{INTRODUCTION}
\label{sec:intro} Laser light is a versatile tool to cool, prepare
and manipulate atoms. Laser cooling
\cite{CohenTannoudji1998Nobel,Chu1998Nobel, Phillips1998Nobel} and
optical pumping \cite{Happer1972Optical} relies on spontaneous
emission, which is particularly important if the laser frequency
is tuned close to the energy of an atomic transition. It is
suppressed if the laser frequency is tuned far from any  internal
excited atomic state. In this limit coherent scattering of photons
dominates and the resulting light force, the dipole force,  can be
derived from an optical potential proportional to the laser
intensity inducing a Stark-shift. This forms the basis for
trapping and the manipulation of cold atoms
\cite{Grimm2000Optical}, Bose-Einstein condensates
\cite{Cornell2002Nobel,Ketterle2002Nobel}, quantum gases
\cite{Bloch2008Manybody,Giorgini2008Theory} and mesoscopic
particles \cite{Gordon1980Motion}, where spontaneous emission has
to be avoided. In free space the back-action of the particles onto
the trapping laser is negligible.  In a microscopic picture, this
means that the probability of a photon to be scattered by a
particle is so small, that the chance for a second scattering
event involving the same photon is negligibly small. Hence, the
modifications of the field are not felt by the particles and the
light forms a conservative optical potential.

The situation changes drastically when the light field is confined
in a high-quality optical resonator. Due to multiple round trips
of intracavity photons not only the dipole force gets strongly
enhanced, but also the back-action of the atoms on the light gets
significant. Since atomic motion and {cavity field dynamics}
influence each other, they have to be treated on equal footing. In
most cases the dipole force then can no longer be derived from a
conservative potential \cite{Horak1997CavityInduced} and the field
dynamics get nonlinear \cite{Vukics2009Cavity}.

To get an intuitive picture, consider for example a moving
point-like atom, or an entire atomic cloud, forming a dielectric
medium with refractive index inside a cavity. This induces a phase
shift on the light field that depends on the position and shape of
the medium relative to the resonator mode structure.
Correspondingly, the cavity resonance frequency is dynamically
shifted with respect to the empty cavity. If this shift is
comparable to the cavity linewidth, the cavity field intensity,
induced by an external pump laser, can undergo a resonant
enhancement and so can the back-action on the motion of the
medium. For several atoms this coupled atom-field dynamics has the
character of a long-range inter-particle interaction. It also
generates a strong nonlinear field response, even if the particles
are linearly polarizable, as e.g. atoms in the low saturation
regime. Coupling to further light modes gives rise to interference
effects, which are the origin of collective instabilities and
self-organization phenomena. Photons leaking out of the cavity
cause a damping of this coupled dynamics. This designable decay
channel can be utilized to cool the motion of the medium
independent of its specific characteristics.

Historically, cavity quantum electrodynamics (QED) was born as a
research field devoted to studying the radiation properties of
atoms when boundaries are present
\cite{Purcell1946Spontaneous,Haroche1992Cqed,Berman1994Cavity}. Advances in
cavity technology over more than 30 years allowed to reach, both
in the microwave \cite{Raimond2001Manipulating,Walther2002Quantum}
and optical \cite{Kimble1998Strong,Mabuchi2002Cavity} frequency
domains, the strong coupling regime where the coherent interaction
between an atomic transition and a single radiation field mode
dominates over all dissipation processes. As a next step, cold and
slow atoms have been integrated successfully within optical cavity
QED experiments, which led to significant coupling of the atomic
motion to the cavity field. It got possible to generate
sufficiently strong forces in order to trap an atom in the field
of a single photon \cite{Hood2000AtomCavity,Pinkse2000Trapping}.
Several experiments achieved strong coupling even in the
dispersive regime of cavity QED where the detuning between the
light field and the internal atomic transitions is large. Although
the resonant energy exchange between atom and field is suppressed
in this regime,  the position-dependent cavity frequency shift
exceeds the cavity linewidth.  Motion-induced changes of the
effective resonator frequency and its back-action on mechanical
motion is also the physical ground of cavity optomechanics
\cite{Kippenberg2008Cavity}, which can be considered as an
extension of dispersive cavity QED towards macroscopic objects.

In this review we survey the recent advancements of cavity QED
systems in which coherent momentum exchange between particles and
radiation field is the dominating effect of the light-matter
interaction. The external degree of freedom of the material
component ranges from the center-of-mass motion of a single atom,
or a cloud of cold atoms, to the {density distribution} of a
continuous medium such as the quantized matter-wave field of an
ultracold gas.

The review illuminates different generic features of the
cavity-generated optical dipole force and is structured in three
main sections. Briefly summarizing, in Sect.~\ref{sec:singleatom}
we discuss the {consequences} of the retardation between atomic
motion and the cavity field dynamics. This time delay leads to an
irreversible dynamics that can be the basis of cooling schemes, as
presented for single atoms in a cavity. Sect.~\ref{sec:coldatoms}
discusses {how the field modification induced by an atom acts back
on the motion of other atoms moving within the cavity.} This
cavity-mediated atom-atom interaction is a source of collective
effects in atomic clouds. Finally, in Sect.~\ref{sec:QuantumGas}
we consider {the collective dynamics of an ultracold gas induced
by its strong coupling to the cavity field.} Owing to the low
temperature, the dynamics involves a reduced set of motional
degrees of freedom, and the system becomes a realization of
various paradigmatic models of quantum many-body physics and
quantum optics \cite{Lewenstein2007Ultracold}.

The most elementary situation that we will discuss
is the dispersive atom-field dynamics of a single atom, or
{polarizable} particle, inside a laser-driven high-finesse cavity
\cite{Pinkse2002Single,Domokos2003Mechanical}. The cavity field
dynamically responds to the position and velocity of the particle,
thereby generating a time-dependent dipole force acting back onto
the particle motion. It is the finite response time of the cavity
field which gives rise to the velocity dependent component of the
force. It can have the character of a friction force shuffling
kinetic energy from the particle to the cavity field and dissipating it via the cavity loss channel.
\cite{Vuletic2000Laser}. This allows for cooling and self-trapping
of particles in the cavity field
\cite{Maunz2004Cavity,Nussmann2005Vacuumstimulated}. Sub-recoil cavity
cooling of an ultracold atomic cloud has been achieved recently
\cite{Wolke2012Cavity}, which paves the way towards reaching
quantum degeneracy without relying on evaporative cooling
techniques. Cavity cooling allows for slowing
of any sufficiently polarizable particle with small absorption,
without the need of a cycling transition.
The possibility of extending the applicability of cavity cooling
beyond atoms has been the subject of extensive research in the
past years. The light field leaking out the cavity carries
information on the trajectory of the particle
\cite{Hood1998RealTime,Maunz2003Emission}. Continous monitoring of the atomic
motion, in turn, can be used for feedback control \cite{Fischer2002Feedback}, which became
by today the standard tool to capture single atoms inside a cavity for
quantum manipulation \cite{Kubanek2009Photonbyphoton}.

For cold atomic ensembles inside a laser-driven cavity
\cite{Kruse2003Cold,Elsasser2004Optical} the atom-field coupling
strength increases and the dynamics becomes more complex. In many
cases the effective coupling strength between particles and cavity
field scales with the square root of the particle number
\cite{Tavis1968Exact,Raizen1989Normalmode,Sauer2004Cavity,Tuchman2006Normalmode}.
As a consequence of this, the cooling of the center-of-mass motion
is correspondingly more effective. Additional complexity arises
from the relative motion of the particles, as the local intensity
of the cavity field experienced by one atom depends on the
position of all other atoms \cite{Horak2001Scaling}. This gives
rise to an effective long-range \cite{Munstermann2000Observation},
or global atom-atom interaction, described by an overall
dispersive shift. The contribution of each particle to this shift
depends on the local field intensity, which is proportional to the
square of the cavity mode function at the position of the atom. In
the low excitation regime this can be captured by a collective
potential. Furthermore, dissipative forces acting on the relative
motion of the particles have been identified
\cite{Chan2003Observation} and interesting correlations between
particles can build up \cite{Asboth2004Correlated}.

The cavity-meditated long-range interactions have a different
character when the mode of the driving field is not identical with
the cavity mode. In this case, the atoms can be considered as
sources for the intracavity field and interference between these
sources becomes crucial. Correspondingly, the effective cavity driving strength
depends on the position of all atoms within the cavity mode
profile and it is the field amplitude rather than its intensity,
which mediates the long-range interaction. For the case of a
transversally laser-driven atomic ensemble in a linear cavity, the
long-range interaction causes a phase transition to a
self-organized phase, in which the atoms arrange themselves in a
checkerboard pattern, thereby maximizing scattering into the
cavity mode \cite{Domokos2002Collective,Black2003Observation}. In
a unidirectionally driven ring cavity geometry, collective
scattering between the two counter-propagating cavity modes results in
a collective instability, referred to as collective atomic recoil
lasing \cite{Kruse2003Observation}. Various mean-field type
theories can be used to describe the non-equilibrium dynamics and
asymptotic behavior of large atomic ensembles, including the
derivation of scaling laws characterizing the above described
critical phenomena
\cite{Asboth2005Selforganization,Griesser2010Vlasov}.

Coupling ultracold atomic ensembles or Bose-Einstein condensates
to the radiation field inside a high-finesse resonator, requires a
quantized description of the atomic motion and reduces the number
of relevant external degrees of freedom
\cite{Gupta2007Cavity,Colombe2007Strong,Brennecke2007Cavity,
Slama2007Superradiant}. In the case of a laser-driven cavity,
situations can be realized where the cavity field couples
dominantly to a single collective motional mode of the atomic
ensemble, providing a direct analogy to cavity optomechanics
\cite{Murch2008Observation,Brennecke2008Cavity,StamperKurn2012Optomechanics}.
Coupling a laser-driven Bose-Einstein condensate to the vacuum
field of a cavity, leads to a quantum phase transition between a
superfluid and a self-organized phase
\cite{Nagy2008Selforganization,Baumann2010Dicke}. This provides an
open-system realization of the Dicke Hamiltonian and its quantum
phase transition
\cite{Dicke1954Coherence,Hepp1973Superradiant,Dimer2007Proposed,Nagy2010DickeModel}.
The self-organized state can also be considered as a supersolid
resulting from a broken Ising-type symmetry. More complex
situations occur in highly degenerate multi-mode cavities
\cite{Gopalakrishnan2009Emergent,Gopalakrishnan2011Frustration,Strack2011Dicke}.

Ultracold gases in optical lattices are one of the most intriguing
systems in which the power of atomic and laser physics can be
exploited to explore generic phenomena of solid-state physics
\cite{Bloch2008Manybody}. The Hubbard model describing the
dynamics of {periodically arranged} bosons or fermions can be \emph{de facto}
realized with adjustable parameters and variable dimensionality.
When the optical lattice potential is created by the field
sustained by an optical high-finesse cavity, the corresponding
cavity Hubbard-model predicts exotic new phases of matter
\cite{Larson2008MottInsulator,Maschler2005Cold}. In many cases the
cavity fields provide for a convenient, built-in real-time
observation tool. Analyzing the emitted fields allows for
dynamical monitoring of quantum phase transitions with minimum and
well controlled measurement back-action \cite{Mekhov2007Probing}.


\section{SINGLE ATOMS IN A CAVITY}
\label{sec:singleatom}

A central objective  of cavity quantum electrodynamics (QED) is
the perfect control of light--matter interaction at the
single-atom and single-photon level in the regime of strong
coupling where atom and cavity field form a single entity. A long
lifetime of such an `atom-photon molecule' requires slow and very
cold atoms to ensure long interaction times and precise control of
the atomic position. At sufficiently small kinetic energies,
however, the light forces induced by even a few intracavity
photons influence the atomic trajectory. The first cavity QED
experiments with cold atoms have already manifested that the
cavity light forces guide or deflect slowly moving atoms.  In
addition, extra diffusion takes place in cavity-sustained dipole
traps which may remove the atom from the interaction volume. Clear
signatures of such effects have been observed  in transmission
spectroscopy experiments
\cite{Mabuchi1996Realtime,Hood1998RealTime,Munstermann1999Dynamics}.
Time-resolved detection of the transmitted light signal allowed
for the reconstruction of atomic trajectories
\cite{Pinkse2000Trapping,Hood2000AtomCavity}.  These experiments
set the stage to include the atomic center-of-mass degrees of
freedom and the optical forces in the cavity QED theory. In the
following decade, the theoretical and experimental efforts
resulted in an extension of the interaction time from the
transit-time range of microseconds to the range of minutes
\cite{Kubanek2011Feedback,Figueroa2011Electromagnetically}.

\subsection{Mechanical effects of light on atoms in a cavity}
\label{sec:model}

The theoretical description of the coupled atom-field
dynamics has been presented in detail by
\textcite{Domokos2003Mechanical}. Here, we recapitulate the notations and methods. Within the vast field
of single-atom cavity QED, we restrict ourselves to the atomic
motion in a cavity, in particular, to the important concept of
cavity cooling. We review the recent experiments demonstrating
cavity cooling of single atoms. It is a manifestation of the
time-delayed action of the electric dipole force on atoms within
the cavity. The understanding at a single-atom level complements
nicely another facet of cavity cooling which we will encounter in
the case of many atom systems, where it appears in the form of the
imaginary part of the collective excitation spectrum.

\subsubsection{A two-level atom  in a cavity}
We consider a single two-level atom with transition frequency
$\omega_A$ coupled to a single mode of the electromagnetic field
inside an optical resonator with resonance frequency $\omega_C$.
These frequencies will be referenced to
the frequency $\omega$ of an external pump laser by defining the
cavity detuning $\DC=\omega-\omega_C$ and the
atomic detuning $\Delta_A=\omega-\omega_A$.  The two relevant
atomic states are the ground state $\ket{g}$ and the excited state
$\ket{e}$. We introduce the atomic raising and lowering operators,
$\sigma^\dagger = \ket{e}\bra{g}$ and $\sigma=\ket{g}\bra{e}$. The
cavity mode variables are the photon creation and annihilation
operators, $a^\dagger$ and $a$, respectively. In the
electric-dipole and the rotating-wave approximations and in a
frame rotating at the angular frequency $\omega$, the
atom-field coupling is described by
\begin{equation}
\label{eq:H_JC}
 H_\sub{JC}/\hbar = -\DC a^\dagger a -\DA(\vecr) \sigma^\dagger \sigma + i g  \lp \sigma^\dagger a \,f(\vecr) -  f^*(\vecr)\, a^\dagger \sigma \rp \,,
\end{equation}
which is usually referred to as the Jaynes--Cummings Hamiltonian
\cite{Jaynes1963Comparison} and, in the quantum optical context,
has been reviewed by \textcite{Shore1993JaynesCummings}. The
emphasis here is on that the position $\vecr$ of the atom is
explicitly taken into account. The spatial dependence of the
atomic detuning, $\DA(\vecr) = \Delta_A -\Delta_S(\vecr)$, may
account for a differential AC-Stark shift $\Delta_S(\vecr)$ which
can be induced by auxiliary, far-detuned optical trapping fields.
The coupling strength in \eref{eq:H_JC} is spatially modulated
according to the intracavity electric field strength which is
proportional to the cavity mode function $f(\vecr)$. For the
effects reviewed in this paper, it is sufficient to consider
modulations on the optical wavelength scale, thus writing
$f(\vecr) = \cos(kx)$ for a standing-wave mode of a Fabry-P\'erot
resonator, or $f(\vecr)=e^{\pm i k x}$ for the running-wave modes
sustained by a ring resonator ($k=\omega/c$ is the optical
wavenumber). The maximum coupling strength is given by the
single-photon Rabi frequency, $g = d \sqrt{\hbar
\omega_C/2\epsilon_0 {\cal V}}$, where $d$ is the atomic dipole
moment along the cavity mode polarization and ${\cal V} = \int d^3
\mathbf{r} |f(\mathbf{r})|^2$ denotes the effective cavity mode
volume (the maximum of $|f(\mathbf{r})|$ is set to 1). The
rotating wave approximation relies on that the characteristic
frequencies of $H_\sub{JC}$ are much smaller than the optical
frequency, \ie, ($|\Delta_A|, |\Delta_C|, g \ll \omega$).  The
atomic center-of-mass (CM) motion is a dynamical component of
the system, which is described by the Hamiltonian
\begin{equation}
\label{eq:H_mech}
 H_\sub{mech} = \frac{\vecp^2}{2 m} + V_\sub{cl}(\vecr) \,,
\end{equation}
where $m$ is the mass of the atom and the term $V_\sub{cl}$ can represent an arbitrary external
trapping potential. For the case of a far off-resonance optical
dipole trap, this term, together with the differential AC-Stark
shift $\Delta_S(\vecr)$ in \eref{eq:H_JC}, fully describe the
effect of the trapping laser.
The characteristic frequency of the CM motion is given by the
kinetic energy of an atom carrying one unit of photon
momentum, $|\vecp| = \hbar k$. We will use throughout the paper
the notion of \emph{recoil frequency} \cite{cct}, with the notation
$$\omega_R \equiv \frac{\hbar k^2}{2m}\,.$$

The system can be excited with a coherent laser field at
frequency $\omega$, which either drives the cavity mode with driving amplitude $\eta$, or directly the atomic internal degree of freedom at Rabi frequency $\Omega$, described by
\begin{equation}
 \label{eq:H_pump}
  H_\sub{pump}/\hbar = i\eta \lp a^\dagger - a \rp +  i \Omega\, h(\vecr) \lp \sigma^\dagger - \sigma \rp \,.
\end{equation}
For the case of pumping the atom with a standing-wave laser
field from a transverse direction perpendicular to the cavity
axis, the spatial mode function is given by $h(\vecr) = \cos(kz)$.
$H_\sub{pump}$ is effectively time-independent since
we work in the frame rotating at the angular frequency $\omega$ of
the monochromatic pump laser.

Cavity QED systems in the optical domain are strongly influenced
by dissipative coupling to the vacuum modes of the electromagnetic
field environment (thermal photons can be neglected at optical
frequencies). Correspondingly, the dynamics of the system is
described by a quantum master equation
\cite{Carmichael2003Statistical}
\begin{equation}
\label{eq:ME}
  \dot{\rho} = -\frac{i}{\hbar}\bigl[H,\rho\bigr] + {\cal L}_\sub{cav}\rho + {\cal L}_\sub{atom}\rho\,,
\end{equation}
with $H = H_\mathrm{JC} + H_\mathrm{mech} + H_\mathrm{pump}$ and $\rho$ denoting the  density operator for the atomic (motional and internal) and cavity degrees of freedom.
The dissipative processes are captured by the Liouville operators
in Born-Markov approximation
\begin{subequations}
\begin{equation}
 \label{eq:L_cav}
{\cal L}_\sub{cav} \rho  = - \kappa \left(a^\dagger a \rho + \rho a^\dagger a - 2 a \rho a^\dagger\right)\,,
\end{equation}
describing decay of the cavity field at rate $\kappa$, and
\begin{multline}
 \label{eq:L_atom}
{\cal L}_\sub{atom} \rho =  - \gamma  \Bigl( \sigma^\dagger \sigma \rho + \rho \sigma^\dagger \sigma
    \\  - 2 \int d^2{\bf u}\; N({\bf u})\sigma e^{-i k_A {\bf u} \vecr}  \rho e^{i k_A {\bf u} \vecr} \sigma^\dagger \Bigr)\,,
\end{multline}
\end{subequations}
describing spontaneous decay of the excited state $\ket{e}$ at
rate $\gamma$ accompanied by the emission of a photon into the
free-space modes of the electromagnetic field environment. This process
involves a recoil of $k_A=\omega_A/c \approx k$ opposite to the direction ${\bf u}$ of the emitted photon,
which is averaged over the directional distribution function
$N({\bf u})$ characterizing the given atomic transition.

In general, the full quantum dynamics of the system defined
by \eref{eq:ME} including all degrees of freedom -- the CM motion,
the internal electronic dynamics and the cavity photon field --
cannot be solved analytically even for a single
atom.

\subsubsection{Dispersive limit}
\label{sec:DispersiveLimit}

For a broad class of cavity QED parameters,
atomic saturation effects are negligible and the atoms can be
considered as linearly polarizable particles. This holds true when
the internal atomic variables $\sigma$, $\sigma^\dagger$ evolve on
a much faster time scale as compared to the other variables due to
a large atomic detuning $\Delta_A$ or a large spontaneous decay
rate $\gamma$. In either case, following the usual technique of
adiabatic elimination, the atomic polarization operator $\sigma$
can be `slaved' to the cavity mode and atomic position `master'
variables. In the absence of direct atom driving, \ie, $\Omega =
0$ in \eref{eq:H_pump}, one obtains
\begin{equation}
\label{eq:SigmaSlaved} \sigma \approx  \frac{g f(\vecr) a}{-i\Delta_A+\gamma}\,.
\end{equation}
This approximation is valid if the population in
the excited atomic state is negligible (low saturation regime). By inserting the slaved variable $\sigma$ into $H_\mathrm{JC}$ and
into the Liouville operator \eref{eq:L_atom}, an effective master equation is obtained.  Of particular interest is the large detuning limit  in which the CM motion and the cavity mode are coupled \emph{dispersively} by
\begin{equation}
\label{eq:H_linpol}
H_\sub{eff}= - \hbar \left(\Delta_C - U_0 |f(\vecr)|^2\right)a^\dagger a \,.
\end{equation}
It captures, on the one hand, the atom-induced dispersive shift of
the cavity mode resonance frequency which depends on the momentary
position of the atom. On the other hand, the cavity field gives
rise to an optical potential $\propto |f(\vecr)|^2$ felt by the
atom whose depth depends on the dynamical photon number.
Dissipation can be treated analogously and the effective Liouville
operator was presented by \textcite{Domokos2001Semiclassical}. The
dispersive and absorptive effects of the atom are expressed in
terms of the parameters
\begin{subequations}
\begin{align}
\label{eq:U0}
U_0 &= \frac{g^2 \Delta_A}{\Delta_A^2+\gamma^2} = - \frac{\omega_{C}}{\cal V} \chi^\prime\,,\\
\label{eq:Gamma0} \Gamma_0 &= \frac{g^2
\gamma}{\Delta_A^2+\gamma^2} = - \frac{\omega_{C}}{\cal V}
\chi^{\prime\prime}\,,
\end{align}
\end{subequations}
respectively. These relations reveal the connection between the
cavity QED parameters and the complex susceptibility
$\chi=\chi^\prime - i \chi^{\prime\prime}$  of a linearly
polarizable object with electric polarization
$\mathbf{P}=\varepsilon_0\chi\mathbf{E}$. With this connection at
hand, the theory presented in this section can be used to
describe a much broader class of particles than only two-level
atoms, and most of the findings can directly be applied to
polarizable particles of sub-wavelength size. In
Sect.~\ref{sec:CoolingMolecules}, the linear polarizability
picture is refined for the case of molecules.

Using the dispersive interaction Hamiltonian, \eref{eq:H_linpol},
the quantized one-dimensional motion of a single atom strongly coupled to
single-mode cavity field has been numerically simulated
\cite{Vukics2005Cavity}. The calculation confirmed
the basic assumption of semiclassical theories (see Sec.~\ref{sec:SemiclassicModel}),
stating that the coherence length of the atomic wavefunction
reduces well below the optical wavelength after a few irreversible
scattering events. This happens although in the dispersive limit the
coupling to the environment is provided by cavity photon loss
rather than spontaneous photon scattering into free space. An
efficient numerical code has been developed providing a general framework for Monte-Carlo wavefunction simulations of
systems composed of the `quantum optical toolbox'
\cite{Vukics2007CQED,Vukics2012CQEDv2}.

If the atom is laser-driven from a transverse
direction, \ie, $\Omega \neq 0$ in \eref{eq:H_pump}, the
adiabatic elimination of the internal degrees of freedom leads to
\begin{equation}
\sigma \approx \frac{g\, f(\vecr)\, a + \Omega\, h(\vecr)}{-i\Delta_A + \gamma}\,.
\end{equation}
Consequently, additional terms appear in the effective adiabatic
Hamiltonian \eref{eq:H_linpol} and the Liouvillean
\eref{eq:L_atom}. In particular, coherent photon scattering
between the transverse laser field and the cavity mode gives rise
to the effective cavity pump term
\begin{equation}
 \label{eq:H_pump_eff}
  H_\sub{pump}/\hbar = \etaeff \, h(\vecr)  \, \lp f^*(\vecr)\, a^\dagger + f(\vecr) a \rp
  \, ,
\end{equation}
with the effective cavity drive amplitude
\begin{equation*}
\etaeff = \frac{\Delta_A g \Omega}{\Delta_A^2+\gamma^2}\,.
\end{equation*}
The atomic recoil accompanied by photon scattering is accounted for by the spatial dependence of this term.

\subsubsection{Semiclassical description of atomic motion}
\label{sec:SemiclassicModel}

In many cavity QED experiments, cold atoms are released from a
magneto-optical trap into the resonator volume. As the temperature
$T$ of the atoms is well above the recoil temperature, $k_B T\gg
\hbar \omega_R$, where $k_B$ is the Boltzmann constant, one can
assume that the reduced density matrix is almost diagonal both in
position and momentum representation. This allows to treat the
position $\vecr$ and momentum $\vecp$ of the atom as stochastic
c-number variables. The regime of ultracold atoms $k_B T\lesssim
\hbar \omega_R$ will be treated in Sec.~\ref{sec:QuantumGas}.

\paragraph{Langevin-type equation}
The separation of the quantized internal and the classical
motional degrees of freedom was developed for the description of
laser-cooling \cite{Gordon1980Motion,Dalibard1985Atomic,cct}. This
approach has been adopted to the cavity QED scenario by extending
the internal degrees of freedom to the combined space of the
atomic polarization and cavity mode
\cite{Horak1997CavityInduced,Hechenblaikner1998Cooling}. By eliminating the internal degrees of freedom, the
dynamics of the CM variables can be formulated in terms of a stochastic
differential equation
\begin{subequations}
\label{eq:Langevin}
\begin{align}
\dot \vecr &= \frac{\vecp}{m} \\ 
\dot \vecp &= \f +  \beta\, \frac{\vecp}{m} + \XXi \,,
\end{align}
\end{subequations}
where $\f$ denotes the classical force and $\beta$ a friction
coefficient in a non-conservative and velocity dependent force
term. In general, $\beta$ can be a tensor in the three-dimensional
space as atomic motion along any direction gives rise to friction
in all three spatial directions \cite{Vukics2004Multidimensional}.
When the eigenvalues of the tensor $\beta$ (or scalar in 1D) are
negative, one encounters \emph{cavity cooling}. The noise term
$\XXi$ induces the stochastic behavior. It has vanishing mean
value and is defined via the diffusion matrix $\mathbf{D}$
according to
\begin{equation}
  \label{eq:noise}
  \langle {\XXi}(t) \circ {\XXi}(t') \rangle = {\bf D} \delta(t-t') \,,
\end{equation}
where $\circ$ denotes the dyadic product. The exact noise
correlation function has a width in the range of the dissipative
parameters $\kappa$ and $\gamma$ of the internal dynamics.
Therefore, it can be approximated by a Dirac-$\delta$ only on the
much slower CM motion time scale set by the inverse of the recoil
frequency, $\omega_R^{-1}\gg \text{min}\{\kappa^{-1},
\gamma^{-1}\}$. The method of calculating the $c$-number
parameters $\f$, $\beta$, and ${\bf D}$ of the Langevin-type
equation from the master equation concerning the internal degrees
of freedom was presented by \textcite{Hechenblaikner1998Cooling}
for the one-dimensional, and by \textcite{Domokos2003Mechanical}
for the three-dimensional case. This method accounts for the
quantum effects of the internal dynamics; hence the full approach
is semiclassical.

The practical use of this method is strongly limited: the
nonlinear quantum master equation for the internal and cavity
degrees of freedom has to be solved numerically and for all atomic
positions $\vecr$. Moreover, the Hilbert space of the photon field
has to be truncated at low photon numbers. This approach was
adopted by \textcite{Doherty2000Trapping,Fischer2001Collective} to
simulate the experiments conducted by
\textcite{Pinkse2000Trapping,Hood2000AtomCavity}.

\begin{figure}[htbp]
\begin{center}
\includegraphics[height=0.45\columnwidth]{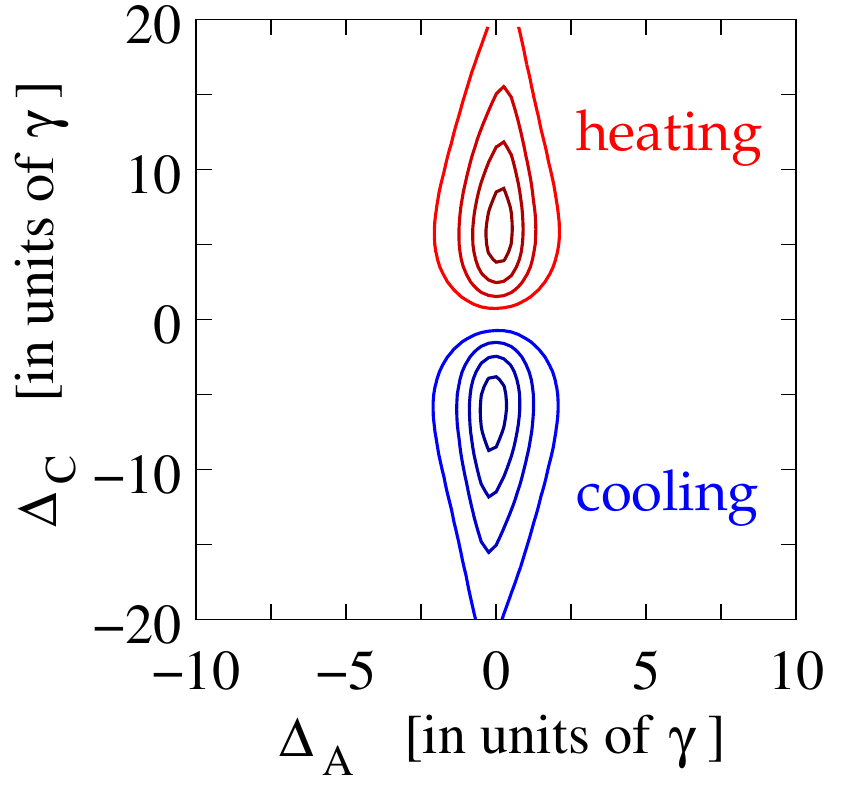}
\includegraphics[height=0.45\columnwidth]{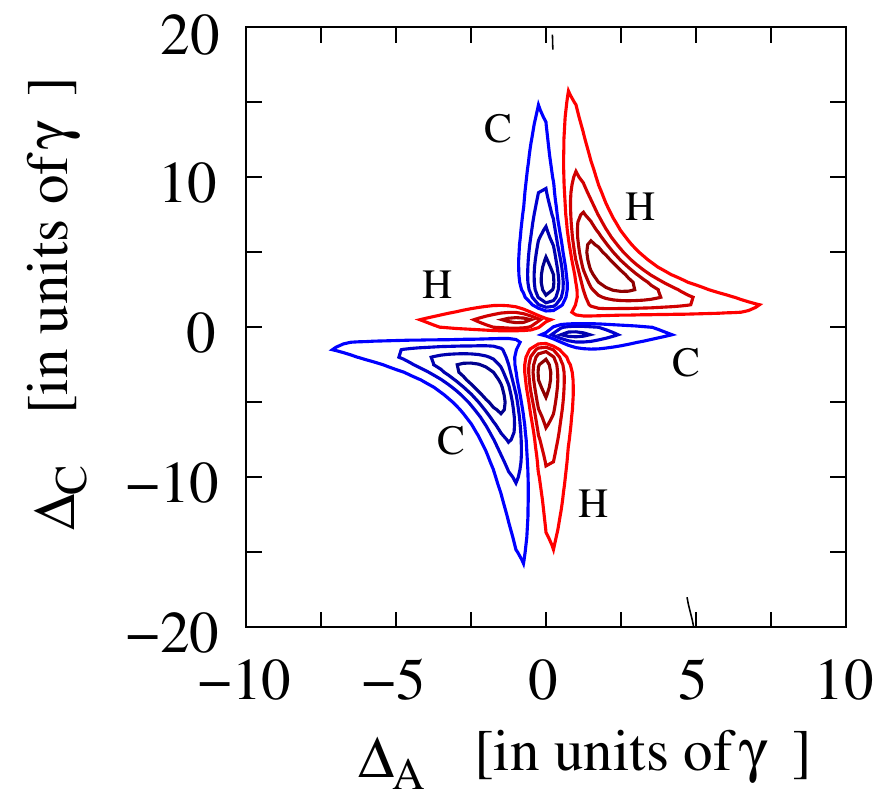}
\caption{(Color online) Cooling and heating regions as a function
of atom and cavity detunings. Shown are contour plots of the
friction coefficient $\beta$ (averaged over an optical wavelength)
acting along the cavity axis on a laser-driven atom. {Left:}
Bad-cavity regime, $g=\gamma/2$, $\kappa=10\gamma$; {Right:} Good
cavity regime, $g=3\gamma$, $\kappa=\gamma$, where the
dressed-state picture can be invoked for interpretation. Blue
contour lines indicate cooling (C, $\beta<0$), red ones heating
(H, $\beta>0$) regions. Note that the spatially averaged friction
coefficient is shown here; tightly confined atoms localized within
a small fraction of the wavelength can follow completely different
behavior depending on their position.} \label{fig:ContourFriction}
\end{center}
\end{figure}

Analytical approximations can be obtained  for the low atomic
saturation regime \cite{Murr2003Suppression}, where the atomic
polarization can be replaced by a bosonic operator and hence the
internal dynamics is described by linear equations of motion. It
is then possible to calculate the friction coefficient $\beta$
and, corresponding to its sign, the cooling versus heating regions
can be mapped as a function of the detunings $\Delta_A$ and
$\Delta_C$, as shown for example in
Fig.~\ref{fig:ContourFriction}.

\paragraph{Semiclassical theory in the dispersive limit}

In the dispersive limit of atom-cavity coupling presented in
Sect.\ \ref{sec:DispersiveLimit}, there is an alternative
semiclassical approach \cite{Domokos2001Semiclassical}. The Wigner
quasi-probability distribution function can be defined in the
joint phase space of the atomic CM motion and the cavity field
amplitude. The quantum master equation translates then into a
partial differential equation for the Wigner function. By dropping
all terms containing higher than second-order derivatives, the
resulting Fokker-Planck equation corresponds to the evolution of
classical stochastic variables associated with the atomic motion
and the cavity field. One can consider this approach as the
construction of a semiclassical model which lies closest to the
true quantum dynamics.  As compared to \eref{eq:Langevin}, here
large intracavity photon numbers are allowed, in fact, the
validity of this approach requires photon numbers larger than 1.

Consider the generic example of a single atom moving in one dimension
along the axis of an externally driven linear cavity with the mode function $f(x)=\cos(kx)$, described by the equations \cite{Domokos2001Semiclassical}
\begin{subequations}
\label{eq:SingleAtomSemiclassic}
\begin{align}
\dot x & =  \frac{p}{m}\,,\\
\dot p & =  -\hbar U_0 |\alpha|^2 \frac{\partial}{\partial x} f^2(x)  + \xi_p \,, \\
\dot \alpha & = \eta -i \left(U_0f^2(x)-\Delta_C\right)\alpha \nonumber\\
&\qquad\qquad\qquad - \left(\kappa + \Gamma_0 f^2(x)\right)\alpha + \xi_\alpha \,.
\end{align}
\end{subequations}
Apart from the noise terms $\xi_p$ and $\xi_\alpha$, these
equations coincide with the classical description in the initial
cavity cooling paper by \textcite{Horak1997CavityInduced}. The
force in (\ref{eq:SingleAtomSemiclassic}b) acting on the atom is
formally identical to the gradient of the optical dipole potential
of the cavity mode. The amplitude $\alpha$, however, depends not
only on the momentary position of the atom but has a memory effect
because of the finite bandwidth $\kappa$ of its linear response.
The actual force has therefore a velocity-dependent character
which can give rise to a viscous friction force, that is, cavity
cooling. Within this approach the friction cannot be determined by
a single coefficient $\beta$, on the other hand, the friction
effect is correctly described for arbitrary velocity.

The noise sources are taken into account in a consistent way as
imposed by quantum mechanics. This results in non-trivial
correlations $\langle \xi_p \xi_\alpha \rangle \neq 0$. The
result for the diffusion matrix as well as the generalization for several atoms was published by
\textcite{Asboth2005Selforganization}. The general model is used for numerically  studying many-body systems, c.f.~Sect.\ \ref{sec:coldatoms},
well above the temperature of quantum degeneracy.

\paragraph{Scattering model}

The semiclassical Langevin-type equation \eqref{eq:Langevin} can
be constructed without mode decomposition of the radiation field.
This approach is required when, instead of a simple
Fabry--P\'erot-type cavity geometry, one considers an atom
interacting with the radiation field of an interferometer which is
composed of an arbitrary one-dimensional configuration of beam
splitters. To deal with this situation a scattering model has been
established by \textcite{Xuereb2009Scattering} and solved for the
force terms acting on a particle as in \eref{eq:Langevin}
\text{by} \textcite{Xuereb2010Optomechanical}. In the scattering
model the atoms and beam splitters are treated
on an equal footing as `scatterers' characterized by a single
polarizability parameter. Thereby, a unified framework is created
to describe opto-mechanical systems in general, which has revealed
the close relationship between cavity cooling of atoms and
radiation-pressure cooling of mirrors
\cite{Metzger2004,Kleckner2006,Arcizet2006,Gigan2006,Schliesser2009}.

\subsection{Cavity cooling}

One of the most promising results from the
understanding of the complex cavity QED dynamics involving atomic
motion is the realization of cavity cooling, \ie , the dissipation
of kinetic energy through the cavity photon loss channel in a
controlled manner.

\begin{figure}[htbp]
\centering
\includegraphics[width=0.375\textwidth]{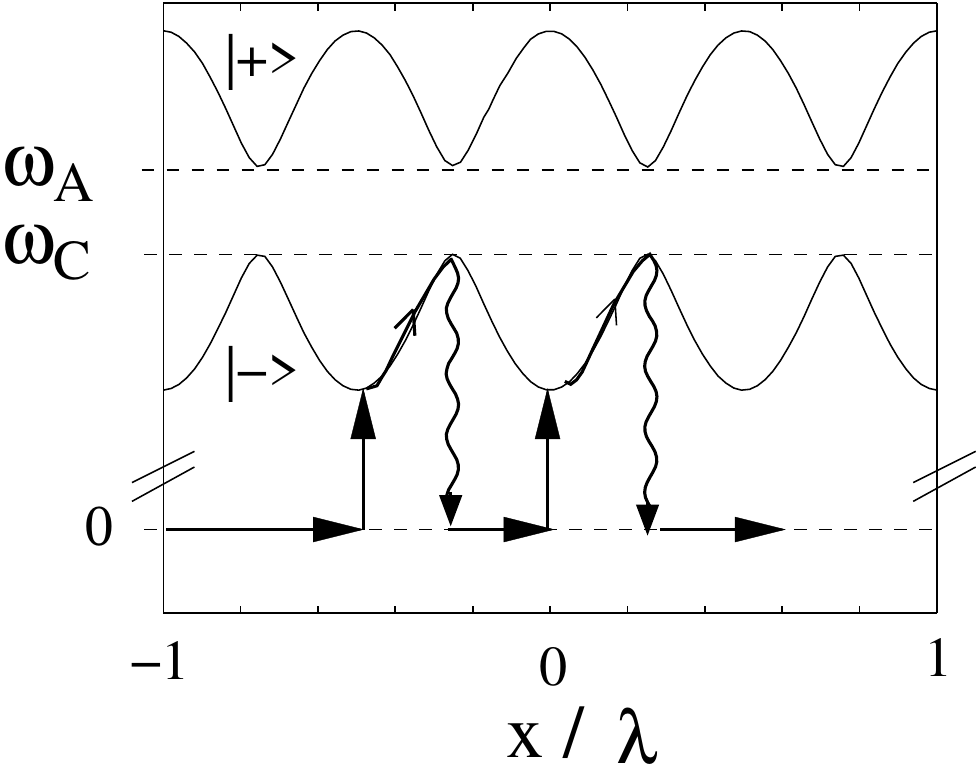}
\caption{Sisyphus-type cooling mechanism underlying the
hyperbolic-shaped cooling region in the right panel of
Fig.~\ref{fig:ContourFriction}. The atomic motion leads to a
modulation of the internal dressed-state energy levels $\ket{\pm}$
which are linear combinations of the $\ket{g,1}$ and $\ket{e,0}$
states with mixing determined by the mode function $f(x)=
\cos(kx)$. This amounts to a state-dependent potential for the CM
degree of freedom. For $\Delta_A < 0$ and $\Delta_C \approx
-\kappa + \frac{g^2}{\Delta_A}$, the transition from the ground
state $\ket{g,0}$ to the lower dressed state $\ket{-}$ is
resonantly excited at an antinode. Thus the excitation happens
more likely at the minimum of the potential wells, whereas
spontaneous or cavity decay transfers the atom-cavity system back
to the ground state homogeneously in space. From
\textcite{Domokos2003Mechanical}.}
    \label{fig:sisyphus}
\end{figure}

Early ideas about using an optical resonator to enhance the
efficiency of laser-cooling relied on the modification of the
spectral mode density of the electromagnetic radiation field in
the presence of spatial boundary conditions
\cite{Mossberg1991Trapping,Lewenstein1993Cooling}. In the most
general form, the notion of cavity cooling in the perturbative
regime has been expressed by \textcite{Vuletic2000Laser}. If an
atom, placed inside an optical cavity, is laser driven at a
frequency below the cavity resonance, $\Delta_C<0$, scattering
favors the emission of photons at frequencies higher than the pump
frequency due to the increased mode density around the cavity
resonance. The energy needed to upshift the photon frequency is
provided by the loss in kinetic energy in processes of inelastic
scattering. With this very simple picture, a robust
three-dimensional cooling effect can be interpreted
\cite{Vuletic2001Threedimensional}. However, the picture holds
only true in the regime of weak atom-photon coupling, see the left
panel of Fig.~\ref{fig:ContourFriction}. When the reabsorption of
a photon starts to become non-negligible, which happens in a
high-finesse cavity, the cooling mechanism substantially changes.
This drastic change is illustrated in
Fig.~\ref{fig:ContourFriction}, where the right panel presents the
friction coefficient for a ratio $g/\kappa$ in the the single-atom
strong-coupling regime. In the case of a standing-wave cavity, the
dynamical cavity cooling effect can be interpreted in the
frequency domain by means of a Sisyphus-type argument
\cite{Horak1997CavityInduced}, using the dressed-state picture of
the strong-coupling regime of cavity QED \cite{Haroche1992Cqed},
see Fig.\ \ref{fig:sisyphus}. For the case of a ring cavity,
interestingly, the intuitive photon scattering picture can be
pursued also in the strong-coupling regime and the full velocity
dependence of the radiation pressure for arbitrary coupling
constant $g$ can be obtained \cite{Murr2006Large}.

In the following we will survey two regimes
where cavity cooling has been demonstrated experimentally, and
present the corresponding intuitive pictures of the cooling
effect. Both regimes are in the dispersive limit of atom-photon interaction keeping the
atomic saturation low.

\subsubsection{Cavity cooling with blue-detuned probe light}
\label{sec:BlueCooling}

\begin{figure}[htbp]
\centering
\includegraphics[width=0.45\textwidth]{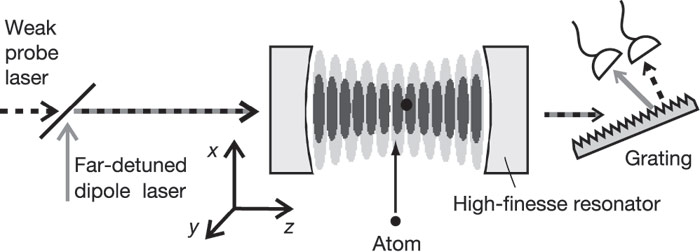}
\caption{Experimental scheme used for the observation of cavity
cooling. Single atoms are captured by an optical dipole trap
formed by far red-detuned light in a longitudinal cavity mode
which is different from the one used for cavity cooling. The
characteristic parameters of the interaction between the weak
probe and the atoms are $(g,\kappa,\gamma) =  2\pi \times (16,
1.4, 3)$ MHz. From \textcite{Maunz2004Cavity}. }
\label{fig:SchemeMaunzExp}
\end{figure}

Cavity cooling of single atoms has been first demonstrated by
\textcite{Maunz2004Cavity} via the observation of extended storage
times and improved localization of single $^{85}$Rb atoms in an
intracavity dipole trap. The trap field was red-detuned with
respect to the atom, however, the cooling was induced by a weak,
blue-detuned probe field. The cooling rate has been estimated to
exceed that achieved in free-space cooling methods by at least a
factor of five, for comparable excitation of the atom.
\textcite{Maunz2004Cavity} present a very intuitive interpretation
of the cooling effect in terms of the classical notion of the
refractive index. Consider a standing-wave optical cavity
resonantly excited by a weak probe laser, $\Delta_C=0$, which is
blue-detuned from the atomic resonance by $\Delta_A = 2\pi \times
35$ MHz $> 0$, see Fig.~\ref{fig:SchemeMaunzExp}. The resulting
light shift parameter,  \eref{eq:U0}, far exceeds the cavity
linewidth, $U_0 > 5 \kappa$, so that even one atom can
significantly influence the optical path length between the cavity
mirrors. Due to \eref{eq:H_linpol}, the atom placed at a node of
the standing-wave mode profile does not couple to the cavity field
and the intracavity intensity is maximum. In contrast, placed at
an antinode, the atom shifts the cavity resonance towards higher
frequency, \ie, out of resonance with the probe laser, resulting
in a reduced intracavity intensity. In a high-finesse cavity,
however, the intensity cannot drop instantaneously when the atom
moves away from a node. The induced blue-shift of the cavity
frequency at almost constant photon number leads to an increase of
the energy stored in the field, before the photons are able to
leak out of the cavity. This occurs at the expense of kinetic
energy of the atoms. The reverse, accelerating effect occurring
when the atom moves from an antinode towards a node is much
weaker, because the cavity is initially  out of resonance with the
probe laser and consequently only a small number of photons are
present and undergo a corresponding red-shift. This argument also
reveals that the delicate correlation between the atomic motion
and the photon number variation, underlying the cooling effect,
imposes an upper bound on the atomic velocity, $kv < \kappa$,
which sets the velocity capture range of cavity cooling.

\begin{figure}[htbp]
\centering
\includegraphics[width=0.45\textwidth]{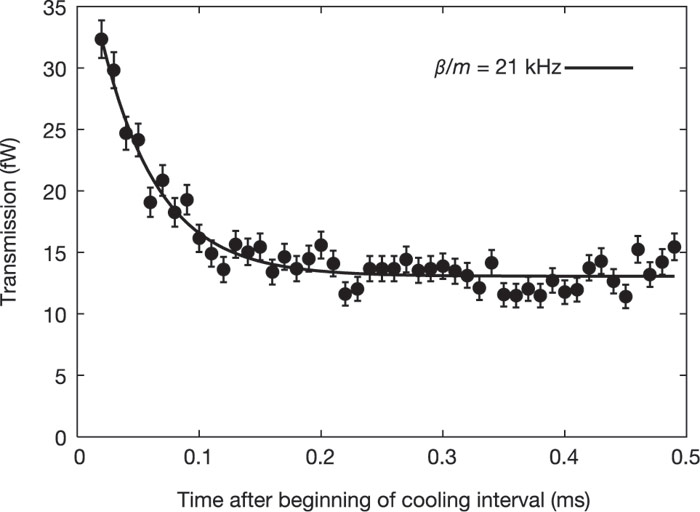}
\caption{Demonstration of cavity cooling. Time-resolved reduction
of the averaged cavity transmission of a weak resonant probe beam
indicating improved localization of the atom at the antinodes of a
far red-detuned trapping field. The closer the atom resides at an
antinode, the larger the detuning of the cavity resonance with
respect to the probe frequency. From \textcite{Maunz2004Cavity}.}
\label{fig:BlueCoolingExp}
\end{figure}

In the experiment, single atoms injected into the cavity are
trapped at the field antinodes of a strong intracavity dipole
trap. To be detectable in cavity transmission of the weak probe
beam, the atoms simultaneously have to be close to an antinode of
the probe field mode. As the probe field induces cavity cooling,
the resulting stronger confinement can be directly read out of the
transmitted signal, as shown in Fig.~\ref{fig:BlueCoolingExp}.
Time-resolved detection of the cavity transmission allowed to
extract a cooling rate of $\beta/m = \unit[21]{kHz}$, which is
large compared to the estimated cooling rate of $\unit[4]{kHz}$
expected for blue-detuned Sisyphus cooling of a two-level atom in
free space, or with the Doppler cooling rate of $\unit[1.5]{kHz}$
at equivalent atomic saturation.

\subsubsection{Cavity cooling and trapping with far red-detuned light}
\label{sec:RedDetuningCooling}

Far-off-resonance dipole traps are commonly used for long-time
capturing and localization of neutral atoms \cite{Grimm2000Optical}. The suppression of
spontaneous emission results in an almost conservative trapping
potential. However, with the elimination of spontaneous emission
($|\Delta_A| \gg \gamma$), any free-space cooling mechanism
also disappears. The far-off-resonant trapping scheme has been revisited for a strongly
coupled atom-cavity system where the cavity mode provides a new
dissipation channel.

Surprisingly, cavity cooling can remain very efficient in the
limit of large atomic detuning $|\Delta_A|\rightarrow \infty$. For
optimal cooling the driving frequency has to be set slightly below
the cavity resonance frequency $\Delta_C \approx -\kappa +U_0$.
Underlying the cooling mechanism is a polariton resonance of the
strongly coupled atom-cavity system (corresponding to the dressed
state $\ket{-}$ in the weak excitation limit where only the lowest
excitations manifold of the Jaynes-Cummings spectrum matters, see
Fig.\ \ref{fig:sisyphus}). Even if $\omega_A$ and $\omega_C$ are
very different, the bare cavity resonance is slightly modified
because the photon excitation mixes with a small amount of the
atomic excitation. In an inhomogeneous system, the mixing leads to
a dependence of the polariton resonance on the atomic position
(see Fig.\ \ref{fig:sisyphus}). Although the modulation is tiny in
amplitude, the resonance is comparably narrow, having a width in
the range of $\kappa$ for the cavity-type polariton. Thus the
system can be very sensitive to the atomic motion and even slow
atom velocities induce large non-adiabatic modulations of the
steady-state field amplitude
\cite{Domokos2004Anomalous,Murr2006Large}.

For demonstration, we consider the simplest case of an atom moving
along the cavity axis in the field generated by an external
driving laser. It has been shown that the standing-wave cavity
field simultaneously traps and cools the atom
\cite{Vukics2005Simultaneous}. For a standing-wave mode,
$f(x)=\cos(kx)$, the cooling rate is given by
\begin{multline}
\label{eq:beta}
\frac{\beta}{2\gamma P_e} = \frac{\omega_R}{\gamma} 4 \sin^2{(kx)}
\times\\\times
\frac{2 g^2 (\DC-U_0 \cos^2{(k x)})
  (\kappa+\Gamma_0 \cos^2{(k x)})}{\lp\lp\DC-U_0 \cos^2{(k x)}\rp^2 +
  \lp\kappa+\Gamma_0 \cos^2{(k x)}\rp^2\rp^2} \,,
\end{multline}
where $P_e$ denotes the mean
population in the excited state $\ket{e}$. Choosing the cavity detuning as
$\DC\approx-\kappa+U_0$  leads to the optimum friction coefficient
which, spatially averaged, reads
\begin{equation}
\label{eq:CoolingScaling}
\frac{\beta}{2\gamma P_e} = \frac{\omega_R}{\gamma}
\left(\frac{g}{\kappa}\right)^2 \,.
\end{equation}
On the left-hand side the cooling rate is normalized to the rate
of spontaneous photon scattering.  This expression shows that the
friction coefficient at a fixed saturation $P_e$ is independent of the atomic detuning, which
gives rise to the perspective  of cooling molecules or other
objects without closed cycling transition (see
Sec.~\ref{sec:CoolingMolecules}).

\begin{figure}[tbhp]
\centering
\includegraphics[width=0.45\textwidth]{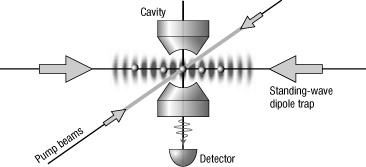}
\caption{Transverse pump scheme of cavity QED. Atoms are
transported into the cavity  using an optical conveyor belt.
Instead of driving the cavity directly, the atoms are transversely
laser-driven giving rise to photon scattering into the cavity
mode. The standing-wave dipole trap yields a large differential AC-Stark shift, \ie , a modulation of the atomic detuning
$\Delta_A(\vecr)$, see \eref{eq:H_JC}. In this geometry the cavity
vacuum field, the weak driving laser tuned according to the optimum
choice in \eref{eq:CoolingScaling}, and the trap laser together
form a very efficient three-dimensional cooling scheme. From
\textcite{Nussmann2005Vacuumstimulated}. }
\label{fig:TransverseSchemeGarching}
\end{figure}
\begin{figure}[htbp]
\centering
\includegraphics[width=0.45\textwidth]{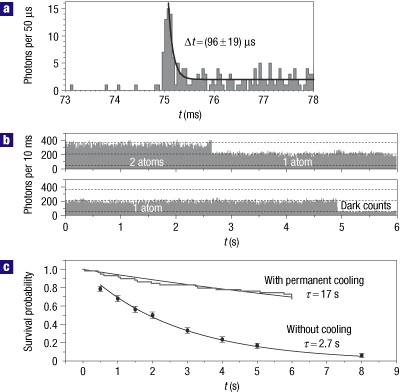}
\caption{Demonstration of cavity cooling and long-time trapping of
a controlled number of atoms inside the cavity. (a) Shown is a
single trace of the recorded photon-count rate indicating the
capture of an atom $\unit[75]{ms}$ after switching on the pump
laser (see Fig.~\ref{fig:TransverseSchemeGarching}). Within
$\unit[100]{\mu s}$, the scattering rate reaches a steady-state
value. (b) The recorded photon-count rate allows for determining
the atom number and the trapping time. (c) The analysis of 50
traces, each of $\unit[6]{s}$ duration and starting with one atom,
yields an average lifetime $\tau$ of $\unit[17]{s}$ (upper curve),
whereas single atoms that are not exposed to the pump laser only
reside for $\unit[2.7]{s}$ in the cavity volume (lower curve).
From \textcite{Nussmann2005Vacuumstimulated}.}
\label{fig:LongCaptureGarching}
\end{figure}

Further studies revealed that the previous setting of detunings
and intensities can be extended to more general geometries,
including the motion perpendicular to the cavity axis, or the
external driving of the atom instead of the cavity.  What is
required for cooling is an inhomogeneity in the system on the
wavelength scale which leads to a position-dependent steady-state
of the coupled atom-cavity system. This inhomogeneity can arise
from the cavity mode function, as for the result in
\eref{eq:beta}, but also from a standing-wave pump field, or from
the spatially modulated AC-Stark shift in a strong standing-wave
laser field \cite{Murr2006Threedimensional}. All these sources
contribute to the cooling efficiency. The resulting cooling effect
has been demonstrated experimentally by
\textcite{Nussmann2005Vacuumstimulated}, making use of an
orthogonal arrangement of a cooling laser, a trapping laser and a
cavity vacuum mode (see Fig.~\ref{fig:TransverseSchemeGarching}).
This combination gives rise to  friction forces  along all three
spatial directions. The achieved cooling efficiency led to
microkelvin temperatures and to an average single-atom trapping
time in the high-finesse cavity as long as 17 seconds, during
which the strongly coupled atom could be observed continuously,
see Fig.~\ref{fig:LongCaptureGarching}.

In the experiment (Fig.~\ref{fig:TransverseSchemeGarching}), a
far-detuned standing-wave dipole trap which is oriented
perpendicular to the cavity axis is used to transport atoms into
the cavity \cite{Kuhr2003Coherence,Dotsenko2005Submicrometer}. The
combination of controlled insertion of single atoms into and
retrieval out of a high-finesse optical resonator with cavity
cooling led to a deterministic strategy for assembling a
permanently bound and strongly coupled atom-cavity system. Long
storage times well above 10 seconds and the controlled positioning
of single or a given small number of atoms on the submicrometer
scale are simultaneously available
\cite{Nussmann2005Submicron,Khudaverdyan2008Controlled}.
\vspace{1cm}

The exploration of cavity cooling was a stimulating and essential
step for the experimental achievement of strongly-coupled cavity
QED systems combined with the control over the atomic motion.
With the implementation of free-space laser cooling and
trapping techniques in cavity experiments sufficiently long
atom-cavity interaction times were demonstrated. These
achievements led to remarkable experimental breakthroughs and
applications in single-atom cavity QED, recently. For example,
high-precision measurements demonstrated the basic cavity QED
model in the optical domain, \ie, by resolving the doublet of the
lowest-lying excitations of the atom-cavity system
\textcite{Maunz2005NormalMode,Boca2004Observation}, as well as the
quantum anharmonic domain of the Jaynes--Cummings spectrum
\cite{Kubanek2008Twophoton,Schuster2008Nonlinear} where squeezed
light can be readily generated \cite{Ourjoumtsev2011Observation}.
Furthermore, the achieved trapping times permitted the development
of a deterministic single-photon source
\cite{Kuhn2002Deterministic,McKeever2004Deterministic}, for having
full polarization control \cite{Wilk2007PolarizationControlled},
and to realize the long time sought atom-photon quantum interface
\cite{Wilk2007SingleAtom,Boozer2007Reversible} and single-atom
quantum memory \cite{Specht2011Singleatom}. Many prosperous
directions can grow out from the realization of the elementary
case of electromagnetically induced transparency with a single
atom
\cite{Kampschulte2010Optical,Figueroa2011Electromagnetically},
such as for example all-optical switching with single photons.

\subsubsection{Temperature limit}

The extra friction term induced by the cavity is closely connected
to the modification of the zero-point field fluctuations. Indeed,
even for large atom-field detuning the diffusion within a
cavity-sustained optical dipole trap can be an order of magnitude
larger than for a free-space field
\cite{vanEnk2001Cooling,Puppe2007Light,Murr2006Momentum}. The
heating rate due to fluctuations can be explicitly calculated in a
semiclassical approach  (see Sec.~\ref{sec:SemiclassicModel})
which allows to estimate the stationary temperature attained by
the atoms. Under optimal conditions one finds the intuitive result
\begin{equation}
\label{eq:CoolingTemp} k_B  T \approx \hbar \kappa \;,
\end{equation}
which is independent of the atomic parameters. This result was
confirmed by numerical simulations
\cite{Domokos2001Semiclassical} and fits very well to experimental
observations.  Interestingly, the result remains largely valid
 in the limit where the temperature reaches the recoil
limit $k_B T \approx \hbar \omega_R$ which is not governed by the
semiclassical description anymore. For a particle trapped in a
harmonic potential with vibrational frequency $\nu
> \kappa$ (resolved-sideband regime) efficient
ground state cooling was proposed by
\textcite{Zippilli2005Cooling}. Quantum interference effects in
the spontaneous emission of a trapped particle in a cavity allow
for ground-state cooling even in the bad-cavity regime where $\nu
< \kappa$ \cite{Cirac1995Laser}. This prediction does not
contradict \eref{eq:CoolingTemp}, since it was made for a strongly
localized trapped atom (Lamb-Dicke regime) at a precisely given
position, whereas the temperature limit above assumes spatial
averaging over the cavity wavelength.

According to \eref{eq:CoolingTemp}, there seems to be no lower
bound on the temperature as long as the cavity finesse can be
increased. However, with decreasing loss rate, the capture range
of the cavity cooling mechanism also shrinks. This relation is
thoroughly discussed by \textcite{Murr2006Large}, based on an
explicit expression for the friction force obtained for arbitrary
velocity. When applying very strong cavity fields, $\alpha \gg 1$,
in close analogy to the mean field treatment of optomechanical
models \cite{Genes2008Groundstate}, it is possible to effectively
enhance the weak atom-field interaction appearing at very large
detunings to an effective strong coupling $g_\sub{eff} =  g_0
\alpha $ at the expense of introducing extra fluctuation terms
\cite{Nimmrichter2010Master}. This setting can considerably speed
up the cooling process and enhance the capture range, while still
leading to a similar final temperature as given by
\eref{eq:CoolingTemp}.

\subsubsection{Cooling in multimode cavities}
\label{single:multimode}

The atom-field dynamics qualitatively changes when invoking
several cavity modes to participate as dynamical degrees of
freedom. In simple terms, not only the magnitude but also the
\emph{spatial shape} of the optical potential and the associated
light forces become a dynamical quantity.

This can be easily demonstrated at the generic example of a ring
cavity geometry \cite{Gangl2000Cold}. In the regime of dispersive
atom-field coupling, the atom not only modifies the resonance
frequencies of the two counter-propagating cavity modes, thereby
tuning their field amplitudes, but also gives rise to phase
locking by coherent photon redistribution between the cavity
modes. This determines the position of the nodes and antinodes of
the emergent standing-wave interference pattern of the cavity
radiation field.  For a red-detuned pump field, $\Delta_A<0$, the particle is
drawn to an antinode of the field which, at the same time, gets
dragged along with the slowly moving atom  (assuming $k v < \kappa,
\gamma$). Due to the delayed response of the intracavity field,
however, the particle is permanently running uphill and thus
experiences a friction force. The two-mode
geometry of a ring cavity has been shown to result in faster cooling and  larger velocity capture
range as compared to a single-mode standing-wave cavity
\cite{Gangl2000Cooling,Schulze2010Optomechanical}. Moreover, the laser pump configuration
used for polarization gradient cooling or velocity-selective coherent population trapping can be envisaged within a ring cavity, for which case very efficient cavity cooling is predicted without fundamental lower limit on the temperature \cite{Gangl2001Cavitymediated}

The more modes in a cavity are available in the vicinity of the
pump frequency, the smaller is the {transverse} length scale on
which the field shape gets modulated in the presence of an atom.
On the one hand, this leads to stronger three-dimensional
localization of atoms around their self-generated intensity
maximum \cite{Salzburger2002Enhanced}. On the other hand, the
cooling time reduces in a highly-degenerate confocal cavity more
or less quadratically, whereas the diffusion increases only
linearly with the effective number of  modes involved
\cite{Domokos2002Dissipative,Nimmrichter2010Master}.

The scope of cavity-mediated optical manipulation of atoms
significantly enlarges also in the case of many-atom systems,
which we will review in Sect. \ref{sec:MultimodeDegenerateCavity}.

\subsection{Extensions of cavity cooling}

The general principle of cavity cooling is expected to be
applicable in a broad range of other systems with different
radiation field geometries or other material components.

\subsubsection{Cooling trapped atoms and ions}

There are several experimental systems in which trapped atoms are
strongly coupled to a high-finesse cavity. Ion trap setups have been combined
with high-finesse cavities in the moderate coupling regime
\cite{Keller2004Continuous,Herskind2009Realization}. There are
all-optical schemes, too, where different longitudinal modes
of a standing-wave cavity are used to separate the optical trap modes
from the cooling ones \cite{Maunz2004Cavity,Schleier2011Optomechanical}.
State-insensitive cooling and trapping of single atoms employing
light field at magic wavelengths, which induces
an almost identical AC-Stark shift of the two relevant
electronic states, has been demonstrated
\cite{McKeever2003StateInsensitive}. Further, trapping of
atoms in low field regions of a blue-detuned intracavity dipole
potential has been investigated experimentally
\cite{Puppe2007Trapping}. In a similar intracavity dipole trap, the axial atomic motion was cooled down to the ground state by way of coherent Raman transitions on the red vibrational sideband, meanwhile the atomic motion was inferred from the recorded Raman spectrum by \textcite{Boozer2006Cooling}.

The cavity cooling mechanism operates also in the case of tightly confined particles.
In the Lamb-Dicke regime for tightly confined particles,
$\sqrt{\omega_R/\nu} \ll 1$ where $\nu$ denotes the
harmonic trap frequency, explicit expressions for the cooling and
heating rates of the CM motion of an atom, trapped in an optical
resonator and driven by a laser field, have been derived both in
the regime of weak and strong atom-cavity coupling. In the
former, a variant of sideband cooling appears
\cite{Vuletic2001Threedimensional,Cirac1995Laser}. Experimentally,
the cavity cooling of a single trapped $^{88}$Sr$^+$ ion in the
resolved-sideband regime has been demonstrated and quantitatively
characterized recently \cite{Leibrandt2009Cavity}. The spectrum of
cavity transmission, the heating and cooling rates, and the
steady-state cooling limit have been measured in perfect
agreement with a rate equation theory. The final temperature
corresponding to 22.5(3) occupied vibrational quanta was limited
by the moderate coupling between the ion and the cavity.

The calculations have been extended to the strong-coupling regime,
where higher-order transitions between eigenstates of the coupled
system have been identified and novel non-trivial parameter
regimes leading to cooling have been revealed
\cite{Zippilli2005Cooling,Blake2011Comparing}.  In the
resolved-sideband regime, $\nu\gg\kappa,\gamma$,  the discreteness
of the vibrational spectrum, which is the same for the electronic
ground and excited states,  gives rise to interference between
different transition paths in analogy to the cooling of trapped
multilevel atoms \cite{Morigi2000Ground}. Ground state cooling is
achievable according to the theoretical predictions
\cite{Zippilli2007Mechanical}.

\subsubsection{Cooling nanoparticles and relation to optomechanics}
\label{sec:CoolingNanoparticles}

The fact that cavity cooling requires only linear polarizability
suggests that it could be directly applicable to large objects,
such as nanobeads \cite{Barker2010Cavity,Chang2009Cavity},  thin
reflective membranes \cite{Genes2009Micromechanical}, or even
small biological objects such as viruses \cite{Romero2010Toward}.
Moreover, since membranes, being macroscopic objects, can have
large static polarizability (refractive index), the cooling can be
much more efficient than for single atoms or molecules. Indeed,
there is a strong connection between cavity cooling of atoms and
dispersive cavity optomechanics \cite{Thompson2008Strong,
Jayich2008Dispersive}, which can  easily be seized in the
framework of the scattering models \cite{Xuereb2009Scattering}.
Cavity cooling of membranes experimentally shows great success
down to the vibrational quantum ground state
\cite{Jayich2011Resolved}.

As the local field strength is strongly enhanced inside a
resonator, optical dipole traps can be operated at very large
detunings, where only the static polarizability of the particle is
relevant \cite{Deachapunya2008Slow,Nimmrichter2010Master}. In such
a setting of coupled optical and mechanical systems, the ring
cavity with degenerate pairs of counterpropagating modes, or other
configurations where degenerate modes are available, can offer the
realization of various effective models.

Consider, for example, a symmetrically pumped ring cavity.
The field can be written as a superposition of the strongly pumped
and thus highly excited cosine mode and the empty sine mode. The
cosine mode fulfills two purposes: (i) it generates the trapping
potential, and (ii) it feeds the sine mode through photon
scattering off the particle (atom, molecule, membrane). The model
Hamiltonian is of the form \cite{Schulze2010Optomechanical}:
\begin{equation}
H=\frac{{p}^{2}}{2m}-\hbar \Delta_C \left( a_{c}^{\dagger} a_{c}+a_{s}^{\dagger }a_{s}\right) -\hbar U({x})+
i\hbar \left( \eta a_{c}^{\dagger }-\eta^{\ast }a_{c}\right)\; ,  \label{ham}
\end{equation}
where $U(x)$ is the dispersive interaction potential, and $a_c$
($a_s$) denote the field amplitudes of the cosine (sine) mode,
respectively.
Linearizing the position around the trap minimum, we can recover the standard optomechanical Hamiltonian,
\begin{multline}
H=\left[ \frac{{p}^{2}}{2m}+\frac{1}{2}m 2\hbar U_{0}  a_c^{\dagger }a_c (k {x})^{2 }\right] \\
-\hbar (\Delta_C - U_0) a_c^{\dagger }a_c - \hbar \Delta_C a_s^{\dagger }a_s \\
 -\hbar U_{0}^{\prime }(a_s+a_s^{\dagger }) {x} \;,
\label{eq:RingOptomech}
\end{multline}%
with quadratic coupling to the cosine trapping mode and linear
coupling to the sine cooling mode. As the particle couples the two
modes there appears an energy splitting which allows to extract
via inelastic scattering kinetic energy from the vibrational
motion in the optical trap \cite{Elsasser2003Collective}. As for
standard cavity cooling the final temperature in the classical
regime is again limited by the cavity linewidth $k_BT \approx
\hbar \kappa$. However, in a very good cavity, when the pump field
is sufficiently strong, one can reach the resolved-sideband
regime, where the trap frequency $\nu$ exceeds the cavity
linewidth, and the final temperature would correspond to less than
a single excitation $k_B T < \hbar \nu$. In this ground-state
cooling limit one has to resort to a quantum description of motion
and the optical fields.  Interestingly, the sine mode
automatically acts as a built-in monitoring system which
continuously observes the vibrational quantum state of the
particle in the cosine mode. Hence close to $T = 0$ one can
observe quantum jumps of the particle via the sine mode
photon counts \cite{Schulze2010Optomechanical}.

\subsubsection{Cooling molecules}
\label{sec:CoolingMolecules}

\paragraph{Cooling the translational motion of molecules}

Molecular structure fundamentally alters and complicates the
picture conceived for laser-cooling two-level atoms. Upon
excitation from the pump field, the molecule can relax by either
Rayleigh scattering back to the ground state $\ket{g}$ at the rate
$\gamma_\sub{Ry}$ or Raman scattering to metastable states at the
rate $\gamma_\sub{Rn}$. There is a multitude of metastable
molecular states (spin-orbit, rotational, and vibrational)
available via inelastic Raman scattering. The generally low
free-space branching ratio $\gamma_\sub{Ry}/\gamma_\sub{Rn}$
results in population shelving after only a few photon scattering
events, thereby prematurely quenching the
cooling process. Because of the prohibitive expense of building
multiple repumping laser systems, optical  cooling of molecules
via free-space dissipative scattering of photons is thought not to
be practicable.

Since cavity-assisted laser-cooling relies on the cavity
dissipation channel, it has been suggested as a potential method
to mitigate Raman loss. Spontaneous photon scattering, in
principle, can be entirely suppressed by using large detuning.
However, as discussed in Sec.~\ref{sec:RedDetuningCooling}, in
order to keep the cooling efficiency constant, one needs to
preserve a given level of excitation in the atom or molecule. Therefore,  merely the large
detuning does not solve the branching ratio problem of molecules
\cite{Lev2008Prospects}. To overcome this severe problem, the
use of an optical cavity with cooperativity parameter much larger
than unity is mandatory, in accordance with
\eref{eq:CoolingScaling}. In this case the enhanced coherent
Rayleigh scattering into a decaying cavity mode can ensure a
vanishingly small probability of the molecule to Raman scatter
during the cooling time. For CN diatomic molecules, the cavity
cooling process has been calculated numerically
\cite{Lu2007Cooling}.

\paragraph{Cooling the rotation and vibration of molecules}
\label{sec:MoleculeRotation}

While theoretical models and experiments have mostly concentrated
on the center-of-mass motion of structureless polarizable
particles or two-level atoms, the complex rovibrational structure
of molecules is one of the central obstacles preventing efficient
laser-cooling of molecules. In many common beam sources the
initial temperature can be designed to be low enough to freeze
most vibrations and only leave few rotational quanta
\cite{Rangwala2003continuous}. Nevertheless, the interaction with
the cooling laser light will in general start to redistribute the
population  within the rovibrational manifolds strongly altering
the optical properties of the molecules and hampering further
cooling. Only a few exceptions of this rule have
been discovered and investigated lately \cite{Shuman2010Laser}.
Cavity cooling, however, can in principle be designed to
counteract this heating process and even further cool the
rovibrational energy of molecule. As an enourmous spread of
transition frequencies is required to facilitate this, it proves
advantageous to simultaneously apply a multitude of different
longitudinal cavity modes
\cite{Kowalewski2007Cavity,Morigi2007Cavity}.  Simulations show
that the rovibrational cooling can be combined
with motional cooling, e.g., in a trap
\cite{Kowalewski2011Cavity}, to get a cold molecular gas in all
degrees of freedom. At this point a practical implementation would
require precooling by other methods, such as the optoelectrical
scheme proposed by \textcite{Zeppenfeld2009Optoelectrical}, to
achieve sufficient interaction times and densities within the
cavity mode volume.

\subsubsection{Cooling and lasing}

Collective coherent emission of a laser-driven atomic ensemble
into the field of an optical cavity accompanied by a very fast and
efficient cooling of atomic motion was observed in an experiment conducted by
\textcite{Chan2003Observation}. Although the effect has not yet been fully
understood, it is attributed to Raman gain within a Zeeman
manifold. The combination of cavity cooling with intracavity gain is an intriguing prospect. It was initially suggested by \textcite{Vuletic2001Cavity} to transform a bad cavity effectively into a good
cavity with fast cooling towards an even lower temperature. While
the principle idea proves to be correct, a more realistic and
detailed modeling, which accounts for fluctuations to consistently
treat the gain, gives a higher limit of the achievable temperature \cite{Salzburger2006Lasing}. This
observation was also confirmed in the optomechanical regime of
cavity cooling, where intracavity gain leads to faster cooling but
a higher final temperature \cite{Genes2009Micromechanical}.

In a standard setup, the intracavity gain could be generated by an
additional inverted medium placed within the cavity.  This would
lead to a technically challenging  setup, if one aims to operate
in the strong-coupling regime. Interestingly, it turns out that in
a conceptually much simpler configuration,  the gain can also be
provided by the same atomic medium which is aimed to be cooled in
the setup. Of course, such a scheme requires a suitable pumping
mechanism which transfers atoms from the lower to the upper level
of the cooling transition, without introducing too much extra
noise. In the ultimate limit one can envisage a single atom, which
is externally pumped within a high-finesse cavity. Stimulated
emission into the cavity mode provides gain to create a trapping
potential for the atom. For a blue-detuned cavity this gain
simultaneously extracts motional energy from the particle and thus
provides cooling \cite{Salzburger2004Atomic}. Fortunately, an
inverted atom is a high-field seeker in the blue-detuned light
field, so that it will be trapped close to optimal gain. Hence
this setup provides for lasing, trapping and cooling of a single
atom within a resonator forming the most minimalistic
implementation of a laser \cite{Salzburger2005Theory}. The system
can be generalized to several particles, which strongly reduces
the requirements on the pump mechanism
\cite{Salzburger2006Lasing}. In the limit of ultracold gases in an
optical lattice, stimulated optical gain occurs concurrent with
Bose enhanced coherent population of the lowest energy band. While
for a pulsed setup this constitutes in principle a very fast and
efficient cooling method, a CW setup could provide a possible
route towards the realization of a CW atom laser
\cite{Salzburger2007Atomphoton,Salzburger2008Twin}.

\subsubsection{Monitoring and feedback control}

Starting  from the early days of cavity QED, a strongly coupled
atom-cavity system was considered as a number-resolving neutral
particle detector \cite{Mabuchi1996Realtime}, a concept which is
still being developed and implemented in  miniaturized devices
\cite{Teper2006ResonatorAided}. Going one step further, the
high-finesse resonator acts as a microscope with which the
trajectory of individual atoms can be reconstructed from the
recorded cavity transmission with high spatial ($< \mu$m) and
temporal  ($< \mu$s) resolution \cite{Hood2000AtomCavity}. The
method can be considerably improved by the use of multimode
cavities. The particle does not only modify the phase and
intensity of the intracavity field, but redistributes light
between the different spatial modes. The output field imaged on a
CCD camera therefore allows to monitor directly and in real time
the motion of the particle
\cite{Horak2002Optical,Maunz2003Emission}. Note that even for
incomplete position information at any given time, the most likely
trajectory of single atoms can be reconstructed with the help of
inversion algorithms based on the coupled equations of motion.

Once the position and motion of the particle are known, it is
straightforward to apply feedback on the motion of a single atom
by adjusting the pump lasers to steer the particle motion within
the cavity and increase its trapping time
\cite{Fischer2002Feedback}. The cavity field both provides
particle detection and mediates the feedback force. This method
has been successfully refined by several groups and resulted in an
increase of single-particle trapping times by several orders of
magnitude \cite{Kubanek2009Photonbyphoton,Kubanek2011Feedback}. By
applying controlled and delayed feedback forces on the particle,
its kinetic energy can be reduced as well. This  kind of feedback
cooling resembles stochastic cooling techniques applied in
high-energy physics. Strongly enhanced  cooling has been predicted
when the feedback scheme, consisting of time-dependent switching
of the trapping field as a function of  the intracavity intensity,
is operated in the dispersive bistability regime
\cite{Vilensky2007Cooling}. This method should also give new
prospects to optomechanical setups.

\newpage
\section{COLD ATOMIC ENSEMBLES IN A CAVITY}
\label{sec:coldatoms}

New research directions opened in cavity QED when cold and
ultracold atomic ensembles were successfully prepared within
high-finesse optical resonators. In the many-body configuration,
the common coupling of atoms to the cavity field creates a wealth
of new possibilities to implement tailored atom-atom interactions
over large distances, an ingredient which usually is absent in
free-space cold atom experiments.

The atom-atom coupling is mediated by the cavity radiation field
between the AC electric dipole moments. However, its nature is
inherently different from the free-space dipole-dipole
interaction.  In a cavity, the interaction strength does not decay
with the interatomic distance and depends only on the local
coupling of the atoms to the cavity field.  Fundamentally, the
interaction is not binary: the ensemble of atoms collectively acts
onto the state of the radiation field which then reacts back on
the individual atoms. This scenario is generally referred to as
\emph{global coupling}. The range of the interaction is given by
the size of the cavity mode, which can be macroscopic.  In cases
where single-atom strong coupling is not achieved, the collective
energy exchange still can be dominated by coherent interaction.

After discussing the nature of the long range atom-atom
interaction mediated by a cavity field in various geometries, we
consider first the many-body influence on the cavity cooling
scheme. Then we address the most spectacular collective effects
realized by cold atoms within linear and ring cavities. Critical
phenomena, instability thresholds, and scaling laws will be
discussed by means of various mean field theories in the end of
this section.

\subsection{Collective coupling to the cavity mode}

Resonant coherent coupling between an ensemble of $N$ two-level atoms
and a single standing-wave cavity mode is described by the
many-body generalization of \eref{eq:H_JC}
\begin{multline}
H/\hbar = -\Delta_C a^\dag a -\sum_j
\Delta_A(\mathbf{r}_j)\sigma_j^\dag \sigma_j\\ + \sum_j i g
f(\mathbf{r}_j)(\sigma_j^\dag a - a^\dag \sigma_j)
\end{multline}
where $j = 1\ldots N$ labels the atoms, and the mode function
$f(\vecr)$, for simplicity, is real. The atomic ensemble can be
represented by a single collective dipole with effective coupling
strength only if (i) the atomic motion can be averaged out (ii)
only the cavity mode is laser-driven, and (iii) the atoms are in
the low saturation regime. In this case the atoms collectively
couple to the cavity mode with an effective strength of $g_{\rm
eff} = g \sqrt{\sum_j f^2(\vecr_j)}$ (the summation index  runs
from 1 to $N$). Correspondingly, an ${N}$-fold enhancement appears
for the many-atom system in terms of the single-atom
cooperativity, ${\cal C} = g^2/(2 \kappa\gamma)$, which measures
the ratio of light scattering into cavity mode versus surrounding
vacuum modes
 \cite{Tuchman2006Normalmode}. For example, the strong
distortions of the single-atom normal mode splitting in the cavity
transmission spectrum induced by a thermal beam of  atoms crossing
the cavity could be interpreted by such a collective mode picture
\cite{Raizen1989Normalmode}. Employing an optical conveyor belt,
an adjustable number, $N = 1\ldots 100$, of cold atoms has been
transported into a microcavity, and large nonlinearities have been achieved as evidenced by the observation of absorptive
optical bistability in the real-time transmission spectrum
\cite{Sauer2004Cavity}.

In general, however,  one has to consider the many-body system composed of a
large number of internal and motional degrees of freedom. We will exhibit this in the following
at the simplest nontrivial case of two atoms in the same mode.

\subsubsection{Cavity-mediated atom-atom interaction}
\label{sec:cavitymediatedatomatomint} Let us discuss the character
of the cavity-mediated atom-atom interaction in two different pump
geometries, namely, pumping the cavity field either directly or
indirectly via light scattering off the laser-driven atoms.

\paragraph{Cavity pumping}

Consider $N$ atoms moving in the field of a laser-driven optical
cavity. The detuning between the driving laser and the
dispersively shifted cavity resonance depends on the position of
all atoms, which in turn experience the optical dipole force of
the intracavity field. For small atomic velocities and in the
low saturation limit, an adiabatic potential can be deduced
\cite{Fischer2001Collective}
\begin{equation}
    \label{eq:ManyAtomPotential}
V(\vecr_1, \ldots \vecr_N) = \frac{\hbar \Delta_A
\abs{\eta}^2}{\Delta_A \kappa + \Delta_C\gamma} \mathrm{atan}
\frac{\gamma\kappa - \Delta_A\Delta_C + g_\mathrm{eff}^2}
{\Delta_A \kappa + \Delta_C\gamma}\,,
\end{equation}
which is analogous to the Born-Oppenheimer approximation used for
describing the motion of nuclei in molecules in the averaged
electronic potential. The potential $V$ depends on the atomic
positions solely via the collective coupling strength
$g_\sub{eff}$, and thus is valid for any number of atoms. This is
not surprising as the adiabatic force is calculated by freezing
the atomic motion. The resulting cavity-mediated long-range
atom-atom interaction gives rise to an asymmetric deformation of
the normal-mode splitting as was observed experimentally by
\textcite{Munstermann2000Observation}.

\begin{figure}
\begin{center}
\includegraphics[angle=0,width=0.48\textwidth]{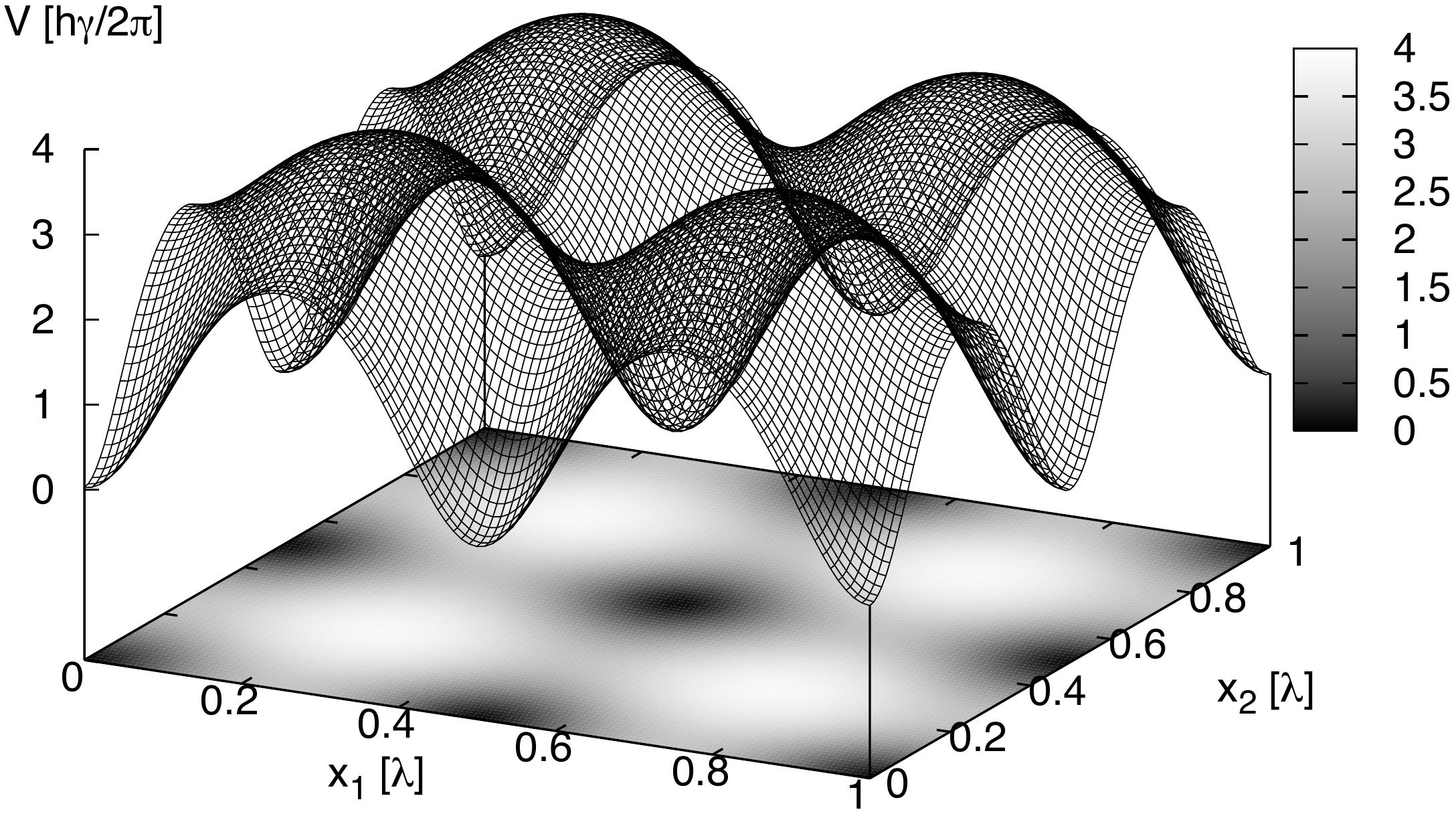}
\includegraphics[angle=0,width=0.48\textwidth]{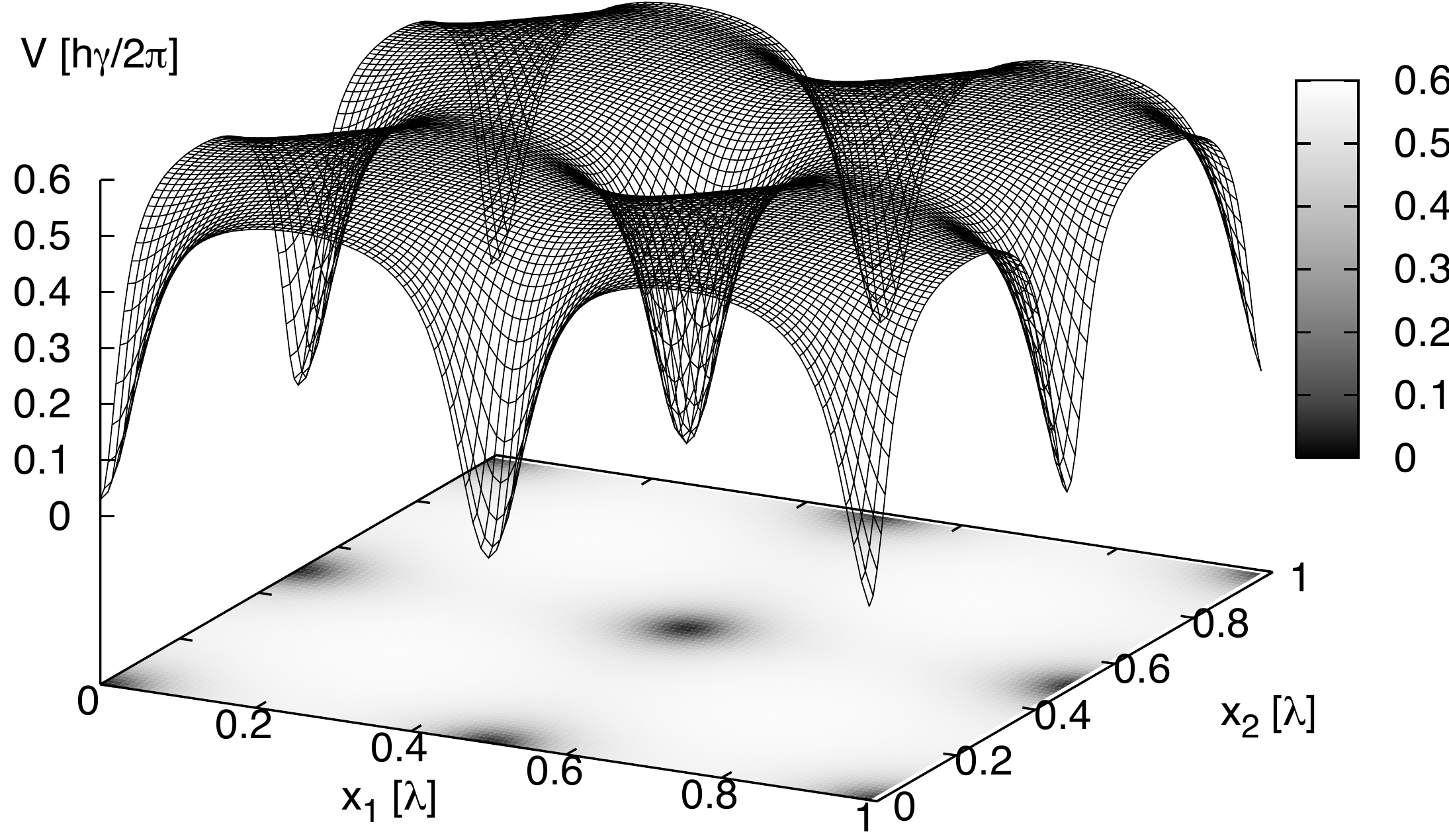}
\caption{The adiabatic cavity potential $V$ as a function of the
atomic positions $x_1$ and $x_2$. The cavity is quasi-resonantly
excited by a pump laser with detuning $\Delta_C=-\kappa +U_0$. The
detuning from the atomic resonance is set to $\Delta_A=-50\gamma$
to ensure the suppression of spontaneous photon scattering. In the
upper graph, typical experimental cavity parameters
 ($\kappa=\gamma/2$,
$g=5\gamma$) \cite{Munstermann2000Observation} have been used,
whereas in the lower graph $g$ was increased fourfold. In the
first case, the potential is well approximated by a sum of two
single-particle potentials. In the second case, either both atoms
are trapped or free. From \textcite{Asboth2004Correlated}.
\label{fig:TwoAtomPotential} }
\end{center}
\end{figure}

In the case of two atoms with positions $x_1$ and $x_2$ the
interaction potential landscape $V(x_1, x_2)$ along the cavity
axis is shown in Fig.\ \ref{fig:TwoAtomPotential} for two
different parameter settings.
The upper graph corresponds to the experimental parameters used in
the Garching group \cite{Munstermann2000Observation}. Although the
single-atom light shift is comparable with the cavity linewidth
$|U_0| \approx \kappa$,  the effective interaction between the
atoms is relatively weak and the potential resembles the familiar
``egg-carton'' surface proportional to $\sin^2(k x_1)+\sin^2(k
x_2)$. For `artificially' enlarged atom-field coupling $g =
20\gamma$, shown in Fig.~\ref{fig:TwoAtomPotential} (lower graph),
the atom-atom interaction strongly affects the potential landscape
felt by the second atom, depending on the position of the first
atom and vice versa. For this parameter setting the single-atom
light shift is sufficiently large, $U_0 \gg \kappa$, that removing
one atom from the cavity antinode makes the potential experienced
by the other atom vanish. Note that the trap is deeper for the
smaller coupling of the upper graph. Interestingly, as was shown
by \textcite{Asboth2004Correlated}, the motion of the two atoms
gets correlated even for the parameter setting of the upper graph
as a consequence of additional non-conservative forces (see Sec.\
\ref{sec:collcoolscaling}).

\paragraph{Atom pumping}

The situation drastically changes if the atoms are laser-driven
from a direction perpendicular to the cavity axis. Intracavity
photons are then created by Rayleigh scattering of laser photons
into the cavity mode. Due to light interference, the scattered
intracavity field exhibits a very sensitive dependence on the
interatomic distance. For two atoms separated by odd integer
multiples of the half-wavelength, the corresponding scattering
amplitudes into the mode have the same magnitude but opposite
sign, resulting in destructive interference and a vanishing cavity
field amplitude. On the other hand, for atoms separated by even
integer multiples of the half-wavelength, the field components
scattered off the two atoms interfere constructively. Compared to
the field intensity created by a single scatterer, the latter case
yields a fourfold enhancement of the intensity, referred to as
\emph{superradiance}
\cite{Dicke1954Coherence,DeVoe1996Observation}.

In the case of pumping directly the atoms the force along the
cavity axis acting on the individual atoms due to light scattering
cannot be expressed as a gradient of a collective potential, at
variance to \eref{eq:ManyAtomPotential}. One can admit this by
checking that $\nabla_i \vec{F}_j \neq \nabla_j \vec{F}_i$, where
$\nabla_i$ is the gradient with respect to the coordinate
$\vecr_i$, and $\vec{F}_j$ is the force acting on atom $j$. If\
there was a potential $V$ such that $\vec{F}_j= - \nabla_j V$, the
two sides should be equal as they are the second derivatives of
the potential  and the order of taking the derivatives is
irrelevant according to Young's theorem. The fact that the force
can not be derived from a potential  is not so surprising, in
hindsight, as we are dealing with an open system with continuous
energy exchange with the environment and an unlimited energy
resource in the form of the pump laser. Actually, the existence of
a potential  \eref{eq:ManyAtomPotential} for the cavity-driving
geometry is the exceptional case.

Approximately, in the limit of $U_0, \Gamma_0 \rightarrow 0$, more
precisely $N^2 U_0 \ll (\kappa, |\Delta_C|)$, the motion of the
atoms is governed by the collective potential
\begin{equation}
\label{eq:ManyAtomPotentialTransverse} V(\vecr_1, \ldots \vecr_N)=
\hbar \frac{\etaeff^2 \Delta_C}{\Delta_C^2 + \kappa^2} \left(
\sum_{j=1}^N \cos(k x_j) \cos(k z_j) \right)^2\,,
\end{equation}
where cosine mode functions were assumed for the cavity and the
pump laser field. The interference effect is manifest: when
scanning the atom-atom distance over a wavelength, the contrast of
the interference in the cavity field intensity is unity regardless
the atom-cavity coupling constant $g$. This is not the case for
cavity pumping, where, in the limit of small coupling constant
$g$, the atoms cause only a small modulation of the cavity
intensity. Therefore, the atom pumping geometry lends itself to
observe spectacular many-body effects even in the weak-coupling
regime.

\begin{figure}
\includegraphics[angle=0,width=0.48\textwidth]{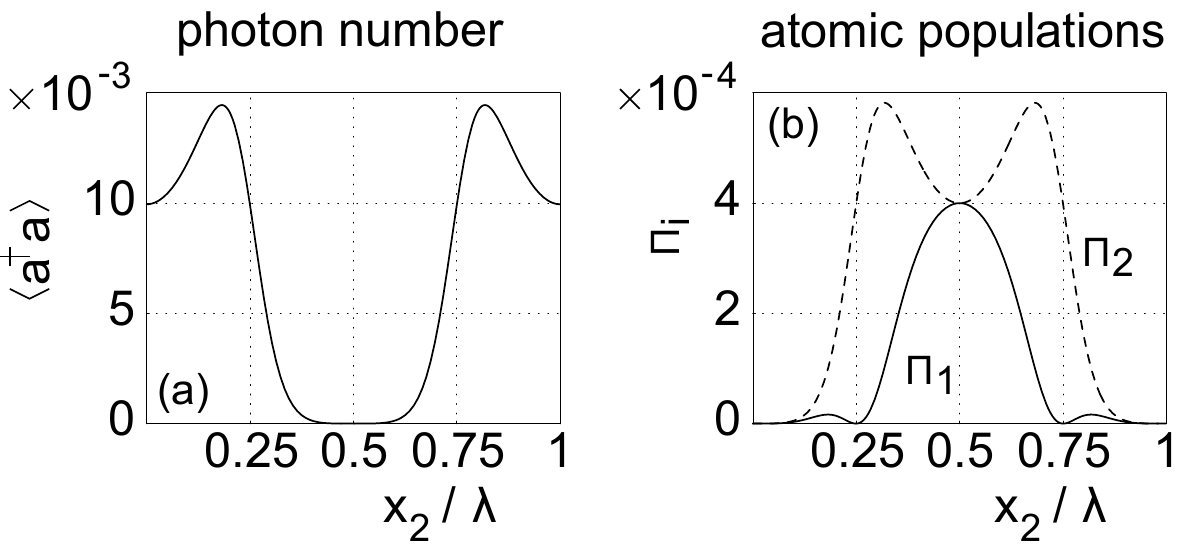}
\caption{Collective scattering of two atoms into the cavity mode
including internal atomic excitation. (a) Mean number of photons
scattered into the cavity mode at pump--cavity resonance,
$\Delta_C = 0$, with coupling strength $g \cos(kx)$. The
atoms are placed at $x_1 = 0$ and $x_2$, respectively. (b)
Excited state populations $\Pi_1$ and $\Pi_2$ of the two atoms (dashed and solid
lines). The parameters are $\kappa = 0.2\gamma, \eta_\mathrm{eff}
= \gamma, g = 10\gamma, \Delta_A = 100\gamma$. From
\textcite{Zippilli2004Collective}.} \label{fig:Two_atom_scatt}
\end{figure}

The superradiant light scattering into the cavity is the basis of
various collective dynamical effects, which have been more
profoundly studied theoretically. Although we mostly neglect
atomic saturation effects in this review, it is important to
reveal modifications of the interference effect in the collective
scattering when a small but finite atomic saturation is taken into
account. Since the saturation also depends on the relative
distance of the particles, new types of nonlinear behavior take
place. For example, as shown in Fig.~\ref{fig:Two_atom_scatt}, the
destructive interference for a separation of half-wavelength
between the atoms is not perfect any more and the photon
scattering generates a nonclassical cavity field with zero
amplitude but finite photon number
\cite{Zippilli2004Forces,Vidal2007Nonlinear}.

\subsubsection{Collective cooling, scaling laws}
\label{sec:collcoolscaling} As was discussed in
Sec.~\ref{sec:BlueCooling}, the cavity cooling force on single
atoms stems from a delicate correlation between the atomic motion
and the retarded dynamics of the cavity field. In a many-atom
system it is at first unclear what happens to these correlations
in the presence of other moving atoms. Furthermore, the
cavity-mediated crosstalk between atoms has a component sensitive
to the atomic velocities \cite{Domokos2003Mechanical}, i.e., atom
1 moving at velocity $v_1$ induces a linear friction force on atom
2, which might yield correlations in velocity space. To answer
this question, one can straightforwardly generalize the
semiclassical model, presented in Sec.~\ref{sec:SemiclassicModel},
for many atoms. In general, however, this leads to an analytically
intractable problem. The dynamics of the many-atom system cannot
be reduced to that of an effective mode, as was the case for the
adiabatic potential \eref{eq:ManyAtomPotential} for atoms at rest.
The two-atom case has been discussed in detail by
\textcite{Asboth2004Correlated}, who found, using the parameter
regime of Fig.~\ref{fig:TwoAtomPotential}a, a buildup of strong
correlations in the motion of two atoms due to the
velocity-dependent cavity forces.

The scaling of the cavity cooling efficiency with the number of
particles has been studied by means of numerical simulations for
$N=1\ldots100$ in the limit of a weakly driven single-mode field,
where the optical dipole potential negligibly perturbs the free
motion of atoms along the cavity axis \cite{Horak2001Scaling}. If
the parameter $U_0$ is chosen sufficiently small so that the
collective light shift is still below the cavity linewidth, $N U_0
< \kappa$, the rate of kinetic energy dissipation is independent
of the number of atoms. This suggests that the individual atoms in
the cloud are cooled independently from each other, although they
are all coupled to the same cavity mode. This holds only in the
weak-coupling limit which is not practical for cooling since the
cooling time is long.

When the collective coupling to the cavity mode is significant
with respect to the linewidth $\kappa$, the scaling behavior of
cooling with the number of atoms has been studied by keeping
$NU_0$ and $\eta/\sqrt{N}$ constant while varying the atom number
$N$. The former ensures an identical maximum collective light
shift induced by the atoms, the latter amounts to a nearly
constant optical potential depth (proportional to $U_0
\eta^2/\kappa^2$). With this rescaling of the parameters, the
effect of individual atoms on the cavity field diminishes as the
number of atoms increases. The final temperature was found
invariant, however, the cooling time increases linearly with the
atom number $N$. As long as the driving $\eta$ is weak enough to
result in a shallow optical potential depth, in which the atoms
move almost freely, all the motional degrees of freedom along the
cavity axis are cooled.

In the limit of tightly confined atoms, both theoretical
calculations \cite{Asboth2004Correlated,Nagy2006Collective} and
experiments \cite{Schleier2011Optomechanical} proved that only the
center-of-mass motion is damped by the cavity-induced friction
force \cite{Gangl1999Collective,Gangl2000Cold}. An
efficient sideband cooling scheme has been proposed by
\textcite{Elsasser2003Collective} for particles
confined in the optical lattice potential generated by two
counter-propagating degenerate modes of a ring cavity. The scheme relies on the collective atom-field coupling
which lifts the degeneracy and creates two standing-wave modes
phase-locked by the back-scattering of light. The lower-lying mode
sustains the optical lattice with an intensity adjusted such that
the upper-lying mode becomes resonant with the vibrational
anti-Stokes Raman transition.  The sideband cooling allows to reach the vibrational
ground state.

A collective enhancement of friction on the center-of-mass
motion has been demonstrated experimentally in the transverse pump
configuration, as an accompanying effect of the self-organization
into a Bragg-scattering lattice (see Sec.~\ref{sec:ColdSelforg}).
Peak decelerations of $-10^3\, \text{m}/\text{s}^2$ have been
observed, and the damping effect has been demonstrated with
light-atom detunings up to $\Delta_A/2\pi = -6$ GHz.
\begin{figure}
\centering
\includegraphics[width=0.9\columnwidth]{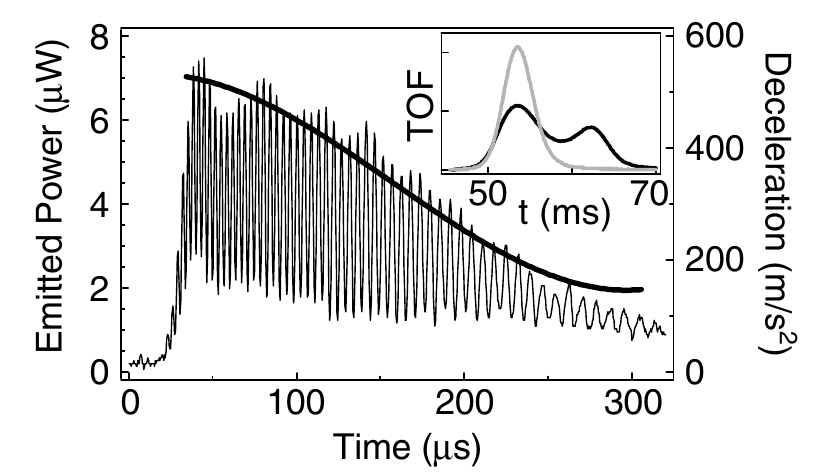}
\caption{Observation of collective friction force on the
center--of-mass motion. The cavity output power (thin line, left
scale) is shown during illumination of a freely falling atomic
cloud (initial velocity 15 cm/s) with a transverse laser beam. The
initial increase signifies the self-ordering into a
Bragg-scattering lattice (see Sec.~\ref{sec:ColdSelforg}).
Deceleration of the center-of-mass motion is recorded via the beat
signal recorded over $300\mu$s. The modulation stems from the
spatial variation of the atom-cavity coupling with period
$\lambda/2$, whereby atoms cannot scatter at a node of the
intracavity standing wave. The changing modulation frequency
indicates the atomic deceleration (thick line, right scale). The
inset shows the density profile (a.u.) of the atomic cloud after
free expansion without (gray line) and with (black line) a 400
$\mu$s exposure to the pump beam. A fraction of about one third of
the atoms is delayed significantly in accordance with the measured
deceleration. Here the pump $I/I_s = 420$ ($I_s=1.1$mW/cm$^2$  is
the D line saturation intensity of Cs), $\Delta_A/2\pi = -1.58$
GHz, and $\Delta_C/2\pi = -10$ MHz. The atom number is $N=2.6
\times 10^7$. From \textcite{Black2003Observation}.}
\label{fig:ExpCollectiveCooling}
\end{figure}
For smaller detunings $\Delta_A/2\pi \approx -160$MHz, similarly
large velocity-dependent friction forces (up to decelerations of
$-1500$ m/s$^2$ and temperatures as low as  $7 \mu$K) have been
observed in another set of experiments \cite{Chan2003Observation}.
Both the large friction and the low temperature cannot be
explained in terms of the interaction between single atoms and the
cavity field.  In this detuning regime a theoretical description
is more involved because the entire hyperfine manifold has to be
taken into account and the interaction can lead to Raman lasing
between different magnetic sublevels.

\subsubsection{Back-action, nonlinear dynamics}

In general, the interplay between the mechanical effect of light
on the atomic motion and light scattering inside the cavity off
the spatial atomic density distribution can lead to highly
non-linear dynamics. Having a large atomic ensemble organized in a
lattice structure, for example, the collective Bragg scattering
much more efficiently redistributes the light between modes than
Rayleigh scattering from the individual atoms. The enhancement
factor, being on the order of the number of atoms, can give rise
to significant sensitivity of light scattering to small variations
of the spatial distribution. As an example, the back scattering
between the counter-propagating modes (denoted by $+$ and $-$) of
a ring cavity was found to depend strongly on the bunching
parameter of the atomic distribution around the trapping sites of
an optical lattice. In experiments performed by the Hamburg group
\cite{Nagorny2003Collective,Elsasser2004Optical}, the amplitude
$\alpha_+$ in one of the modes was actively stabilized by a
feedback loop, $\dot{\alpha}_+ = 0$, and thus the other mode
obeyed the nonlinear equation of motion
\begin{equation}
\label{eq:RingCavityFeedbackBistability} \dot \alpha_- = i N U_0
\, B \, \frac{\alpha_-^2}{\alpha_+}-\kappa \alpha_- -i NU_0
\alpha_+ \, B^* + \eta_-\;.
\end{equation}
Here, $\eta_-$ denotes the driving amplitude of this mode and
$B=\langle e^{-2ikz}\rangle$ the bunching parameter (see also
Sec.~\ref{sec:CARL}). For a thermal cloud the atomic bunching
follows approximately $B \propto \alpha_-^*/|\alpha_-| \,
\exp\left\{- \mbox{const} / \sqrt{|\alpha_-|}\right\}$. The
resulting nonlinear dynamics was exemplified by a new kind of
optical bistability in the dispersive atom-field coupling regime,
which is outside the range of optical bistability effects relying
on the nonlinearity of the internal atom-field coupling
\cite{Lugiato1984II}. Subsequent experiments
\cite{Klinner2006Normal} also revealed the mechanical effect of
light on the atomic distribution in the dispersive regime through
the normal-mode splitting.

\subsection{Non-equilibrium phase transitions and collective
instabilities}

The nonlinear collective dynamics of thermal atoms in a
high-finesse resonator can give rise to non-equilibrium phase
transitions and collective instabilities. In the following we
present two, experimentally evidenced examples which have been
theoretically studied both in the thermodynamic limit and by means
of microscopic models.

\subsubsection{Spatial self-organization into a Bragg-crystal}
\label{sec:ColdSelforg}

A thermal cloud of cold atoms interacting with a single mode of a
high-finesse Fabry-P\'erot cavity undergoes a phase transition upon
tuning the power $P$ of a far-detuned laser beam (wavelength
$\lambda$) which illuminates the atoms from a direction
perpendicular to the cavity axis
\cite{Domokos2002Collective,Asboth2005Selforganization}. Below a
threshold power $P_\mathrm{cr}$, the thermal fluctuations
stabilize the homogeneous density distribution of the atomic
cloud, and light which is scattered off the atoms into the cavity
destructively interferes, rendering the mean cavity field
amplitude to be zero. Above threshold, $P>P_\mathrm{cr}$, the
atoms self-organize into a $\lambda$-periodic crystalline
checkerboard order which is bound by the interference between
the pump field and the macroscopic cavity field, resulting from
Bragg scattering into the cavity mode.

This self-organization effect can be described in terms of a
semiclassical model similar to \eref{eq:SingleAtomSemiclassic},
generalized to many atoms. A set of variables ${\bf p}_j$ and
${\bf r}_j$ is introduced, the index $j=1, \ldots, N$ labeling the
atoms. For simplicity, the atomic motion is considered in two
dimensions spanned by the cavity axis and the pump laser
direction, with coordinates $x$ and $z$, respectively. The
equation of motion for the coherent cavity field amplitude
$\alpha$ is given by
\begin{subequations}
\label{eq:sde}
\begin{multline}
\label{eq:sde_al}
  \dot\alpha = i \Big[\Delta_C - U_0 \sum_j \cos^2(k x_j)\Big] \alpha
  - \Big[ \kappa + \Gamma_0 \sum_j \cos^2(k x_j)\Big] \alpha
  \\
  - i \etaeff \sum_j \cos(k x_j) \cos(k z_j) + \xi_\alpha \,,
\end{multline}%
where the effective pumping strength of the cavity mode is denoted
by $\etaeff = \frac{\Omega g \Delta_A}{\Delta_A^2+\gamma^2}$, see
\eref{eq:H_pump_eff}.  Due to the interference term
$\sum_j\cos(kx_j) \cos(kz_j)$ light scattering into the cavity
vanishes for a homogeneous atomic density distribution. It can be
small even if all the atoms are maximally coupled but the signs of
the summands alternate. The light forces exerted on the individual
atoms along the cavity and pump direction are given by
\begin{multline}
\label{eq:sde_pz}
  \dot {p_x}_j = - \hbar U_0 |\alpha|^2 \frac{\partial}{\partial x_j}
  \cos^2(k x_j)
  \\
  - \hbar \etaeff (\alpha + \alpha^*)
  \frac{\partial}{\partial x_j} \cos(k z_j) \cos(k x_j) + {\xi_x}_{j}\; ,
\end{multline}
\begin{multline}
\label{eq:sde_px} \dot {p_z}_j = - \hbar U_0 (\Omega/g)^2
\frac{\partial}{\partial z_j} \cos^2(k z_j)
\\
- \hbar \etaeff ( \alpha +\alpha^*)
\frac{\partial}{\partial z_j} \cos(k x_j) \cos(k z_j)+
{\xi_z}_{j}\; ,
\end{multline}
\end{subequations}

These equations include Langevin noise terms $\xi_\alpha$, ${\xi_x}_j$,
and ${\xi_z}_j$, defined by the non-vanishing second-order correlations,
\begin{subequations}
 \label{eq:noise_sde}
\begin{align}
  \langle \xi_\alpha^* \xi_\alpha \rangle &= \kappa  + \sum_{j=1}^N
  \Gamma_0 \cos^2(k x_j) \;, \\
  \langle {\xi}_{n} \xi_\alpha \rangle &= i \hbar \Gamma_0 \partial_n{\cal
  E}({\bf r}_j) \cos(k x_j) \;,\\
  \langle \xi_{n} \xi_{m} \rangle &= 2 \hbar^2 k^2 \Gamma_0 |{\cal E}({\bf
  r}_j)|^2 \overline{ u_n^2} \delta_{nm} + \hbar^2 \Gamma_0  \nonumber \\
 &  \Bigl[\partial_n {\cal E}^*({\bf r}_j) \, \partial_m{\cal
  E}({\bf r}_j) + \partial_n {\cal E}({\bf r}_j) \, \partial_m{\cal
  E}^*({\bf r}_j) \Bigr] \; ,
\end{align}
\end{subequations}
with indices $n,m=x_j, z_j$. The noise terms associated with
different atoms are not correlated. The complex dimensionless
electric field is given by ${\cal E}({\bf r}) = \cos(k x) \alpha +
\cos(k z) \Omega/g$.

\begin{figure}[htbp]
\centering
\includegraphics[width=0.95\columnwidth]{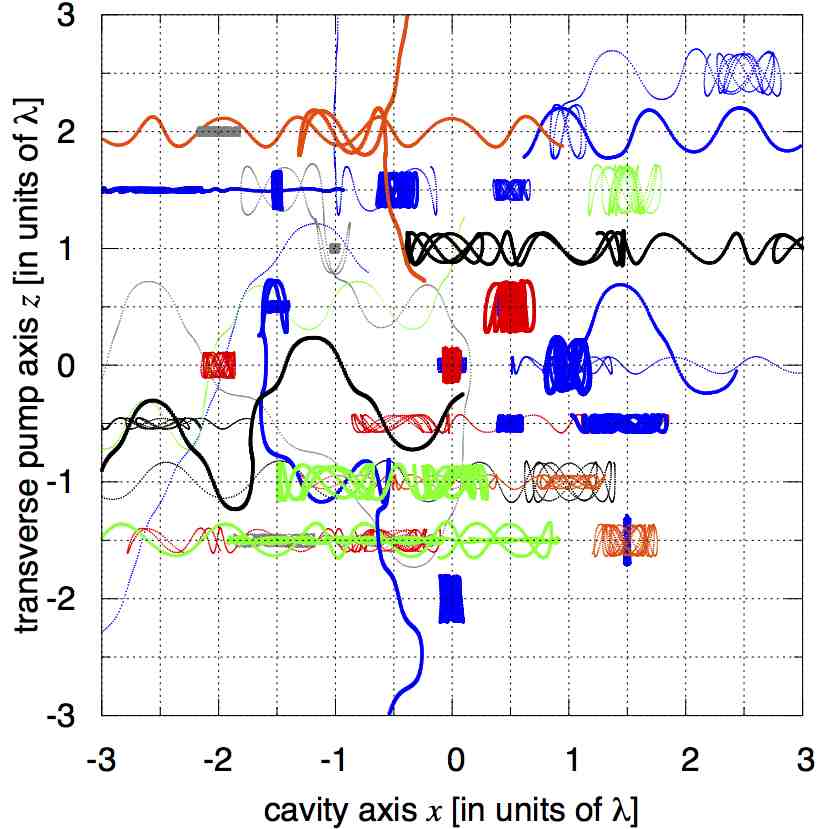}
\caption{(Color online) Self-organization of laser-driven atoms in
a cavity. Numerically simulated two-dimensional trajectories
during the first 50 $\mu$sec of transverse illumination. A
checkerboard pattern of trapped atoms emerges, in which the
occupied trapping positions are separated by even multiples of
$\lambda/2$ (= 1 edge). The grid lines indicate points of maximum
coupling to the standing-wave cavity or pump field. There is a
possible complementary configuration with atoms occupying the
other set of intersections. Parameters: $\gamma=20/\mu$sec, $(g,
\kappa)= (2.5, 0.5)\gamma$, atomic detuning $\Delta_A =-500
\gamma$, cavity detuning $\Delta_C =- \kappa+N U_0$, and the
pumping strength $\Omega=50\gamma$. From
\textcite{Asboth2005Selforganization}.} \label{fig:selforg2D}
\end{figure}
\begin{figure}[t!]
\centering
\includegraphics[width=0.91\columnwidth]{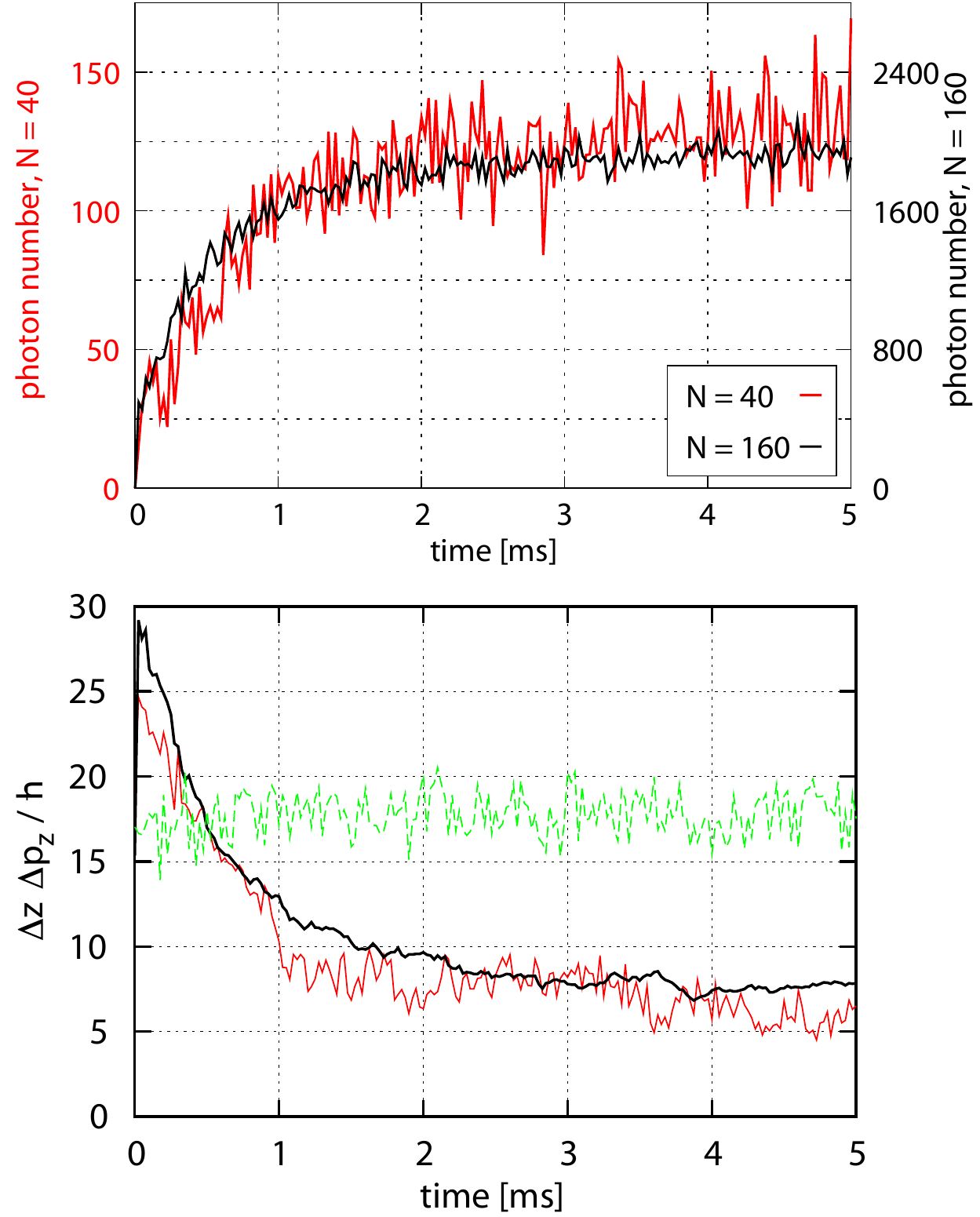}
\caption{(Color online) Cavity cooling in the self-organized
phase. The time evolution of the photon number in the cavity
(left) and the phase space density of the atoms (right) on a long
time scale, for $N=40$ and $N=160$ atoms (different vertical
scalings are used for illustrating the superradiance $|\alpha|^2
\propto N^2$). In the right panel, the green curve fluctuating
around a constant value corresponds to a uniformly distributed
$N=40$ atoms driven below the self-organization threshold. The
parameters are the same as in Fig.~\ref{fig:selforg2D}. From
\textcite{Asboth2005Selforganization}.} \label{fig:slowselforg}
\end{figure}

In Fig.\ \ref{fig:selforg2D}  a numerical simulation of the
trajectories of 40 atoms during the first 50 \(\mu\)s of
illumination are shown. The initial configuration is given by an
ensemble of thermal atoms with random positions from a uniform,
and velocities from a thermal distribution. The cavity mode
initially is in the vacuum state ($\alpha=0$). With the right
choice of parameters the emergence of a periodic pattern in the
spatial density distribution of the atoms is observed, accompanied
by the buildup of a coherent cavity field amplitude
(Fig.~\ref{fig:slowselforg}). In the emerging configuration, assuming
a red-detuned pump laser,  the trapped atoms are oscillating about intensity maxima of the interfering pump-cavity field.
Along the cavity and pump direction these are separated by even multiples of the optical wavelength.
Since only the black or white fields of the underlying checkerboard lattice pattern are occupied,
constructive interference leads to efficient Bragg scattering of pump photons into the cavity.

As will be shown in more detail  later, the process of self-organization relies on the right choice of
the detuning $\delta_C = \Delta_C- N U_0/2$ between the pump laser
and the dispersively shifted cavity resonance $\Delta_C$. For the
case $\delta_C<0$, the potential term $\cos(kx_j)\cos(kz_j)$ in
Eqs.~(\ref{eq:sde}b,c) attracts atoms towards the ``majority''
sites and repels them from the ``minority'' sites, providing
positive feedback. Initiated by density fluctuations, one of the
two possible Bragg lattices is then formed in a runaway process.
For the case $\delta_C>0$, the scattered cavity field creates
potential maxima (minima) at the positions of the majority
(minority) sites, counteracting the amplification of density
fluctuations and preventing a dynamical instability. Furthermore,
in this regime the delayed cavity response causes cavity heating
of the atomic motion which obscures an equilibrium situation in
the lack of other dissipative process.

For $\Delta_C-NU_0<0$, the initial fast buildup of a coherent
cavity field continues over a longer
timescale.  The kinetic energy of the oscillating and the
untrapped atoms dissipates owing to the cavity cooling mechanism, which leads to an increase of the number of trapped atoms and a stronger
localization in the potential wells. This further improves
coherent scattering into the cavity, as indicated by the slow
increase in the cavity field intensity shown in
Fig.~\ref{fig:slowselforg}. Comparing the time evolution of the
intracavity photon number for the self-organization process of 40
and 160 atoms (rescaled in Fig.~\ref{fig:slowselforg} by a factor
of 16) demonstrates the superradiance effect, i.e.~the field
intensity scales cooperatively as the square of the particle
number. The right panel of the figure shows that the cooling rate, described by the decrease of the phase space density of the atoms, is also similar for $N=40$ and $N=160$ and that the self-organization leads to smaller phase space densities than the homogeneous distribution below threshold.

\begin{figure}[b!]
\begin{center}
\includegraphics[width=0.9\columnwidth]{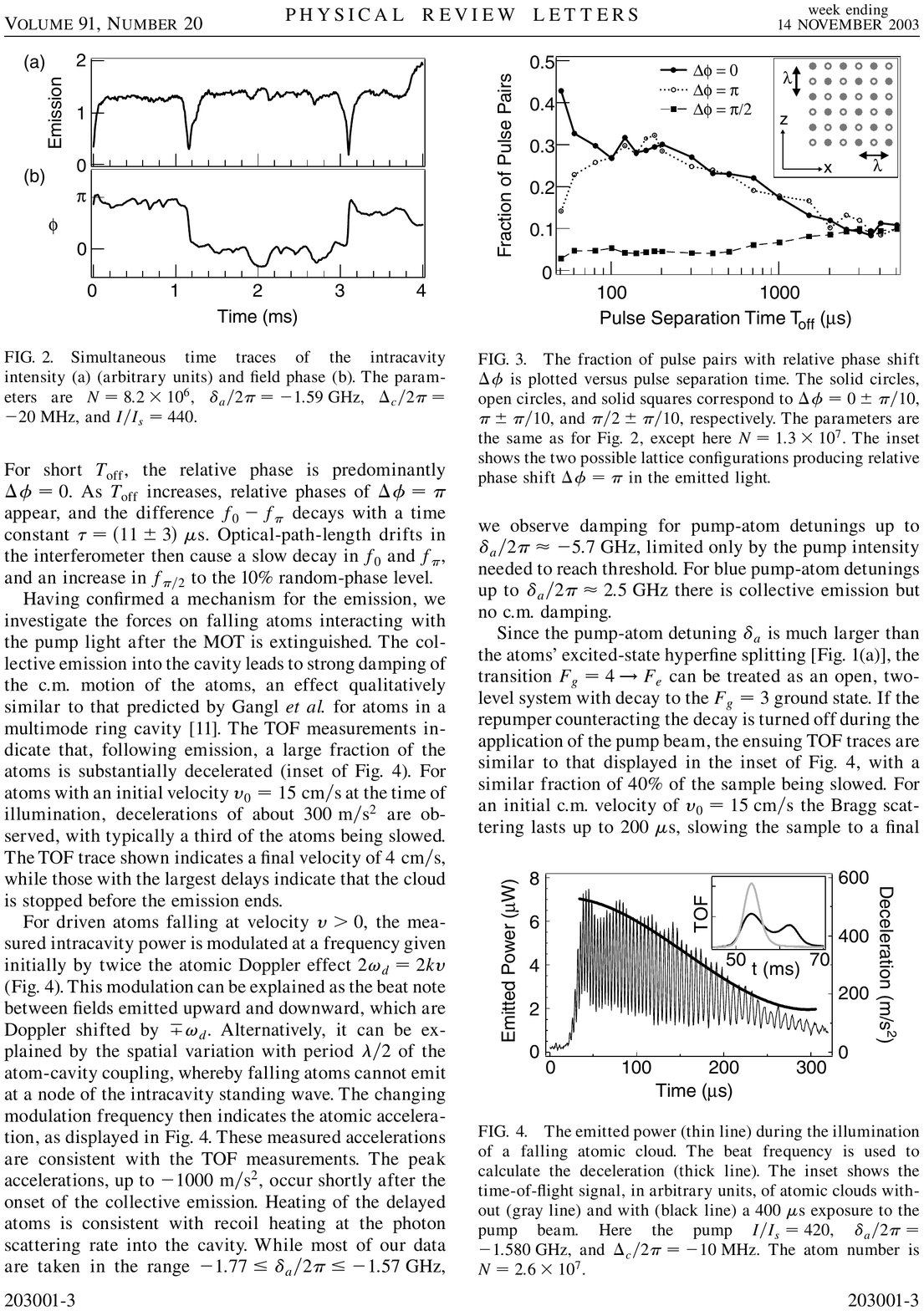}
\caption{Observation of spontaneous symmetry breaking in the
self-organization phase transition. Simultaneous time traces of
the  (a) intracavity intensity (in arbitrary units) and (b) the
relative pump-cavity phase. Drops in the intensity correspond to
time intervals during which the beams of the magneto-optical trap (MOT) are switched on forcing
the atomic density distribution to randomize. After switching off
the MOT beams, the atoms self-organize again into one of the two
possible checkerboard patterns, as indicated by the relative phase
signal. Experimental parameters were $N =8.2 \times 10^6$,
$\Delta_A= -2 \pi \times \unit[1.59]{GHz}$, $\Delta_C = - 2\pi
\times \unit[20]{MHz}$, and $I/I_\sub{sat} = 440$. From
\textcite{Black2003Observation}.} \label{fig:Selforg_Exp}
\end{center}
\end{figure}
Self-organization of laser-cooled atoms has been observed in
experiments of the MIT group with $N \approx 10^7$ Cs atoms
prepared at a temperature of $\unit[6]{\mu K}$ in a nearly
confocal Fabry-P\'erot cavity \cite{Black2003Observation}. Above a
threshold intensity of the transverse pump beam, collective
emission of light into the cavity was observed at a rate which
exceeds the free-space single-atom Rayleigh scattering rate by a
factor of up to $10^3$. This experiment clearly demonstrated the
process of spontaneous symmetry breaking by measuring $\pi$-jumps
in the phase of the emitted cavity field relative to the
transverse pump field, corresponding to self-organization into
the black or white lattice sites of a checkerboard pattern
(Fig.~\ref{fig:Selforg_Exp}). Retardation between the cavity field
and the atomic motion resulted in a collective friction force on
the center-of-mass degree of freedom. A deceleration of up to
$\unit[1000]{m/s^2}$ has been achieved with atom-cavity detunings
as large as $\Delta_A = -2\pi\times \unit[1.58]{GHz}$.

\begin{figure}[htbp]
\centering
\includegraphics[width=0.9\columnwidth]{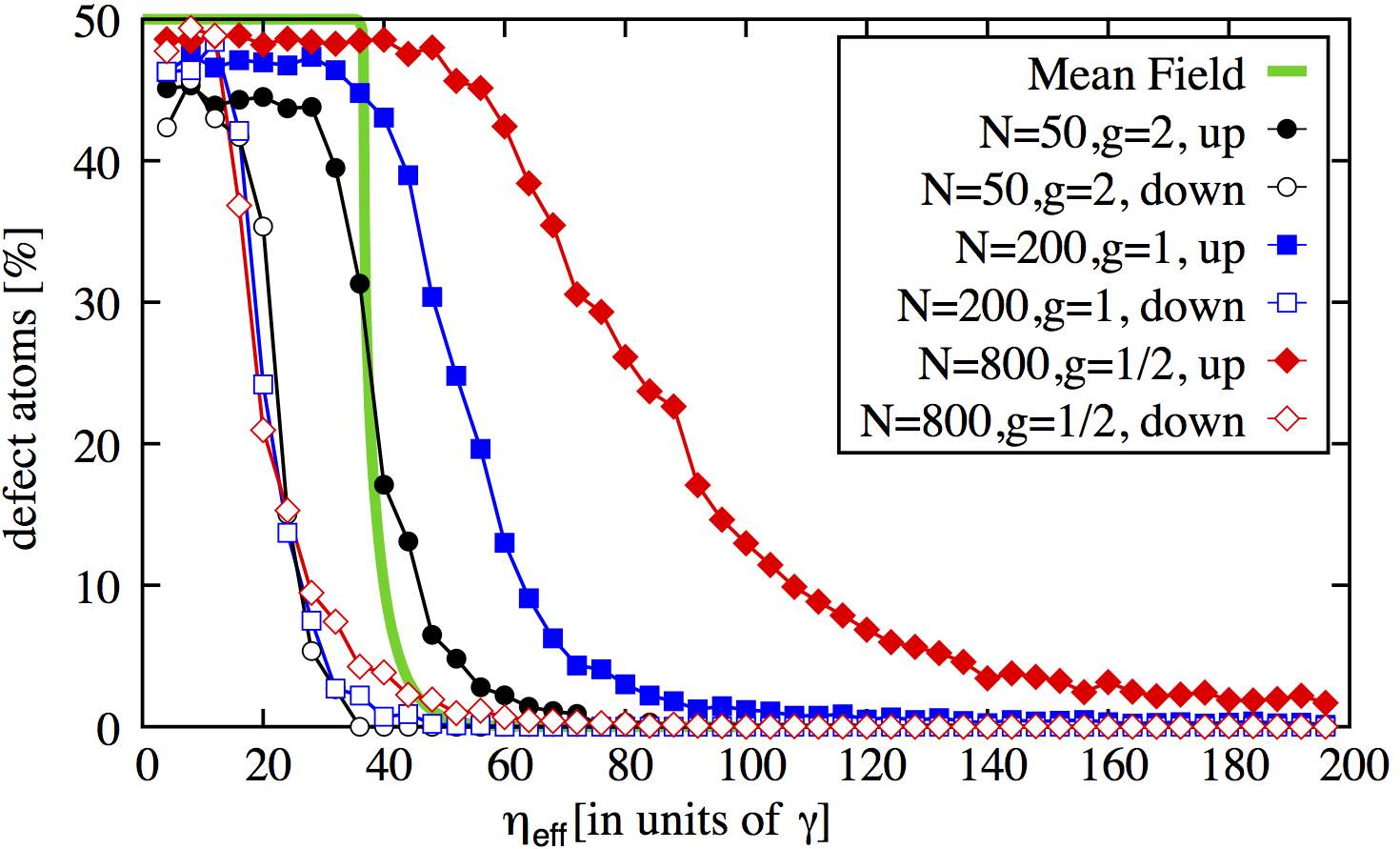}
\caption{(Color online) Hysteresis effect for finite measurement
time. Ratio of atoms in the ``defect'' positions against pumping
strength $\eta$, 4ms after the loading of the trap with a uniform
(``up'') or organized (``down'') gas of atoms. The different
curves show the approach towards the thermodynamic limit. The
parameters are $\kappa=\gamma/2, \Delta_A= -500 \gamma$, $N g^2 =
200 \gamma^2$ $\Delta_C=-\kappa-Ng^2/|\Delta_A|$, $k_B T = \hbar
\kappa$. From \textcite{Asboth2005Selforganization}.}
\label{fig:threshold_Ng2}
\end{figure}
For finite atom number $N$ and finite measurement time, an
interesting hysteresis effect accompanies self-organization, as
shown in Fig.~\ref{fig:threshold_Ng2}. The thermodynamic limit
$N\to \infty$ is approached by simulations of Eq.~(\ref{eq:sde})
with the atomic density  $N/V \propto N g^2$ and the cavity loss
rate kept constant.  The percentage of defect atoms after $4$ ms
of simulation time as a function of the pumping laser strength
clearly shows the transition. However, the transition  point is
dependent on $N$ and on whether the initial positions were
uniformly distributed (``up'') or at ``odd'' points of maximal
coupling (``down''). The breadth of the hysteresis increases with
the atom number, but decreases with the measurement time. This behavior can be explained by
taking into account statistical fluctuations arising from the finite atom number $N$.
Assuming that the self-organization from a uniform distribution (``up'' curves) is triggered when the fluctuating energy difference between the even and odd sites momentarily exceeds the mean kinetic energy, a scaling $N g^4$ of the threshold with the pump intensity was found in accordance with the numerical results of the plot. The disappearance of the lattice pattern for decreasing pump power  (``down'' curves) when the system is started from the ordered phase occurs at the half of the mean-field threshold (see \eref{eq:threshold_selforg} in the next subsection), independently of the atom number $N$.

Self-organization of laser-driven atoms occurs also in a
transversally driven ring-resonator geometry supporting two
running-wave modes \cite{Nagy2006Selforganization}. In contrast to
the linear single-mode cavity case, here the transition from the
homogeneous to the organized density distribution involves
spontaneous breaking of a continuous (rather than a discrete)
translational symmetry (see also
\ref{sec:MultimodeDegenerateCavity}).

\subsubsection{Collective atomic recoil laser}
\label{sec:CARL}

Collective atomic recoil lasing (CARL) is the prominent many-body
instability effect in a ring cavity, originally predicted by \textcite{Bonifacio1994Exponential}. An ensemble of cold atoms
couples to two counter-propagating modes of a unidirectionally
pumped high-finesse ring cavity. Light scattering off the atomic
ensemble between these cavity modes leads to a collective
instability corresponding to an exponential gain for the
back-propagating field mode amplitude in conjunction with an
atomic bunching at the antinodes of a self-organized optical
lattice. In the presence of dissipation of the atomic kinetic
energy, a steady-state operation of CARL can be achieved with a
self-determined atomic drift velocity and back-reflected light
frequency.

The CARL scheme involves an interplay between the influence of the
atomic motion on the Rayleigh scattering of light and, reversely,
the mechanical effect of light upon the atomic motion. The former
effect has been previously seen, for example, as the so-called
recoil induced resonance (RIR) in the transmission spectrum of a
probe beam making a small angle with a one-dimensional
lin$\perp$lin optical molasses \cite{Courtois1994Recoil}, and also
in a optical dipole trap formed by counter-propagating modes of a
ring cavity \cite{Kruse2003Cold}. This narrow, dispersion-like
resonance around the pump field frequency originates from a
two-photon Raman transition between different momentum states of
the atoms. Different populations of the corresponding momentum
states lead to gain or attenuation of a probe beam. For a thermal
velocity distribution, a probe frequency tuned slightly below the
pump frequency gives rise to gain, whereas for
negative detuning there is an attenuation of the probe.
The probe transmission spectrum measurement provides information
about the temperature, and even more about the velocity
distribution \cite{Brzozowska2006Bound}.

\begin{figure}[htbp]
\begin{center}
\includegraphics[width=0.9\columnwidth]{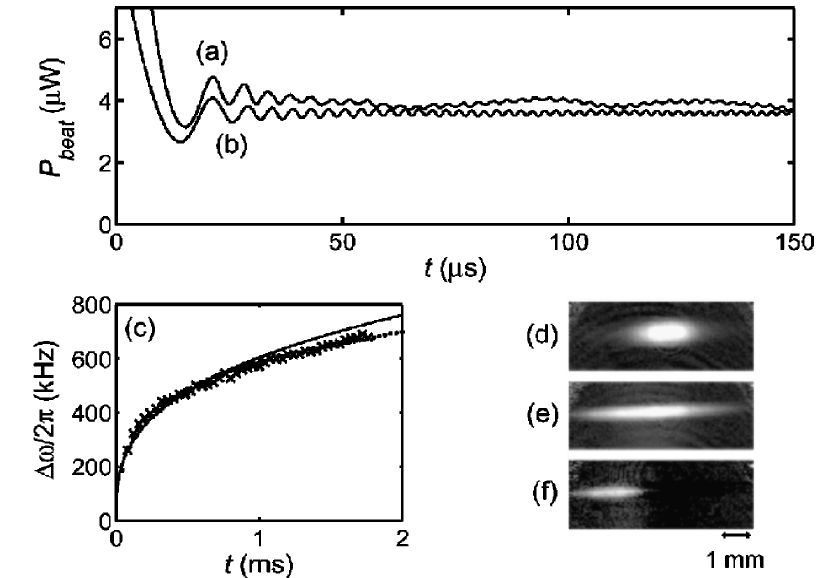}
\caption{Observation of lasing mediated by collective atomic
recoil. (a) Recorded time evolution of the observed beat signal
between the $\alpha_+$ and $\alpha_-$ cavity modes of a ring
cavity. Initially both modes are pumped to form an optical lattice
of  $N=10^6$ atoms. The initial drop is due to the decay of the
unpumped mode after switching off the $\alpha_-$ pump at $t=0$.
Well beyond the ring-down time of about $\unit[10]{\mu s}$, the
persisting oscillations demonstrate the coherent backscattering of
the pumped mode.  (b) Numerical simulation with the temperature
adjusted to $\unit[200]{\mu K}$. (c) The symbols ($\times$) trace
the evolution of the beat frequency after switch-off (dotted line
is from numerical simulation). The increase of the beat frequency
corresponds to the acceleration of the Bragg lattice of back
reflecting atoms, which lasts until the Doppler-shifted frequency
drops out of cavity resonance.  Absorption images of a cloud of $6
\times 10^6$ atoms recorded at (d) $\unit[0]{ms}$ and (e)
$\unit[6]{ms}$ after switching off the probe beam pumping. All
images are taken after a $\unit[1]{ms}$ free expansion time. (f)
This image is obtained by subtracting from image (e) an absorption
image taken $\unit[6]{ms}$ after switch-off with low cavity
finesse for which no collective recoil is expected. The
intracavity power has been adjusted to the same value as in the
high finesse case. From \textcite{Kruse2003Observation}.}
\label{fig:TransientCarl}
\end{center}
\end{figure}

It has been predicted that, based on the RIR gain effect, a weak
probe field injected in the direction opposite to the strong pump
field would exponentially amplify due to the self-bunching of a
part of the atoms into a lattice which Bragg reflects the strong
pump beam more efficiently than single atoms
\cite{Bonifacio1994Exponential}. However, one needs  long enough
interaction time so that back-action of the light scattering on
the velocity distribution can have a significant effect
\cite{Berman1999Comparison}. This can be accomplished, for
example, by confining the light modes into a cavity. Lasing
mediated by the collective atomic recoil between the
counter-propagating modes of a unidirectionally ring cavity has
been observed \cite{Kruse2003Observation}. Back-action on the
atomic motion has been demonstrated by detecting the displacement
of the atoms accelerated by the momentum transfer process. The
self-consistent solution is an accelerating Bragg-lattice of atoms
co-moving with the standing wave formed by the pump and the
back-reflected component having a Doppler-shifted frequency
$\Delta \omega = 2 k v$ with respect to the pump. The phase
dynamics  of the counter-propagating modes can be monitored as a
beat signal between the outcoupled beams, which reveals the
acceleration by an increasingly red-detuned probe as a  function
of time (Fig.~\ref{fig:TransientCarl}).

The runaway process can be counteracted by introducing some
external friction force on the atom motion. The dissipation gives
rise to a steady-state solution which involves a constant drift of
the entire atomic cloud at a speed $v$. Accordingly, we transform
the atomic position variables such as $z_j = \tilde z_j +  vt$,
and the coherent field amplitude of the running-wave field mode
propagating opposite to the pumped field mode as $\alpha_- =
\tilde\alpha_- e^{2ikvt}$. The drift velocity $v$ is to be
determined in a self-consistent manner. The
semiclassical equations describing atomic motion are given by
\begin{subequations}
\label{eq:CARL}
\begin{equation}
\dot{p}_j = -\frac{\beta}{m} p_j + \hbar U_0 \, 2 i k \, \left(\alpha_+^* \tilde{\alpha}_- e^{-2 i k \tilde{z}_j} - \tilde{\alpha}_-^* \alpha_+ e^{2 i k \tilde{z}_j} \; ,\right)
\end{equation}
where $\beta$ represents the linear friction arising, e.g., from
collisions with a buffer gas
\cite{Bonifacio1996Doppler,Perrin2001Phase} or laser-cooling in an
optical molasses \cite{Kruse2003Observation}. The cavity field
amplitudes evolve as
\begin{align}
\dot \alpha_+ &= \left( i \delta_C -\kappa \right) \alpha_+ -i N U_0 B  \tilde\alpha_- + \eta \;,\\
\dot {\tilde\alpha}_- &= \left[ i\left(\delta_C - 2 k v \right) -\kappa \right] \tilde\alpha_- -i N U_0 B^* \alpha_+ \;,
\end{align}
where $\delta_C= \Delta_C - N U_0$ is the effective detuning of
the pump frequency from the atoms-shifted cavity resonance. The
atomic positions enter through the bunching parameter
 \begin{equation}
B= \frac{1}{N} \sum_{j=1}^N e^{-2 i k \tilde{z}_j} \equiv b e^{-i \varphi} \; .
\end{equation}
\end{subequations}
A closed set of equations can be formed in which the atomic cloud
is characterized by the three real parameters $v$, $b$, and
$\phi$. A trivial solution of these equations corresponds to the
case where the atoms are uniformly distributed in space ($b=0$,
$v=0$) and the counter-propagating field mode amplitude vanishes
($\alpha_- =0)$. A non-trivial steady-state solution can be
obtained numerically from the coupled algebraic equations (time
derivatives set to zero). This solution can be approximated
analytically by assuming perfect bunching, $b=1$ and $\varphi=0$.
The only remaining free parameter is then given by the steady-state
drift velocity $v$ which obeys the algebraic equation
\begin{equation}
\label{eq:CARL_v}
2 k v = 8 \frac{m\, \omega_R}{\beta} \, N U_0^2 \frac{\kappa |\eta|^2}{|D|^2} \;,
\end{equation}
where $D=(i \delta_C -\kappa) [i(\delta_C-2 kv)-\kappa] + N^2U_0^2
b^2$. In order to gain insight into the solution, the following
simplifications can be made:  (i) neglect in $D$ the last $U_0^2$
term originating from (second-order) scattering of the $\alpha_-$
mode back into the pumped $\alpha_+$ one, and (ii) consider
resonance $\delta_C=0$. Then one can identify the typical
solutions of two different regimes. In the limit of a large
Doppler shift, $kv \gg \kappa$, the drift velocity and the back
reflected power scale with the atom number as $v\propto N^{1/3}$
and $|\alpha_-|^2 \propto N^{4/3}$, respectively. This is referred
to as the CARL limit in the literature. In the opposite limit of a
small Doppler shift, $kv \ll \kappa$, the velocity obeys $v\propto
N$ and the intensity exhibits superradiant behavior, $|\alpha_-|^2
\propto N^2$, corresponding in this geometry to Bragg
retro-reflection. The relation between these two kinds of
superradiant instabilities in the collective interaction of light
with an atomic gas has been established experimentally by
\textcite{Slama2007Superradiant, Slama2007Cavityenhanced}. While
superradiant Rayleigh scattering from atomic clouds is normally
observed only at very low temperatures, i.e., well below
$\unit[1]{\mu K}$, \cite{Inouye1999Superradiant}, the presence of
the ring cavity enhances cooperativity and allows for
superradiance with thermal clouds as hot as several $\unit[10]{\mu
K}$.

\begin{figure}[htbp]
\begin{center}
\includegraphics[width=0.9\columnwidth]{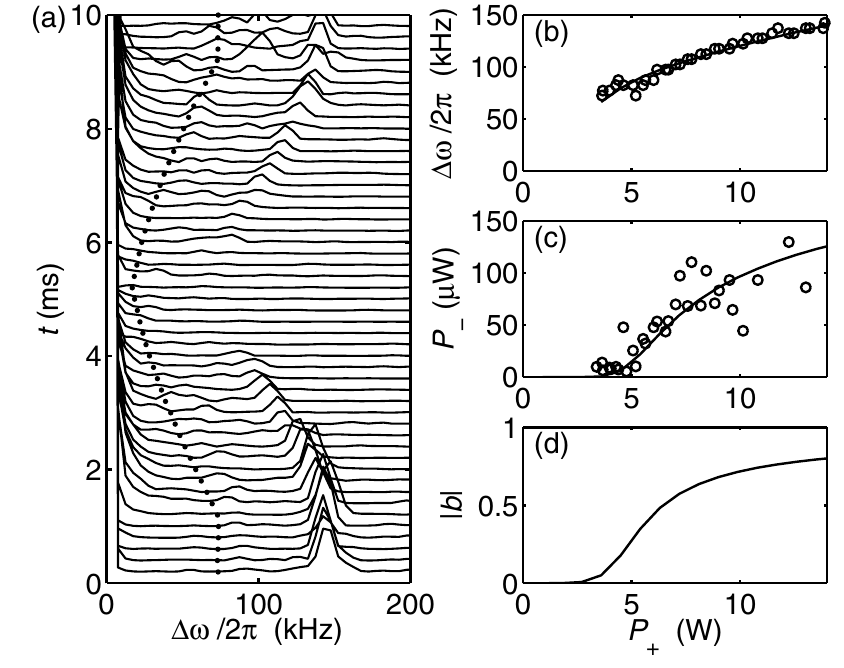}
\caption{Pump power threshold for the collective atomic recoil
lasing. (a) Sectionwise Fourier-transform of the interference
signal $P_\sub{beat}$ with the pump laser power being ramped down
and up (the dotted line is proportional to the pump laser power).
At $t=0$ the system is in the ordered CARL phase, by gradually
decreasing the power the drift velocity decreases (the peaks in
the Fourier spectra shift downward) until the back reflection
ceases at a threshold pump power. Ramping up the pump power from
about $t=5$ ms, the appearance of a peak is delayed and occurs at
about the same threshold value of the pump power.  (b)
Dependence of the CARL frequency on the intracavity pump power.
(c) Dependence of the probe field intensity on the pump power. The
CARL laser threshold is around $P_+ = 4$ W intracavity power.  The
fitted curves are based on a Fokker-Planck theory outlined in
Sec.~\ref{sec:ThermalMeanField}. (d) Calculated bunching
parameter. The parameters are $\beta=4 \kappa$, $N=10^6$ atoms,
$\Delta_A = - 2\pi \times 1.7$ THz and $T=200 \mu$K. From
\textcite{Cube2004SelfSynchronization}.} \label{fig:steadyCarl}
\end{center}
\end{figure}
In the experiments of the T\"ubingen group
\cite{Kruse2003Observation,Cube2004SelfSynchronization}, an
optical molasses has been used to impose a motional damping force
on the atoms. In such a viscous CARL system,  there is a
steady-state operation with a self-consistent drift velocity,
according to \eref{eq:CARL_v}, at which the friction compensates
the acceleration due to back scattering of photons. The Fourier
spectrum of the beat signal between the pumped mode and the
counter-propagating one  in steady-state is shown in
Fig.~\ref{fig:steadyCarl}a for various pump strengths. The drift
velocity can be deduced from the Doppler-shift of the
back-reflected field as  shown in Fig.~\ref{fig:steadyCarl}b, the
corresponding range of velocities is from 7 to $\unit[13]{cm/s}$.
The beat frequency as well as the drift frequency vary as a
function of the pump power. However, the dramatic feature of
Fig.~\ref{fig:steadyCarl} is  the clear appearance of a threshold:
the self-bunching and backscattering starts only above a
well-defined threshold pump power. This measurement provides thus
an experimental evidence of a phase transition to a state of
synchronized atomic motion \cite{Cube2004SelfSynchronization}.
Underlying the critical behavior is the diffusion accompanying
friction: the spontaneous photon scattering of the molasses laser
beams leads to a random heating force which stabilizes the
homogeneous distribution and thereby can prevent the formation of
a Bragg-lattice for weak pump power \cite{Robb2004Collective}.
Above threshold, the dissipation and the fluctuations  together
lead to a position distribution which exhibits a finite bunching
parameter. These effects are discussed in the framework of various
mean-field theories for the atomic position distribution (see
Sect.\ \ref{sec:MeanField}).

We note that the CARL system and its phase transition has been
studied also in the case where the atomic transition frequency is
close-to-resonance with the pump laser field
\cite{Perrin2001Phase}. Then the atomic polarization plays a
dynamical role and the transition does not require spatial
bunching but the emergence of coherent polarization grating
\cite{Perrin2002Microscopic}.

The viscous CARL transition exhibits analogy with that of the
generic Kuramoto model which describes the  self-synchronization
of coupled oscillators with different frequencies
\cite{Kuramoto1975Selfentrainment,Strogatz2000From}. To reveal the
analogy, one can transform the CARL equations using the following
assumptions: (i) the motion is overdamped ($\dot p_j =0$); (ii)
the pumped field amplitude $\alpha_+$ is a constant of time
($\alpha_+\approx \eta/\kappa$); (iii) the counter-propagating
mode amplitude is stationary, oscillating at frequency $\omega_0$,
\ie,  effectively $\dot{\tilde{ \alpha}}_- = - i\omega_0
\tilde{\alpha}_-$ with the frequency $\omega_0$ being determined
by the constant drift velocity; and (iv) $\kappa\ll \omega_0$.
With these assumptions and by using the notation $\theta_j=2 k
\tilde{z}_j$, Eqs.~(\ref{eq:CARL}) simplify to
\begin{equation}
\dot \theta_j = \frac{2k}{m\, \beta} \xi + K b \,\sin{(\varphi-\theta_j)}\; ,
\end{equation}
which is formally
equivalent with the Kuramoto model. The Langevin-type random noise  $\xi$, associated with the
friction term $-\beta p$ in Eq.~(\ref{eq:CARL}a),
introduces the random frequencies present in the Kuramoto-type
systems.  The coupling strength is $K= 2\omega_R\, N U_0^2 \,
|\alpha_+|^2/(\omega_0 \,\beta)$. The mean-field character is
obvious: each oscillator couples only to the mean-field quantities
$b$ and $\varphi$. The phase $\theta_j$ is pulled toward the
mean-field phase $\varphi$, which increases the order parameter
$b$. The coupling is proportional to $b$, which sets a positive
feedback loop. With the increasing coherence $b$, even more
oscillators can be recruited to the synchronized pack (those being
within the bandwidth $K b$), further increasing $b$. Such a
runaway process starts only above a critical coupling $K$.

\subsection{Phase-space and mean-field descriptions for large particle numbers}
\label{sec:MeanField}

Due to the nonlinearity of the coupled atom-field dynamics, exact
analytic results for atomic ensembles coupled to optical cavities
are rather sparse and the computational demand often hinders
simulations at realistic particle numbers
\cite{Deachapunya2008Slow,Salzburger2009Collective}.  In the
following, we will present mean-field methods which allow to
predict instability thresholds and to model the effective dynamics
in  the thermodynamic limit. Starting with the assumption of a
quasi-thermal distribution, the critical behavior can be revealed
and the threshold for criticality can be approximated. We will
continue by adopting phase-space methods that can account for
general position and velocity distributions. A Vlasov-type
equation allows for a more accurate estimation of the threshold,
and also for performing stability analysis and setting up a phase
diagram. This analysis leads to such intriguing predictions that
the cavity-mediated interaction combined with the cavity cooling
effect can be the basis of a generally applicable sympathetic
cooling scheme. Finally, a Fokker-Planck-type equation can be
constructed in order to determine the steady-state of the
non-equilibrium systems in the thermodynamic limit. The methods
show similarities with  plasma physics where equally complex,
coupled dynamics of particles and fields occurs
\cite{Montgomery1971Theory}.

\subsubsection{Critical point}
\label{sec:ThermalMeanField}

The simplest mean-field model is based on the assumption that the
atomic motion is overdamped and the distribution function of the
atomic positions $\rho(x,t)$ is a thermal distribution. The
critical point of the CARL instability (Sec.~\ref{sec:CARL}) and
that of the self-organization (Sec.~\ref{sec:ColdSelforg}) can be
calculated with this approach.

\paragraph{Dynamical equations}
The motional damping is characterized by a linear friction coefficient $\beta$ (half of the kinetic energy damping rate is $\beta/m$ with $m$ being the atomic mass) and a temperature $T$.  Then the mean atom density distribution obeys the Smoluchowski-equation
\begin{equation}
\label{eq:Smoluchowski} \frac{\partial \rho(x,t)}{\partial t} = -
\frac{1}{\beta} \frac{\partial}{\partial x} \left[ F(x) \rho(x,t)
- k_B T\frac{\partial \rho(x,t)}{\partial x} \right]\;.
\end{equation}

In the ring cavity geometry of CARL \cite{Kruse2003Observation}, for example, the
force $F(x)$ is given by the last term on the right-hand side of
Eq.~(\ref{eq:CARL}a). It contains the field mode amplitudes which
couple back to the atomic density distribution via the
bunching parameter $\mathcal{B}$,
\begin{subequations}
\begin{align}
\label{eq:AlphaMeanField} \dot {\tilde\alpha}_-  &= -\kappa
{\tilde\alpha}_- -i U_0 \alpha_+\, \mathcal{B} \;,\\
\text{with} \quad \mathcal{B} &=  \int_0^\lambda dx\, \rho(x,t) \,e^{2i k x}\,.
\end{align}
For simplicity, the center-of-mass velocity and the detuning $\delta_C$
were set to zero. Linear perturbation calculus leads to the instability threshold of the homogeneous solution
\cite{Robb2004Collective}. Since the spatial coupling functions
are sinusoidal, only a few Fourier components are involved in
the initial dynamics. In particular, in order to determine the
instability of the homogeneous distribution, only the single mode
function $e^{-i k x}$ of the counter-propagating cavity mode
needs to be taken into account. In the so-called CARL limit (see
Sec.~\ref{sec:CARL}), one obtains the following threshold condition for the
cavity pump amplitude:
\begin{equation}
\label{eq:CARL_threshold} \eta^2 \geq
\left(\frac{k_BT}{\hbar}\right)^{{3}/{2}} \,
\sqrt{\frac{m \omega_R}{\beta}} \; \frac{\kappa^{5/2}}{N\,U_0^2} \, .
\end{equation}
\end{subequations}

\paragraph{Canonical distribution}
A further simplification can be made in the mean-field approach
if $\kappa$ is the largest rate in the dynamics. Adiabatic
elimination of the cavity field dynamics then results in a self-consistent
optical potential $V(x)$ in which the spatial density $\rho(x)$ of
the atoms is determined by a canonical distribution,
\begin{equation}
\label{eq:rho_mf} \rho(x) = \frac{1}{Z} \exp(-V(x)/(k_B T))\; ,
\end{equation}
with the partition function $Z=\int \exp(-V(x)/(k_B T)) dx$
ensuring normalization of $\rho(x)$ to unity.  The temperature
could be identified with the one which is achieved in
cavity cooling, $k_B T \approx \hbar \kappa$, but in general it
can be set by other means, e.g., by laser-cooling in an external
optical molasses. The nonlinearity enters the equations
through the dependence of the potential $V(x)$ on the atomic
density $\rho(x)$ itself, \ie, $V=V(x, \rho(x))$.

In principle, the optical force acting on the atoms in the cavity does
not derive from a potential when the back-action of the
atomic motion on the radiation field amplitude is significant. As was noted by
\textcite{Asboth2007Comment}, dynamical equations based on forces, such as
\eref{eq:Smoluchowski} and \eref{eq:AlphaMeanField} have to be used.  However, in the spirit of the
mean-field approach, the effect of an individual atom on the field
amplitude is negligible with respect to the summed effect of all
the others. In the limit of many atoms with small single-atom coupling, the motion of a single atom is very well
described by an effective potential determined by the many-body ensemble.

For the example of self-organization in a standing-wave
cavity, see Sec.~\ref{sec:ColdSelforg}, the light potential along the cavity axis is given by
\begin{equation}
\label{eq:AdiabaticPotential} V(x) = U_2 \cos^2 (k x) + U_1 \cos
(kx)\, ,
\end{equation}
which is composed of the sum of a \(\lambda/2\)-periodic potential
stemming from the cavity field and a \(\lambda\)-periodic one
arising from the interference between cavity and pump fields. The
depths of these potentials are
\begin{subequations}
\begin{align}
\label{eq:prefactors}
U_2 &= N^2 \langle \cos(kx) \rangle^2 \, \hbar I_0\,U_0\\
U_1 &= 2 N  \langle \cos(kx) \rangle \, \hbar I_0 \, (\Delta_C - N
U_0 \langle \cos^2(kx) \rangle)\;,
\end{align}
\end{subequations}
where $I_0$ is a dimensionless single-atom scattering parameter,
$I_0 \propto \eta^2$. Equation \eqref{eq:rho_mf} has to be solved in a self-consistent manner by iteration. In
\textcite{Asboth2005Selforganization}, the threshold has been analytically determined to be
\begin{equation}
\label{eq:threshold_selforg} \eta_\mathrm{eff,c}^2 = {\frac{k_B
T}{\hbar}} {\frac{(\kappa^2 + \delta_C^2) }{N \abs{\delta_C}}} \;.
\end{equation}

\subsubsection{Stability analysis and phase diagram}
\label{sec:Vlasov}

In the following we will account explicitly for the effect of the velocity distribution on the dynamics and on the instability
threshold. A mean-field model based on the Vlasov-equation for the phase space distribution $f(x,v,t)$ has been derived from a microscopic theory for the infinite system size by \textcite{Griesser2010Vlasov}. For one-dimensional motion along the cavity axis, the dynamical equation reads
\begin{equation}
\label{eq:VlasovEq}
\frac{\partial f}{\partial t}+ v\frac{\partial f}{\partial x}-\partial_x\phi(x,\alpha)\frac{\partial f}{\partial v}=0\; ,
\end{equation}
where $\phi(x,\alpha)$ is the potential corresponding to a momentary field amplitude $\alpha$.
For the generic example of a laser-driven cold atomic cloud in a single-mode standing-wave resonator with mode function $\cos(kx)$, the potential is
\begin{equation}
\label{eq:VlasovPotential} \phi(x,\alpha)=\frac{2\hbar}{m}
\left(\frac{U_0}{4}|\alpha|^2\cos(2kx)+\eta_\sub{eff}
\mathrm{Re}\left(\alpha\right)\cos(kx)\right)\; .
\end{equation}
Similarly to the adiabatic potential,
\eref{eq:AdiabaticPotential}, there is a $\lambda/2$ and a
$\lambda$-periodic term, this latter originates from interference
between the transverse pump laser and the intracavity field.
However, in this more general approach the cavity field amplitude
is kept dynamical obeying the self-consistent equation
\begin{multline}
\label{eq:FieldWithVlasov}
\dot\alpha=(-\kappa+i\delta_C)\alpha + \eta \\
-i\alpha\frac{NU_0}{2}\int_{-\infty}^\infty
dv\int_0^\lambda \cos(2kx) f(x,v,t)dx\\
- i N \eta_\sub{eff} \int_{-\infty}^\infty dv\int_0^\lambda \cos(kx)
f(x,v,t)dx  \; ,
\end{multline}
where $\delta_C=\DC-NU_0/2$. This approach based on the
Vlasov-equation is well suited to study the mean-field dynamics at
short times in order to test the stability of stationary states.
Over longer time scales, statistical fluctuations have to be taken
into account in the framework of a Fokker--Planck equation for the
velocity distribution, which is presented in
Sec.~\ref{sec:FokkerPlanck}.

\begin{figure}[t!]
\centering
\includegraphics[width=0.98\columnwidth]{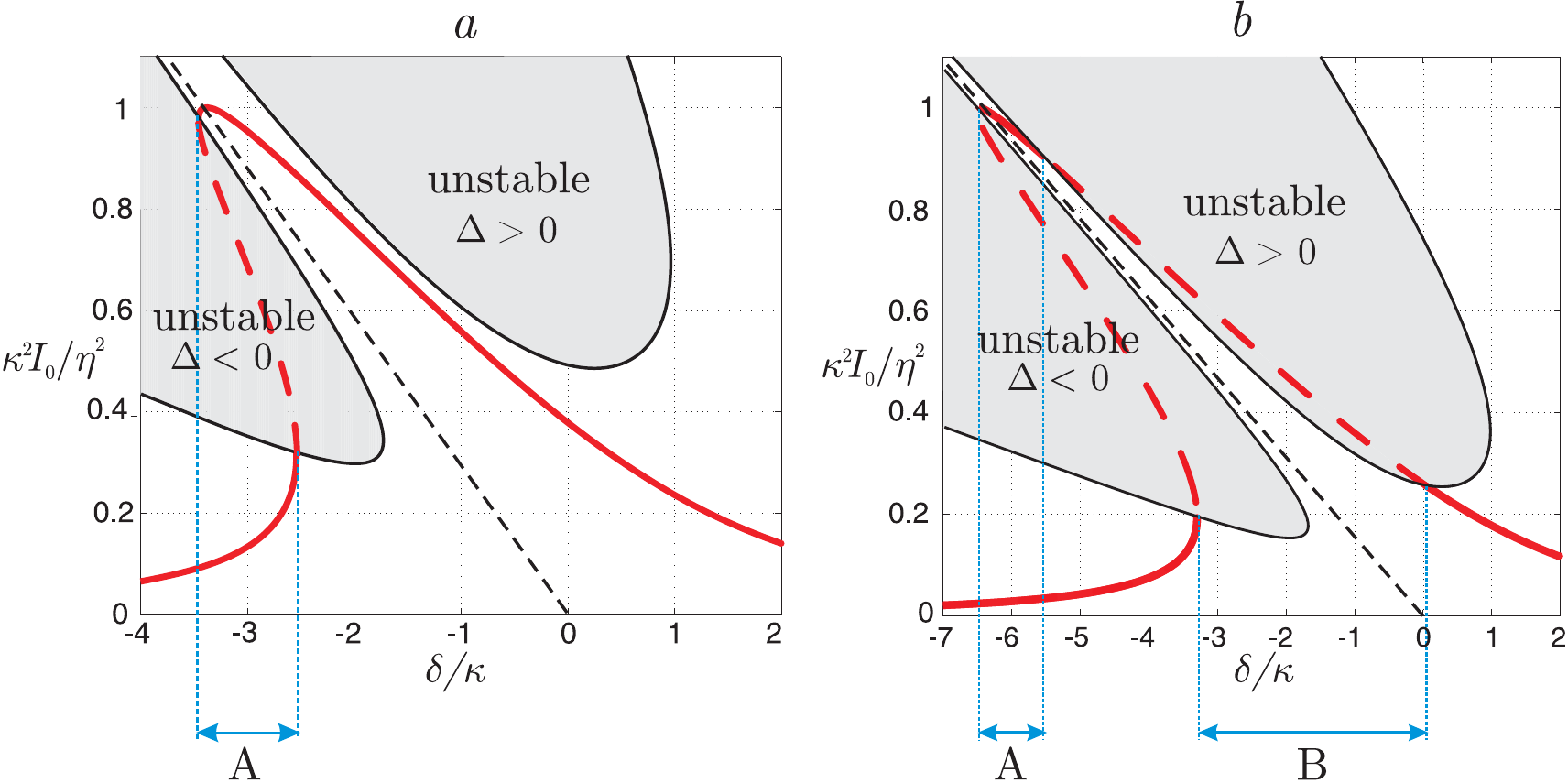}
\caption{(Color online) Normalized solutions for the steady-state
photon number (red) $I_0 = |\alpha|^2$ versus effective cavity
detuning $\delta = \delta_C-N U_0$ for a thermal gas in a driven
standing-wave cavity. The driving strength is $\eta=13\kappa$ (a)
and $\eta=18\kappa$ (b). Those parts of the response curve that
lie inside the instability region (shaded area) are depicted in
dashed and correspond to linearly unstable steady states. The
intervals designated A correspond to bistability, the interval
designated B supports no stable steady state at all. The
parameters are $N=10^5$, $U_0 = 0.04 \kappa$, $\eta = 18 \kappa$,
$\kappa = 2000 \omega_R$ and $k_B T= \hbar \kappa$. From
\textcite{Griesser2011Nonlinear}.} \label{fig:stability}
\end{figure}

\paragraph{Nonlinear response of a cold atomic cloud in a driven Fabry-P\'erot cavity}
For cavity pumping only $(\eta_\sub{eff}=0,\eta \neq 0)$,  the self-consistent
steady-state solution exhibits a strong nonlinear optical
response. Underlying the nonlinearity,  the particle distribution
and thus the effective refractive index of the cloud depends on
the cavity pump intensity.  Above a sufficient pump strength,
multiple stationary solutions appear, reminiscent of optical
bistability.One can perform a
systematic stability analysis of these solutions by studying the dynamics of small fluctuations
of the field and the particle distribution \cite{Griesser2011Nonlinear}. As shown in
Fig.~\ref{fig:stability}, the stability analysis reveals regions
of bistability as well as parameter ranges where no stable
solutions exist.

In the parameter regions of instability, the  numerical solution of the dynamical Vlasov equation reveals a limit cycle behavior
with subsequent appearance of higher frequencies than the fundamental cycle \cite{Griesser2011Nonlinear}.
In the quantum regime essentially the same behavior is retrieved with the recoil frequency determining the
oscillation frequencies $\nu \approx 4\omega_R$ \cite{Ritter2009Dynamical} (see
Sec.~\ref{sec:Optomechanics}).

\paragraph{Self-organization of a laser-driven cloud of atoms}
For a purely transverse pump geometry $(\eta=0, \eta_\sub{eff}\neq
0)$, one can systematically recalculate the critical pump
amplitude $\eta_\mathrm{eff,c}$  that marks the transition from
the stable regime to the unstable one, where small fluctuations
are amplified and grow exponentially.

The Vlasov equation \eref{eq:VlasovEq} together with the equation
for the coherent cavity field amplitude ${\alpha}$ possesses an
infinite number of stationary solutions with a spatially
homogeneous density distribution and zero cavity field but
different velocity distribution, which, however, are not
necessarily stable against fluctuations. Indeed, any symmetric
velocity distribution $g(v/\vT)=L\vT f(v)$ for $\delta_C<0$ is
stable only if
\begin{equation}\label{eq_critical_value}
\frac{N|\eta_\sub{eff}|^2}{\kB
T}\vp\int_{-\infty}^\infty\frac{g^\prime(\xi)}{-2\xi}
d\xi<\frac{\delta_C^2+\kappa^2}{\hbar|\delta_C|}\;,
\end{equation}
where $\vp$ denotes the Cauchy principal value. Here we have
defined the thermal velocity $\vT^2=2\kB T/m$; $L$ denotes the
cavity length. For a Gaussian distribution the integral evaluates
to one, and the condition is equivalent to
\eref{eq:threshold_selforg}.

\begin{figure}[t!]
\centering
$\delta_C=-\kappa$
  \includegraphics[width=0.98\columnwidth]{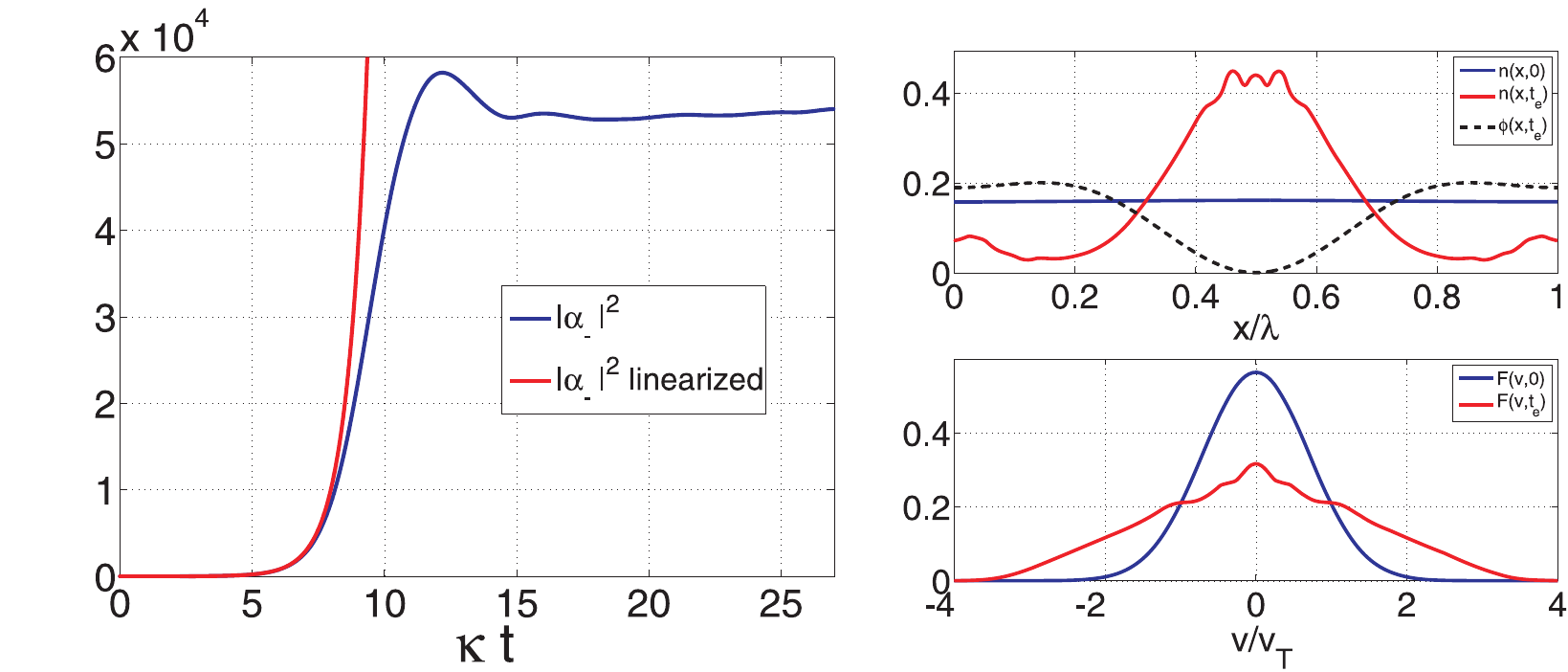}\\[0.5\baselineskip]
$\delta_C=\kappa$
  \includegraphics[width=0.98\columnwidth]{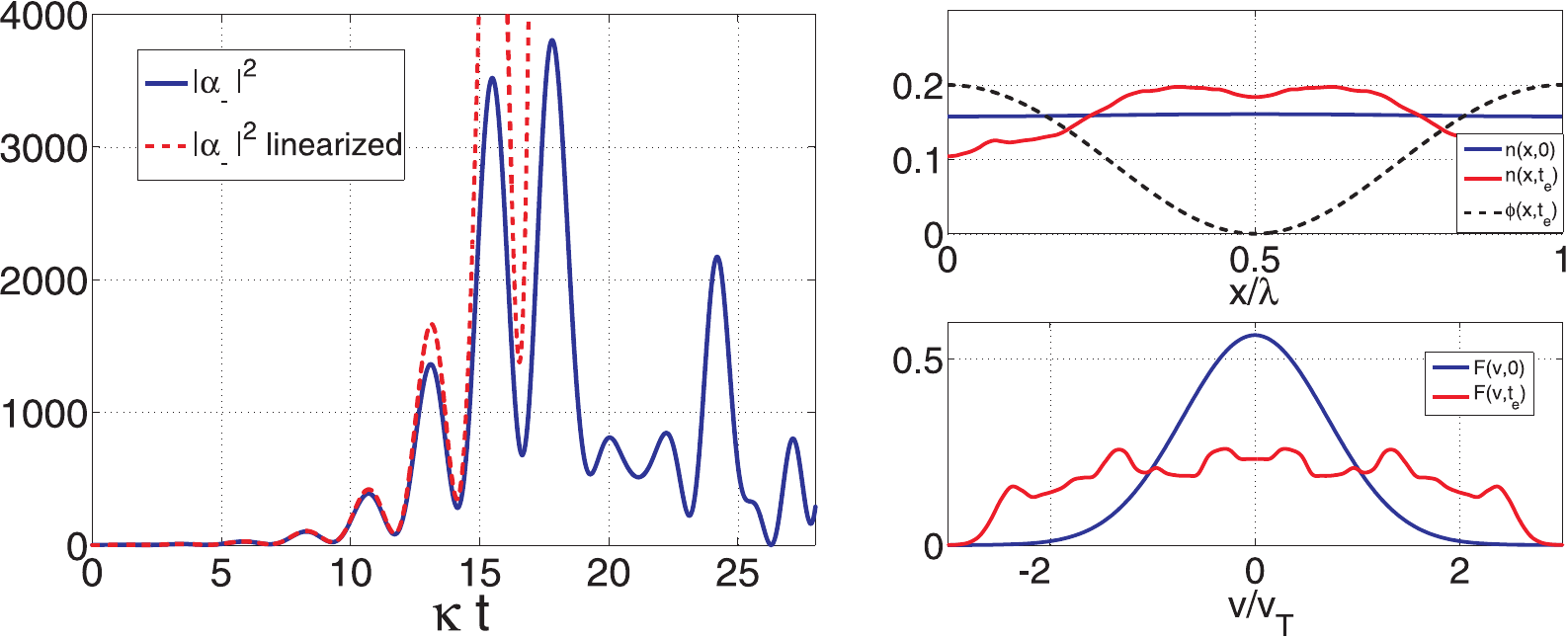}
\caption{(Color online) Laser illuminated cold gas in a ringcavity: time evolution of the intra-cavity field
intensity (left), and the instantaneous spatial $n(x)$ (upper
right) and velocity $F(v)$ (lower right) distributions of the
particles along the cavity axis at times $t =0$ and $t_e = 28/\kappa$ for $\delta_C =
-\kappa$ (upper row) and $\delta_C = \kappa$ (bottom row). For
$\delta_C<0$, after a transient exponential growth, the field
intensity saturates accompanied by the trapping of atoms in the
self-organized pattern. By contrast, for $\delta_C>0$ the
transient exponential growth is followed by oscillations.
Parameters are $N=10^4$, $U_0=-\kappa/N$,
$\eta_\mathrm{eff}=0.05\kappa$, $kv_T=\kappa$, and $v_R=v_T/5$.
From \textcite{Griesser2010Vlasov}.} \label{Kombi3}
\end{figure}
Figure~\ref{Kombi3} shows the results of a numerical simulation
of Eqs.~(\ref{eq:VlasovEq}), (\ref{eq:VlasovPotential}) and
(\ref{eq:FieldWithVlasov}), initialized with a perturbed gaussian
distribution
\begin{equation}
 \label{pert}
f(x,v,0)=\frac{1}{\lambda\sqrt{\pi}v_T}e^{-v^2/v_T^2}(1-\epsilon\cos(kx))\;,
\end{equation}
with $\epsilon\ll 1$, for the case of a transversely pumped ring cavity, where the light can be scattered into a superposition of two resonant cavity modes. Apart from possessing continuous translational symmetry, the dynamics is qualitatively very similar to the single mode case \cite{Griesser2010Vlasov} shown in Fig.~\ref{fig:slowselforg} of the previous section.  One clearly recognizes the striking difference in the dynamical behavior for positive and negative values of $\delta_C$. While we have an instability in both cases, self-organization is only found for $\delta_C<0$.

\paragraph{Sympathetic self-ordering and cooling}
\label{sub_sec_cold_vlasov_symp}

Self-organization and collective coherent light scattering into a
high-finesse cavity in principle allows for trapping and cooling
of any kind of polarizable particles. In practice, however, the
required phase-space densities and laser intensities to initiate
the ordering process are hard to achieve for atomic species,
molecules or nanoparticles which cannot be efficiently optically
precooled \cite{Deachapunya2008Slow,Lev2008Prospects}. As an
alternative approach, the self-organization threshold can be
achieved  by inserting different species simultaneously into an
optical resonator.

\begin{figure}[t!]
\includegraphics[width=8.5cm]{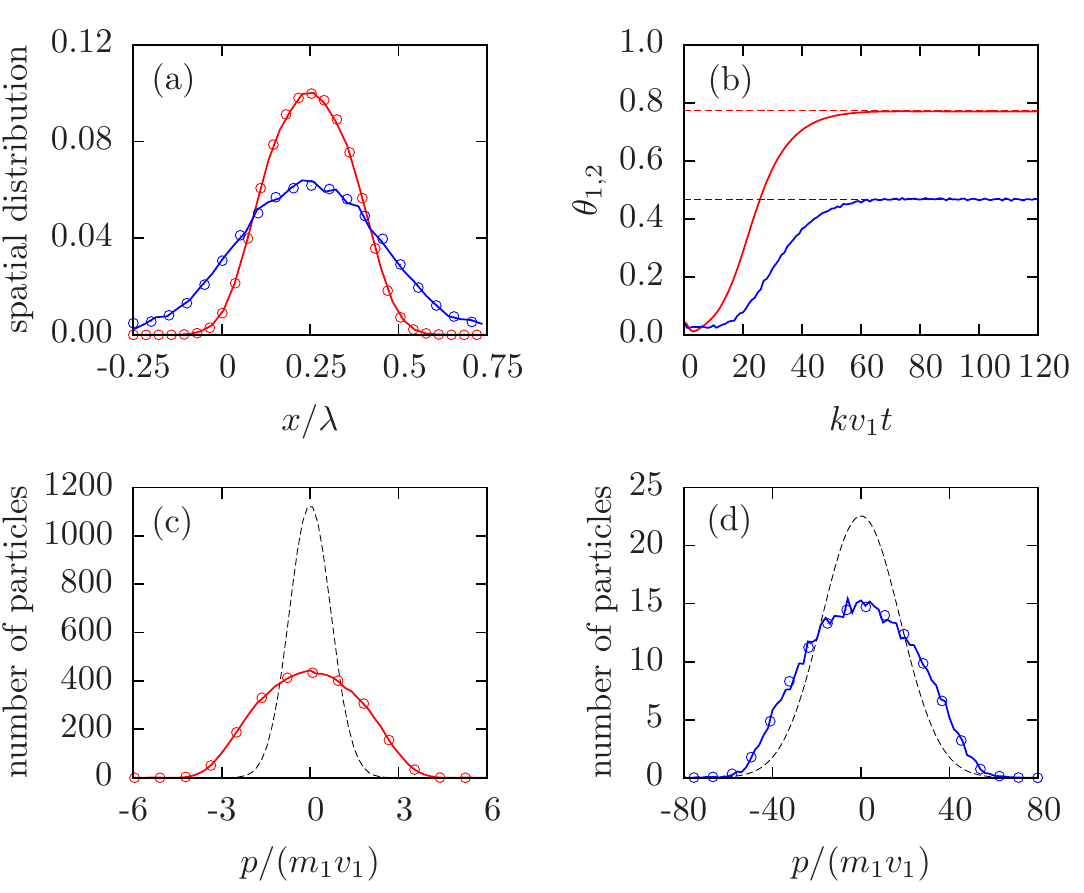}
\caption{(Color online) Simultaneous self-organization of two
species. The system is started from a perturbed uniform state above
the instability threshold, \eref{threshold}, in such a way that
species one (two) itself would be pumped six times above (far
below) the critical point. Figure (a) shows the position
distributions in the final state, (c) and (d) the momentum
distributions initially (dashed lines) and after self-organization
(solid line).  Solid lines depict the results of stochastic
trajectory simulations for ensembles of particles as in
Eq.(\ref{eq:sde_pz}), while open circles show the predictions of
the corresponding Vlasov model. Figure (b) shows the time
evolution of the two order parameters $\theta_{1(2)} $ approaching
theoretical steady-state values. Parameters are $N_1=10^4$,
$N_2=500$, $m_2=10m_1$, $\kB T_1=10^4\hbar\kappa$, $\kB
T_2=2.5\times 10^5\hbar\kappa$, $\eta_1=2.4\kappa$,
$\eta_2=27.4\kappa$ and $\omrec=10^{-2}\kappa$. From
\textcite{Griesser2011Selforganisation}.}
\label{fig:TwoSpeciesSelforg}
\end{figure}

The Vlasov-type model can be generalized to a dilute gas of
various kinds of $N_s$ polarizable point particles of mass $m_s$
illuminated by a single transverse standing-wave laser field.  For
the threshold value, a condition analogous to
\eref{eq_critical_value} can be obtained
\cite{Griesser2011Selforganisation}. The homogeneous distribution
is unstable if and only if
\eq{\label{threshold}\sum_{s=1}^S\frac{\hbar\, N_s\,
\eta_{s}^2}{\kB T_s}\left(\mathrm{vp}\int_{-\infty}^\infty
\frac{g_s'(u)}{-2 u}
du\right)>\frac{\kappa^2+\delta_C^2}{|\delta_C|},} where $\kB T_s
= m_sv_s^2/2$. Note that the right-hand side of \eref{threshold}
depends only on cavity parameters, and all terms in the sum on the
left-hand side are positive and proportional to the pump
intensity. This guarantees that inserting any additional species
into the cavity always increases the total light scattering rate
and thus lowers the minimum power needed to start the
self-organization process, regardless of temperature and
polarizability or density of the additional particles. Moreover,
the different species can be located at different regions within
the cavity. Assisted self-organization of a species which, alone,
would be pumped below threshold is shown in
Fig.~\ref{fig:TwoSpeciesSelforg}.

\begin{figure}[t!]
\includegraphics[width=8cm]{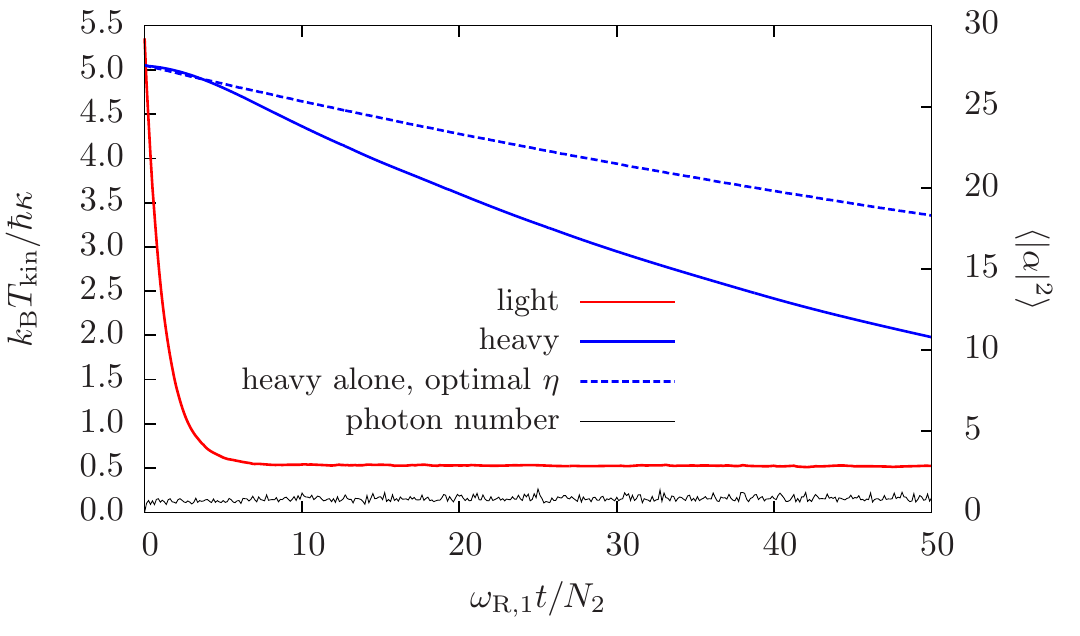}
\caption{(Color online) Sympathetic cavity cooling.  Time
evolution of the kinetic temperatures of a heavy and a light
species.  The blue dashed line represents the heavy particle alone
and the blue solid line the enhanced cooling in the presence of a
lighter species below self-organization threshold. Parameters are
$m_2=200m_1$, $N_1=200$, $N_2=200$, $\sqrt{N_1}\eta_1=134\omrec$,
$\sqrt{N_2}\eta_2=134\omrec$, $\kappa=200 \omrec$ and
$\delta_C=-\kappa$. From \textcite{Griesser2011Selforganisation}.}
\label{fig:Sympathetic}
\end{figure}
Below the self-organization threshold  cooling occurs {thereby}
equalizing the stationary momentum distributions for all species.
Fig.~\ref{fig:Sympathetic} exhibits the enhanced decay of the
kinetic energy of the heavy particles in presence of cavity field
and a cold species. Although the distributions get independent in
stationary equilibrium, the cooling process itself involves energy
exchange between different  species.  Thus if any of the species
is cold or can be cooled by different means, the other components
are sympathetically cooled in parallel.

\subsubsection{Non-equilibrium steady-state distributions}
\label{sec:FokkerPlanck}

Over longer time scales, diffusion has to be accounted
for in terms of a nonlinear Fokker--Planck equation for the
statistically averaged velocity distribution. This allows for the
calculations of cooling time scales and the unique steady-state
distribution.

\paragraph{Transverse pump configuration below threshold}
Below the instability threshold, \eref{eq_critical_value}, the
mean spatial distribution $f$ is homogeneous, i.e., independent of
$x$. The statistical fluctuations of the potential and the actual
atomic distribution gets important.  For the spatially averaged
distribution, a lengthy calculation \cite{Niedenzu2011Kinetic}
leads to a nonlinear Fokker--Planck equation for the velocity
distribution ${F(v,t)}=\overline{f(x,v,t)}$,
\begin{equation}\label{eq_nonlin_fpe}
\frac{\partial}{\partial t}{F}+\frac{\partial}{\partial v}\Big(A[{F}] {F}\Big)=\frac{\partial}{\partial v}\left(B[ {F}]\frac{\partial}{\partial v} {F}\right)\, ,
\end{equation}
with coefficients
\begin{subequations}
\begin{align}
A[F]& =\frac{2\hbar k\delta_C\kappa \eta_\mathrm{eff}^2}{m}\frac{kv}{|D(ikv)|^2}\\
B[F]& =\frac{\hbar^2k^2\eta_\mathrm{eff}^2\kappa}{2m^2}\frac{\kappa^2+\delta_C^2+k^2v^2}{|D(ikv)|^2}\,.
\end{align}
\end{subequations}
These functionals depend on $\langle F\rangle$ via the dispersion
relation
\begin{multline}\label{eq_dispersion}
D(s)=(s+\kappa)^2+\delta_C^2-\\-i\hbar
k\delta_C\frac{NL\eta_\mathrm{eff}^2}{2m}\int_{-\infty}^\infty
dv\left(\frac{F^\prime(v)}{s+ikv}-\frac{F^\prime(v)}{s-ikv}\right)\,
,
\end{multline}
which encodes all cavity-mediated long-range particle
interactions. Far below threshold the dispersion relation reduces
to $D(ikv)\simeq(ikv+\kappa)^2+\delta_C^2$, which corresponds to
the case of independent particles.

Steady-state solutions of \eref{eq_nonlin_fpe} exist only for
negative detuning $\delta_C<0$, where light scattering is
accompanied by kinetic energy extraction from atomic motion. Below
threshold one obtains non-thermal $q$-Gaussian velocity
distribution functions \cite{deSouza1997Students}:
\begin{equation}\label{eq_qgauss}
{F(v)}\propto\left(1-(1-q)\frac{mv^2}{2\kB T}\right)^\frac{1}{1-q}\; ,
\end{equation}
with $q = 1+\omrec/|\delta_C|$ and the effective temperature
\begin{equation}\label{eq_temperature}
\kB T=\hbar\frac{\kappa^2+\delta_C^2}{4|\delta_C|}\geq
\frac{\hbar\kappa}{2}\; .
\end{equation}
The minimum temperature is reached for $\delta_C=-\kappa$. The
magnitude of the detuning $|\delta_C|/\omrec$ determines the shape
of the distribution. For $|\delta_C|=\omrec$, it is a Lorentzian
distribution,  whereas for $|\delta_C|/\omrec\rightarrow\infty$,
i.e.\ $q\rightarrow 1$, it converges to a Gaussian distribution
with kinetic temperature $\kB \Tkin = m\ew{v^2}$ .

Inserting the steady-state $q$-Gaussian~\eqref{eq_qgauss}
distribution into the threshold
condition~\eqref{eq_critical_value} gives a self-consistent
stability criterion. As a result, the homogeneous distribution is
stable only if
\begin{equation}\label{eq_selfconsistent_threshold}
\sqrt{N}\eta_\sub{eff} \leq \kappa\sqrt{\frac{2}{3-q}}\; ,
\end{equation}
where the equality is reached for $\delta_C=-\kappa$, i.e., for optimum cavity cooling given by \eref{eq:beta}.  The
stability criterion can be rewritten in the intuitive form
\begin{equation}
N|U_0|V_p \leq \kappa^2,
\end{equation}
where $V_p = \Omega^2/\Delta_A$ is the optical potential depth
created by the pump laser, and $N U_0$ is the total dispersive
shift of the cavity resonance. Note that even if the initial
temperature is too high for the homogeneous distribution to be
unstable, cavity cooling induced self-organization is possible.

\paragraph{Transverse pump configuration above threshold}
Above the self-organization threshold, the inhomogeneous spatial
distribution can still be derived from  a Fokker--Planck equation
similar to \eref{eq_nonlin_fpe} by using action-angle variables
\cite{Luciani1987Kinetic,Chavanis2007Kinetic}. In the limit of
deep trapping and a harmonic approximation for the potential, the steady state
is a thermal distribution with a temperature depending both on the
effective trap frequency $\omega_0$  and on the cavity linewidth
$\kappa$,
\begin{equation}\label{eq_temperature_so}
\kB T = \hbar\frac{\kappa^2+\delta_C^2+4\omega_0^2}{4|\delta_C|} \stackrel{\delta_C=-\omega_0}{\approx} \hbar \omega_0\;.
\end{equation}
The effective trap frequency  $\omega_0$  can be approximated by
\begin{equation}
\omega_0^2\simeq\sqrt{N}\eta_\mathrm{eff}\omrec\left(\frac{\eta_\mathrm{eff}}{\eta_\mathrm{eff,c}}+\sqrt{\frac{\eta_\mathrm{eff}^2}{\eta_\mathrm{eff,c}^2}-1}\right) \; ,
\end{equation}
which is valid in the regime $|\delta_C|\gg\omrec$ with $\eta_\mathrm{eff,c}$
given by the equality of \eref{eq_selfconsistent_threshold}. As the
temperature depends explicitly on the pump strength, a stronger
pump laser beam results in more confined particles with increased kinetic
energy.  The system has the interesting property that the more
particles we add, the deeper the optical potential gets, which
shows analogy to self-gravitating
systems \cite{Posch2005Stellar}.

\begin{figure}
  \centering
  \includegraphics[width=\columnwidth]{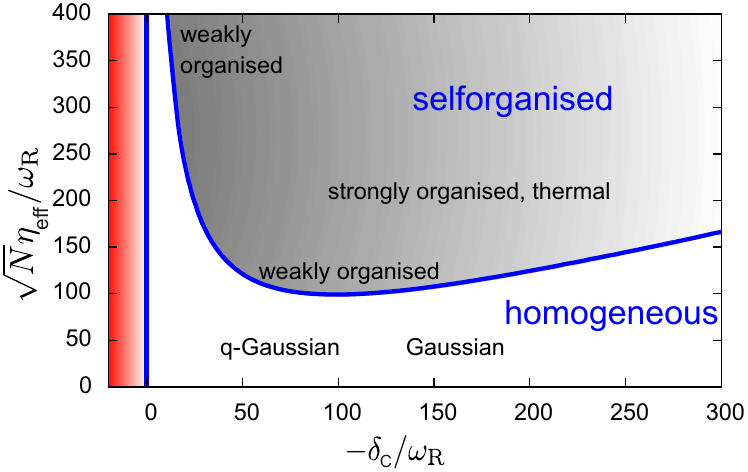}
  \caption{(Color online) Schematic view of the phase diagram in the weak-coupling limit
($N|U_0|\ll\kappa$) for $\kappa=100\omrec$. Equilibrium solutions
exist only for $\delta_C<-\omrec/2$, a Lorentzian steady-state
velocity distribution is realized for the case
$|\delta_C|=\omrec$. For large negative values of the detuning
$\delta_C$, strongly organized equilibrium solutions exist already
for pump strengths slightly above the critical value. Adapted from
\textcite{Niedenzu2011Kinetic}
  }\label{fig_phasediagram}
\end{figure}

Finally, putting all this together one obtains the self-consistent
phase diagram for self-organization which accounts for cavity cooling, see Fig.~\ref{fig_phasediagram}.

\newpage
\section{QUANTUM GASES IN OPTICAL CAVITIES}
\label{sec:QuantumGas}

Quantum gases are considered as ideal model systems to study
quantum many-body phenomena under well-controlled experimental
conditions. The possibilities which arise from loading ultracold
atomic ensembles of different particle statistics into various
optical potential landscapes and to tune the strength of the
contact atom-atom interaction, make these system well suited for
quantum information and simulation research \cite{Bloch2008Manybody}. The merger of the
field of ultracold gases with that of cavity QED provides a set of
additional possibilities. Cavity-mediated atom-atom interactions can be tailored by
choosing different resonator and pump geometries and give rise to novel quantum phases. Closely related, the atomic
back-action upon the cavity-generated lattice potentials can be
significant, which paves the way to study phonon or soft-condensed
matter physics with ultracold gases \cite{Lewenstein2007Ultracold}. Further, coherent scattering into the cavity field can be used for non-destructive and real-time probing of different many-body phases.

The coupling between a Bose-Einstein condensate and an optical cavity is conceptually fundamental since a single mode of a matter wave
field interacts with a single mode of the light field: as all
atoms occupy the same motional quantum state they couple
identically to the optical cavity field. This situation can
substantially reduce the number of degrees of freedom necessary to
describe the system. Therefore the experimental situation can
often be almost perfectly described by fundamental Hamiltonians of
matter-light interaction. These include the Tavis-Cummings or
Dicke model, as well as the generic model for
cavity optomechanics.

\subsection{Experimental realizations}
\label{sec:UltracoldExp} Experimentally, there have been different
approaches to realize and study Bose-Einstein condensates or
bosonic atomic ensembles close to quantum degeneracy in optical
high-finesse cavities
\cite{Slama2007Superradiant,Brennecke2007Cavity,Colombe2007Strong,Gupta2007Cavity,Purdy2010Tunable}.
So far, all groups used $^{87}$Rb atoms. In the T\"ubingen group
\cite{Slama2007Superradiant} a Bose-Einstein condensate was loaded
for the first time into a ring cavity with large mode volume using
magnetic trapping and transport. This experiment extended prior
work on the collective atomic recoil laser (see
Sect.~\ref{sec:CARL}) with laser-cooled atoms
\cite{Kruse2003Observation,Cube2004SelfSynchronization} into the
ultracold regime.

\begin{figure}[htbp]
\begin{center}
\includegraphics[width=\columnwidth]{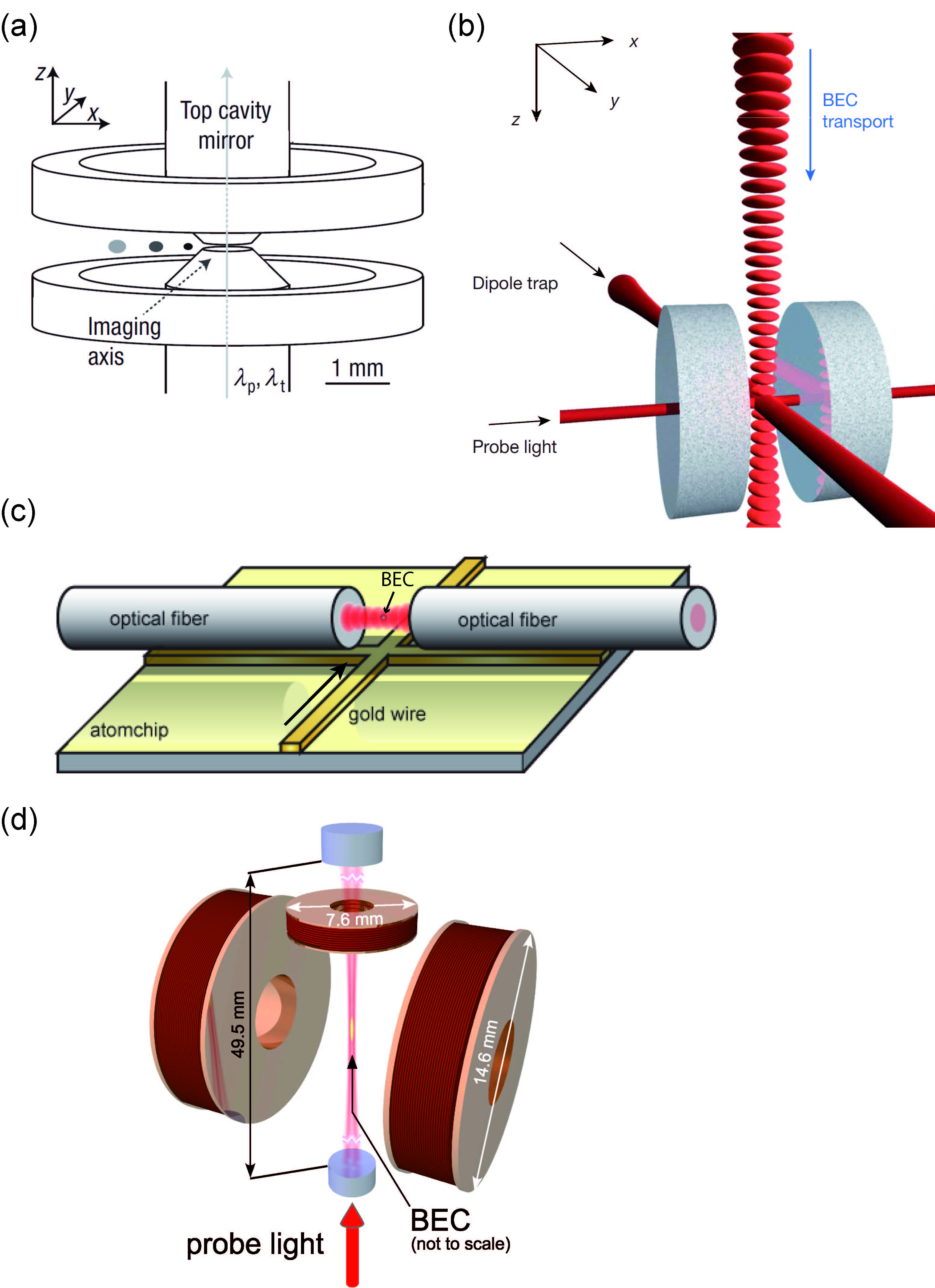}
\caption{(Color online) Different experimental schemes for
preparing ultracold atoms and Bose-Einstein condensates inside
high-finesse optical Fabry-P\'erot resonators. (a) Ultracold atoms
are prepared in a magnetic trap, formed using electromagnets
coaxial with the vertically oriented high-finesse cavity (length =
$\unit[194]{\mu m}$) and delivered along the $x$-axis towards the
cavity center. Once overlapping with the cavity mode, the atoms
are loaded into a deep intracavity lattice potential provided by a
far-detuned cavity pump field. Adapted from
\textcite{Murch2008Observation}. (b) Ultracold atoms, prepared in
a magnetic trap placed above the optical resonator, are loaded
into a vertically oriented optical lattice potential and
transported into the cavity by controlled frequency chirping the
counter-propagating laser beams. Once in the cavity (length  =
$\unit[176]{\mu m}$), the atoms are loaded into a crossed-beam
harmonic dipole trap where Bose-Einstein condensation is achieved.
From \textcite{Brennecke2007Cavity}. (c) A Bose-Einstein
condensate is prepared in an atom-chip-based magnetic trap and
positioned afterwards with subwavelength precision in the mode of
a fibre-based Farby-P\'erot cavity which has a length of
$\unit[39]{\mu m}$. Figure courtesy of J.~Reichel. (d) A magnetic
QUIC trap is used to prepare and to transfer a Bose-Einstein
condensate into the field of a vertically oriented Fabry-P\'erot
resonator with a length of $\unit[5]{cm}$. Figure courtesy of
A.~Hemmerich.} \label{fig:expschemes}
\end{center}
\end{figure}
Loading ultracold quantum gases or Bose-Einstein condensates into
ultra-high finesse optical cavities of \emph{small} mode volume,
which operate in the single-atom strong coupling regime of cavity
QED, has been achieved by applying different concepts. The
Berkeley group \cite{Gupta2007Cavity} prepared an ultracold gas of
up to $10^5$ atoms within an ultra-high finesse Fabry-P\'erot
resonator by loading it into a vertically oriented, deep
intracavity optical lattice potential, see
Fig.~\ref{fig:expschemes}a. A cavity with similar parameters was
used in the approach of the Z\"urich group
\cite{Brennecke2007Cavity}. Here, Bose-Einstein condensates of
typically $2 \times 10^5$ atoms were transported into the mode
volume of the optical cavity using an optical elevator formed by
two counter-propagating laser beams with controlled frequency
difference, see Fig.~\ref{fig:expschemes}b. The Paris group
\cite{Colombe2007Strong} used an atomic chip to produce
Bose-Einstein condensates of up to 3000 atoms and control its
position on a sub-wavelength scales within a novel type of
fibre-based Fabry-P\'erot cavity with high mirror curvature and
reduced mode volume, see Fig.~\ref{fig:expschemes}c. Also the
Berkeley group \cite{Purdy2010Tunable} achieved sub-wavelength
positioning of Bose-Einstein condensates of a few thousands of
atoms inside a conventional small-volume high-finesse optical
cavity using an atomic chip.

A novel BEC-cavity system operating in an interesting and so far
unexplored parameter regime was presented recently by the Hamburg
group \cite{Wolke2012Cavity}. Here, Bose-Einstein condensates of
typical $2 \times 10^5$ atoms are prepared and transferred
magnetically into the field of a $\unit[5]{cm}$ long
near-concentric Fabry-P\'erot resonator resulting in a large
single-atom cooperative $\mathcal{C} \gg 1$
 and a very narrow cavity bandwidth on the order of the recoil frequency
$\omega_R$, see Fig.~\ref{fig:expschemes}d.

\begin{figure}[t!]
\begin{center}
\includegraphics[width=\columnwidth]{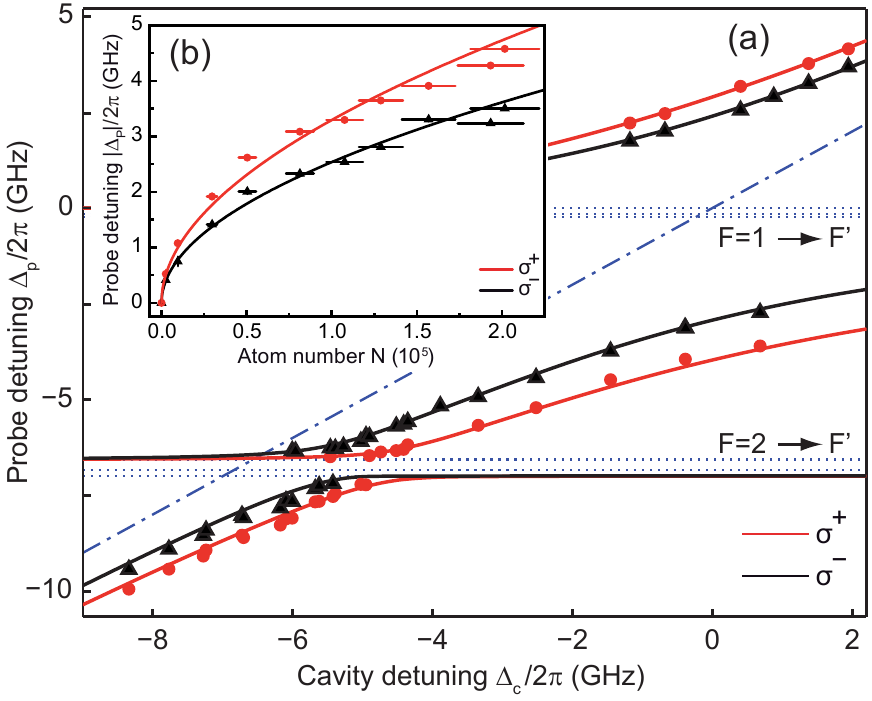}
\caption{(Color online) Collective vacuum Rabi-splitting of a
coupled BEC-cavity system. The displayed data was obtained by
cavity transmission spectroscopy using a weak probe laser beam
\cite{Brennecke2007Cavity}. The detuning of the probe beam with
respect to the frequency $\omega_{A}$ of the $|F=1\rangle
\rightarrow |F^\prime=1\rangle$ transition of the D$_2$ line of
$^{87}$Rb is denoted by  $\Delta_p$. Two orthogonal circular
polarizations of the transmitted light were recorded and are
displayed as red circles ($\sigma^+$) and black triangles
($\sigma^-$). (a) Position of the probed resonances as a function
of the detuning $\Delta_{c}$ between the empty cavity resonance
and the atomic transition frequency $\omega_{A}$ for $2.2 \times
10^5$ atoms. Bare atomic resonances are shown as dotted lines,
whereas the empty cavity resonance of the $\mathrm{TEM}_{00}$ is
plotted as a dashed-dotted line. The solid lines are the result of
a theoretical model including the influence of higher-order cavity
modes. (b) Shift of the lower resonance of the coupled BEC-cavity
system from the bare atomic resonance as a function of atom number
for $\Delta_{c}=0$. The solid lines are fits of the square-root
dependence on the atom number $N$. Adapted from
\textcite{Brennecke2007Cavity}.}
\label{fig:EnergySpectrumBrennecke}
\end{center}
\end{figure}

The lowest electronic excitation spectrum of degenerate and
non-degenerate atomic samples strongly coupled to the cavity field
has been studied by
\textcite{Brennecke2007Cavity,Colombe2007Strong}. The presence of
$N$ atoms which collectively couple to the cavity field results in
an enhanced collective coupling which scales as $\sqrt{N}$. A
correspondingly large vacuum Rabi-splitting was measured in the
experiments \cite{Brennecke2007Cavity,Colombe2007Strong}. The
energy spectrum of a coupled BEC-cavity system together with the
square-root dependence of the energy splitting on the atom number
are shown in Fig.~\ref{fig:EnergySpectrumBrennecke}.

The electronic excitation spectrum is sensitive to the effective
number of atoms coupled maximally to the cavity mode, i.e.~it
depends on the density distribution of the atoms integrated over
the cavity mode profile. However, the above described measurements
of the electronic excitation spectrum have an energy resolution
given by the excited state and cavity lifetimes, which is too
large to probe the low-energy excitations of the external degree
of freedom of a Bose-Einstein condensate. Probing quantum
statistics and quantum correlations in atomic many-body states
using the dispersive coupling to far-detuned laser and cavity
fields is discussed in Sect.~\ref{sec:TrappedUltracoldQND}.

\subsection{Theoretical description}
\label{sec:ManyBodyTheory} This section provides the theoretical
basis for a quantum many-body description of a coupled and
laser-driven BEC-cavity system at zero temperature. For
simplicity, we consider two different pump laser fields with equal
frequency $\omega$ which propagate along and transverse to the
axis of a Fabry-P\'erot resonator. A similar many-body description
for the case of a BEC in a driven ring-cavity has been presented
e.g.~in \textcite{Moore1999Quantum}. The Hamiltonian is composed
of an atomic, a cavity and an atom-field interaction part
\begin{equation}\label{eq:ManyBodyH}
H=H_A+H_C+H_{AC}\;.
\end{equation}
In the following, we will assume a sufficiently large detuning of
the cavity frequency $\omega_C$ and the pump laser frequency
$\omega$ from the atomic transition frequency $\omega_A$, so that
the atom-field interaction is of purely dispersive nature, see
Sec.~\ref{sec:DispersiveLimit}. In this case all excited states
can be adiabatically eliminated and the atom resides most of the
time in its electronic ground state. Correspondingly, the motional
degree of freedom is captured by a scalar matter-wave field
operator $\Psi(\vecr)$.

The atomic many-body Hamiltonian is given by
\begin{equation}
\label{eq:H_A} H_A=\!\int d^3\vecr \, \Psi^\dagger(\vecr)\Big[
{\cal H}^{(1)} +
\frac{u}{2}\Psi^\dagger(\vecr)\Psi(\vecr)\Big]\Psi(\vecr) \; ,
\end{equation}
where $u = 4 \pi \hbar^2 a_s/m$ denotes the strength of the
short-range s-wave collisions with scattering length $a_s$
\cite{Pitaevskii2003Bose-Einstein}. Here, the single-atom
Hamiltonian
\begin{equation}
\label{eq:H_A1} {\cal H}^{(1)} = \frac{\vecp^2}{2m} +
V_\mathrm{cl} (\vecr)\; ,
\end{equation}
includes an external trapping potential $V_\mathrm{cl}(\vecr)$
which also incorporates the potential caused by the transverse
pump laser field.

The dynamics of a single, coherently laser-driven cavity mode with
mode function $\cos(kx)$ and resonance frequency $\omega_C$ is
described in a frame rotating at the pump laser frequency $\omega$
by the Hamiltonian
\begin{equation}
\label{eq:H_C} H_C = - \hbar\Delta_C\, a^\dagger a  + i \hbar
\eta\, (a^\dagger - a) \; .
\end{equation}
As before, the detuning between the pump laser frequency and the
cavity resonance frequency is denoted by $\Delta_C = \omega -
\omega_C$. The generalization to multimode cavities will be
discussed in Sec.~\ref{sec:MultimodeDegenerateCavity}.

The dispersive interaction between the pump and cavity radiation
fields and the atoms reads (in the frame rotating at $\omega$)
\begin{multline}\label{eq:H_AC}
H_{AC} = \int d^3\vecr \,\Psi^\dagger(\vecr)\Big[ \hbar U_0
\cos^2(k x) \, a^\dagger a \\ +  \hbar \eta_\sub{eff}\, \cos(k x)
\cos(k z)\, \left(a^\dagger +  a\right)   \Big] \Psi(\vecr)\;.
\end{multline}
The first term arises from the absorption and stimulated emission
of cavity photons, with $U_0 = \frac{g^2}{\Delta_A}$ denoting the
maximum atomic light-shift for a single intracavity photon. As
before, $g$ denotes the maximum atom-cavity coupling strength and
$\Delta_A = \omega -\omega_A$ the pump-atom detuning. The second
term corresponds to the coherent redistribution of photons between
the standing-wave transverse pump laser (with mode function
$\cos(kz)$) and the cavity field. The maximum scattering rate for
a single atom is given by the two-photon (vacuum) Rabi frequency
$\etaeff = \frac{\Omega g}{\Delta_A}$, where $\Omega$ is the Rabi
frequency of the transverse pump laser. Both interaction terms in
Eq.~\eqref{eq:H_AC} can be viewed as a four-wave mixing of light
and matter wave fields \cite{Rolston2002Nonlinear}.

The system is subject to dissipation due to photon leakage through
the cavity mirrors. The corresponding irreversible evolution can
be modeled by the Liouville terms in the master equation,
\eref{eq:L_cav}, or, equivalently, by a Heisenberg-Langevin
equation \cite{Gardiner1991Quantum} for the cavity field operator
$a$,
\begin{equation}
\label{eq:HL_Equation} \frac{d}{dt}a = -i [a\,,H] -\kappa{}a +
\xi\,,
\end{equation}
with cavity field decay rate $\kappa$.\footnote{It is important to
note that the original problem is intrinsically time-dependent
because of the external laser driving, although this time
dependence has been formally eliminated by going into a rotating
frame. Nevertheless, the coupling to the reservoir occurs at the
high, optical frequency range and thus the simple form of the loss
description can be used irrespective of the low-frequency dynamics
imposed by the  effective Hamiltonian.} The Gaussian noise
operator $\xi$ maintains the commutation relation for the photon
operators in the presence of cavity decay. In the optical domain,
the temperature of the bath of electromagnetic field modes can be
set to zero. Correspondingly, $\xi$ has zero mean value and the
only non-vanishing correlation function reads
\begin{equation}
\label{eq:NoiseCorrelation} \langle \xi(t) \xi^\dagger(t')\rangle
= 2 \kappa \delta(t-t')\;,
\end{equation}
according to the fluctuation-dissipation theorem. Further possible dissipation channels
can, for example, act directly on the atomic cloud.

As the cavity field mediates a global coupling among all atoms, a
mean-field approach is well suited to solve the above set of
equations
\cite{Horak2000Coherent,Horak2001Dissipative,Nagy2008Selforganization}.
The mean-field description assumes the presence of a
macroscopically populated matter wave field $\varphi(\vecr) =
\langle \Psi(\vecr)\rangle$ (\emph{condensate wavefunction}) and a
coherent cavity field with amplitude $\alpha = \langle a \rangle$
which can be separated from the quantum fluctuations according to
\begin{subequations}
\label{eq:MeanField}
\begin{align}
 a &\rightarrow \alpha + \delta a\;,\\
 \Psi(\vecr) & \rightarrow \sqrt{N_c} \varphi(\vecr) + \delta
\Psi(\vecr) \;.
\end{align}
\end{subequations}
Here, $N_c$ denotes the number of condensate atoms with
$\varphi(\vecr)$ being normalized to 1. Quantum fluctuations are
assumed to be small and their mean values vanish by definition,
i.e.~$\langle\delta a \rangle =0$ and $\langle\delta
\Psi(\vecr)\rangle =0$. The dynamical equations of motion
resulting from Eqs.~\eqref{eq:MeanField} contain a hierarchy of
terms according to the different powers of the fluctuation
operators. To zeroth order in the fluctuations, one obtains a
Gross--Pitaevskii-type equation for the condensate wavefunction
$\varphi(\vecr,t)$ coupled to an ordinary differential equation
for $\alpha(t)$:
\begin{subequations}
\label{eq:motion}
\begin{multline}
i \hbar \frac{\partial}{\partial{}t}\varphi(\vecr,t) =
\Big[- \frac{\hbar^2 \nab^2}{2\,m} + V_0(\vecr)  + N_c u \, |\varphi(\vecr,t)|^2 \\
+ \hbar U_0|\alpha(t)|^2  \cos^2(k x)  \\ +  2\, \hbar
\eta_\sub{eff} \re\{\alpha(t)\}\, \cos(k x)\cos(k z)
\Big]\varphi(\vecr,t) \; . \label{eq:CavityGPE}
\end{multline}
\begin{multline}
i\frac{\partial}{\partial{}t}\alpha(t) = [-\Delta_C + N_c U_0
\langle \cos^2(k x)\rangle - i\kappa]\,\alpha(t) + i \eta \\ + N_c
\eta_\sub{eff} \langle \cos(k x) \cos(k z)\rangle\,,
\label{eq:CavityMeanField}
\end{multline}
\end{subequations}
where we used the notation $\langle f(\vecr) \rangle = \int d^3r
f(\vecr) |\varphi (\vecr, t) |^2$. The Gross--Pitaevskii equation
contains potential-like terms which depend on the amplitude
$\alpha$ and intensity $|\alpha|^2$ of the cavity field, and
express the mechanical effect of the cavity light upon the atoms.
The dynamics of the cavity field involves spatial averages over
the condensate density distribution.  Because of cavity decay, the
time evolution leads to a self-consistent stationary solution for
the mean fields, which usually is obtained only numerically
\cite{Nagy2008Selforganization}.

For a given condensate wave function and coherent cavity field
amplitude in steady state, the quantum fluctuations to leading
order form a linear system, which provides the energy spectrum of
excitations. With the notation $\vec{R}=[\delta  a,\delta
a^\dagger,\delta \Psi(\vecr), \delta \Psi^\dagger(\vecr)]$, the
time evolution of the fluctuation operators takes the compact form
\begin{equation}
\label{eq:LinearFluctuations} \frac{\partial}{\partial
t}\vec{R}=\mat{M}\vec{R}+\vec{\Xi}\,,
\end{equation}
where $\mat{M}$ is the linear stability matrix of the mean field
solution \cite{Nagy2008Selforganization}, and the term
$\vec{\Xi}=[\xi,\xi^\dagger,\ 0,0]$ accounts for the quantum input
noise of the cavity field. In general, the matrix $\mat{M}$ is
non-normal, i.e.~it does not commute with its hermitian adjoint.
Therefore it has different left and right eigenvectors, denoted by
$\lev{k}$ and $\rev{k}$, that form a bi-orthogonal system with
scalar product $(\lev{k},\rev{l})=\delta_{k,l}$. The decoupled
quasi-normal excitation modes defined by
$\rho_k=(\lev{k},\vec{R})$ are mixed excitations of the photon and
the matter wave fields.

The spectrum of excitations was analyzed first from a cavity
cooling point of view in the cavity pumping geometry ($\eta \neq
0$, $\eta_\sub{eff} = 0$). The imaginary part of the spectrum
revealed that excitations of the ultracold atomic gas can be
damped through the cavity loss channel
\cite{Horak2001Dissipative,Gardiner2001Cavityassisted}, provided
the decay rate $\kappa$ is on the order of the recoil frequency
$\omega_R$. The excitation spectrum was used in further studies to
describe critical phenomena, such as the dispersive optical
bistability in the cavity pumping geometry
\cite{Szirmai2010Quantum} and the self-organization phase
transition \cite{Nagy2008Selforganization,Konya2011Multimode} in
the atom pumping geometry ($\eta_\sub{eff} \neq 0$, $\eta=0$), see
Sec.~\ref{sec:Optomechanics} and Sec.~\ref{sec:DickeQPT}.

In the stable regime, second-order correlation functions can be
derived from \eref{eq:LinearFluctuations}. Importantly, there can
be a non-vanishing population of the atomic excited modes,
$\langle \delta \Psi^\dagger(\vecr) \delta \Psi(\vecr) \rangle
\neq 0$, even at zero temperature. This \emph{quantum depletion}
of the condensate is independent of collisional interactions,
which are known to cause a finite population of the Bogoliubov
modes \cite{Pitaevskii2003Bose-Einstein}. At variance, here the
condensate depletion arises from the cavity-mediated atom-atom
interactions as well as the dissipative process associated with
cavity decay. The quantum noise accompanying the photon loss
process couples into the atomic system and excites atoms out of
the condensate mode. Formally, it stems from the term containing
the photon creation operator $a^\dagger$ in the equations of
motion of $ \Psi(\vecr)$. This noise amplification mechanism is
analogous to the Petermann excess noise factor in lasers with
unstable cavities \cite{Grangier1998Simple}. It was shown by
\textcite{Szirmai2009Excess} that a depletion on the order of
${\sqrt{\Delta_C^2+\kappa^2}}/{\omega_R}$ can be expected rather
independently of the atom--field coupling, even for $U_0
\rightarrow 0$ and $\etaeff = 0$. It is a signature of the global
coupling that the quantum depletion is independent of the total
atom number $N$. In most of the experiments with linear cavities,
the ratio $\kappa/\omega_R$ is large ($\sim 10^3$). Since cavity
decay can also be interpreted as a continuous weak measurement of
the cavity photon number, the depletion can be attributed to
quantum back-action upon the atomic many-body state, as discussed
in Sec.~\ref{sec:QuantumBackaction}. It is also interesting to
note that the second-order correlation functions reveal an
entanglement between the matter-wave and the cavity field modes
\cite{Szirmai2010Quantum}.

\subsection{Cavity opto-mechanics with ultracold atomic ensembles}
\label{sec:Optomechanics}

In this section we focus on the dispersive interaction between the
collective motion of a quantum gas and a single-mode Fabry-P\'erot
cavity, which is coherently driven with amplitude $\eta$ by a
laser field at frequency $\omega$. In this case the dispersive
matter-light interaction, \eqref{eq:H_AC}, is given by
\begin{multline}\label{eq:H_AC_onaxis}
H_{AC} = \int d^3\vecr \,\Psi^\dagger(\vecr)\Big[ \hbar U_0
\cos^2(k x) \, a^\dagger a \Big] \Psi(\vecr)\;.
\end{multline}
On the one hand, the atomic medium experiences a periodic
potential, whose depth is proportional to the intracavity photon
number $ a^\dag a$. The potential depth for a single cavity photon
is $U_0 = g^2/\Delta_A$ and can be tuned in the experiment via the
detuning $\Delta_A$ between the cavity pump frequency $\omega$ and
the atomic transition frequency. On the other hand, the atom-light
interaction causes a dispersive shift of the empty cavity
frequency, which is determined by the spatial overlap between the
atomic density $\Psi^\dag(\mathbf{r})\Psi(\mathbf{r})$ and the
cavity intensity mode function $\cos^2(kx)$. A change in the
atomic density distribution caused by the intracavity dipole force
can therefore dynamically act back on the intracavity field
intensity by shifting the cavity resonance with respect to the
driving field.

In general, the interplay of these two effects results in a highly
nonlinear evolution of the coupled atoms-cavity system. For
certain limiting situations, however, the system can effectively
be described in the framework of cavity optomechanics
\cite{Kippenberg2008Cavity}, which studies the radiation-pressure
interaction between a harmonically suspended mechanical element
and the field inside an electromagnetic resonator. In a frame
rotating at $\omega$ this is described by the generic cavity
optomechanics Hamiltonian
\begin{equation}
\label{eq:H_OM} H_\mathrm{OM} = \hbar \omega_m c^\dag c - \hbar
(\delta_C - G X)a^\dag a +i \hbar \eta (a^\dag -a)
\end{equation}
where $c^\dag$ and $c$ denote creation and annihilation operators
of the mechanical oscillator at frequency $\omega_m$. The
mechanical element couples via its position quadrature $X = (c +
c^\dag)/\sqrt{2}$ with coupling strength $G$ to the intracavity
photon number $a^\dag a$. The detuning between the driving laser
and the cavity resonance frequency for zero displacement $X$ is
denoted by $\delta_C$. Configurations in which the harmonic
oscillator couples quadratically in $X$ to the cavity field have
recently been realized \cite{Thompson2008Strong,Purdy2010Tunable}.
This offers the possibility to detect phonon Fock states of the
mechanical element and to prepare squeezed states of the
mechanical oscillator or the optical output field.

\subsubsection{Experimental realizations}
Particular experimental situations allow to realize the
optomechanics Hamiltonian \eref{eq:H_OM} with an atomic ensemble
dispersively coupled to the field inside an optical cavity. This
relies on the fact, that the cavity field affects and senses
predominantly a single collective motional mode, which matches the
spatial cavity mode profile and plays the role of the harmonically
suspended mechanical element. Two different approaches for
realizing cavity optomechanics with ultracold atoms have so far
been realized experimentally.

\paragraph{Collective center-of-mass motion in the Lamb-Dicke regime}

\begin{figure}[t!]
\begin{center}
\includegraphics[width=\columnwidth]{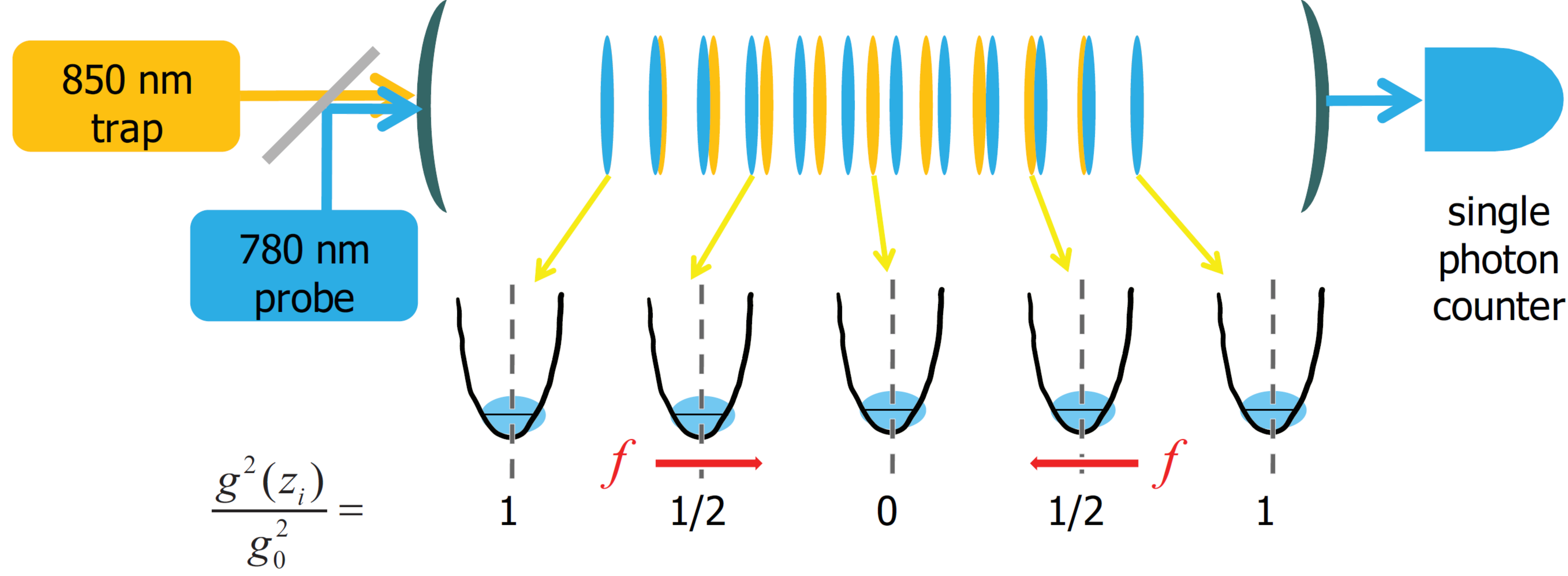}
\caption{(Color online) Scheme for cavity optomechanics with
ultracold atoms confined in the Lamb-Dicke regime. A high-finesse
cavity supports two longitudinal modes: one with wavelength of
about \unit[780]{nm} that is near the $\mathrm{D}_2$ line of
$^{87}$Rb, and another with wavelength of about \unit[850]{nm}.
The latter produces a one-dimensional optical lattice potential,
with trap minima indicated in orange, in which ultracold $^{87}$Rb
atoms are confined within the lowest vibrational band. The atomic
clouds induce, depending on their trapping position $z_i$,
dispersive frequency shifts on the \unit[780]{nm} cavity
resonance, as shown in the bottom line. In turn, the cavity field
exerts a position dependent force $f$, as indicated by the red
arrows. In the Lamb-Dicke regime, the collective atoms-cavity
interaction reduces to the generic optomechanics Hamiltonian
wherein a single collective mode of harmonic motion linearly
couples to the cavity field. From \textcite{Botter2009Quantum}.}
\label{fig:OMscheme_LambDicke}
\end{center}
\end{figure}

In experiments performed by the Berkeley group
\cite{Gupta2007Cavity,Murch2008Observation}, ultracold atoms are
loaded into the lowest band of a far-detuned intracavity lattice
potential, forming a stack of hundreds of tightly confined atom
clouds (see Fig.~\ref{fig:OMscheme_LambDicke}). Each atom cloud is
harmonically suspended with oscillation frequency $\omega_m$ and
extends along the cavity axis by only a fraction of the optical
wavelength, thus realizing the Lamb-Dicke regime. A cavity mode,
whose periodicity differs from that of the trapping lattice
potential, couples strongly to a single collective center-of-mass
mode of the atomic stack. All remaining collective modes decouple
from the cavity field and can be considered as a heat bath to
which the distinguished collective mode is only weakly coupled via
e.g.~collisional atom-atom interactions.
The system realizes the linear optomechanics Hamiltonian
\eref{eq:H_OM}, with the optomechanical coupling strength being
given by $G = \sqrt{N_\mathrm{eff}} k U_0 X_\mathrm{ho}$. Here,
$k$ is the cavity wave vector, $X_\mathrm{ho} = \sqrt{\hbar/2 m
\omega_m}$ denotes the harmonic oscillator length with atomic mass
$m$, and $N_\mathrm{eff}\approx N/2$ with total atom number $N$.

The quadratic coupling regime of optomechanics with ultracold
atoms was realized in an atom-chip-based setup
\cite{Purdy2010Tunable} which allows for subwavelength positioning
of a tightly confined ultracold ensemble. By preparing as low as
two atomic clouds, tightly confined at adjacent lattice sites of a
far-detuned intracavity lattice potential, and controlling their
center-of-mass position along the cavity axis both linear and
quadratic optomechanical coupling can be realized, providing an
atoms-based realization of the "membran-in-the-middle" approach
\cite{Thompson2008Strong}.

\paragraph{Collective density oscillations in a Bose-Einstein condensate}

\begin{figure}[htbp]
\begin{center}
\includegraphics[width=\columnwidth]{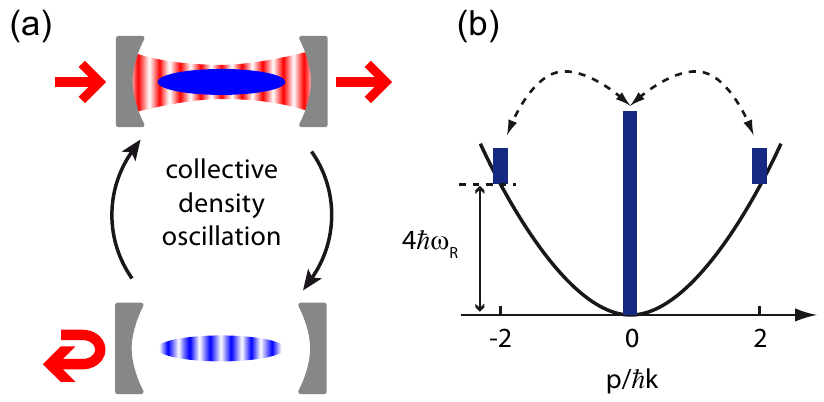}
\caption{(Color online) Cavity-optomechanics with a weakly
confined Bose-Einstein condensate dispersively coupled to the
field of an optical high-finesse resonator. (a) A collective
density excitation of the condensate (blue) with periodicity
$\lambda/2 = \pi/k$ acts as a mechanical oscillator with
oscillation frequency $4 \omega_R$. Optomechanical coupling to the
cavity field is provided by the dependence of the optical path
length on the atomic density distribution within the spatially
periodic cavity mode structure. (b) Condensate atoms initially
prepared close to zero momentum, $p=0$, are scattered off the
intracavity lattice potential into the symmetric superposition of
states with momentum $p = \pm 2 \hbar k$. Matter-wave interference
with the macroscopic zero-momentum component results in a harmonic
density oscillation evolving at frequency $4 \omega_R$. Adapted
from \textcite{Brennecke2008Cavity}.} \label{fig:OMscheme_Zurich}
\end{center}
\end{figure}

A different route to realize linear cavity optomechanics with an
ultracold atomic ensemble was experimentally explored in the
Z\"urich group \cite{Brennecke2008Cavity}. A Bose-Einstein
condensate of typically $10^5$ atoms is prepared in an external
harmonic trapping potential, extending over several periods of the
cavity standing-wave mode structure (see
Fig.~\ref{fig:OMscheme_Zurich}). In contrast to the Lamb-Dicke
regime considered before, here a momentum picture is more
appropriate. Initially, all condensate atoms are prepared --
relative to the recoil momentum $\hbar k$ -- in the zero-momentum
state $|p = 0\rangle$. The dispersive interaction with the cavity
field diffracts atoms into the symmetric superposition of momentum
states $|\pm 2\hbar k\rangle$ along the cavity axis. Matter-wave
interference between the macroscopically occupied zero-momentum
component and the recoiling component results in a spatial
modulation of the condensate density with periodicity $\lambda/2$,
which oscillates in time at the frequency $4 \omega_R = 2\hbar
k^2/m$. As long as diffraction into higher-order momentum modes
can be neglected, the dynamics of the coupled system is again
captured by the simple optomechanics Hamiltonian \eref{eq:H_OM}.
Here, collective excitations of the recoiling momentum mode play
the role of phonon excitations of a mechanical mode with
oscillation frequency $\omega_m = 4 \omega_R$. The coupling rate,
$G = \sqrt{N}U_0/2$, again scales with the square root of the
atoms number, indicating the collective nature of the atom-light
interaction. The realization of an optomechanical system employing
the collective motion of a Bose-Einstein condensate triggered
subsequent theoretical studies along this direction
\cite{Zhang2009Nonlinear,Chen2010Classical,Chen2011Alloptical,DeChiara2011Entanglement}.

Further realizations of cavity optomechanics with atomic ensembles
have been proposed theoretically by dispersively coupling a
quantum-degenerate Fermi gas \cite{Kanamoto2010Optomechanics} or
the internal spin-degrees of freedom of a quantum gas
\cite{Brahms2010Spin,Jing2011Quantum} to the field of an optical
cavity. The latter system has been shown to exhibit a formal
analogy with a torsional oscillator coupled quadratically to the
cavity mode. It provides an ideal nondestructive tool for the
control of quantum spin dynamics, and was proposed to resolve the
quantum regime of an antiferromagnetic spin-1 condensate.

The realization of an optomechanical system using ultracold atoms
offers direct access to the quantum regime of cavity
optomechanics. In contrast to solid-state realizations of
optomechanics, evaporative cooling techniques available for atomic
gases allow for a natural and very pure preparation of the
mechanical oscillator mode in its quantum ground state.
Correspondingly, these systems pave the way to directly study
quantum effects of the optomechanical interaction
\cite{Murch2008Observation,Brooks2012Nonclassical,Brahms2012Optical}.

The easy tunability of system parameters like e.g.~the mechanical
oscillator frequency $\omega_m$ (via the external confining
potential), the optomechanical coupling strength $G$ (via the atom
number or the pump-atom detuning) or the initial temperature of
the mechanical oscillator allows to explore the transition between
different regimes of optomechanics. Most important, the coupling
strengths $G$ achievable with atomic systems open access to the
'granular' regime of optomechanics
\cite{Ludwig2008optomechanical,Murch2008Observation}, where single
excitations in either of the two subsystems have a non-negligible
effect upon the dynamics of the other. This can be measured by the
granularity (or quantum) parameter which is defined as $\zeta =
G/\kappa$. For $\zeta = 1$, already a single excitation of the
mechanical mode shifts the cavity resonance by half its linewidth,
and already a single photon entering the cavity imparts one
excitation quantum in the mechanical oscillator. In future
research, this might allow the generation and detection of quantum
correlations between the mechanical and light degrees of freedom.
Further research possibilities based on atoms-based realizations
of optomechanics, are given by possible implementations of
precision sensors of forces, or devices to manipulate light fields
on a quantum level.

\subsubsection{Nonlinear dynamics and bistability for low photon number}
\label{sec:Bistability} The optomechanical interaction,
\eref{eq:H_OM}, being intrinsically nonlinear gives rise to
dispersive optical bistability and nonlinear dynamics of the
coupled system. Optical bistability \cite{Lugiato1984II}, a well
studied phenomenon in nonlinear optics, refers to the co-existence
of two stable steady-state solutions when e.g.~driving an optical
cavity filled with a medium whose refractive index depends on the
light intensity. In typical nonlinear Kerr media and solid-state
realizations of optomechanics, the occurrence of bistability
typically requires large intracavity power in order to
significantly alter the system's optical properties. The large
coupling strength achieved in the atomic-ensemble realizations of
optomechanics induces optical bistability at a mean-intracavity
photon level below one \cite{Gupta2007Cavity,Ritter2009Dynamical}.
This achievement is desirable for applications ranging from
optical communication to quantum computation
\cite{Cirac1997Quantum,Imamoglu1997Strongly}.

The occurrence of bistability in optomechanical systems can be
understood from the corresponding semiclassical equations of
motion for the oscillator displacement $X$ and the coherent
intracavity field amplitude $\alpha$ derived from Hamiltonian
\eref{eq:H_OM}
\begin{eqnarray}\label{eq:motion_OM}
\ddot{X} + \omega_m X &=& -\omega_m G |\alpha|^2\\\notag
\dot{\alpha} &=& \left(i(\delta_C - G X) - \kappa \right)\alpha +
\eta\,.
\end{eqnarray}
In the bad cavity regime, $\kappa \gg \omega_R$, the atoms move on
a timescale which is large compared to the lifetime
$(2\kappa)^{-1}$ of intracavity photons. Correspondingly, the
cavity field adiabatically follows the atomic dynamics according
to
\begin{equation}\label{eq:meanphotonnumber}
|\alpha|^2 = \frac{\eta^2}{\kappa^2 + (\delta_C - GX)^2}\,.
\end{equation}
Retardation effects resulting in dynamical back-action cooling or
heating of the mechanical element are neglected in this
approximation. Inserting this expression into the equation of
motion for $X$, \eref{eq:motion_OM}, yields $\ddot{X} =
-\frac{d}{dX}V_\mathrm{OM}(X)$. The optomechanical potential given
by
\begin{equation}\label{eq:V_OM}
V_\mathrm{OM}(X) = \frac{1}{2}\hbar\omega_m X^2 - \frac{\hbar
\eta^2}{\kappa} \arctan \left( \frac{\Delta(X)}{\kappa}\right)\, ,
\end{equation}
captures the combination of the harmonic confinement and the
cavity dipole forces. Here, $\Delta(X) = \delta_C - G X$ denotes
the detuning between the driving laser and the atoms-shifted
cavity resonance.

\begin{figure}[htbp]
\begin{center}
\includegraphics[width=\columnwidth]{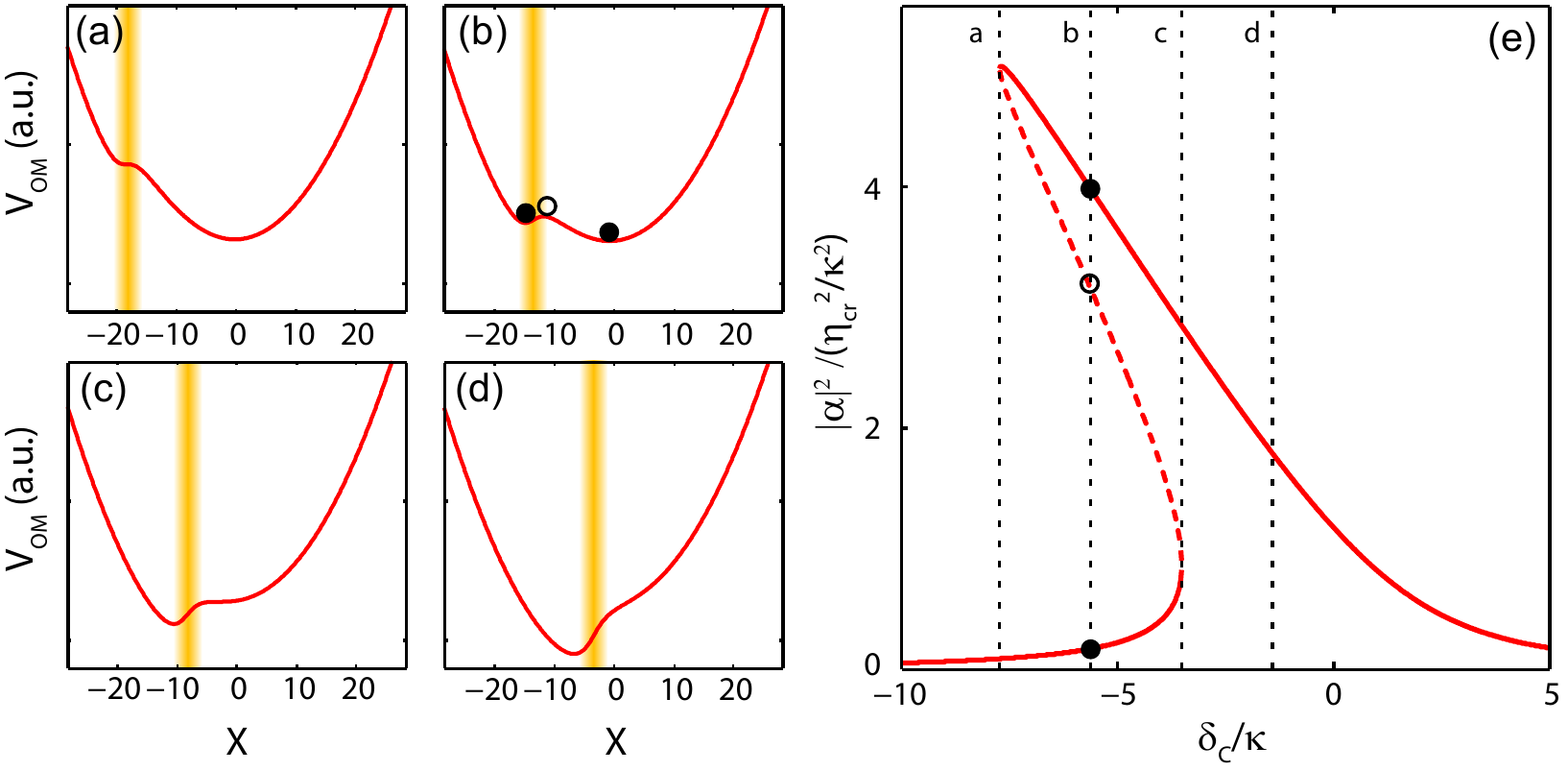}
\caption{(Color online) Optomechanical potential and bistability.
(a-d) Optomechanical potential landscape $V_\mathrm{OM}(X)$ for
different pump-cavity detunings $\delta_C$, indicated by the
dashed lines in (e). The shaded regions show the resonance profile
of the cavity. (e) Mean intracavity photon number $|\alpha|^2$ in
steady state. Open and close circles correspond to the situation
shown in (b). Parameters are $G = 0.42 \kappa$ and $\eta =
\sqrt{5} \eta_\mathrm{cr}$.} \label{fig:V_OM}
\end{center}
\end{figure}
The optomechanical potential provides an intuitive picture to
understand the steady-state as well as the dynamical behavior of
the coupled system, see Fig.~\ref{fig:V_OM}. Above a critical
cavity pump strength $\eta_\mathrm{cr}$, determined by
$\eta_\mathrm{cr}^2 = \frac{8}{3\sqrt{3}}\frac{\omega_m
\kappa^3}{G^2}$, the optomechanical potential exhibits (within a
certain detuning range) two local minima, which correspond to
different intracavity light intensities as shown in the bistable
resonance curve, Fig.~\ref{fig:V_OM}e.

Depending on the direction into which $\delta_C$ is adiabatically
tuned, the system remains in either of the two local minimum of
$V_\mathrm{OM}$, following the upper or lower bistable resonance
branch. When reaching the critical detuning, where one of the
local minima turns into a saddle point, the system starts to
perform transient oscillations in the remaining potential minimum,
which translate into a periodically modulated cavity light
intensity. Due to damping of the collective atomic motion, the
system finally relaxes to the steady-state in the remaining
potential minimum.

\begin{figure}[htbp]
\begin{center}
\includegraphics[width=\columnwidth]{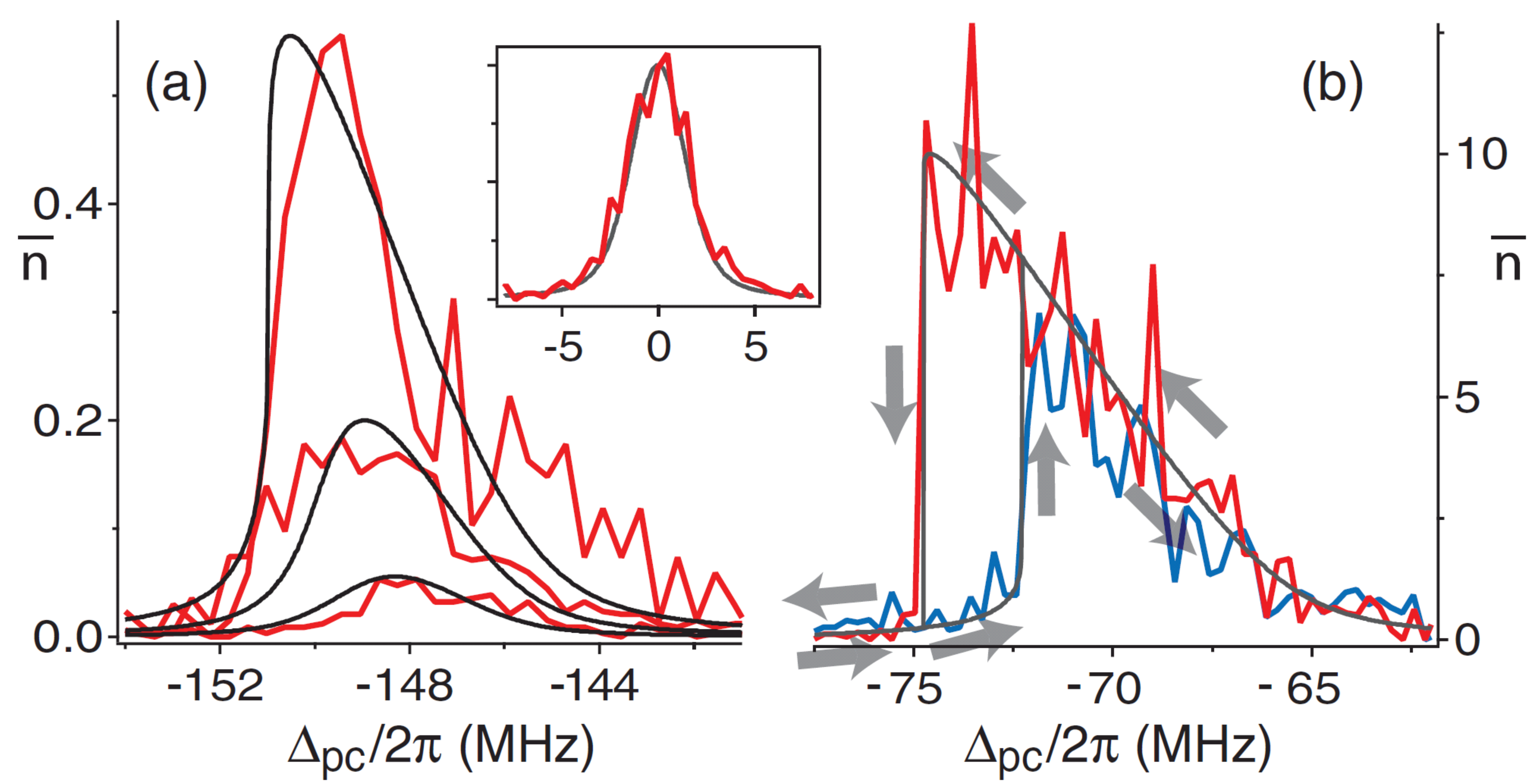}
\caption{(Color online) Dispersive optical bistability with
collective atomic motion. (a) Observed cavity line shapes (red)
for increasing cavity input power at low intracavity photon number
$\bar{n}$ and model line shapes (black), based on the
Voigt-profile of the bare cavity line (inset). $\Delta_{pc}$
denotes the detuning between the probe laser frequency and the
empty cavity frequency. (b) Lower (blue) and upper (red) branches
of optical bistability observed in single sweeps across resonance.
From \textcite{Gupta2007Cavity}.} \label{fig:bistability}
\end{center}
\end{figure}

Optical bistability induced by collective atomic motion was
observed at low intracavity photon number both in the Berkeley
group \cite{Gupta2007Cavity} and the Z\"urich group
\cite{Ritter2009Dynamical}. The lower and upper bistability
branches were observed in single experimental runs by slowly
sweeping the frequency of the driving laser twice across
resonance, first with increasing detuning and then with decreasing
detuning, see Fig.~\ref{fig:bistability}. Upon increasing the
probe strength, the cavity transmission profile becomes more and
more asymmetric and exhibits hysteresis.

Dispersive optical bistability with collective atomic motion was
also studied in the regime of quadratic optomechanical coupling
\cite{Purdy2010Tunable}. Here, instead of displacing the
center-of-mass motion of the mechanical element, the intracavity
dipole forces increase or decrease the rms width of the
compressible atomic ensemble, depending on whether the atoms are
confined at a maximum or a minimum of the intracavity probe
lattice potential. The corresponding change of the dispersive
cavity shift leads again to bistable resonance curves as observed
in the experiment \cite{Purdy2010Tunable}.

\begin{figure}[htbp]
\begin{center}
\includegraphics[width=0.8\columnwidth]{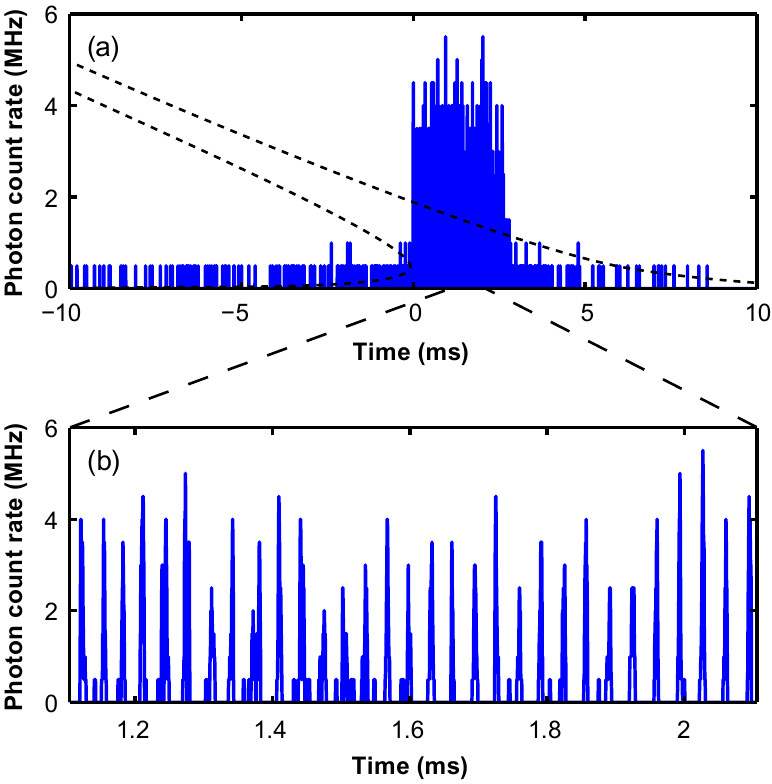}
\caption{Nonlinear dynamics of a driven BEC-cavity system. (a)
Sweeping the driving laser frequency across the bistable resonance
curve (indicated in dashed, scaled by a factor of 4) excites
large-amplitude density oscillations in the condensate. (b) The
magnified cavity transmission signal indicates how the density
oscillations tune the cavity frequency periodically in an out of
resonance with the driving laser. The mean intracavity photon
number on resonance was 7.3 corresponding to a photon count rate
of \unit[5.8]{MHz}. Adapted from \textcite{Brennecke2008Cavity}.}
\label{fig:transient_oscillations}
\end{center}
\end{figure}

Dynamical optomechanical effects arise in small-amplitude
oscillations of the system around steady state. As a result of the
optomechanical interaction the frequency of such oscillations is
shifted with respect to the bare oscillation frequency $\omega_m$,
in the literature often referred to as the ''optical spring
effect''. In the case of linear optomechanical coupling this can
be inferred from a quadratic expansion around the steady state
minima of the optomechanical potential $V_\mathrm{OM}$,
Fig.~\ref{fig:V_OM}. Experimentally, the optomechanical frequency
shift for the collective atomic motion was observed and quantified
in agreement with theory both in the linear and the quadratic
coupling regime \cite{Purdy2010Tunable}. Highly nonlinear
oscillations in the optomechanical potential with relatively large
amplitude have been excited and observed in cavity transmission
either by a sudden displacement of the optomechanical potential
\cite{Gupta2007Cavity} or by crossing the instability point of the
bistable curves \cite{Brennecke2008Cavity}, see
Fig.~\ref{fig:transient_oscillations}.

\subsubsection{Quantum-measurement back-action upon collective atomic motion}
\label{sec:QuantumBackaction}

The accuracy of any position measurement of a mechanical element
is limited by quantum mechanics. Referred to as the
\emph{standard quantum limit}, this has been extensively studied in
connection with the development of gravitational-wave detectors
\cite{Caves1980QuantumMechanical}. In a generic optomechanical
setup, which allows for high-precision measurements of the
position of a mechanical element, the standard quantum limit
arises from the balance between two noise terms: (i) detection
shot noise, given by the random arrivals of photons on the
detector and (ii) radiation-pressure induced displacement noise
caused by the quantum fluctuations of the intracavity photon
number. Whereas detection noise can be decreased by increasing the
light power, this comes at the expense of increased radiation
pressure force fluctuations. The optimal sensitivity is achieved
if these two noise contribution are balanced. A direct
experimental observation of the intracavity photon number
fluctuations, requires large optomechanical coupling strengths
between the intracavity field and the mechanical element in
combination with the suppression of thermal or technical noise
sources which perturb the mechanical motion.

The utilization of collective atomic motion of an ultracold gas
strongly coupled to the field inside a Fabry-P\'erot resonator,
allowed for the first observation of measurement-induced
back-action upon a macroscopic mechanical element formed of $10^5$
atoms, caused by intracavity quantum force fluctuations
\cite{Murch2008Observation}. In the non-granular regime $\zeta =
G/\kappa \ll 1$, the spectral density of intracavity photon number
fluctuations \cite{Marquardt2007Quantum,Nagy2009Nonlinear} agrees with that in an
empty driven cavity, and reads
\begin{equation}\label{eq:S_nn}
S_{nn}(\omega) = \frac{2 \bar{n} \kappa}{\kappa^2 + (\Delta(X) +
\omega)^2}.
\end{equation}
Here, $\bar{n} = |\alpha|^2$ denotes the steady-state
mean-intracavity photon number given in
Eq.~\eqref{eq:meanphotonnumber}. These photon number fluctuations
are transmitted into the momentum of the mechanical element via
the optomechanical interaction, giving rise to a diffusion-like
increase of the phonon number
\begin{equation}
\frac{d}{dt}\langle c^\dag c\rangle  = \kappa^2 \zeta^2
S_{nn}(-\omega_m) \; ,
\end{equation}
as can be derived e.g.~from an effective master equation for the
mechanical oscillator \cite{Nagy2009Nonlinear}.

\begin{figure}[htbp]
\begin{center}
\includegraphics[width=0.9\columnwidth]{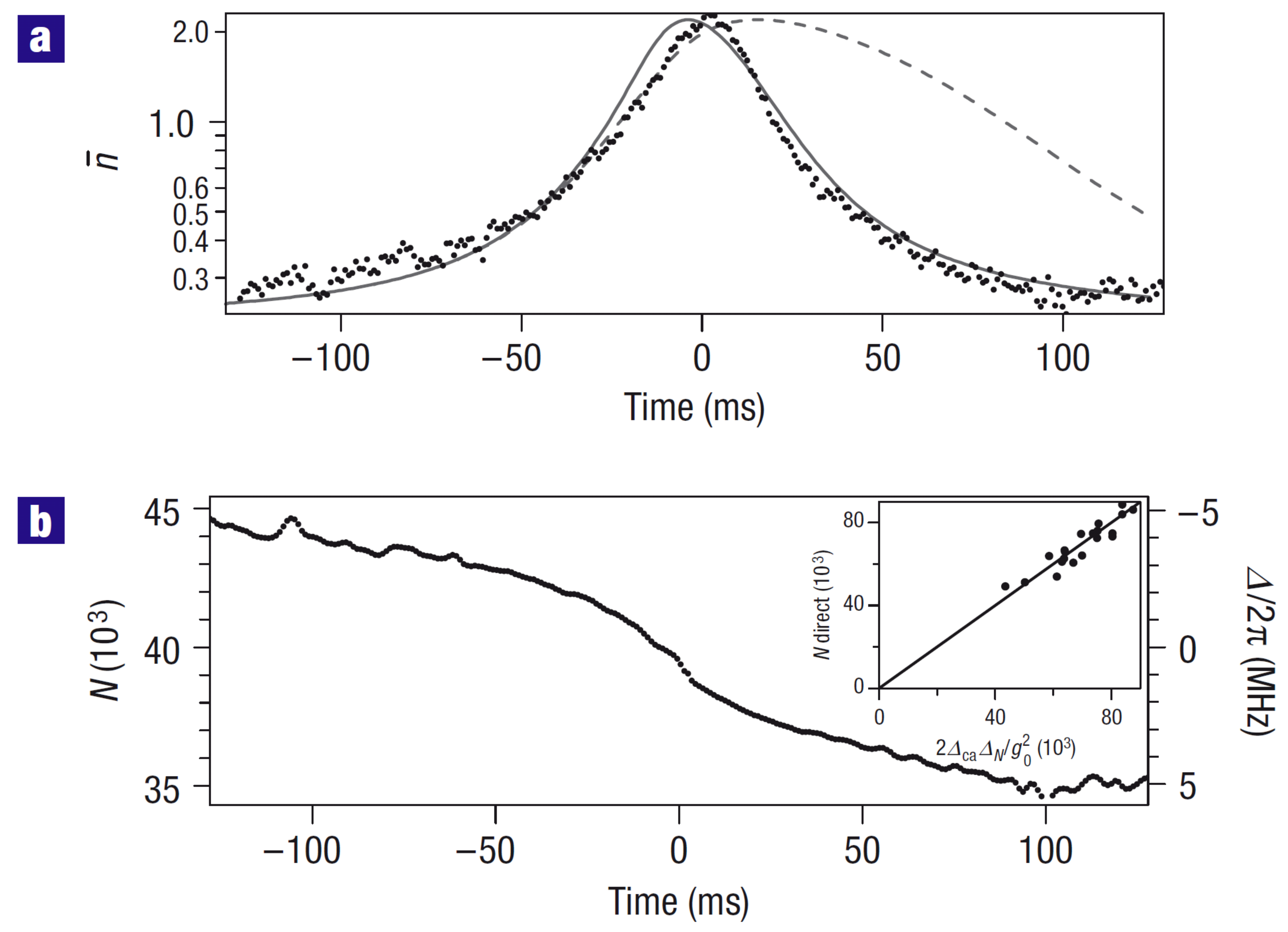}
\caption{Observing measurement-induced back-action upon the
collective motion of an ultracold atomic ensemble. (a)
Mean-intracavity photon number $\bar{n}$ (points), monitored as
the system is brought across the cavity resonance due to
evaporative atom loss. The expected photon number including (solid
line) and excluding (dashed line) measurement back-action is shown.
(b) Total atom number $N$ as a function of time as inferred from
data shown in (a), using the empty cavity line shape and the
linear relation between atom number and dispersive cavity shift
(inset). From \textcite{Murch2008Observation}.}
\label{fig:Murch2008Fig2}
\end{center}
\end{figure}
\textcite{Murch2008Observation} measured the corresponding heating
rate of the atomic ensemble in a bolometric way by quantifying the
evaporative atom loss, see Fig.~\ref{fig:Murch2008Fig2}. After
preparing the mechanical oscillator close to its ground state, the
cavity transmission of a weak probe beam at fixed frequency is
monitored on a single-photon counting module. Continuous
background atom loss tunes the atoms-shifted cavity frequency in
resonance with the driving laser. The atom loss rate is deduced
from the comparison between the recorded transmission curve and
the empty-cavity resonance curve. The corresponding single-atom
heating rate is found to exceed the free-space spontaneous heating
rate, which was deduced by measuring the atom loss rate far from
the cavity resonance, by a factor of 40, in agreement with the
theoretical expectation. As cavity-mediated coherent amplification
and damping of the mechanical oscillator is negligible in the
experiment, the observation of back-action heating can be
interpreted as a direct measurement of photon number fluctuations
in a coherently driven cavity.

\begin{figure}[htbp]
\begin{center}
\includegraphics[width=.85\columnwidth]{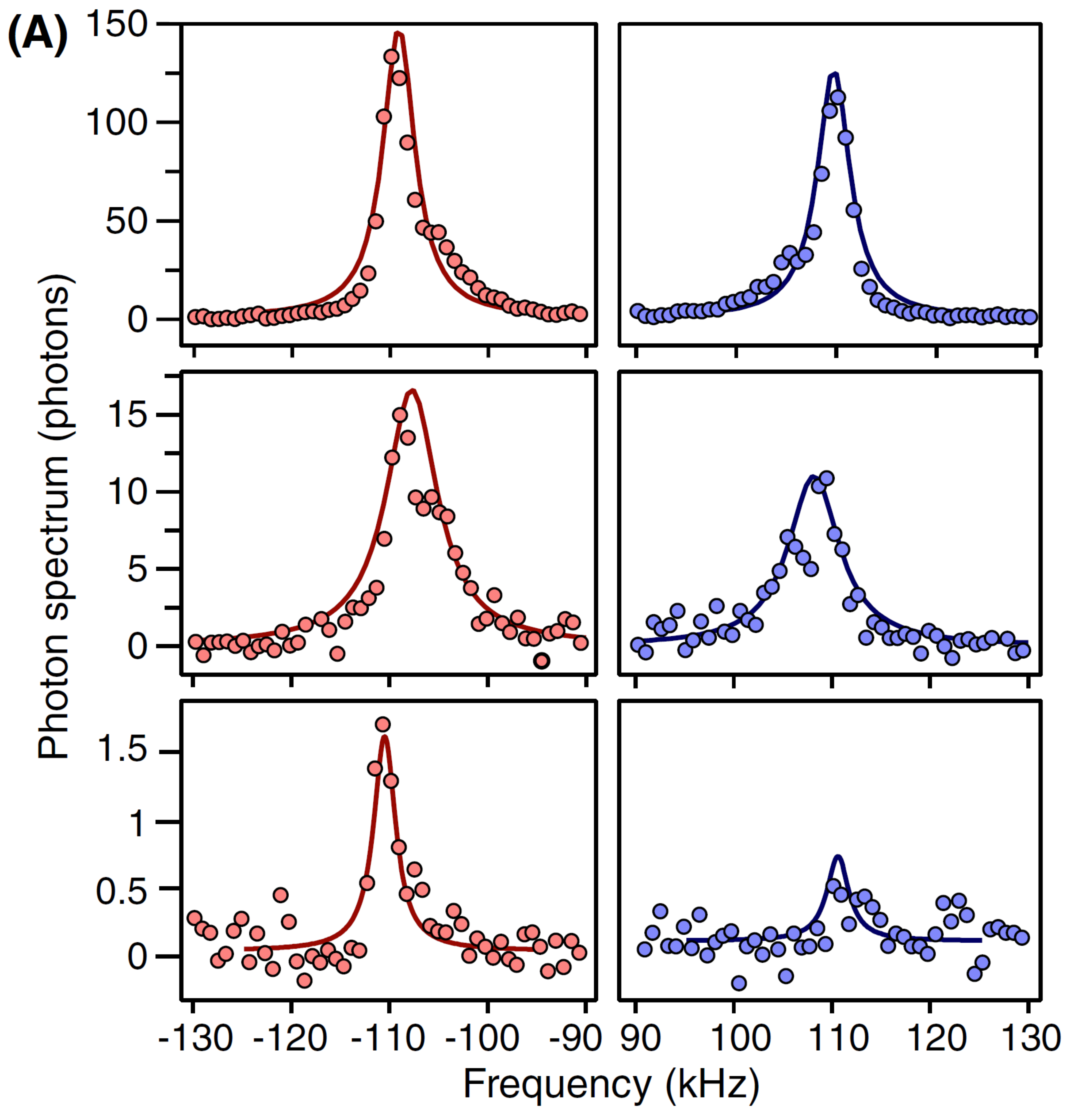}
\caption{Optical detection of quantization of collective atomic
motion in the cavity output spectrum. Shown are measured Stokes
sidebands (left panels) and anti-Stokes sidebands (right panels)
for increasing intracavity photon number (bottom to top) together
with the theoretical prediction (solid lines). The observed Stokes
asymmetry provides a calibration-free measure for the mean
occupation number of the mechanical oscillator, which was deduce
for the lowest graph to be $0.49$. The mechanical oscillation
frequency was $\omega_m = 2\pi\times \unit[110]{kHz}$. From
\textcite{Brahms2012Optical}.} \label{fig:Brahms2012Fig2}
\end{center}
\end{figure}

Another direct signature of quantum back-action of light upon
collective atomic motion was obtained by monitoring the Stokes and
anti-Stokes sidebands of the cavity transmission subsequent to the
preparation of the collective motional degree of freedom close to
its ground state \cite{Brahms2012Optical}, see
Fig.~\ref{fig:Brahms2012Fig2}. The observed sideband asymmetry
provides a direct measurement of the quantized collective motion
and serves as a record of the energy exchanged between motion and
the light in agreement with a continuous back-action limited
quantum position measurement.

The disturbance of collective atomic motion via the intracavity
quantum force fluctuations acts back again onto the intracavity
light field. In particular, the resulting motional-induced
modulation of the cavity field can interfere with the coherent or
vacuum cavity input field giving rise to nonlinear optical
parametric amplification and -- for negligible technical or
thermal fluctuations -- to ponderomotive squeezing
\cite{Mancini1994Quantum,Fabre1994Quantumnoise}. Only recently,
these effects where observed for the first time in the Berkeley
group utilizing the optomechanical coupling between collective
atomic motion and an optical cavity field
\cite{Brooks2012Nonclassical}.

\subsubsection{Cavity cooling in the resolved sideband regime}

For the small-volume cavities which were employed in the
experiments performed by the Berkeley and Zurich group, the cavity
decay rate $\kappa$ exceeds the mechanical oscillation frequency
$\omega_m$ of the collective atomic degree of freedom by more than
one order of magnitude. In this \emph{non-resolved sideband
regime} of cavity optomechanics cooling of the mechanical
oscillator into its ground state utilizing cavity dissipation is
not possible \footnote{Indeed, in those experiments the
preparation of the mechanical degree of freedom close to its
ground state was achieved directly by evaporative cooling.}.
Rather, the minimal steady-state occupation number when driving
the cavity field with a laser field which is red-detuned by
$\sqrt{\omega_m^2+\kappa^2}$ from the cavity resonance is given by
$\frac{\kappa}{2 \omega_m}\gg 1$ for weak optomechanical coupling
strength \cite{Marquardt2007Quantum}.

Ground-state cooling becomes possible only in the \emph{resolved
sideband regime} where $\omega_m \gg \kappa$
\cite{Kippenberg2008Cavity}. Here, the cavity is able to resolve
the Stokes resp.~anti-Stokes sidebands which correspond to adding
resp.~removing motional quanta from the mechanical degree of
freedom. The large asymmetry between these processes which is
achieved by driving the cavity near the anti-Stokes sideband
results in a steady-state phonon occupation number of
$(\frac{\kappa}{2 \omega_m})^2 \ll 1$ \cite{Marquardt2007Quantum}.
Optomechanical cooling in the resolved sideband regime is
equivalent to optical Raman sideband cooling of tightly confined
atoms or ions.

\begin{figure}[htbp]
\begin{center}
\includegraphics[width=0.8\columnwidth]{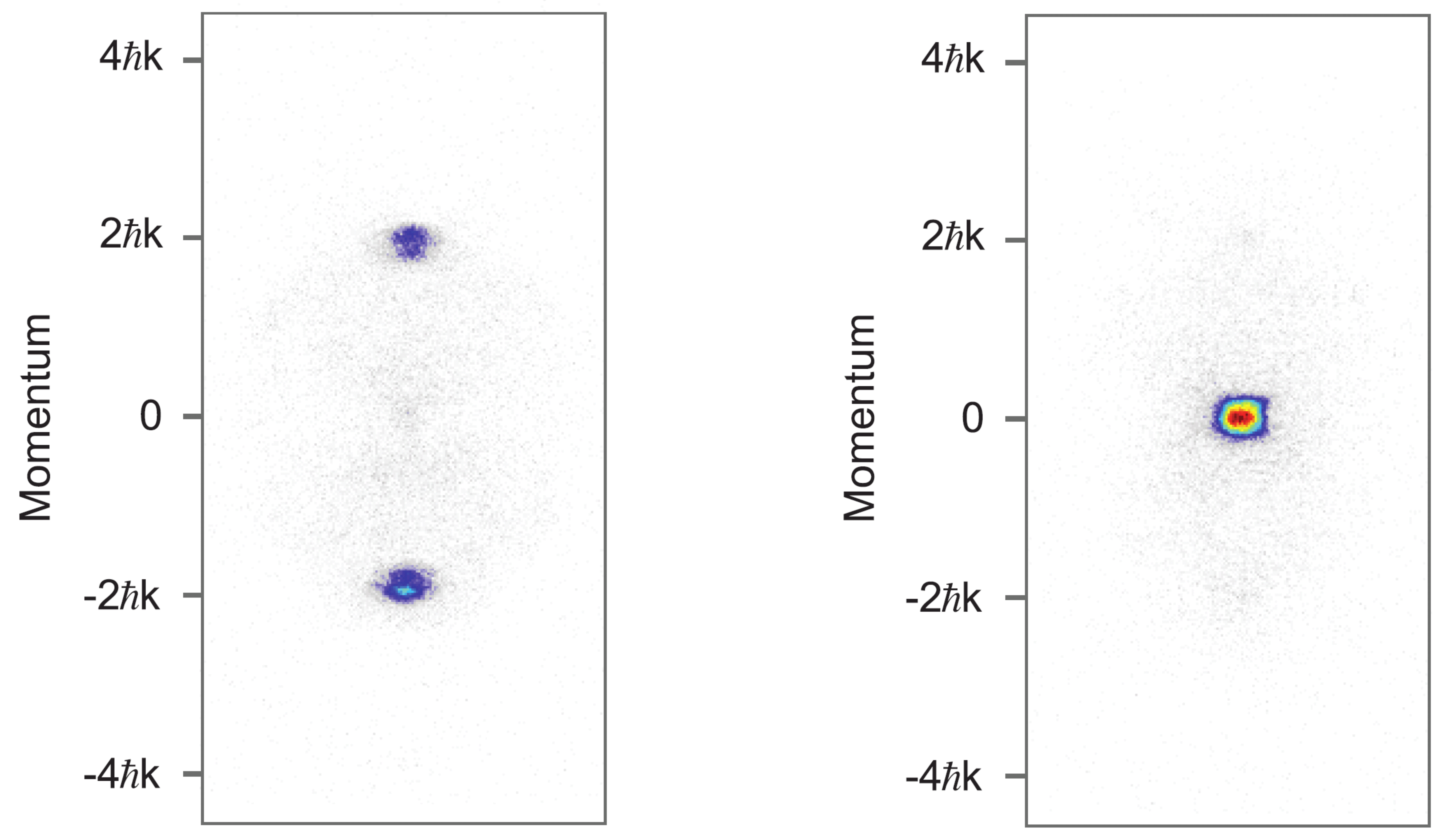}
\caption{Observation of sub-recoil cavity cooling with a $^{87}$Rb
BEC in a narrow-bandwidth Fabry-P\'erot resonator. Shown are
atomic momentum distributions after driving the cavity field with
a far-detuned laser field at $\unit[803]{nm}$. First, a
$\unit[400]{\mu s}$ long pulse, blue-detuned from the cavity
resonance, transfers atoms into the momentum states $|\pm 2\hbar k
\rangle$ (left), subsequently a $\unit[200]{\mu s}$ long
red-detuned pulse transfers the atoms back into the zero-momentum
state. Binary atomic collisions result in a substantial depletion
of the $\pm 2\hbar k$-momentum state populations visible as a
diffusive halo. From \textcite{Wolke2012Cavity}.
\label{fig:Wolke2012Fig3}}
\end{center}
\end{figure}

Cavity cooling in an optomechanical-type BEC-cavity system which
ranges in the good cavity regime, $\kappa < \omega_m = 4
\omega_R$, was demonstrated recently by
\textcite{Wolke2012Cavity}. By driving the cavity field
selectively close to the Stokes or anti-Stokes sidebands atoms
where transferred via cavity-stimulating backward scattering from
the macroscopically populated zero-momentum state into a
superposition of momentum states $|\pm 2 \hbar k \rangle$ and
back, see Fig.~\ref{fig:Wolke2012Fig3}. This experiment paves the
way towards the achievement of quantum degeneracy starting from a
thermal gas without relying on evaporative cooling techniques.

\subsection{Non-equilibrium phase transitions}
\label{sec:DickeQPT}

Self-organization of a laser-driven atomic ensemble inside an
optical resonator, as was considered for thermal atoms in Section
\ref{sec:ColdSelforg}, was extended both theoretically and
experimentally into the ultracold regime, where atomic motion
becomes quantized. Correspondingly, the transition point to the
self-organized phase is not determined anymore by thermal density
fluctuations, rather by the competition between kinetic energy
cost and potential energy gain associated with a spatial
modulation of the atomic matter-wave in the cavity-induced lattice
potential. In case of a weakly interacting Bose-Einstein
condensate, the reduced number of momentum states accompanying in
the dynamics allows for a simplified description in terms of a
collective spin degree of freedom, providing a direct link between
self-organization and an open-system realization of the Dicke
quantum phase transition.

\subsubsection{Self-organization of a Bose-Einstein condensate}

\begin{figure}[b!]
\begin{center}
\includegraphics[angle=0,width=0.9\columnwidth]{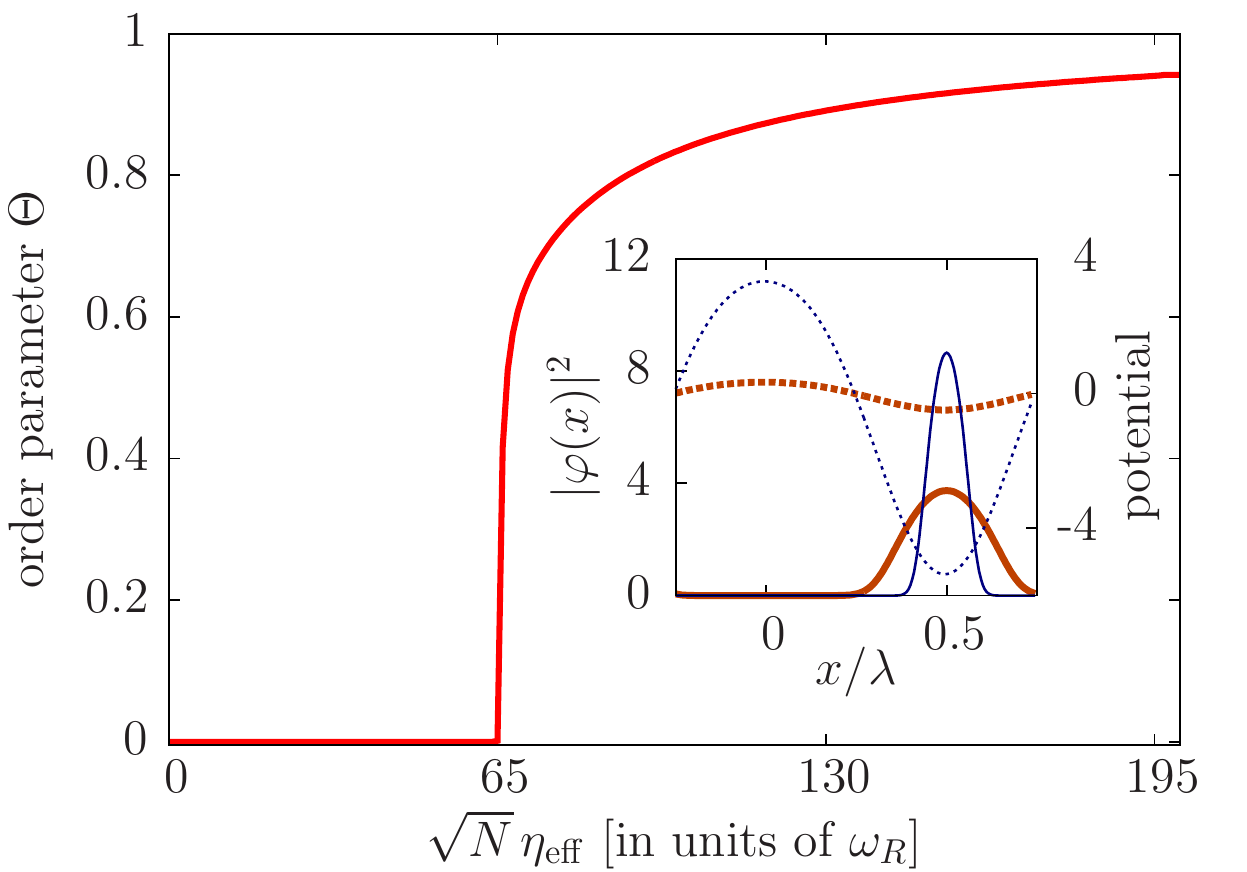}
\caption{Self-organization of a driven Bose-Einstein condensate in
a standing-wave cavity. Plotted is the steady-state order
parameter $\Theta$ as a function of the effective cavity pump
strength $\eta_\mathrm{eff}$, as obtained from a numerical
solution of the mean-field equations. Parameters are $N U_0 = -100
\omega_R$, $\Delta_C = -300 \omega_R$, $\kappa = 200 \omega_R$ and
$\mu_0 = 10 \hbar \omega_R$ in the homogeneous phase. According to
\eref{eq:MFthreshold} the homogeneous phase is stabilized in this
parameter regime dominantly by collisional interaction energy. The
inset shows the condensate wave functions (solid lines) for
$\sqrt{N}\etaeff = 100 \omega_R$ (broader, brown solid) and $300
\omega_R$ (narrower, blue solid) and the corresponding optical
dipole potentials (dashed lines). Adapted from
\textcite{Nagy2008Selforganization}.\label{fig:Nagy2008Fig1}}
\end{center}
\end{figure}

Self-organization of a dilute Bose-Einstein condensate (BEC),
which is located in a single-mode optical cavity and illuminated
transversally to the cavity axis by a far-detuned laser field, was
studied theoretically by \textcite{Nagy2008Selforganization}. In
terms of a mean-field description, the steady-state of the system
was obtained from the equations of motion for the coherent cavity
field amplitude $\alpha$ and the atomic mean-field
$\varphi(\mathbf{r})$, see Eq.~\eqref{eq:CavityMeanField} and
\eqref{eq:CavityGPE}, setting the on-axis pump strength $\eta$ to
zero. For simplicity only atomic motion along the cavity axis was
taken into account. The numerical solution for the steady-state
order parameter $\Theta = \langle
\varphi|\cos(kx)|\varphi\rangle$, obtained by numerically
propagating the equations of motion into imaginary time, is shown
in Fig.~\ref{fig:Nagy2008Fig1}. Above a critical two-photon Rabi
frequency $\eta_\mathrm{eff}$, the order parameter takes a
non-zero value indicating self-organization of the atoms in a
$\lambda$-periodic density pattern.
A stability analysis of the non-organized steady state, $\Theta =
0$, yields the following analytic expression for the critical
point
\begin{equation}\label{eq:MFthreshold}
\sqrt{N}\eta_\mathrm{eff,c} =
\sqrt{\frac{(\delta_C^2+\kappa^2)(\omega_R + 2
\mu_0/\hbar)}{-2\delta_C}}
\end{equation}
where $\mu_0$ denotes the chemical potential of the homogeneous
condensate and $\delta_C$ the detuning of the pump laser from the
dispersively shifted cavity resonance. In contrast to the thermal
case \eref{eq:threshold_selforg}, the critical transverse pump
power scales in the zero-temperature limit with the recoil
frequency (and the chemical potential), which reflects the fact
that the homogeneous phase is stabilized by the kinetic energy
(and atom-atom collisions).

\begin{figure}[t!]
\begin{center}
\includegraphics[angle=0,width=0.9\columnwidth]{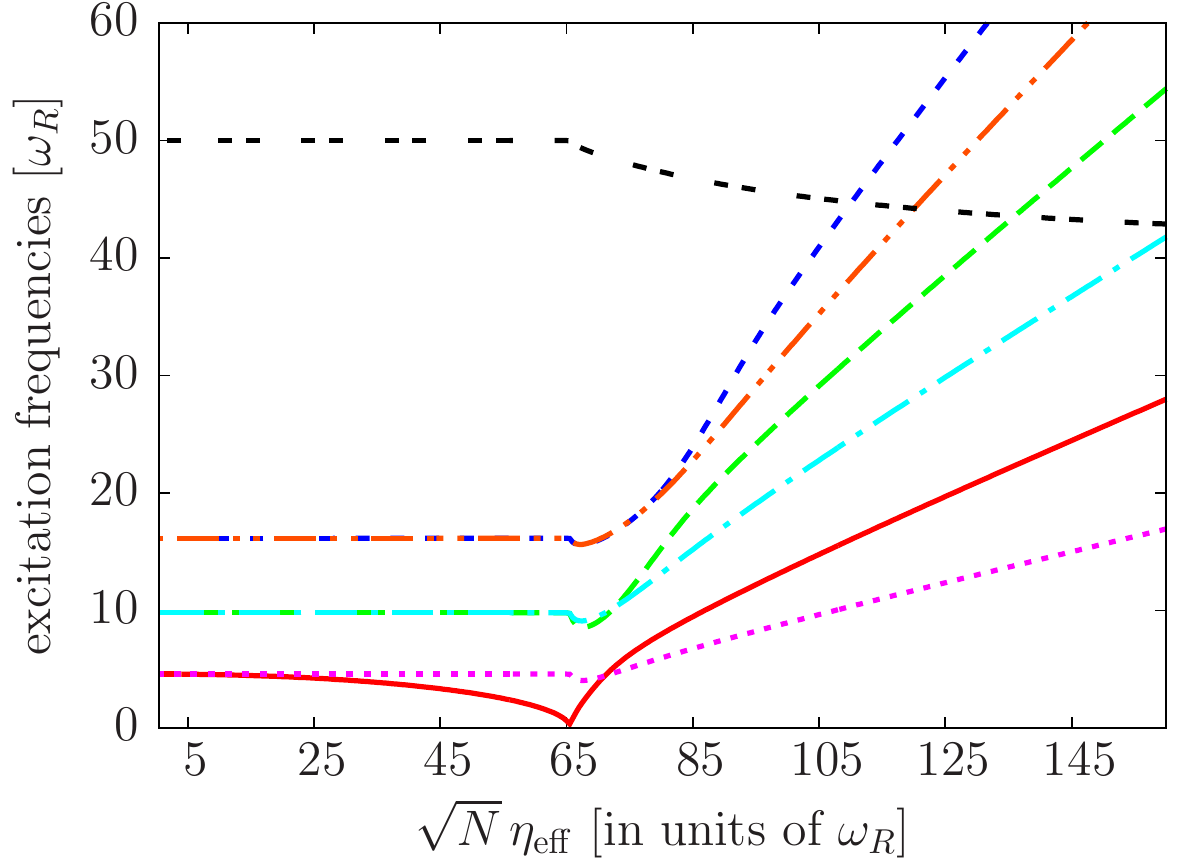}
\caption{Collective excitation spectrum of the transversally
driven condensate-cavity system. Shown are the eigenfrequencies of
the lowest six collective atomic-like and the first cavity-like
(divided by 5) excited states as a function of the transverse pump
amplitude. For vanishing pump amplitude the Bogoliubov spectrum
$\Omega_n = \sqrt{n^2 \omega_R (n^2 \omega_R + 2 \mu_0)}$ for a
condensate in a 1D box potential of size $\lambda$ is retained.
Self-organization is indicated by the softening of the lowest
lying collective mode towards the critical pump amplitude
$\sqrt{N} \eta_\mathrm{eff,c} \approx 65.5 \omega_R$. Parameters
are the same as in Fig.~\ref{fig:Nagy2008Fig1}. From
\textcite{Nagy2008Selforganization}. \label{fig:Nagy2008Fig4}}
\end{center}
\end{figure}
A deeper understanding of the process of self-organization is
gained from the collective excitation spectrum on top of the
steady-state mean-field solution. This was calculated in
\textcite{Nagy2008Selforganization,Konya2011Multimode} using a
Bogoliubov-type approach based on the separation ansatz
Eq.~\eqref{eq:MeanField}. The eigenvalues of the linearized
equations for condensate and cavity fluctuations,
Eq.~\eqref{eq:LinearFluctuations}, yield the energy spectrum of
excitations (polaritons) shown in Fig.~\ref{fig:Nagy2008Fig4}. For
the considered case where the pump-cavity detuning $\delta_C$ is
large compared to the recoil frequency $\omega_R$, the excitations
separate into two classes, according to whether they are
dominantly cavity-like or atom-like. The occurrence of
self-organization is recognized in a characteristic softening of
the atom-like excitation mode which matches the spatial
interference pattern between cavity and transverse pump mode
(Fig.~\ref{fig:Nagy2008Fig4}, red solid line).

\begin{figure}
\begin{center}
\includegraphics[angle=0,width=0.9\columnwidth]{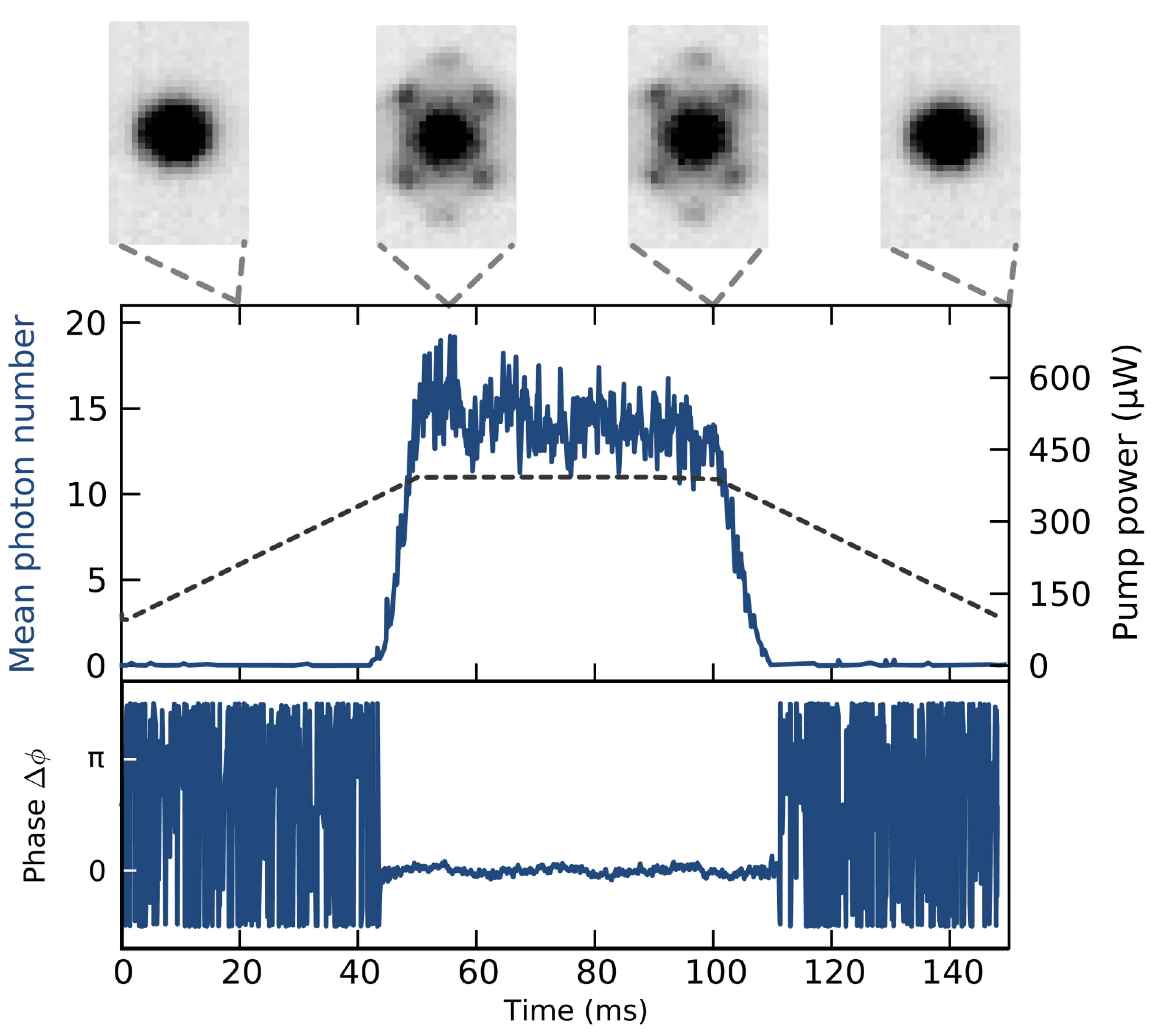}
\caption{Observation of self-organization with a Bose-Einstein
condensate. Simultaneous time traces of the mean-intracavity
photon number (middle panel) and the relative pump-cavity phase
$\Delta \phi$ (lower panel) while ramping the transverse pump
power twice across the critical point at $\approx
\unit[0.35]{mW}$. The absorption images (upper panel) show the
atomic momentum distribution for the indicates times. The line of
sight is perpendicular to the pump-cavity plane. Parameters are
$\Delta_C = -2\pi \times \unit[20]{MHz}$, $\kappa = 2\pi \times
\unit[1.3]{MHz}$ and $N = 10^5$. Adapted from
\textcite{Baumann2010Dicke, Baumann2011Exploring}.
\label{fig:Baumann2010fig3} }
\end{center}
\end{figure}

Self-organization of a Bose-Einstein condensate was observed by
the Zurich group \cite{Baumann2010Dicke}. A BEC of about $10^5$
atoms, harmonically confined inside a high-finesse optical
Fabry-P\'erot resonator, was illuminated by a far red-detuned
standing-wave laser beam. By gradually increasing the power of the
transverse laser beam, the transition to the self-organized phase
was observed in a sharp raise of the intracavity light intensity
accompanied by the build-up of macroscopic populations in the
momentum states $(p_x, p_z) = (\pm\hbar k, \pm\hbar k)$, see
Fig.~\ref{fig:Baumann2010fig3}. Above the critical pump power, the
relative phase $\Delta\phi$ between pump field and cavity field is
observed to stay constant, which demonstrates that the system
reached a steady state. By controlling the transverse pump power,
the system can be transferred repeatedly from the normal in the
self-organized phase and back (Fig.~\ref{fig:Baumann2010fig3}).

\begin{figure}
\begin{center}
\includegraphics[angle=0,width=0.9\columnwidth]{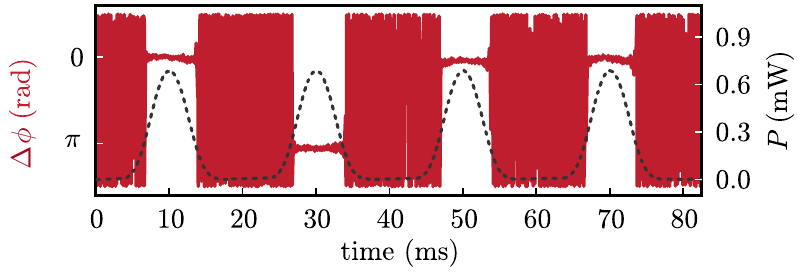}
\caption{Observation of symmetry breaking at the self-organization
transition with a BEC. Shown is in red the relative pump-cavity
phase $\phi$ monitored on a heterodyne detector while repeatedly
entering the self-organized phase by tuning the transverse pump
power $P$ (dashed). The system organizes into one out of two
possible checkerboard patterns corresponding to the two observed
phase values differing by $\pi$. From \textcite{Baumann2010Dicke}.
\label{fig:Baumann2011fig2}}
\end{center}
\end{figure}

The process of symmetry breaking at the transition point was
studied in \textcite{Baumann2011Exploring}. In repeated
realizations of the self-organized phase, two possible values of
the relative phase $\Delta \phi$ with a difference of $\pi$ were
observed, according to self-organization into either the even
($u(x,z)>1$) or odd ($u(x,z)<1$) sites of the underlying mode
interference mode profile $u(x,z) = \cos(kx)\cos(kz)$
(Fig.~\ref{fig:Baumann2011fig2}). The finite spatial extent of the
atomic cloud results in a tiny imbalance between the even--odd
populations in the non-organized phase. This effectively acts as a
symmetry breaking field, which favors the realization of one
particular organized pattern, as was observed in the experiment.
The influence of the symmetry breaking field could be overcome by
increasing the speed at which the transition is crossed, in
accordance with a simple model description based on the
adiabaticity condition \cite{Baumann2011Exploring}.

\begin{figure}
\begin{center}
\includegraphics[angle=0,width=0.9\columnwidth]{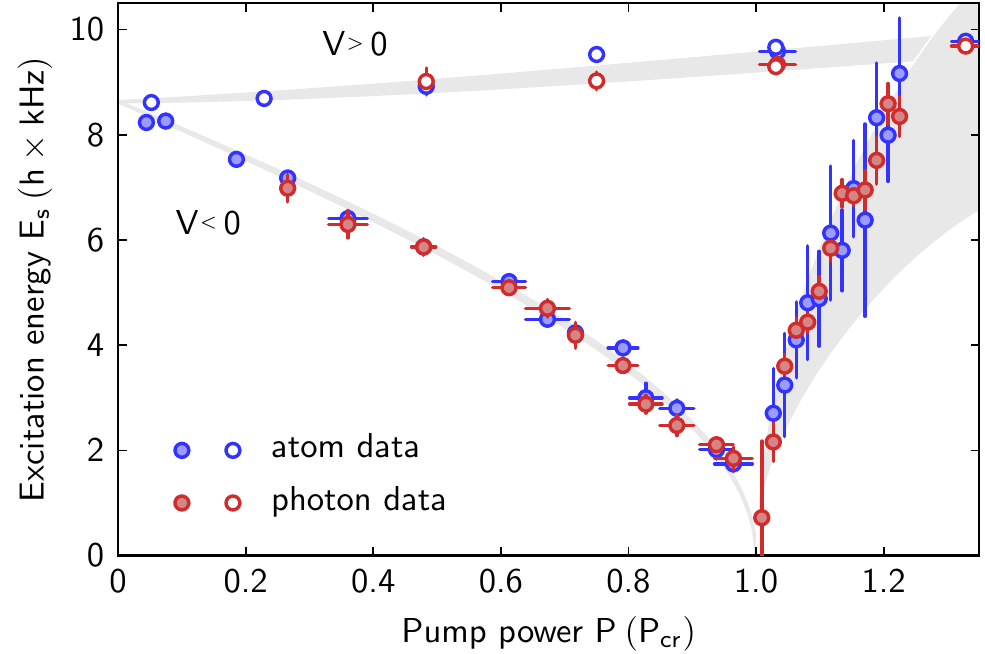}
\caption{(Color online) Observation of mode softening induced by cavity-mediated
atom-atom interactions in a Bose-Einstein condensate. Shown is the
motional atomic excitation energy at momenta $(\pm \hbar k, \pm
\hbar k)$ along the cavity and pump direction as a function of the
transverse laser power $P$ , which sets the modulus $|V|$ of the
cavity-mediated atom-atom interaction. The sign of $V$ is
determined by the sign of $\delta_C$. For negative interaction
strength $V$, the system organizes at the critical pump power
$P_\mathrm{cr}$, while for positive interaction an increased
excitation energy is observed in accordance with the absence of a
phase transition. From \textcite{Mottl2012RotonType}.
\label{fig:Mottl2012Fig3}}
\end{center}
\end{figure}

In the limit where the cavity field adiabatically follows the
atomic motion, the process of self-organization can also be
understood as a result of the cavity-mediated atom-atom
interactions, see \ref{sec:cavitymediatedatomatomint}. On a
microscopic level, these are induced by the virtual exchange of
cavity photons between different laser-driven atoms, accompanied
by the creation of atom pairs recoiling with momentum $\hbar k$
along the pump and cavity direction. The resulting
$\lambda$-periodic density correlations in the atomic cloud
energetically compete with the cost in kinetic energy, which gives
rise to a characteristic roton-type softening in the dispersion
relation of the condensate at momenta $(\pm \hbar k, \pm \hbar
k)$, see also Fig.~\ref{fig:Nagy2008Fig4}. Once the softened
excitation energy reaches the ground state energy, the system
self-organizes by macroscopically occupying those momentum states.
Such mode softening was observed by the Zurich group
\cite{Mottl2012RotonType} using a variant of Bragg spectroscopy
\cite{Stenger1999Bragg} where the cavity field was probed with a
weak laser pulse whose frequency was detuned by a variable amount
from the transverse pump laser field. The observed excitation
spectrum as a function of sign and modulus of the cavity-mediated
atom-atom interaction strength $V$ is depicted in
Fig.~\ref{fig:Mottl2012Fig3}. The vanishing of the excitation gap
at the transition point towards the organized phase is accompanied
by a diverging susceptibility of the system to $\lambda$-periodic
density perturbations \cite{Mottl2012RotonType}. As was
theoretically considered by \textcite{Oztop2012Excitations}, the
softened excitation spectrum can also be probed parametrically via
amplitude modulation of the transverse pump laser.

Conceptually, the self-organized BEC can be regarded as a
supersolid \cite{Leggett1970Can,Gopalakrishnan2009Emergent},
similar to those proposed for two-component systems
\cite{Buchler2003Supersolid}. Non-trivial diagonal long-range
order is induced by the cavity-mediated long-range interactions,
which restricts the periodic density modulation to two possible
checkerboard patterns, in contrast to traditional optical lattice
experiments with laser fields propagating in free space.
Simultaneously, the organized phase exhibits off-diagonal
long-range order which is not destroyed while crossing the phase
transition. Only when deeply entering the organized phase,
tunneling between different sites of the optical checkerboard
potential gets suppressed and phase coherence is lost
\cite{Vidal2010Quantum}.

\subsubsection{Open-system realization of the Dicke quantum phase transition}
\label{sec:Open-SystemDickeQPT}

Self-organization of a laser-driven BEC in an optical resonator
can be considered as an open-system realization of the Dicke
quantum phase transition, where the quantized atomic motion acts
as a macroscopic spin which strongly couples to the cavity field.
The Dicke model goes back to the pioneering work of R.~W.~Dicke
\cite{Dicke1954Coherence} and describes the collective interaction
between matter and the electromagnetic field. Consider $N$
two-level systems with transition frequency $\omega_0$, forming a
collective spin variable $\mathbf{J}$, which couple identically to
a single resonator mode at frequency $\omega$. This system can be
described in terms of the Dicke Hamiltonian (also referred to as
the Tavis-Cummings model \cite{Tavis1968Exact})
\begin{equation}\label{eq:DickeHamiltonian}
H_\mathrm{Dicke}/\hbar = \omega a^\dag a + \omega_0 J_z +
\frac{\lambda}{\sqrt{N}}(J_+ + J_-)(a + a^\dag)
\end{equation}
with the collective coupling strength denoted by $\lambda \propto
\sqrt{N}$. The ladder operators $J_\pm = J_x \pm i J_y$ describe
creation and annihilation of collective atomic excitations.

According to \textcite{Dicke1954Coherence}, a collectively excited
medium, which carries correlations among the different atomic
dipoles, decays within a much shorter time into its ground state
than a single atom. This phenomenon, termed superradiance (or
superfluorescence), originates from spontaneous phase-locking of
the different radiators resulting in a short radiation burst whose
intensity is proportional to the number of atoms squared.
Superradiant emission of laser-excited media has been studied
extensively in the past \cite{Gross1982Superradiance}.

In contrast to this transient phenomenon, the Dicke Hamiltonian
Eq.~\eqref{eq:DickeHamiltonian} was shown in 1973 to exhibit also
a ground-state version of superradiance
\cite{Hepp1973Superradiant,Wang1973Phase,Carmichael1973Higher,Lambert2004Entanglement}.
When the collective coupling strength $\lambda$ reaches the
critical value $\lambda_\mathrm{cr} = \sqrt{\omega \omega_0}/2$,
the Dicke model undergoes a quantum phase transition from a normal
into a superradiant phase, which is characterized by a macroscopic
cavity field amplitude $\langle a\rangle$ and a macroscopic
polarization $\langle J_-\rangle$ of the atomic medium. Apart from
its fragility upon the inclusion of the $A^2$ term originating
from the minimal coupling Hamiltonian \cite{Rzazewski1975Phase},
the experimental realization of the superradiant Dicke phase
transition with direct dipole transitions was obscured in the past
due to practical limitations in the available dipole coupling
strengths.

The proposal by \textcite{Dimer2007Proposed} circumvents these
issues by considering a pair of stable atomic ground states which
are coupled via two different Raman transitions involving a single
ring cavity mode and external laser fields. This scheme realizes
the Dicke model through an effective Hamiltonian in an open-system
dynamics, including external driving and cavity loss, where the
critical coupling strength can be reached for realistic
experimental parameters.

The transversally driven BEC-cavity system is formally equivalent
to this proposal upon replacing the electronic atomic states by a
pair of motional atomic states, as was shown in
\textcite{Baumann2010Dicke,Nagy2010DickeModel}. The two motional
states are given by the flat condensate mode $|p_x, p_z\rangle =
|0,0\rangle$ and the coherent superposition of the four momentum
states $|\pm \hbar k, \pm \hbar k\rangle$, where $x$ and $z$
denote the cavity and pump direction, respectively. Coherent light
scattering between the transverse pump beam and the cavity mode
couples these two momentum states via two distinguishable Raman
channels, resulting in a dipole-type interaction between cavity
mode and the corresponding collective spin degree of freedom, see
Eq.~\eqref{eq:DickeHamiltonian}. The parameters $(\omega_0,
\omega, \lambda)$ of the corresponding realization of the Dicke
Hamiltonian are given by the energy difference $2\omega_R$ between
the two momentum modes (neglecting atom-atom collisions), the
effective detuning $-\delta_C$ between the pump laser frequency
and the dispersively shifted cavity mode frequency, and the
collective two-photon Rabi frequency $\sqrt{N}\eta_\mathrm{eff}/2$
between pump laser and cavity mode, respectively. In the
experiment \cite{Baumann2010Dicke}, $\delta_C$ exceeds the recoil
frequency by three orders of magnitude, thus realizing the
dispersive regime of the Dicke model. Higher-order momentum modes
do not contribute in the phase transition dynamics itself and are
populated only when deeply entering the self-organized phase
\cite{Konya2011Multimode}.

\begin{figure}
\begin{center}
\includegraphics[angle=0,width=0.9\columnwidth]{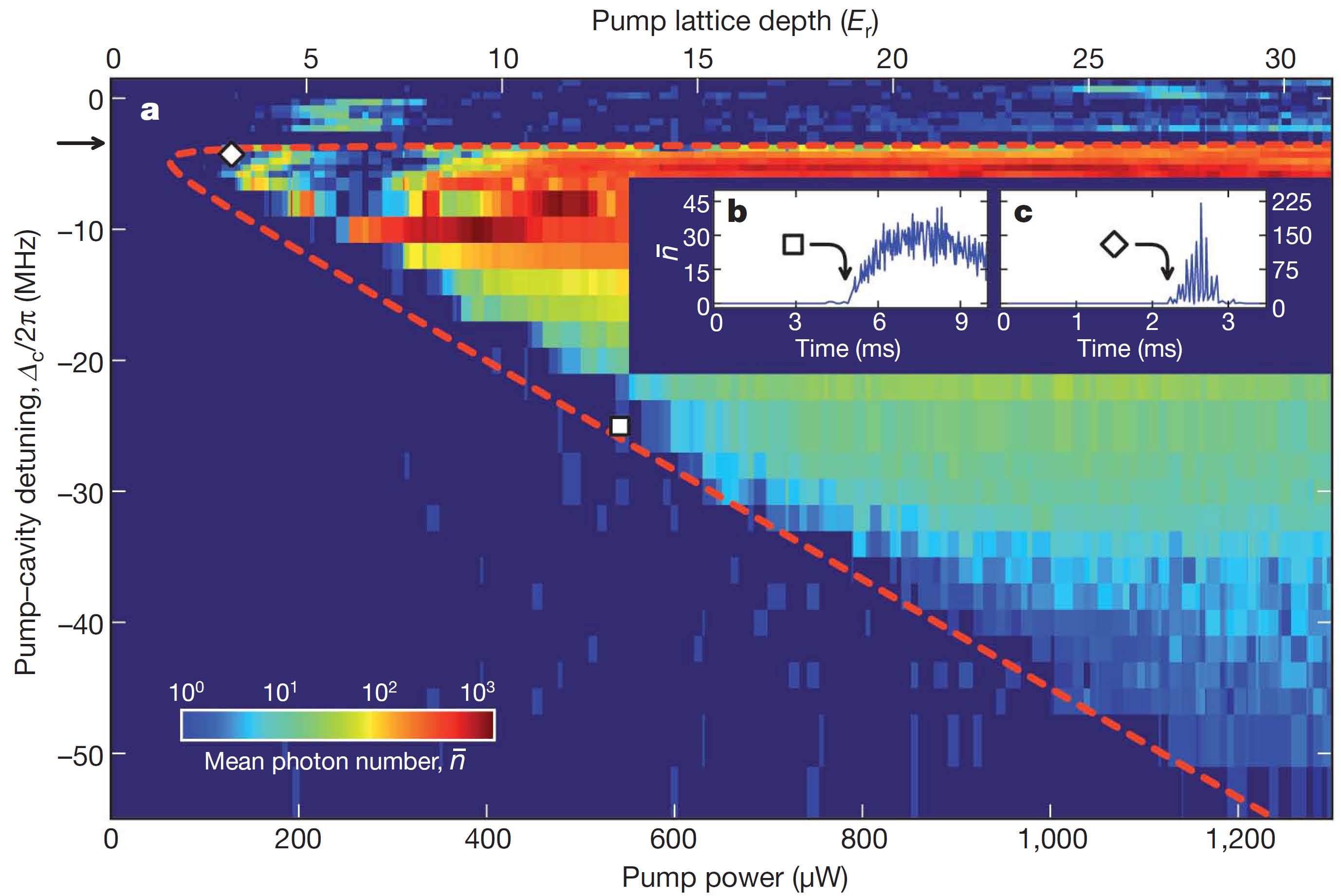}
\caption{(Color online) Dicke model phase diagram. (a) Displayed is in color the
recorded mean intracavity photon number $\bar{n}$ as a function of
the transverse pump power $P$ and the pump-cavity detuning
$\Delta_C$. A sharp phase boundary is observed in agreement with a
mean-field description (dashed line). The dispersively shifted
cavity resonance for the non-organized BEC is indicated by a
horizontal arrow. (b,c) Time traces of $\bar{n}$ while gradually
increasing the pump power to $\unit[1.3]{mW}$ for the indicated
pump-cavity detuning. From \textcite{Baumann2010Dicke}.
\label{fig:Baumann2010fig5}}
\end{center}
\end{figure}

From the analogy to the Dicke model the following expression for
the critical coupling strength is obtained upon including cavity
decay \cite{Dimer2007Proposed}
\begin{equation}
\lambda_\mathrm{cr} =
\sqrt{\frac{(\kappa^2+\omega^2)\omega_0}{4\omega}}\, .
\end{equation}
In the absence of atom-atom collisions, this condition agrees with
the result obtained from the stability analysis of the mean-field
equations, Eq.~\eqref{eq:MFthreshold}. Experimentally
\cite{Baumann2010Dicke}, the phase boundary was mapped out as a
function of pump-cavity detuning $\Delta_C$ in agreement with the
theoretical prediction, see Fig.~\ref{fig:Baumann2010fig5}.

It is instructive to contrast the Dicke quantum phase transition,
realized with a BEC in a single-mode cavity, with the occurrence
of free-space superradiant Rayleigh scattering off an elongated
BEC which is driven by off-resonant laser light
\cite{Inouye1999Superradiant}. In this experiment, a superradiant
light pulse was emitted along the axial direction of the atomic
cloud accompanied by the creation of recoiling matter-wave
components, once the pump intensity exceeded a critical value.
This dynamical effect is equivalent to Dicke superradiance of a
collectively excited medium \cite{Dicke1954Coherence}, where
matter-wave amplification phase-locks the spontaneous emission
events into the continuum of optical field modes. The minimal pump
intensity required for superradiance to occur is determined by the
balance between loss and gain processes. In contrast, light
scattering off the BEC into a single cavity mode is a reversible
process and the critical pump strength dominantly results from the
finite pump-cavity detuning.

In the self-organized phase the detuning $\delta_C$ between pump
laser and the dispersively shifted cavity resonance becomes a
dynamic quantity. This effect is not taken into account by the
description in terms of the Dicke model \eref{eq:DickeHamiltonian}
which is a valid approximation as along as the maximum dispersive
cavity shift $U_0 N$ is small compared to the pump-cavity detuning
$|\Delta_C|$. In the case where $U_0 N$ exceeds $|\Delta_C|$ the
system can exhibit dynamically frustrated behavior characterized
by a periodic sign change of the effective pump detuning from the
dispersively shifted cavity resonance, as was observed in
\textcite{Baumann2010Dicke}, see Fig.~\ref{fig:Baumann2010fig5}c.
Theoretically the influence of the additional nonlinear dispersive
term $\sim U_0 J_z a^\dag a$ appearing in the Dicke model
\eref{eq:DickeHamiltonian} upon the dynamics of the coupled
BEC-cavity system was investigated in great detail by
\textcite{Keeling2010Collective,Bhaseen2012Dynamics,Liu2011Lightshiftinduced}.
Employing a semiclassical description,
\textcite{Bhaseen2012Dynamics} reveal a rich phase diagram
including distinct superradiant fixed points, bistable and
multistable coexistence phase and regimes of persistent
oscillations, and explore the timescales for reaching these
asymptotic states. It is emphasized that the behavior of the open
system is controlled by the stable attractors, which do not
necessarily coincide with the points of minimal free energy. As
such, there is a crucial distinction between the $\kappa
\rightarrow 0$ limit of the dynamical system and the equilibrium
behavior at $\kappa=0$. Similar conclusion can be drawn in the
quantum case, as discussed in the following.

The coupling of the cavity field to the electromagnetic field
environment, causing cavity decay, amounts to a weak measurement
of the coupled BEC-cavity system. The corresponding quantum
back-action results in a diffusion-like depletion of the ground
state of the Dicke Hamiltonian, even at zero temperature. The
underlying physics is similar to that described in section
\ref{sec:QuantumBackaction}, with the important difference that
the system gets increasingly susceptible to quantum back-action
when approaching the critical point.

\begin{figure}
\begin{center}
\includegraphics[angle=0,width=0.4\textwidth]{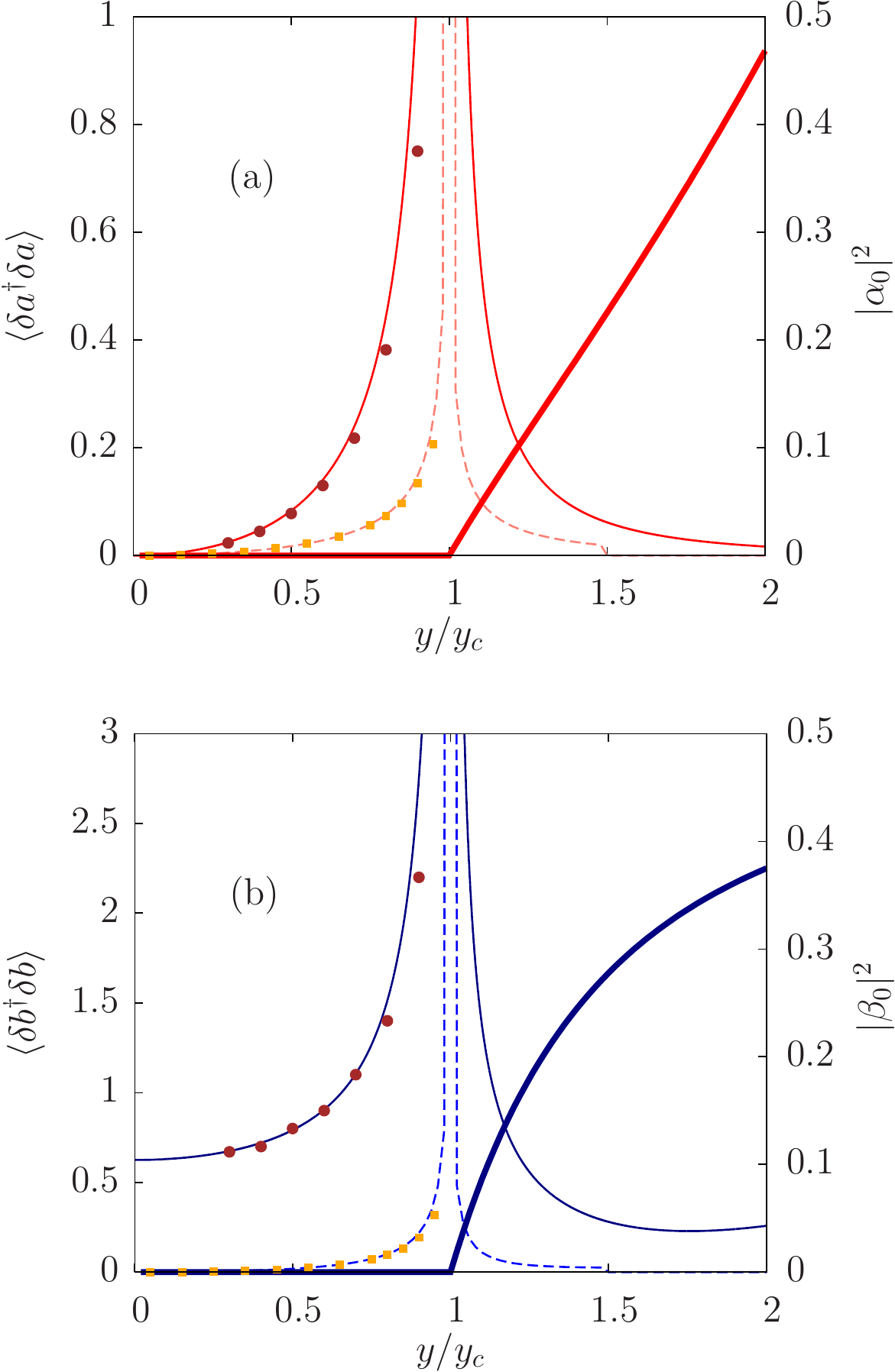}
\caption{(Color online) Criticality in the closed and open-system Dicke phase
transition. The mean values (thick, right axes) and the incoherent
excitation numbers (thin, left axes) of the photon (a) and atomic
(b) fields are plotted as a function of relative coupling strength
$y/y_\mathrm{cr} = \lambda/\lambda_\mathrm{cr}$. The incoherent
excitation numbers in steady-state for the open system (thin
solid) diverge at the critical point with exponent $-1$ in
contrast to the ground-state number of excitations for the closed
system (thin solid), which diverges with the mean-field exponent
$-1/2$. From \textcite{Nagy2011Critical}.
\label{fig:Nagy2011Fig2}}
\end{center}
\end{figure}

The rate at which the ground state of the Dicke Hamiltonian
initially gets depleted due to cavity decay was calculated in
\textcite{Nagy2010DickeModel} based on the Langevin equation
approach \eref{eq:LinearFluctuations}. In the dispersive regime
$|\delta_C|\gg \omega_R$, the ground state depletion happens
mostly in the atomic space and the corresponding diffusion rate
can be approximated below threshold by the expression $\omega_R
\kappa/|\delta_C| (\lambda/\lambda_\mathrm{cr})^2$. Per atom, this
corresponds for $|\delta_C| \gg \kappa$ to a heating rate of
$\kappa \frac{\eta_\mathrm{eff}^2}{\delta_C^2}$. Note, the formal
equivalence of this result with the spontaneous heating rate in a
far-detuned dipole trap. Importantly, the use of a large detuning
$|\delta_C|$ removes the time limitation imposed by
measurement-induced back-action.

Measurement-induced back-action drives the BEC-cavity system into
a steady state which is a dynamical equilibrium between diffusion
and damping. It is very interesting that this limiting state is
not the same as the equilibrium state of the system, i.e., for
$\kappa = 0$ the ground state at $T = 0$. Namely, the order of the
two limiting procedures, $t\rightarrow\infty$ and
$\kappa\rightarrow 0$ cannot be interchanged. The steady-state
($t\rightarrow \infty$ limit) occupation of the cavity field and
the excited momentum state were calculated in
\textcite{Nagy2011Critical}. Flux and second-order time
correlations of the cavity output signal were investigated
theoretically in \textcite{Oztop2012Excitations}. The mean-field
obtained in the thermodynamic limit is a smooth function of
$\kappa$ and the steady-state solution tends to that of the ground
state of \eref{eq:DickeHamiltonian} for $\kappa \rightarrow 0$. By
contrast, the comparison of the quantum fluctuations present in
the ground state of the Dicke model and in the steady-state of the
damped-driven system exhibit a significant difference. In the
ground-state of \eref{eq:DickeHamiltonian} the second-order
correlation functions diverge towards the critical point with the
exponent $-1/2$, indicating a mean-field-type transition, whereas
in the non-equilibrium case the quantum fluctuations exhibit a
divergence with exponent $-1$ (see Fig.~\ref{fig:Nagy2011Fig2}).
At the same time, the singularity of the ground-state entanglement
\cite{Lambert2004Entanglement} between cavity and atomic subsystem
is regularized at the critical point by the quantum noise
associated with cavity decay. The non-vanishing entanglement,
however, shows that the quantum character of the Dicke quantum
phase transition (self-organization at zero temperature)  is not
fully destroyed in case of an open-system dynamics, and it cannot
be exactly mapped to a thermal noise-driven phase transition.

\subsubsection{Phases in highly degenerate cavities}
\label{sec:MultimodeDegenerateCavity}

Self-organization of polarizable particles into periodic
structures induced and stabilized by the intracavity light field
resembles the process of crystallization. In a cavity with only a
single standing-wave mode tuned into resonance with the pump
field, only the amplitude of the cavity field is a dynamical
quantity. The formation of a periodic crystal from a
spatially homogeneous distribution breaks the discrete symmetry
corresponding to the even and odd antinodes of the standing-wave
mode profile. Already in the two-mode setting of a  ring cavity
sustaining two degenerate counter-propagating modes, the
self-organization is accompanied by spontaneous breaking of a
continuous translational symmetry
\cite{Nagy2006Selforganization,Niedenzu2010Microscopic}, which
induces rigidity against lattice deformations. In the case of
highly degenerate multimode cavities the field has much more
freedom to adjust locally to the particle distribution.

\begin{figure}[htbp]
    {\includegraphics[width=0.95\columnwidth]{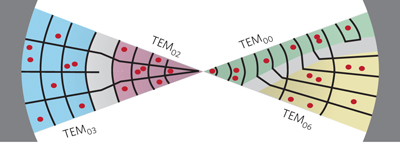}}
\caption{(Color online) Self-ordered states in a concentric multimode cavity
forming two-dimensional patterns. The diagram shows a regime near
threshold, with domains locally populating distinct
$\mathrm{TEM}_{xy}$ cavity modes in the equatorial plane. Domains
can be punctuated by dislocations (shown in the left half of the
figure), but might also show textural variation in space (right
half of the figure). The black lines represent nodes of the cavity
field, which separate 'even' and 'odd' antinodes. As the atoms are
Bose-condensed, the atomic population per site is not fixed. From
\textcite{Gopalakrishnan2009Emergent}.}
  \label{fig:MultimodeDomain}
\end{figure}

The general structure of the resulting complex phase diagram was
studied in
\textcite{Gopalakrishnan2009Emergent,Ritsch2009Crystals}. In
general such setups allow to realize conceptually novel systems
and to explore and discover properties of crystalline and
liquid-crystalline ordering, including intrinsic effects as
dislocations, the growth and arrangement of crystal grain
boundaries (see Fig.~\ref{fig:MultimodeDomain}), and the nature of
the phonon spectrum. Multimode cavities also offer a natural
connection to models developed in the field of neural networks as
the Hopfield model or similar spin models with infinite range
statistical couplings. First ideas about this relation have been
raised very recently in
\textcite{Strack2011Dicke,Gopalakrishnan2011Frustration,Gopalakrishnan2011Exploring}.
Extensions to fermionic atoms in multimode cavities have been
considered in \textcite{Muller2012Quantum}.

Gopalakrishnan and coworkers generalized and adapted a
field-theoretical framework, also successfully used in solid-state
physics, to describe many-body systems coupled to a multitude of
degenerate modes of a high-finesse cavity
\cite{Gopalakrishnan2010Atomlight,Keeling2010Liquid}. For a
quasi-two-dimensional cloud of atoms confined in the equatorial
plane of a concentric optical cavity, the transition from the
homogeneous distribution into a spatially modulated one is of the
Brazovskii type \cite{Brazovskii1975Phase}, which describes the
phase transitions from isotropic to striped structures in liquid
crystals. The description is based on an effective equilibrium
theory which is valid when the effective cavity loss rate $\kappa
\eta_\mathrm{eff}^2/\Delta_C^2$ is smaller than the recoil
frequency $\omega_R^{-1}$, see \ref{sec:Open-SystemDickeQPT}.
Here, the dispersive cavity shift was assumed to be much smaller
than the pump-cavity detuning $\Delta_C$.
Unlike the Landau theory of crystallization, here the free energy
of the system does not have a cubic term that breaks the symmetry
at the phase transition. The transition persists at zero
temperature, hence it realizes a quantum phase transition of an
unusual university class. The non-equilibrium extension of this
theory, which includes the effect of photon leakage out of the
cavity as a perturbation, leads to the conclusion that the photon
loss corresponds to an effective temperature and quantum
correlations are washed out by decoherence on timescales longer
than the cavity decay time.  Note that the Bogoliubov-type mean-field model description of the open-system Dicke-model in a single mode cavity also predicts the depletion of  the ground state
\cite{Nagy2010DickeModel}  due to measurement-induced back-action
\cite{Murch2008Observation}, or, in other words, due to the
diffusion induced by fluctuations accompanying cavity photon loss
\cite{Nagy2009Nonlinear}. However, unlike the Brazovskii transition, the
Dicke-model system is driven into a  steady-state, significantly different from the ground
state, which has a critical point \cite{Nagy2011Critical}.

\begin{figure}[htbp]
    {\includegraphics[width=0.95\columnwidth]{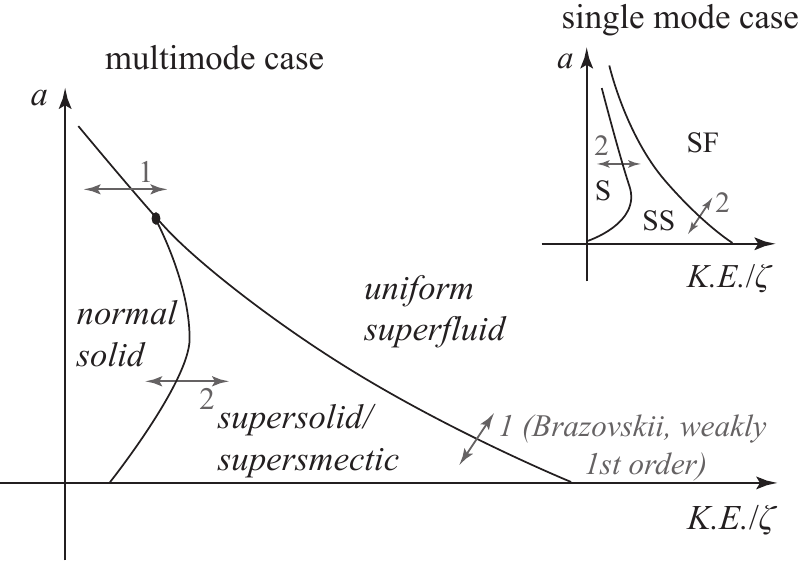}}
  \caption{Schematic zero-temperature phase diagram for a BEC in a
concentric cavity. The control parameters are the atomic
scattering length $a$ and the inverse effective atom-cavity
coupling $\zeta^{-1}$ with $\zeta=\eta_\mathrm{eff}^2/\Delta_C$.
For weak, repulsive interactions and increasing atom-cavity
coupling, the superfluid first undergoes self-organization via the
Brazovskii transition, thus forming a supersolid. If the
transverse laser intensity is increased further, the supersolid
undergoes a transition into a normal solid (i.e., a Mott
insulator). For strong, repulsive interactions, the uniform BEC
can lose phase coherence concurrently with a first-order
self-organization transition. This situation is to be contrasted
with that for the case of a single-mode cavity (inset), in which
there should always be a supersolid (SS) region separating the
uniform fluid (SF) and normal solid (S) regions. First- and
second-order transitions are marked by 1 and 2, respectively. From
\textcite{Gopalakrishnan2010Atomlight}.}
  \label{fig:MultimodePhaseDiagram}
\end{figure}

Similarly to the case of the single-mode experiment performed in
the Zurich group \cite{Baumann2010Dicke}, the emergent crystalline
state in a transversally driven multimode cavity can be considered
as a supersolid phase where crystalline order and off-diagonal
long-range order (long-range phase coherence) coexist. The phase
diagram, schematically shown in Fig.\
\ref{fig:MultimodePhaseDiagram} for a multimode cavity is
strikingly different from the single-mode case in that a region
with direct uniform superfluid-to-normal phase transition occurs,
whereas in the single-mode cavity there is always a supersolid
state between the uniform and the normal solid phases
\cite{Vidal2010Quantum}. It is also observed that, for a strongly
layered three-dimensional structure, the inter-layer frustration
precludes global ordering and the system breaks  up into
inhomogeneous domains.

\subsection{Extended Hubbard-type models for ultracold atoms in cavities}
\label{sec:TrappedUltracold}

The theoretical description of the quantum many-body dynamics of
ultracold atoms confined in optical lattices and strongly
interacting with a quantized cavity field can be based on a
sophisticated extension of the Bose-Hubbard (BH) model
\cite{Fisher1989Boson}. In static optical lattices, the BH model
properly accounts for the quantum statistical properties of
bosonic atoms at the lattice sites, as well as the inter-particle
quantum correlations \cite{Bloch2008Manybody}. The basic
assumption, valid in the limit of very low temperature, is that
the dynamics can be restricted to the lowest (or lowest few) Bloch
bands of the periodic optical potential. Correspondingly, the
many-body wavefunction can be expressed in terms of Wannier
functions localized at individual lattice sites.

However, if the optical lattice potential is sustained by
the mode of a high-finesse cavity, thus becoming a dynamical degree of freedom, it gets a highly nonlinear
problem to determine the Wannier functions themselves and thereby
the ground state of the many-body system. For a laser-driven
cavity, e.g., the atoms dispersively shift the cavity resonance,
thus affecting the intracavity field amplitude which itself
determines the lattice depth. Hence, the optical
lattice potential and the state of the atoms have to be evaluated
in a self-consistent way
\cite{Maschler2005Cold,Maschler2008Ultracold,Larson2008MottInsulator,Vidal2010Quantum},
as will be presented below.  We focus on the most studied particular case of
spinless bosons, and mention only that the  cases of fermions and spin
particles are expected to lead to interesting novel effects \cite{Larson2008Cold,Sun2011Dynamics}.

\subsubsection{Bose-Hubbard model with cavity-mediated atom-atom interactions}

Consider an ensemble of $N$ bosonic particles subject to an
optical lattice potential which is generated by the field of an
optical resonator and possibly by an additional, far off-resonant
standing-wave laser field. The latter is represented by the
external potential term $V_\sub{cl}(\mathbf{r})$ in the
single-atom Hamiltonian \eref{eq:H_A1}. Restricting the motional
dynamics to the lowest energy band (lowest vibrational state), we
expand the atomic field operator
\begin{equation}
\label{eq:Wannier}
\Psi({\bf r})=\sum_{i=1}^{M}{b_i w({\bf r}-{\bf r}_i)}\,,
\end{equation}
in the Wannier basis of atomic states localized at sites
$i=1\ldots M$, where $b_i$ denotes the associated annihilation
operator. Upon inserting this expansion into
Eqs.~(\ref{eq:ManyBodyH}), one obtains
\begin{multline}
\label{eq:CavityBH} H= \sum_m \left(-\hbar\Delta_{C,m} a^\dag_m
a_m + i \hbar
\eta_m\, (a^\dagger_m - a_m)\right) \\+\sum_{i,j=1}^M{\left( E_{i,j} + V_\sub{cl} J_{i,j}^{\text {cl}} \right) b_i^\dag b_j}\\
+ \frac{\hbar}{\Delta_{A}} \sum_{l,m}{g_l g_m a^\dag_l
a_m}\left(\sum_{i,j=1}^M{J_{i,j}^{lm}b_i^\dag
b_j}\right) \\
+\frac{U}{2}\sum_{i=1}^M{b_i^\dag b_i(b_i^\dag b_i-1)}\,,
\end{multline}
where several cavity modes with mode functions $f_m({\bf r})$
and corresponding photon annihilation operators $a_m$ are considered. The
coefficients $E_{i,j}$ and $J_{ij}^{\text {cl}}$ are defined as in
the standard BH Hamiltonian
\begin{subequations} \label{eq:BHcoefficients}
\begin{align}
E_{i,j} &=  \int{d^3{\bf r}}\,w({\bf r}-{\bf r}_i)\, \left(- \frac{\hbar^2\nabla^2}{2m}\right) \, w({\bf r}-{\bf r}_j)\,, \\
J_{i,j}^{\text {cl}} &= \int{d^3{\bf r}}\,w({\bf r}-{\bf r}_i) \,
f_{\text {cl}} ({\bf r}) \,w({\bf r}-{\bf r}_j)\,,
\end{align}
where we separated the characteristic magnitude $V_\sub{cl}$ of
the classical trapping potential (difference between maximum and
minimum) from its  spatial form $f_{\text {cl}}({\bf r})$. The
last term of Eq.~(\ref{eq:CavityBH}) describes the on-site
interaction with $U= ({4\pi a_s\hbar^2}/{m}) \int{d^3{\bf
r}|w({\bf r})|^4}$. Of primary interest are the extra couplings
generated by the cavity modes with matrix elements
\begin{equation}\label{eq:JCav}
J_{i,j}^{lm}=\int{d^3{\bf r}}\,w({\bf r}-{\bf r}_i) f_l^*({\bf r})
f_m({\bf r})  w({\bf r}-{\bf r}_j)\,.
\end{equation}
\end{subequations}
From the cavity field point of view, the diagonal elements, $l=m$,
correspond to the atomic state-dependent dispersive shifts of the
cavity mode frequency $\omega_{C,m}$, whereas the off-diagonal
elements, $l \neq m$, describe photon scattering between different
cavity modes. The Wannier functions $w({\bf r}-{\bf r}_i)$
appearing in these integrals, in principle depend on the dynamic
potential terms generated by the cavity field. This renders the
problem highly non-trivial.

In the most general case, the Wannier functions have to be
calculated for each photon number state to define a corresponding
manifold of parameters in the BH model, \eref{eq:CavityBH}. In
other words, the couplings $J_{i,j}^{lm}$ are replaced by
operators which can be easily expressed in a Fock basis. Such a
brute force approach is necessary if the effect of the cavity
field on the trapping potential is significantly different for
adjacent Fock states \cite{Horak2001Manipulating}. Typically,
numerical simulations have to be performed to study, e.g.,
microscopic processes underlying many-body effects that are
understood in the mean-field limit
\cite{Vukics2007Microscopic,Maschler2007Entanglement,Niedenzu2010Microscopic}.
Obviously, this approach is limited to small system sizes of a few
particles moving in a few cavity modes. Most of the works,
however, used approximations to treat the cavity-generated optical
potential in which the localized Wannier functions are defined in
a self-consistent manner.

\subsubsection{Cavity-enhanced light scattering for quantum measurement and preparation}
\label{sec:TrappedUltracoldQND}

Before addressing the problem of dynamical cavity-induced
potentials within the framework of the BH model, we note that a
lot of applications have been developed based on the coupling of
quantized cavity field modes to trapped, ultracold atomic systems
in a simple scattering regime, as was exhaustively reviewed by
\textcite{Mekhov2012Quantum}. In the scattering scenario, the
external lattice potential $V_\sub{cl}(\vecr)$ is taken strong
enough to define solely the localized Wannier functions, and their
modification due to the cavity light forces is negligible. The
quantized cavity field modes are a perturbative probe which can
yield a mapping between quantum properties of atomic many-body
states and light observables.  This system gives, for example, a
means to determine the quantum state of ultracold atoms by light
scattering \cite{Miyake2011Bragg}.

\begin{figure}
\includegraphics[width=0.95\columnwidth]{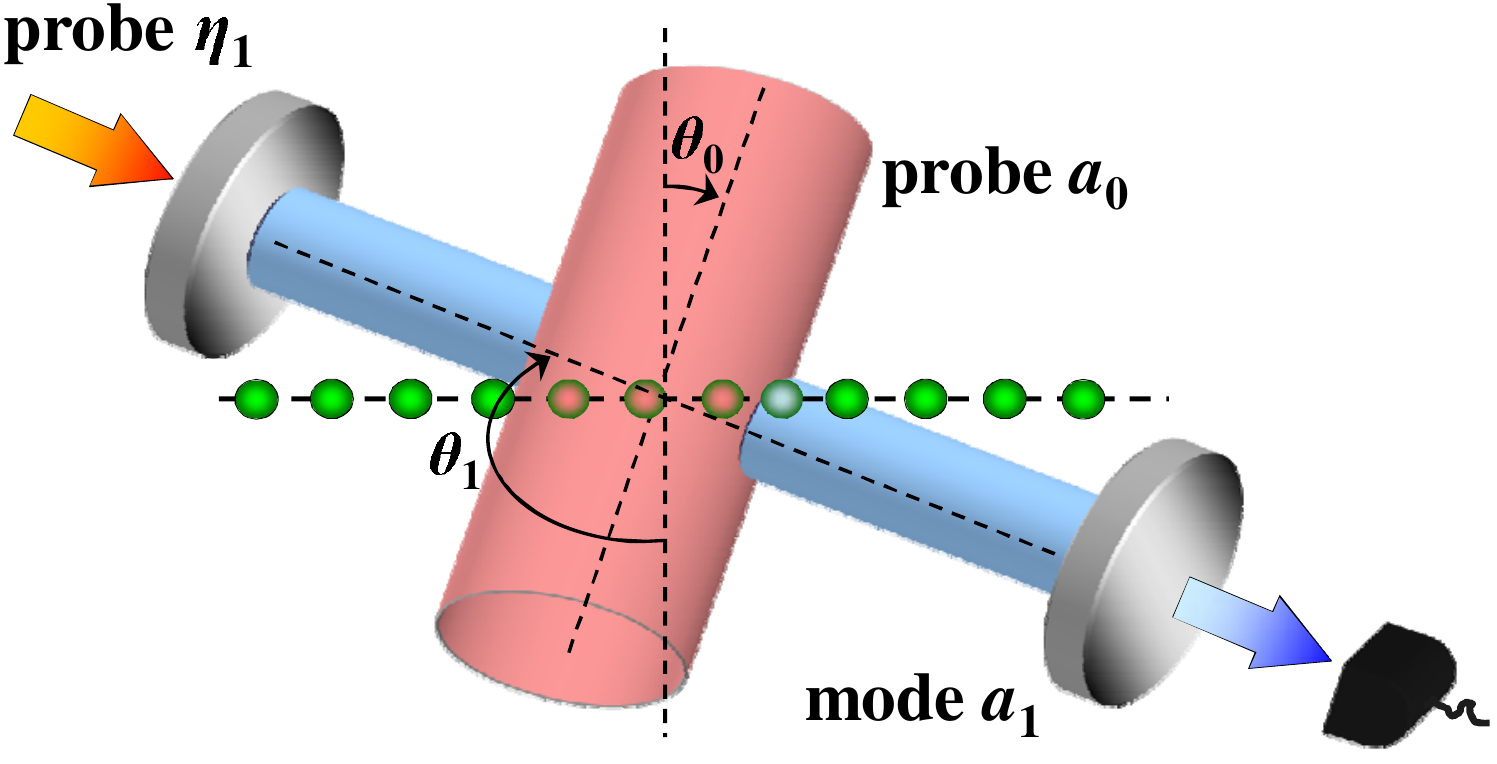}
\caption{(Color online) Scheme for quantum non-demolition
measurement of atomic many-body states in an optical lattice. $N$
atoms are trapped in a one-dimensional lattice potential (green,
$M$ lattice sites) which partially overlaps with a cavity mode
$a_1$ (blue) and a transverse probe mode $a_0$ (red). The number
of illuminated lattice sites is denoted by $K$. Depending on their
many-body state the atoms act as a quantum refractive index whose
statistical distribution with respect to the probe and cavity
modes can be mapped out via transmission or diffraction
spectroscopy as a function of the probe-cavity detuning or the
angles $\Theta_0$ and $\Theta_1$.
} \label{fig:QNDscattering}
\end{figure}

\begin{figure}
{\includegraphics[width=0.95\columnwidth]{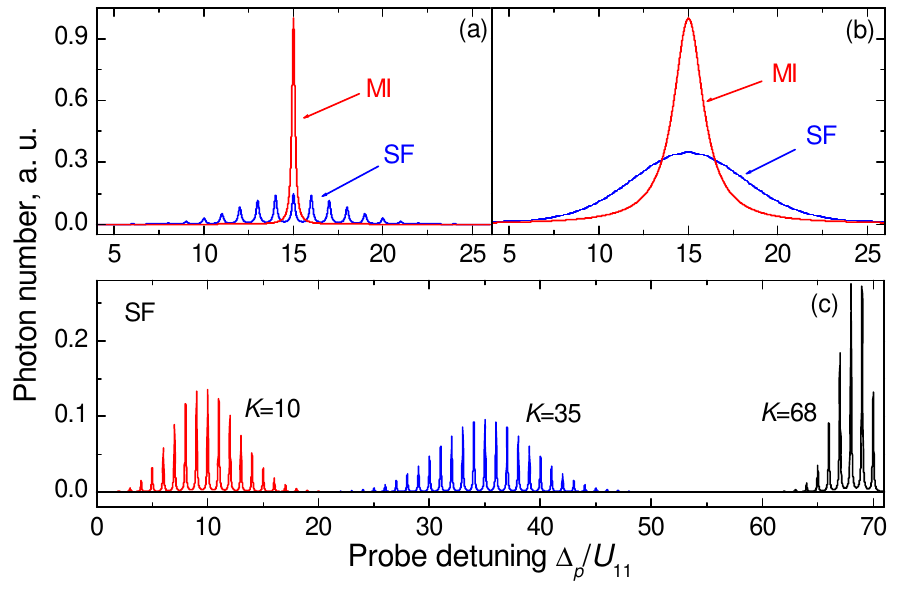}} \caption{(Color online) Cavity
transmission spectra showing the atom number distribution of an
ultracold gas in the intracavity part of an optical lattice, see
Fig.~\ref{fig:QNDscattering}. Shown are transmission profiles
as a function of the probe-cavity detuning $\Delta_p$ of a MI
(red) and a SF (blue) state for (a) a good cavity with $\kappa
=0.1 U_{11}$ and (b) a bad cavity with $\kappa = U_{11}$, where $U_{11}=g_1^2/\Delta_A$. In (b)
the satellites are not resolved but the spectra for SF and MI
states are still different. Parameters are $N = M = 30$ and
$K=15$. (c) Spectra for a SF state with $N = M = 70$ and different
number of illuminated lattice sites $K$.  From
\cite{Mekhov2007Probing}.} \label{fig:cavtransmission}
\end{figure}

A typical quantum measurement scheme involving a single cavity
mode is depicted in Fig.~\ref{fig:QNDscattering}.
It was shown that various quantum states of ultracold bosons
trapped in the lowest band of an optical lattice and having equal
mean densities can be distinguished
\cite{Mekhov2007CavityEnhanced,Chen2007Cavity}. As a
characteristic example, the very different transmission spectra of
the Mott insulator (MI) and the superfluid (SF) states are
exhibited in Fig.~\ref{fig:cavtransmission}. In contrast to
standard techniques this measurement is non-destructive, limited
only by quantum measurement back-action.
Depending on the chosen geometry, light scattering is sensitive to the global and local atom number fluctuations
\cite{Chen2009Cavity,Bhattacherjee2010Probing,Mekhov2007Probing}, or to long-range correlations
between two or more lattice sites (e.g. four-point correlations) \cite{Bux2011CavityControlled,Mekhov2007Light}.

The back-action of repeated quantum non-demolition measurements
drives the atomic many-body state towards specific states
\cite{Mekhov2009Quantum}. This is in full analogy to microwave
cavity QED experiments
\cite{Brune1990Quantum,Guerlin2007Progressive} where,
complementarily, the field state is driven into nonclassical Fock
states by measuring the state of a train of atoms crossing the
cavity.

\subsubsection{Self-consistent Bose-Hubbard models in cavity mean-field approximation}

Genuine cavity-induced dynamical effects appear when the excited
cavity modes significantly modify the trapping with respect to the
external potential $V_\mathrm{cl}(\mathbf{r})$. Even without
noticeably reshaping the Wannier-functions at the trapping sites,
the perturbative light probe can modify the tunneling rates as was
shown for free-space Bragg scattering by \textcite{Rist2010Light}.
In a cavity, local changes of the atomic distribution influence
the whole cavity-sustained optical lattice potential. Thereby a
new type of long-range interaction between the particles appears
and gives rise to resonant nonlocal co-tunneling or momentum space
pairing \cite{Mekhov2007Light}, effects which go far beyond the
standard Bose-Hubbard model.

When the Wannier functions themselves are dynamically influenced
by the cavity, a self-consistent mean-field approach, similar to
the one in Sec.~\ref{sec:ManyBodyTheory}, has been broadly adopted
to describe the nonlinear dynamics of trapped, ultracold atoms in
a cavity
\cite{Maschler2005Cold,Maschler2008Ultracold,Larson2009Dilute,Larson2008Quantum,Vidal2010Quantum,Chen2009Bistable,Nimmrichter2010Master}.
Splitting the cavity field amplitude $a=\langle a \rangle +\delta
a$ into its mean value and fluctuations, the main assumption is
that only the highly excited mean field can modify the trapping
potential, whereas the fluctuations amount to a perturbative
probe. Since the cavity mean-field amplitude $\langle a \rangle$
depends on the momentary atomic quantum state, so does the depth
and shape of the on-site potential. Therefore, the Wannier
functions in \eref{eq:Wannier} and hence the coefficients
\eref{eq:BHcoefficients} in the Hubbard-type Hamiltonian
\eref{eq:CavityBH}, have to be determined self-consistently in
conjunction with the proper cavity mean field $\langle a \rangle$.
It is noteworthy that, often, the self-consistent calculation does
not lead to a unique solution.

%
\paragraph{Phases in dynamical optical lattices}

Consider a linear cavity with only a single mode being driven and
overlapping with a static optical lattice potential
$V_\sub{cl}({\bf r})$. Assuming only nearest-neighbor hopping to
be relevant we keep adjacent $\langle i,j \rangle$ pairs from the
double sum over indices $i$ and $j$ in \eref{eq:CavityBH}.  The
corresponding many-body Hamiltonian is given by
\begin{multline} \label{eq:BHhamCP}
H = E_0 \hat{N}+E\hat{B}+\left( \hbar U_0 a^\dag
a+V_{\textrm{cl}}\right)\left( J_0\hat{N}+ J\hat{B}\right) \\
-\hbar\Delta_C a^\dag a-i\hbar\eta\left(
a-a^\dag\right)+\frac{U}{2}\hat{C}.
\end{multline}
The relevant atomic degrees of freedom are the total atom number
$\hat{N}$ and the collective nearest-neighbor coherence $\hat{B}$ defined as
\begin{equation}
\hat N = \sum_{i=1}^M{b^\dag_i b_{i}} \,,\quad
\hat{B}=\sum_{i=1}^M{b^\dag_i b_{i+1}} +  {b^\dag_{i+1} b_i}\,,
\end{equation}
respectively, and the operator $\hat{C}=\sum_i b_i^\dag
b_i\left(b_i^\dag b_i-1\right)$ for the two-body on-site
interaction. The coefficients $E_0$, $E$, $J_0$ and $J$ derive
from \eref{eq:BHcoefficients} contracted to a single cavity mode
and assuming uniform coupling along the lattice.

To exhibit the underlying physics one may neglect
the photon number dependence of the Wannier functions and
adiabatically eliminate the cavity field via the Heisenberg
equation of motion
\begin{equation}\label{eq:heisenbergCP}
\dot{a}=\left\{i\left[
\Delta_C-U_0\left(J_0\hat{N}+J\hat{B}\right)\right]-\kappa\right\}a+\eta
\,.
\end{equation}
This is a good approximation as long as the cavity field decays
fast compared to the timescale of atomic motion. Since tunneling
in deep lattice potentials is slow compared to the recoil
frequency, this applies widely in experimental setups.

An effective atomic Bose-Hubbard model, formally identical to the
usual one but with coefficients $J_0$ and $J$ depending on the
many-body state, has been considered systematically \emph{in the
thermodynamic limit}. By evaluating the stability of the Mott
insulator states,  the phase diagram has been constructed.
\begin{figure}
\includegraphics[width=0.75\columnwidth]{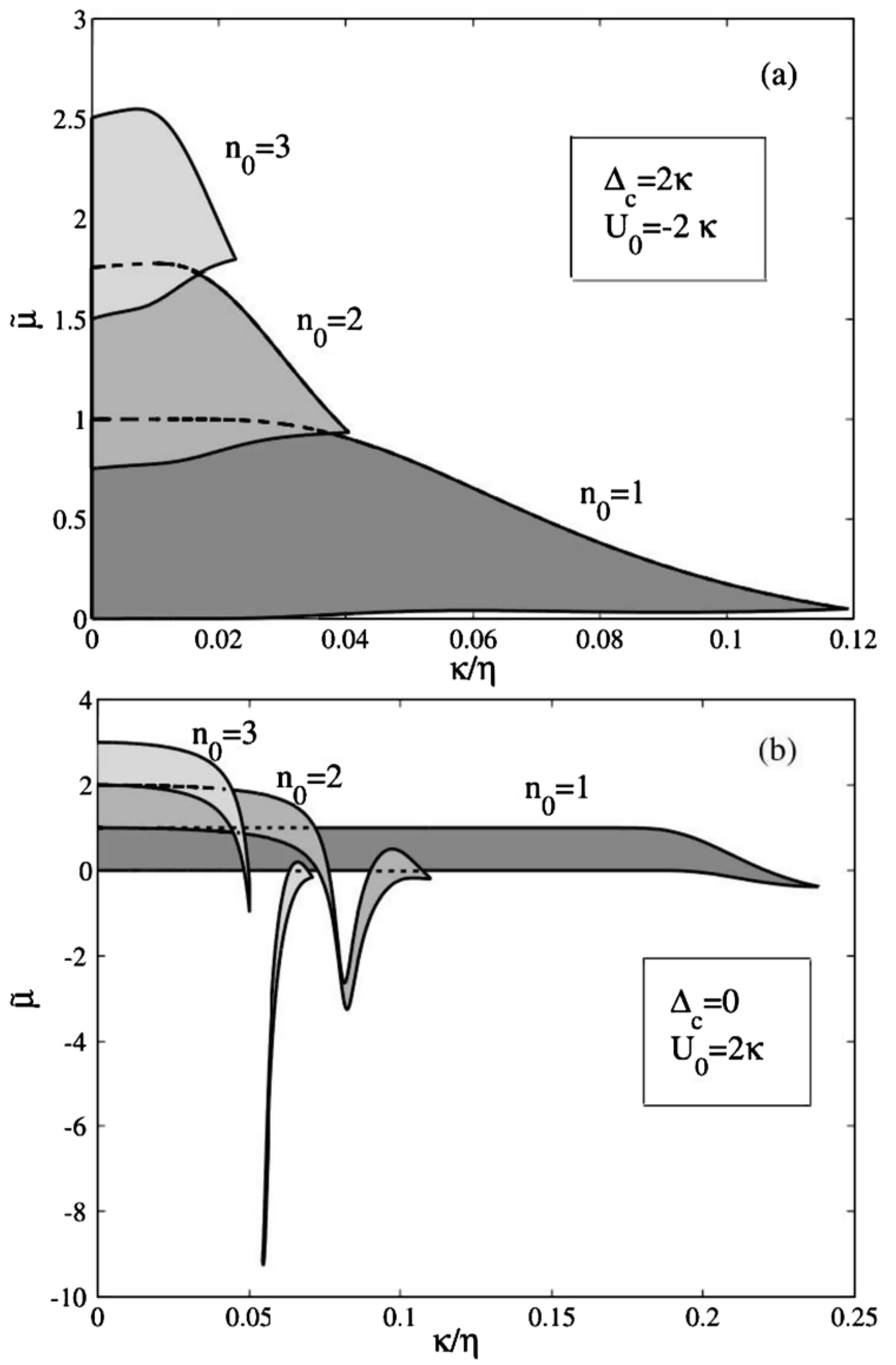}
\caption{Phase diagram with
overlapping Mott insulator states. Boundaries of different Mott
lobes (shaded regions) as a function of the rescaled chemical
potential $\tilde \mu$ and the inverse of the pump strength $\eta$
(in units of $\kappa$) in the 1D cavity lattice potential of $K
=50$ sites. Parameters are (a) $(\Delta_C, U_0) = (2 \kappa, -2
\kappa)$ and (b) $(\Delta_C, U_0) = (0, 2\kappa)$. The Mott lobes
are labeled by the number of atoms per site $n_0$. The dashed
lines show the boundaries of zones which are hidden. Outside
the shaded parameter regions, the state of the system is
superfluid in most cases. From
\cite{Larson2008MottInsulator}.}
\label{fig:MottOverlap}
\end{figure}
As shown in Fig.~\ref{fig:MottOverlap}, the model predicts the
existence of competing Mott insulator states
\cite{Larson2008MottInsulator,Larson2008Quantum}. The overlapping
Mott lobes indicate the possibility of bistability in this
laser-driven, nonlinear system, see \ref{sec:Bistability}. The
state of the system can be controlled by fine tuning the pump
parameters near the shifted cavity resonance. For certain
parameters a state with two atoms per site can lead to a much
higher photon number and thus deeper optical potential, so that
its energy falls below that of the state with unity filling.

In order to gain insight into the nature of atom-atom coupling via the cavity field, a simple effective Hamiltonian can be constructed. The adiabatic field amplitude
can be expanded to second order in the small tunneling matrix element $J$
\begin{equation}\label{eq:elimfield}
a\approx\frac{\eta}{\kappa-i\delta_C}\left[\mathbf{1}-i
\frac{U_0J}{\kappa-i\delta_C}\hat{B}-\frac{(U_0J)^2}{(\kappa-i\delta_C)^2}\hat{B}^2\right]\,,
\end{equation}
where the effective detuning $\delta_C=\Delta_C - U_0 J_0 N$ was
introduced, and the atom number was set to $N$. Inserting this
solution back into the Hamiltonian \eref{eq:BHhamCP} and the
Liouville operator \eref{eq:L_cav}, which accounts for cavity
damping, leads to an effective adiabatic model. It comprises the
nonlinear Hamiltonian
\begin{subequations}
\begin{multline}\label{eq:effHam0}
H_{\textrm{ad}}=(E+JV_{\textrm{cl}})\hat{B}+\frac{U}{2}\hat{C}\\+\frac{\hbar
U_0 J \eta^2}{\kappa^2+{\delta_C}^2}\left(\hat{B}+\frac{U_0J
\delta_C
}{\kappa^2+{\delta_C}^2}\frac{\kappa^2-3{\delta_C}^2}{\kappa^2+{\delta_C}^2}\hat{B}^2\right)\,,
\end{multline}
and Liouville operator
\begin{equation}\label{eq:adhamproc3}
\mathcal{L}_{\textrm{ad}}\varrho = \frac{\kappa
U_0^2J^2\eta^2}{\left(\kappa^2+{\delta_C}^2\right)^2}\left(2\hat{B}\varrho\hat{B}-\hat{B}^2\varrho-\varrho\hat{B}^2\right)\,,
\end{equation}
\end{subequations}
which describes decoherence in the basis of the eigenstates of the
operator $\hat{B}$. Note that the adiabatic elimination procedure
described above is not rigorous mathematically, since we
adiabatically approximated the solution of a nonlinear dynamical
equation in \eref{eq:elimfield}, which appears also as an ordering
ambiguity of the involved operators
\cite{Maschler2008Ultracold,Larson2008Quantum}.

\begin{figure}
{\includegraphics[width=0.66\columnwidth]{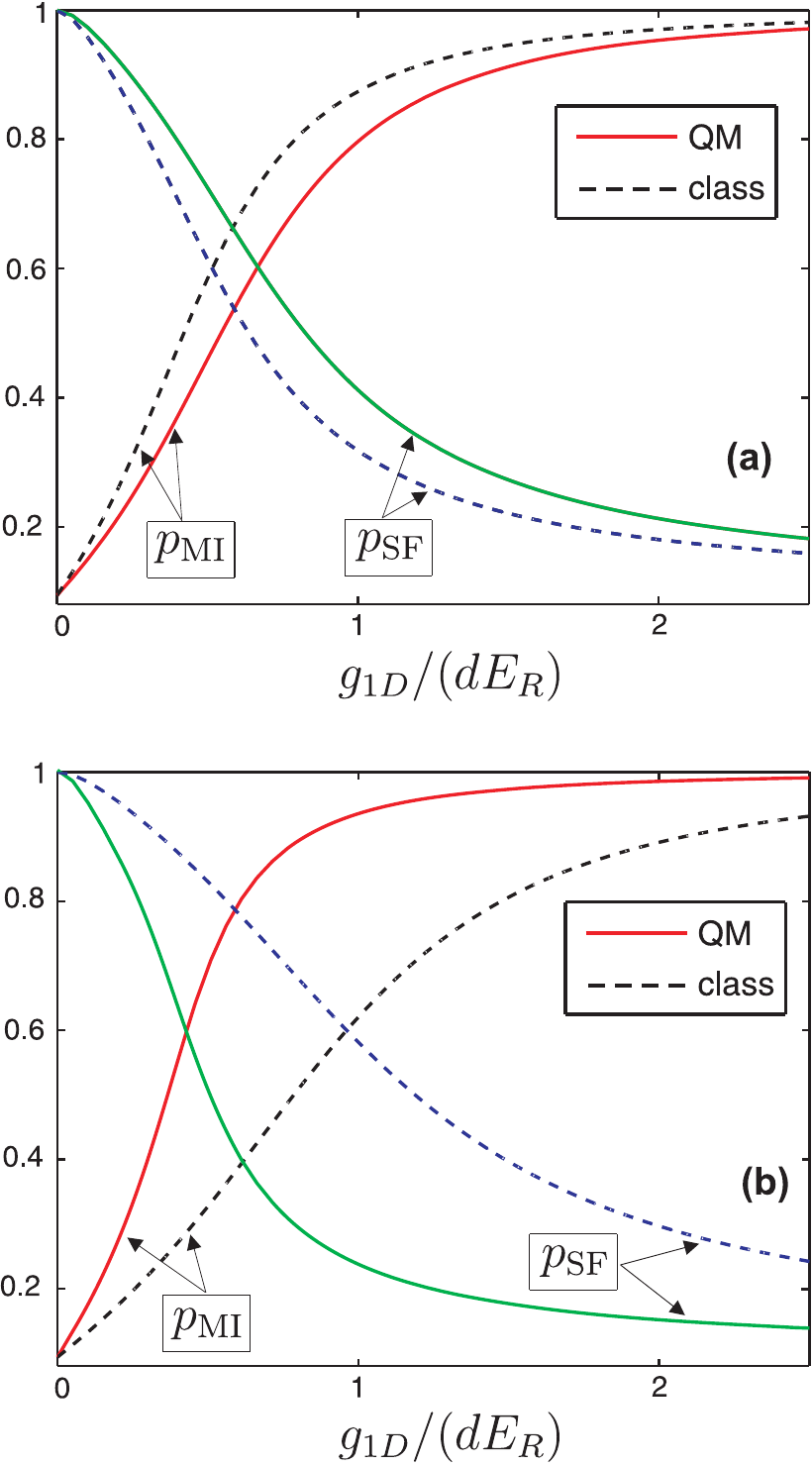}} \caption{Mott
insulator (MI) to superfluid (SF) phase transition in a cavity
optical lattice.  The probabilities $p_{\textrm{MI}}$ and
$p_{\textrm{SF}}$ to find the atoms in the states
$|\Psi_\text{MI}\rangle$ and $|\Psi_\text{SF}\rangle$ as a
function of the dimensionless 1D on-site interaction strength
$g_{1D}/(dE_R)$ ($d$ is the lattice constant, $E_R = \hbar
\omega_R$ is the recoil energy) are compared for two cases: first,
for an optical lattice sustained by the quantum field of a cavity
mode ($V_\sub{cl}=0$), and second, for a purely classical optical
lattice ($\eta_1=0$). We choose $\eta$ such that in each of the
two examples (a and b) both potentials have equal depth for zero
on-site interaction $g_{\textrm{1D}}$. The quantum (QM) and
classical (class) cases are depicted with solid and dashed lines,
respectively. Parameters are
$(U_{0},\kappa,\eta)=(-1,1/\sqrt{2},\sqrt{5.5})\omega_R$. The
detuning between the probe and the dispersively shifted cavity
frequencies affects the position of the phase transition. In (a)
resp.~(b) this detuning is positive, $\Delta_C-U_{0}N=\kappa$,
resp.~negative $\Delta_C-U_{0}N=-\kappa$ and the transition point
is shifted towards smaller resp.~higher interaction strengths in
comparison to that in a classical lattice. From
\textcite{Maschler2008Ultracold}.} \label{fig:cavity_phase}
\end{figure}

For a small, numerically tractable system, the lowest energy
eigenstate of $H_\mathrm{ad}$ can be calculated.  As a key
example, the SF-to-MI quantum phase transition in an optical
lattice sustained entirely by a quantized standing-wave cavity
field was analyzed \cite{Maschler2008Ultracold}. The dynamical
response of the photon number to the atomic motion is able to
strongly modify atomic number fluctuations and hence to drive the phase
transition. Depending on the cavity parameters (e.g.\ the detuning
between cavity and external pump laser), the photon fluctuations
can either suppress or enhance atomic fluctuations and hopping,
therefore pushing the system towards or outwards the MI or SF
states. Accordingly, as depicted in Fig.~\ref{fig:cavity_phase},
the position of the SF-to-MI phase transition in a cavity optical
lattice potential can be shifted (keeping the mean potential depth
constant) towards either smaller or larger values of the
collisional atom-atom interaction strengths, depending on whether
the pump-cavity detuning is chosen positive (a) or negative (b).

\begin{figure}
{\includegraphics[width=0.9\columnwidth]{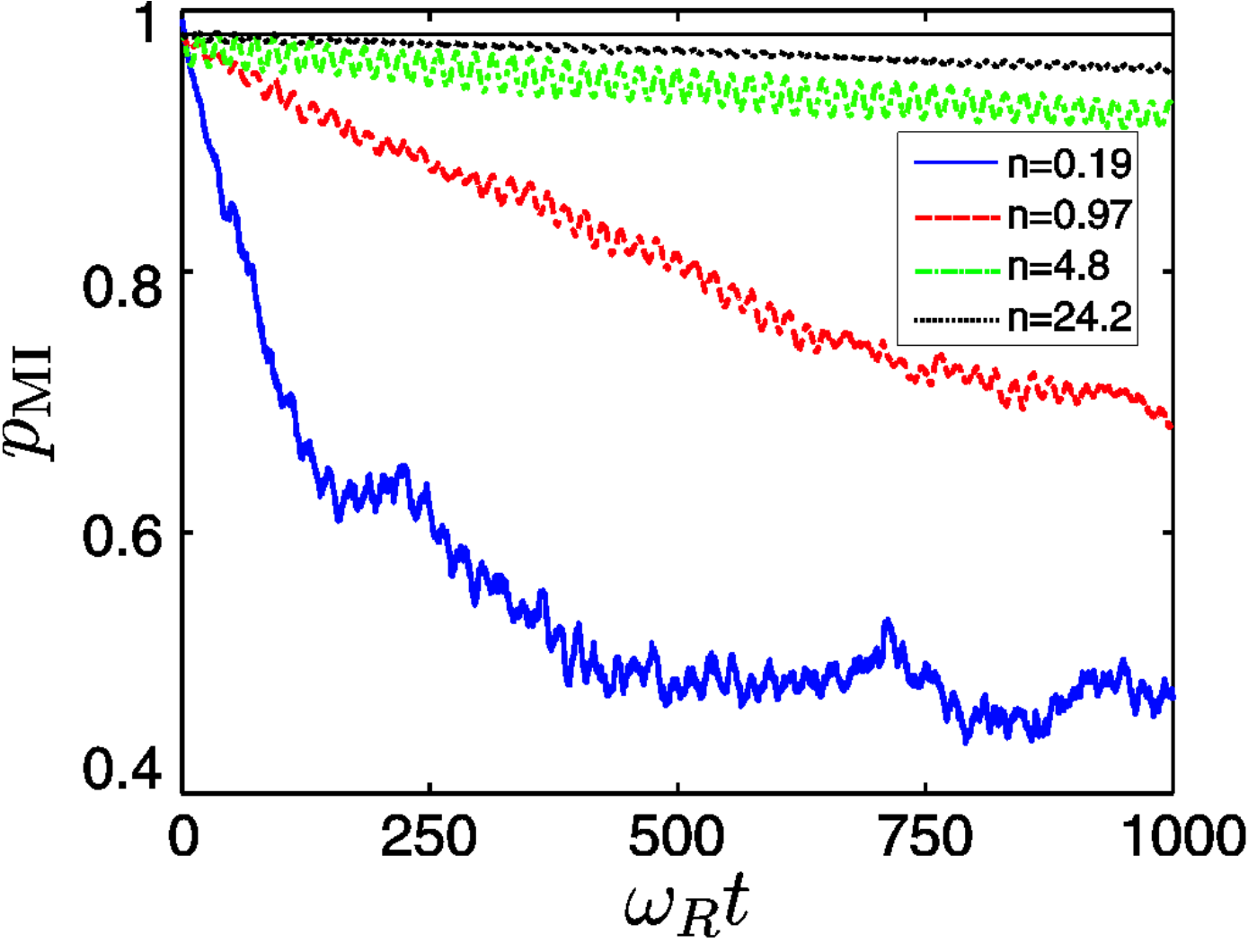}}
\caption{\label{fig:MIdepletion} (Color online) Effect of photon
number granularity upon the Mott insulator (MI) state for two
atoms in a cavity-sustained optical lattice. The time evolution of
the occupation probability $p_\text{MI}$ of the MI state is shown
for various mean intracavity photon numbers $n$. The atom-cavity
coupling $g$ is adjusted such that the average potential depth
($8E_R$) is identical for all curves. Whereas for a static optical
lattice potential of equal depth the system remains in the
initially prepared MI state (solid line), photon number
fluctuations for low $n$ deplete the MI state. From
\textcite{Maschler2008Ultracold}.}
\end{figure}

Figure \ref{fig:MIdepletion} demonstrates the importance of the
photon number fluctuations in a quantum potential by testing the
stability of the Mott phase in a potential of fixed average depth
but different mean photon number. While in an almost classical
field (highly excited coherent state with many photons) the Mott
phase is stable, photon number fluctuations (uncertainty) inherent
in a weak coherent state of few photons enhance tunneling and
decay of the perfect order.

This is explicitly shown in Fig.\ \ref{fig:MIdepletion} depicting
the decay of an initially prepared perfectly ordered atomic state.
We choose different mean intracavity photon numbers and keep the
average depth of the potential constant by readjusting the
coupling strength $U_0$. In the classical limit (very large photon
number and small atom-cavity coupling), the system remains in the
initially prepared MI state. For smaller photon numbers
$\bar{n}\sim 20$,
 the initial MI state only slowly degrades in time. However, when the
photon number fluctuations become comparable to the mean,
 i.e.~for mean photon numbers as small as $\bar{n}\sim1$, the system quickly
escapes from the MI state via fluctuation induced tunneling. Note
that in order to keep the average optical potential constant,
lower photon numbers are connected to a larger potential per
photon, so that the potential fluctuations are additionally
enhanced at low photon numbers. The classical limit is also
approached in the bad cavity limit, $\kappa \gg \omega_R$, where
number fluctuations occur so fast, that particles only see the
average and do not have the time to tunnel during an intensity
fluctuation.

The quantum properties of the cavity light become predominant if already single intracavity photons
create an optical potential of considerable depth, capable of trapping numerous atoms. As quantum mechanics allows
for the existence of superpositions of photon number states, one may obtain superpositions of several potentials with different depths \cite{Horak2001Manipulating}.

The complete phase diagram of ultracold atoms in \emph{two-band}
BH models coupled to a cavity light field has been
calculated by \textcite{Silver2010BoseHubbard} by means of a
variational approach and the analogy to the Dicke-model superradiant
phase transition has been pointed out, see Sec.~\ref{sec:Open-SystemDickeQPT}.

\paragraph{Self-organization within the Hubbard-model approach}

\begin{figure}
{\includegraphics[width=0.9\columnwidth]{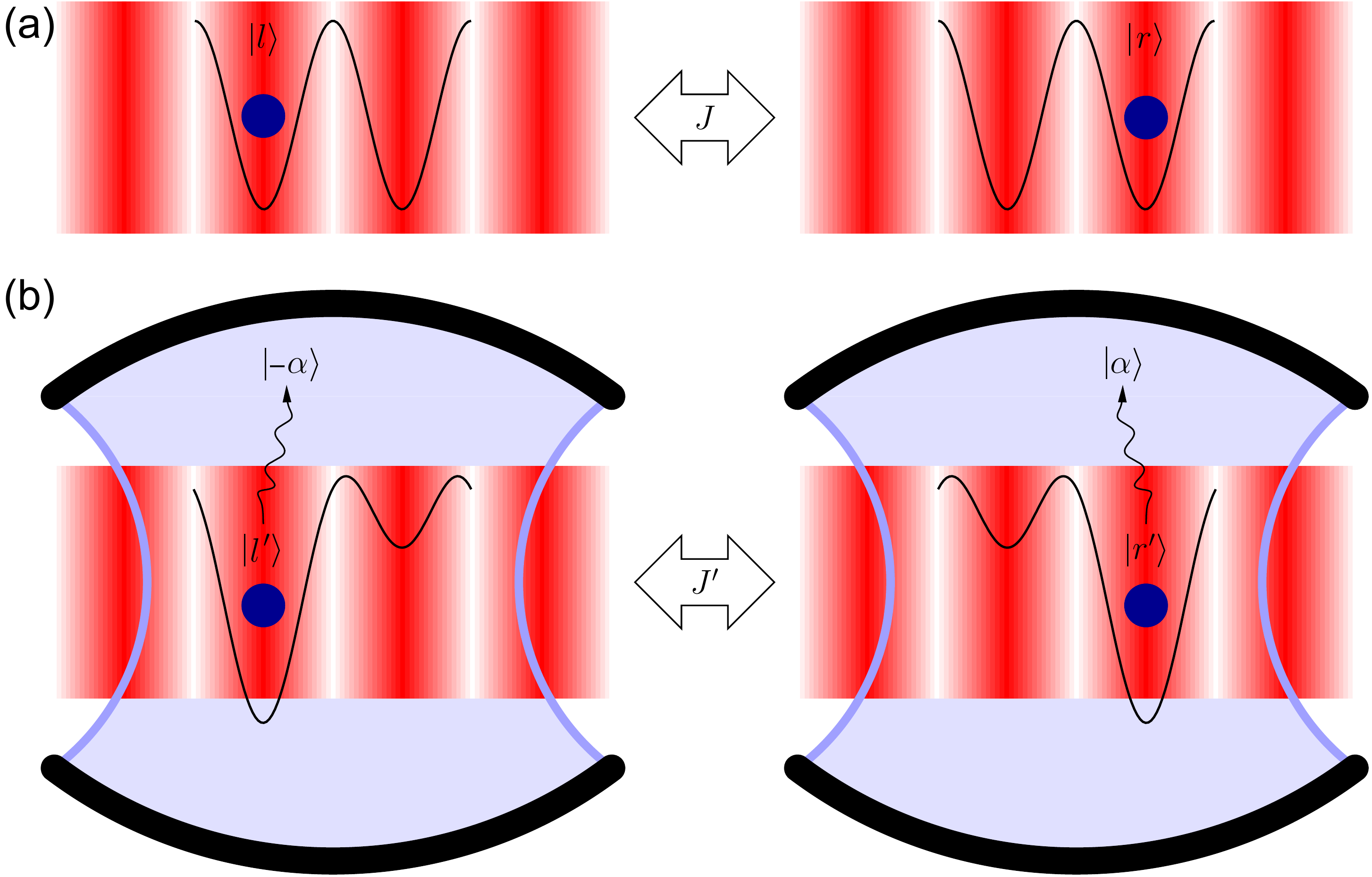}}
  \caption{Self-organization as a quantum seesaw effect. (a)
Atoms which are trapped in a free-space 1D lattice potential with
two adjacent sites (left and right) tunnel with rate $J$ between
the corresponding Wannier states $\ket{l}$ and $\ket{r}$. (b)
Coupling the atoms in addition to the field of a cavity whose axis
is perpendicular to the optical lattice induces light scattering
between the optical lattice laser and the cavity field with
opposite phases from the two sites. The modified potential
resulting from the interference of the lattice field and the
cavity field discriminates the two sites and causes positive
feedback and atomic ordering into one of them. The process starts
by spontaneous symmetry breaking, and depends on the quantum
statistics of the initially prepared many-body state.}
  \label{fig:seesaw}
\end{figure}

A possible influence of quantum statistical properties on the
spatial self-organization process, described in
Sec.~\ref{sec:ColdSelforg} and \ref{sec:DickeQPT}, can be studied
within the framework of an extended BH model. For simplicity, the
geometry is modified in comparison with the generic case of
self-organization, as depicted schematically in
Fig.~\ref{fig:seesaw}.
Atoms are confined in a static optical lattice potential which is
oriented perpendicular to the cavity axis. As before, large
atom-laser detuning and negligible atomic saturation are assumed.
The laser fields providing the optical lattice potential are
considered to be tuned close to resonance with a cavity mode,
therefore inducing coherent cavity driving via Rayleigh scattering
off the atoms. The single-atom Hamiltonian corresponding to this
geometry reads~\cite{Maschler2007Entanglement}
\begin{multline}
\label{eq:Hameff} H=\frac{p^2}{2m}+ V_\sub{cl}\cos^2(kx)
-\hbar\left(\Delta_C-U_0\right)a^\dagger a
\\+\sqrt{\hbar V_\sub{cl} U_0}\cos(kx)\left(a+a^\dagger\right)\; ,
\end{multline}
where $V_\sub{cl}$ denotes the depth of the static lattice potential, $\Delta_C$ the
detuning between the lattice laser and the cavity resonance, and
$U_0$ the light shift of the cavity resonance frequency per atom.
In a simple and intuitive picture the dynamic cavity field plays
the role of a \emph{``seesaw''} potential. Interference between
the $\cos(kx)$ potential, generated through photon scattering, and
the static $\cos^2(kx)$ potential determines the overall potential
felt by the atoms. The spatial symmetry of the system allows for
the emergence of two possible ordered configurations, with all
atoms residing at either odd ($\cos(kx)=1$) or even
($\cos(kx)=-1$) lattice sites.

The many-body Bose-Hubbard Hamiltonian, \eref{eq:CavityBH},
adapted to this scheme reads
\begin{multline}
\label{BHham} H=\sum_{i,j}J_{i,j}b_i^\dagger b_j-
\hbar\left(\Delta_C - U_0\sum_i b_i^\dagger b_i\right)a^\dagger
a\\+
\left(a+a^\dagger\right)\sum_{i,j}\hbar\tilde{J}_{i,j}b_i^\dagger
b_j.
\end{multline}
Here, the standard matrix elements for the kinetic and potential
energy $p^2/2m + V_\mathrm{cl}\cos^2(kx)$ between sites $i$ and $j$ are
denoted by $J_{i,j}$, whereas $\tilde{J}_{i,j}$ gives the matrix
elements of the interference term $\sqrt{U_0 V_\mathrm{cl}/\hbar}\cos(kx)$.
On-site interactions are neglected at this point and, for
consistency, we require weak coupling per atom, i.e.\ $\hbar \, |U_0| \ll
|V_\mathrm{cl}|$.

\begin{figure}
    \includegraphics[width=0.9\columnwidth]{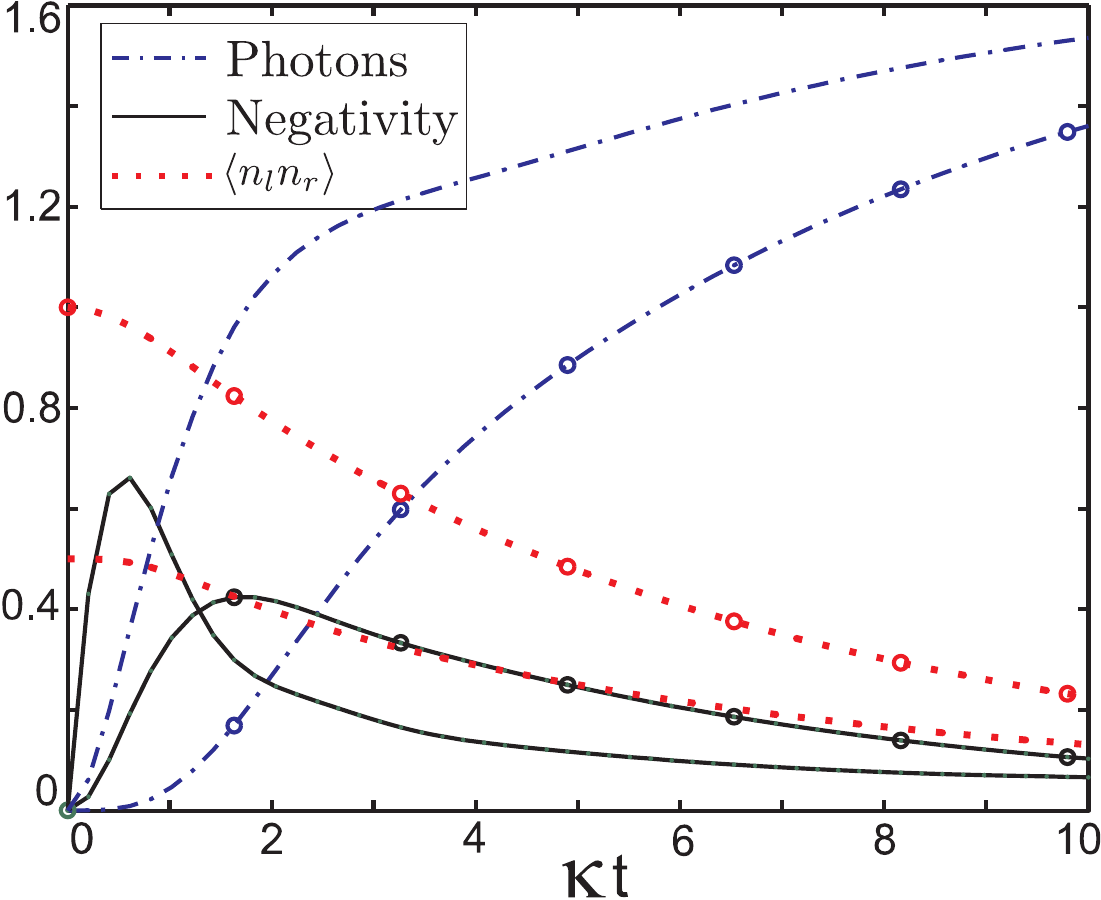}
\caption{Entanglement-assisted self-organization in a quantum
optical lattice potential, see Fig.~\ref{fig:seesaw}. Shown are
atom-light entanglement (solid lines), mean cavity photon number
(dotted-dashed lines), and the two-site atom-atom correlation
function (dotted lines) for two atoms in two adjacent lattice
wells, denoted by left ($l$) and right ($r$). Lines with extra
circles show the case of exactly one atom in each well at start
(MI state), while the other lines show the evolution for an
initially symmetric superposition state for each atom (SF state).
The parameters are $U_0=-2 \kappa, \Delta_C=-6\kappa,
J=\kappa/100$ and $\tilde{J}=1.6\kappa $. From
\textcite{Maschler2007Entanglement}.}
  \label{fig:TimeEv2}
\end{figure}

Essential dynamical properties of this system beyond the
mean-field approximation become evident already for two atoms.
Monitoring the microscopic physics of self-organization as shown
in Fig.~\ref{fig:TimeEv2}, the process resembles the decay of a
homogeneously filled lattice with one particle per site on average
to the self-ordered state, where both particles occupy even or odd
sites. Following the decay of the probability for the two atoms
sitting in different wells ($\langle n_l n_r \rangle$), we first
note that the formation of the self-organized state is accompanied
by a fast growth of atom-field entanglement. Most importantly,
however, one finds a striking dependence of the self-organization
dynamics on the initial quantum fluctuations. A SF state with both
atoms prepared in the symmetric superposition of the two wells,
$1/2 \, (b_l^\dagger+b_r^\dagger)^2 |0\rangle$, self-organizes
much faster than a perfectly ordered MI state, $b_l^\dagger
b_r^\dagger |0\rangle$, with exactly one atom per well. In the
latter case, the cavity field remains in the vacuum state until a
tunneling event induces atomic coherence between the left and
right lattice site, triggering the decay of the MI state towards
the self-organized state.

Under some approximations this model can be extrapolated to the
thermodynamic limit where quantum phase transitions similar to the
one predicted in the mean-field approach can be studied. Among
various other properties this leads to the coexistence of diagonal
long-range order and long-range coherence \cite{Vidal2010Quantum},
indicating new phases to appear in the gaps between Mott-like
states with different integer filling factors.

\paragraph{Ring cavity}

When several independent cavity modes are dynamically interacting
with the atoms, not only the depth but also the shape and the
spatial periodicity of the potential can change. In the generic
case of a ring cavity the depth and the longitudinal position of
the lattice is dynamical.  While, already in standard optical
lattices the validity of the lowest-band assumption is often
doubtful and corrections are necessary,  this approximation loses
its meaning in a ring cavity. Expansions based on a single set of
Wannier functions cannot be consistently formulated since the
lowest-band Wannier functions for a given position contain
contributions from a large number of higher bands for a slightly
shifted position. Hence small lattice shifts immediately involve
many higher-order bands.

\begin{figure}[h!]
\begin{center}
\includegraphics[width=0.9\columnwidth]{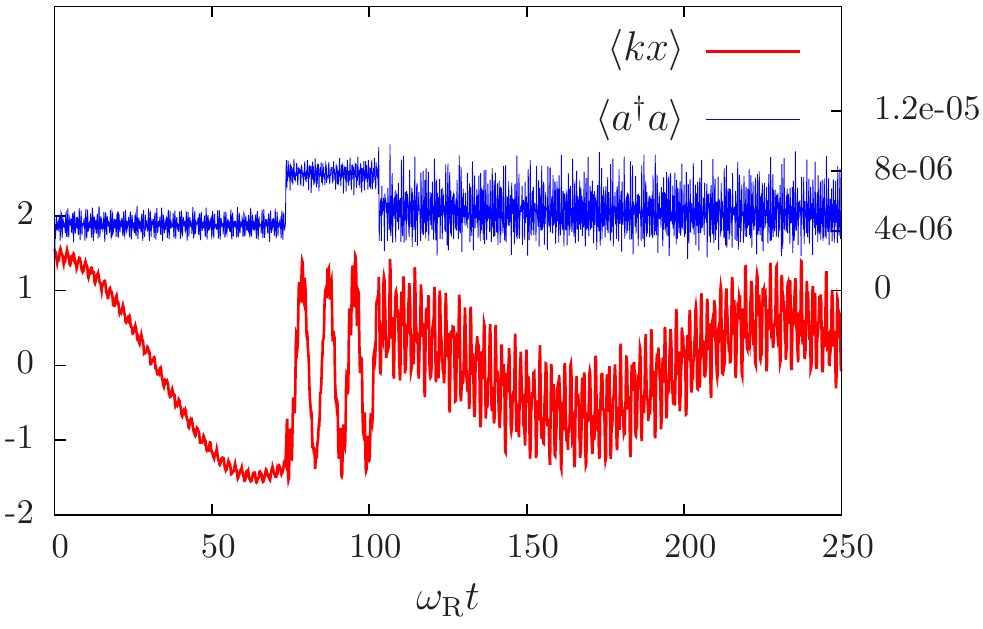}
\caption{(Color online) Correlated photon jumps and tunneling of an atom in a
symmetrically driven ring cavity. The sample trajectory, showing
the position expectation value  $\langle k x \rangle$ (left axis)
and the mean photon number $\langle a^\dag a \rangle$ of the
undriven sine mode (right axis), presents two quantum jumps
occurring at $\omrec t\approx 70$ and $\omrec t\approx 100$. They
lead to a simultaneous change of the photon number and the atomic
band excitation corresponding to different tunnel oscillation
frequencies. In a higher band much faster oscillations are
obtained between neighboring sites. This generates a higher
effective hopping amplitude and heating on average. The parameters
are $U_0=-2\omega_R$, $\alpha_c=\sqrt{6}$ (field amplitude of the
driven cosine mode), $\Delta_C=U_0-\kappa$ and
$\kappa=500\omega_R$. From \textcite{Niedenzu2010Microscopic}.}
\label{fig:SampleTrajectory}
\end{center}
\end{figure}
The simplest example of two cavity modes sustained by a ring
cavity reveals, that a naive crude truncation of the Bose-Hubbard
model with respect to the pumped cosine mode  at the lowest band
decouples the atoms from the associated sine mode. This
immediately eliminates central dynamical effects of the system as
overall momentum conservation and nonlocal correlated hopping
\cite{Niedenzu2010Microscopic}.

As an example, Fig.\ \ref{fig:SampleTrajectory} shows that a quantum jump in the lattice
photon number is accompanied by a sudden change of the tunneling
rate between adjacent sites. After the second jump in the trajectory shown in the Figure, the system returns to the original mean values, however,  both the position and photon number quantities exhibit much larger noise,  which demonstrates the effect of heating stemming from the photon number fluctuations.

\section{OUTLOOK}

After almost two decades of active research in optical cavity QED
with cold and ultracold atoms, initially dominated by theoretical
investigations, the field presently exhibits fast growth with
several experimental groups demonstrating spectacular effects.
Single atoms are routinely cavity-cooled and trapped over seconds
within optical high-finesse resonators, providing a
well-controlled quantum system for quantum information science.
Ultracold quantum gases prepared in magnetic or optical traps are
now reliably coupled to high-quality cavity modes. Even in the far
dispersive regime, these systems are governed by strong
back-action effects of the collective atomic motion on the cavity
field degrees of freedom. Cavity decay offers a unique channel to
monitor the complex coupled atom-light dynamics non-destructively
and in real time. Many central atomic variables can be accessed by
quantum non-demolition measurements which minimize quantum
back-action.

With the aim of trapping and cooling ensembles, nano- and even
microscopic particles or arrays of thin membranes, the research
field of cavity QED, reviewed in this paper, overlaps and unifies
more and more with the rapidly growing field of optomechanics
\cite{StamperKurn2012Optomechanics}. Trapping and cooling arrays
of membranes in cavities is only one very striking
example \cite{Xuereb2012Strong}. Practical applications as
ultra-sensitive detectors of mass, acceleration or magnetic fields
or even tests of general relativity seem within range of current
technology.

Initiated by theoretical studies and early experiments with cold
atomic ensembles, a non-equilibrium quantum phase transition
between a superfluid and a supersolid phase has been investigated
experimentally. Recently, theoretical investigations opened new
directions and possibilities towards controlled preparation and
investigations of the physics of spin glasses, more complex
supersolid and superglass
phases \cite{Gopalakrishnan2011Frustration,Strack2011Dicke}. Further prominent
solid-state Hamiltonians involving phonons or polarons could be
studied with unprecedented control and observation
possibilities \cite{Mekhov2012Review}.

A recent breakthrough experiment demonstrating sub-recoil cavity
cooling towards quantum degeneracy \cite{Wolke2012Cavity} opens
the prospect of replacing evaporative cooling techniques by cavity
cooling and direct preparation of exotic quantum states from a
thermal gas. This also paves the way towards implementing a
continuous atom-laser as a new tool in ultracold atom physics
\cite{Salzburger2007Atomphoton}.

Still, important challenges as cooling and trapping of molecular
samples or large suspended objects have not been experimentally
demonstrated. The prospects of multi-species implementations in
multimode cavity environments still have to be fully evaluated.
Apart from this point, experiments seem ahead of theoretical and
numerical simulation possibilities, where theory has to be
improved and better suited models need to be developed.

In a long term vision, cavity-sustained light fields allow to
couple hybrid systems of very different physical nature like
superconducting qubits, cold quantum gases and micromechanical
oscillators without destroying quantum coherence of the systems,
brought about by any classical coupling of such systems. In this
way cavity-based setups with ultracold gases could develop into an
important building block for quantum information processing or
other quantum-based future technologies \cite{Henschel2010Cavity}
or a route to an even better atomic lattice clock
\cite{Nicholson2012Cavity}.\\

{\bf Acknowledgments}

\vspace{1cm} We thank D.~Nagy, G.~Szirmai, G.~K\'onya, A.~Vukics,
J.~Asb\'oth, I.~Mekhov, C.~Genes, K.~Baumann, R.~Mottl, T.~Donner
and R.~Landig for stimulating discussions. H.~R.~acknowledges
support from the Austrian Science Fund FWF (grant F 4013-N16).
P.~D.\ acknowledges the financial support from the Hungarian
National Office for Research and Technology under the contract
ERC\_HU\_09 OPTOMECH, and from the Hungarian Academy of Sciences
(Lend\"ulet Program, LP2011-016). F.~B.\ and T.~E.\ acknowledge
financial funding from SQMS (ERC advanced grant), NAME-QUAM (EU,
FET open), NCCR-QSIT and ESF (POLATOM).

\bibliography{cqedreview2011}

\begin{thebibliography}{278}%
\makeatletter
\providecommand \@ifxundefined [1]{%
 \@ifx{#1\undefined}
}%
\providecommand \@ifnum [1]{%
 \ifnum #1\expandafter \@firstoftwo
 \else \expandafter \@secondoftwo
 \fi
}%
\providecommand \@ifx [1]{%
 \ifx #1\expandafter \@firstoftwo
 \else \expandafter \@secondoftwo
 \fi
}%
\providecommand \natexlab [1]{#1}%
\providecommand \enquote  [1]{``#1''}%
\providecommand \bibnamefont  [1]{#1}%
\providecommand \bibfnamefont [1]{#1}%
\providecommand \citenamefont [1]{#1}%
\providecommand \href@noop [0]{\@secondoftwo}%
\providecommand \href [0]{\begingroup \@sanitize@url \@href}%
\providecommand \@href[1]{\@@startlink{#1}\@@href}%
\providecommand \@@href[1]{\endgroup#1\@@endlink}%
\providecommand \@sanitize@url [0]{\catcode `\\12\catcode `\$12\catcode
  `\&12\catcode `\#12\catcode `\^12\catcode `\_12\catcode `\%12\relax}%
\providecommand \@@startlink[1]{}%
\providecommand \@@endlink[0]{}%
\providecommand \url  [0]{\begingroup\@sanitize@url \@url }%
\providecommand \@url [1]{\endgroup\@href {#1}{\urlprefix }}%
\providecommand \urlprefix  [0]{URL }%
\providecommand \Eprint [0]{\href }%
\providecommand \doibase [0]{http://dx.doi.org/}%
\providecommand \selectlanguage [0]{\@gobble}%
\providecommand \bibinfo  [0]{\@secondoftwo}%
\providecommand \bibfield  [0]{\@secondoftwo}%
\providecommand \translation [1]{[#1]}%
\providecommand \BibitemOpen [0]{}%
\providecommand \bibitemStop [0]{}%
\providecommand \bibitemNoStop [0]{.\EOS\space}%
\providecommand \EOS [0]{\spacefactor3000\relax}%
\providecommand \BibitemShut  [1]{\csname bibitem#1\endcsname}%
\let\auto@bib@innerbib\@empty
\bibitem [{\citenamefont {Arcizet}\ \emph {et~al.}(2006)\citenamefont
  {Arcizet}, \citenamefont {Cohadon}, \citenamefont {Briant}, \citenamefont
  {Pinard},\ and\ \citenamefont {Heidmann}}]{Arcizet2006}%
  \BibitemOpen
  \bibfield  {author} {\bibinfo {author} {\bibnamefont {Arcizet}, \bibfnamefont
  {O.}}, \bibinfo {author} {\bibfnamefont {P.~F.}\ \bibnamefont {Cohadon}},
  \bibinfo {author} {\bibfnamefont {T.}~\bibnamefont {Briant}}, \bibinfo
  {author} {\bibfnamefont {M.}~\bibnamefont {Pinard}}, \ and\ \bibinfo {author}
  {\bibfnamefont {A.}~\bibnamefont {Heidmann}}} (\bibinfo {year} {2006}),\
  \href@noop {} {\bibfield  {journal} {\bibinfo  {journal} {Nature}\ }\textbf
  {\bibinfo {volume} {444}}~(\bibinfo {number} {7115}),\ \bibinfo {pages}
  {71}}\BibitemShut {NoStop}%
\bibitem [{\citenamefont {Asb\'{o}th}\ and\ \citenamefont
  {Domokos}(2007)}]{Asboth2007Comment}%
  \BibitemOpen
  \bibfield  {author} {\bibinfo {author} {\bibnamefont {Asb\'{o}th},
  \bibfnamefont {J.~K.}}, \ and\ \bibinfo {author} {\bibfnamefont
  {P.}~\bibnamefont {Domokos}}} (\bibinfo {year} {2007}),\ \href {\doibase
  10.1103/PhysRevA.76.057801} {\bibfield  {journal} {\bibinfo  {journal}
  {Physical Review A (Atomic, Molecular, and Optical Physics)}\ }\textbf
  {\bibinfo {volume} {76}}~(\bibinfo {number} {5}),\ \bibinfo {pages}
  {057801+}}\BibitemShut {NoStop}%
\bibitem [{\citenamefont {Asb\'{o}th}\ \emph {et~al.}(2004)\citenamefont
  {Asb\'{o}th}, \citenamefont {Domokos},\ and\ \citenamefont
  {Ritsch}}]{Asboth2004Correlated}%
  \BibitemOpen
  \bibfield  {author} {\bibinfo {author} {\bibnamefont {Asb\'{o}th},
  \bibfnamefont {J.~K.}}, \bibinfo {author} {\bibfnamefont {P.}~\bibnamefont
  {Domokos}}, \ and\ \bibinfo {author} {\bibfnamefont {H.}~\bibnamefont
  {Ritsch}}} (\bibinfo {year} {2004}),\ \href {\doibase
  10.1103/PhysRevA.70.013414} {\bibfield  {journal} {\bibinfo  {journal}
  {Physical Review A}\ }\textbf {\bibinfo {volume} {70}}~(\bibinfo {number}
  {1}),\ \bibinfo {pages} {013414+}}\BibitemShut {NoStop}%
\bibitem [{\citenamefont {Asb\'{o}th}\ \emph {et~al.}(2005)\citenamefont
  {Asb\'{o}th}, \citenamefont {Domokos}, \citenamefont {Ritsch},\ and\
  \citenamefont {Vukics}}]{Asboth2005Selforganization}%
  \BibitemOpen
  \bibfield  {author} {\bibinfo {author} {\bibnamefont {Asb\'{o}th},
  \bibfnamefont {J.~K.}}, \bibinfo {author} {\bibfnamefont {P.}~\bibnamefont
  {Domokos}}, \bibinfo {author} {\bibfnamefont {H.}~\bibnamefont {Ritsch}}, \
  and\ \bibinfo {author} {\bibfnamefont {A.}~\bibnamefont {Vukics}}} (\bibinfo
  {year} {2005}),\ \href {\doibase 10.1103/PhysRevA.72.053417} {\bibfield
  {journal} {\bibinfo  {journal} {Physical Review A}\ }\textbf {\bibinfo
  {volume} {72}}~(\bibinfo {number} {5}),\ \bibinfo {pages}
  {053417+}}\BibitemShut {NoStop}%
\bibitem [{\citenamefont {Barker}\ and\ \citenamefont
  {Shneider}(2010)}]{Barker2010Cavity}%
  \BibitemOpen
  \bibfield  {author} {\bibinfo {author} {\bibnamefont {Barker}, \bibfnamefont
  {P.~F.}}, \ and\ \bibinfo {author} {\bibfnamefont {M.~N.}\ \bibnamefont
  {Shneider}}} (\bibinfo {year} {2010}),\ \href {\doibase
  10.1103/PhysRevA.81.023826} {\bibfield  {journal} {\bibinfo  {journal}
  {Physical Review A}\ }\textbf {\bibinfo {volume} {81}}~(\bibinfo {number}
  {2}),\ \bibinfo {pages} {023826+}}\BibitemShut {NoStop}%
\bibitem [{\citenamefont {Baumann}\ \emph {et~al.}(2010)\citenamefont
  {Baumann}, \citenamefont {Guerlin}, \citenamefont {Brennecke},\ and\
  \citenamefont {Esslinger}}]{Baumann2010Dicke}%
  \BibitemOpen
  \bibfield  {author} {\bibinfo {author} {\bibnamefont {Baumann}, \bibfnamefont
  {K.}}, \bibinfo {author} {\bibfnamefont {C.}~\bibnamefont {Guerlin}},
  \bibinfo {author} {\bibfnamefont {F.}~\bibnamefont {Brennecke}}, \ and\
  \bibinfo {author} {\bibfnamefont {T.}~\bibnamefont {Esslinger}}} (\bibinfo
  {year} {2010}),\ \href {\doibase 10.1038/nature09009} {\bibfield  {journal}
  {\bibinfo  {journal} {Nature}\ }\textbf {\bibinfo {volume} {464}}~(\bibinfo
  {number} {7293}),\ \bibinfo {pages} {1301}}\BibitemShut {NoStop}%
\bibitem [{\citenamefont {Baumann}\ \emph {et~al.}(2011)\citenamefont
  {Baumann}, \citenamefont {Mottl}, \citenamefont {Brennecke},\ and\
  \citenamefont {Esslinger}}]{Baumann2011Exploring}%
  \BibitemOpen
  \bibfield  {author} {\bibinfo {author} {\bibnamefont {Baumann}, \bibfnamefont
  {K.}}, \bibinfo {author} {\bibfnamefont {R.}~\bibnamefont {Mottl}}, \bibinfo
  {author} {\bibfnamefont {F.}~\bibnamefont {Brennecke}}, \ and\ \bibinfo
  {author} {\bibfnamefont {T.}~\bibnamefont {Esslinger}}} (\bibinfo {year}
  {2011}),\ \href {\doibase 10.1103/PhysRevLett.107.140402} {\bibfield
  {journal} {\bibinfo  {journal} {Physical Review Letters}\ }\textbf {\bibinfo
  {volume} {107}},\ \bibinfo {pages} {140402+}}\BibitemShut {NoStop}%
\bibitem [{\citenamefont {{Berman}}(1994)}]{Berman1994Cavity}%
  \BibitemOpen
  \bibfield  {author} {\bibinfo {author} {\bibnamefont {{Berman}},
  \bibfnamefont {P.~R.}}} (\bibinfo {year} {1994}),\ \href@noop {} {\emph
  {\bibinfo {title} {Advances in atomic, molecular, and optical physics, New
  York: Academic Press, edited by Berman, Paul R.}}}\BibitemShut {Stop}%
\bibitem [{\citenamefont {Berman}(1999)}]{Berman1999Comparison}%
  \BibitemOpen
  \bibfield  {author} {\bibinfo {author} {\bibnamefont {Berman}, \bibfnamefont
  {P.~R.}}} (\bibinfo {year} {1999}),\ \href {\doibase 10.1103/PhysRevA.59.585}
  {\bibfield  {journal} {\bibinfo  {journal} {Phys. Rev. A}\ }\textbf {\bibinfo
  {volume} {59}},\ \bibinfo {pages} {585}}\BibitemShut {NoStop}%
\bibitem [{\citenamefont {Bhaseen}\ \emph {et~al.}(2012)\citenamefont
  {Bhaseen}, \citenamefont {Mayoh}, \citenamefont {Simons},\ and\ \citenamefont
  {Keeling}}]{Bhaseen2012Dynamics}%
  \BibitemOpen
  \bibfield  {author} {\bibinfo {author} {\bibnamefont {Bhaseen}, \bibfnamefont
  {M.~J.}}, \bibinfo {author} {\bibfnamefont {J.}~\bibnamefont {Mayoh}},
  \bibinfo {author} {\bibfnamefont {B.~D.}\ \bibnamefont {Simons}}, \ and\
  \bibinfo {author} {\bibfnamefont {J.}~\bibnamefont {Keeling}}} (\bibinfo
  {year} {2012}),\ \href {\doibase 10.1103/PhysRevA.85.013817} {\bibfield
  {journal} {\bibinfo  {journal} {Physical Review A}\ }\textbf {\bibinfo
  {volume} {85}},\ \bibinfo {pages} {013817+}}\BibitemShut {NoStop}%
\bibitem [{\citenamefont {Bhattacherjee}\ \emph {et~al.}(2010)\citenamefont
  {Bhattacherjee}, \citenamefont {Kumar},\ and\ \citenamefont
  {Mohan}}]{Bhattacherjee2010Probing}%
  \BibitemOpen
  \bibfield  {author} {\bibinfo {author} {\bibnamefont {Bhattacherjee},
  \bibfnamefont {A.~B.}}, \bibinfo {author} {\bibfnamefont {T.}~\bibnamefont
  {Kumar}}, \ and\ \bibinfo {author} {\bibfnamefont {M.}~\bibnamefont {Mohan}}}
  (\bibinfo {year} {2010}),\ \href {\doibase 10.2478/s11534-009-0158-x}
  {\bibfield  {journal} {\bibinfo  {journal} {Central European Journal of
  Physics}\ }\textbf {\bibinfo {volume} {8}}~(\bibinfo {number} {5}),\ \bibinfo
  {pages} {850}}\BibitemShut {NoStop}%
\bibitem [{\citenamefont {Black}\ \emph {et~al.}(2003)\citenamefont {Black},
  \citenamefont {Chan},\ and\ \citenamefont
  {Vuleti\'{c}}}]{Black2003Observation}%
  \BibitemOpen
  \bibfield  {author} {\bibinfo {author} {\bibnamefont {Black}, \bibfnamefont
  {A.~T.}}, \bibinfo {author} {\bibfnamefont {H.~W.}\ \bibnamefont {Chan}}, \
  and\ \bibinfo {author} {\bibfnamefont {V.}~\bibnamefont {Vuleti\'{c}}}}
  (\bibinfo {year} {2003}),\ \href {\doibase 10.1103/PhysRevLett.91.203001}
  {\bibfield  {journal} {\bibinfo  {journal} {Physical Review Letters}\
  }\textbf {\bibinfo {volume} {91}}~(\bibinfo {number} {20}),\ \bibinfo {pages}
  {203001+}}\BibitemShut {NoStop}%
\bibitem [{\citenamefont {Blake}\ \emph {et~al.}(2011)\citenamefont {Blake},
  \citenamefont {Kurcz},\ and\ \citenamefont {Beige}}]{Blake2011Comparing}%
  \BibitemOpen
  \bibfield  {author} {\bibinfo {author} {\bibnamefont {Blake}, \bibfnamefont
  {T.}}, \bibinfo {author} {\bibfnamefont {A.}~\bibnamefont {Kurcz}}, \ and\
  \bibinfo {author} {\bibfnamefont {A.}~\bibnamefont {Beige}}} (\bibinfo {year}
  {2011}),\ \href {\doibase 10.1080/09500340.2010.543957} {\bibfield  {journal}
  {\bibinfo  {journal} {Journal of Modern Optics}\ }\textbf {\bibinfo {volume}
  {58}}~(\bibinfo {number} {15}),\ \bibinfo {pages} {1317}}\BibitemShut
  {NoStop}%
\bibitem [{\citenamefont {Bloch}\ \emph {et~al.}(2008)\citenamefont {Bloch},
  \citenamefont {Dalibard},\ and\ \citenamefont {Zwerger}}]{Bloch2008Manybody}%
  \BibitemOpen
  \bibfield  {author} {\bibinfo {author} {\bibnamefont {Bloch}, \bibfnamefont
  {I.}}, \bibinfo {author} {\bibfnamefont {J.}~\bibnamefont {Dalibard}}, \ and\
  \bibinfo {author} {\bibfnamefont {W.}~\bibnamefont {Zwerger}}} (\bibinfo
  {year} {2008}),\ \href {\doibase 10.1103/RevModPhys.80.885} {\bibfield
  {journal} {\bibinfo  {journal} {Reviews of Modern Physics}\ }\textbf
  {\bibinfo {volume} {80}}~(\bibinfo {number} {3}),\ \bibinfo {pages}
  {885}}\BibitemShut {NoStop}%
\bibitem [{\citenamefont {Boca}\ \emph {et~al.}(2004)\citenamefont {Boca},
  \citenamefont {Miller}, \citenamefont {Birnbaum}, \citenamefont {Boozer},
  \citenamefont {McKeever},\ and\ \citenamefont
  {Kimble}}]{Boca2004Observation}%
  \BibitemOpen
  \bibfield  {author} {\bibinfo {author} {\bibnamefont {Boca}, \bibfnamefont
  {A.}}, \bibinfo {author} {\bibfnamefont {R.}~\bibnamefont {Miller}}, \bibinfo
  {author} {\bibfnamefont {K.~M.}\ \bibnamefont {Birnbaum}}, \bibinfo {author}
  {\bibfnamefont {A.~D.}\ \bibnamefont {Boozer}}, \bibinfo {author}
  {\bibfnamefont {J.}~\bibnamefont {McKeever}}, \ and\ \bibinfo {author}
  {\bibfnamefont {H.~J.}\ \bibnamefont {Kimble}}} (\bibinfo {year} {2004}),\
  \href {\doibase 10.1103/PhysRevLett.93.233603} {\bibfield  {journal}
  {\bibinfo  {journal} {Physical Review Letters}\ }\textbf {\bibinfo {volume}
  {93}}~(\bibinfo {number} {23}),\ \bibinfo {pages} {233603+}}\BibitemShut
  {NoStop}%
\bibitem [{\citenamefont {Bonifacio}(1996)}]{Bonifacio1996Doppler}%
  \BibitemOpen
  \bibfield  {author} {\bibinfo {author} {\bibnamefont {Bonifacio},
  \bibfnamefont {R.}}} (\bibinfo {year} {1996}),\ \href {\doibase
  10.1016/0030-4018(95)00585-4} {\bibfield  {journal} {\bibinfo  {journal}
  {Optics Communications}\ }\textbf {\bibinfo {volume} {124}}~(\bibinfo
  {number} {5-6}),\ \bibinfo {pages} {469}}\BibitemShut {NoStop}%
\bibitem [{\citenamefont {Bonifacio}\ \emph {et~al.}(1994)\citenamefont
  {Bonifacio}, \citenamefont {De~Salvo}, \citenamefont {Narducci},\ and\
  \citenamefont {D'Angelo}}]{Bonifacio1994Exponential}%
  \BibitemOpen
  \bibfield  {author} {\bibinfo {author} {\bibnamefont {Bonifacio},
  \bibfnamefont {R.}}, \bibinfo {author} {\bibfnamefont {L.}~\bibnamefont
  {De~Salvo}}, \bibinfo {author} {\bibfnamefont {L.~M.}\ \bibnamefont
  {Narducci}}, \ and\ \bibinfo {author} {\bibfnamefont {E.~J.}\ \bibnamefont
  {D'Angelo}}} (\bibinfo {year} {1994}),\ \href {\doibase
  10.1103/PhysRevA.50.1716} {\bibfield  {journal} {\bibinfo  {journal}
  {Physical Review A}\ }\textbf {\bibinfo {volume} {50}}~(\bibinfo {number}
  {2}),\ \bibinfo {pages} {1716}}\BibitemShut {NoStop}%
\bibitem [{\citenamefont {Boozer}\ \emph {et~al.}(2006)\citenamefont {Boozer},
  \citenamefont {Boca}, \citenamefont {Miller}, \citenamefont {Northup},\ and\
  \citenamefont {Kimble}}]{Boozer2006Cooling}%
  \BibitemOpen
  \bibfield  {author} {\bibinfo {author} {\bibnamefont {Boozer}, \bibfnamefont
  {A.~D.}}, \bibinfo {author} {\bibfnamefont {A.}~\bibnamefont {Boca}},
  \bibinfo {author} {\bibfnamefont {R.}~\bibnamefont {Miller}}, \bibinfo
  {author} {\bibfnamefont {T.~E.}\ \bibnamefont {Northup}}, \ and\ \bibinfo
  {author} {\bibfnamefont {H.~J.}\ \bibnamefont {Kimble}}} (\bibinfo {year}
  {2006}),\ \href {\doibase 10.1103/PhysRevLett.97.083602} {\bibfield
  {journal} {\bibinfo  {journal} {Physical Review Letters}\ }\textbf {\bibinfo
  {volume} {97}}~(\bibinfo {number} {8}),\ \bibinfo {pages}
  {083602+}}\BibitemShut {NoStop}%
\bibitem [{\citenamefont {Boozer}\ \emph {et~al.}(2007)\citenamefont {Boozer},
  \citenamefont {Boca}, \citenamefont {Miller}, \citenamefont {Northup},\ and\
  \citenamefont {Kimble}}]{Boozer2007Reversible}%
  \BibitemOpen
  \bibfield  {author} {\bibinfo {author} {\bibnamefont {Boozer}, \bibfnamefont
  {A.~D.}}, \bibinfo {author} {\bibfnamefont {A.}~\bibnamefont {Boca}},
  \bibinfo {author} {\bibfnamefont {R.}~\bibnamefont {Miller}}, \bibinfo
  {author} {\bibfnamefont {T.~E.}\ \bibnamefont {Northup}}, \ and\ \bibinfo
  {author} {\bibfnamefont {H.~J.}\ \bibnamefont {Kimble}}} (\bibinfo {year}
  {2007}),\ \href {\doibase 10.1103/PhysRevLett.98.193601} {\bibfield
  {journal} {\bibinfo  {journal} {Physical Review Letters}\ }\textbf {\bibinfo
  {volume} {98}}~(\bibinfo {number} {19}),\ \bibinfo {pages}
  {193601+}}\BibitemShut {NoStop}%
\bibitem [{\citenamefont {{B}otter}\ \emph {et~al.}(2009)\citenamefont
  {{B}otter}, \citenamefont {{B}rooks}, \citenamefont {{G}upta}, \citenamefont
  {{Zhao}-{Yuan}}, \citenamefont {{M}oore}, \citenamefont {{M}urch},
  \citenamefont {{Purdy}},\ and\ \citenamefont {{S}tamper
  {K}urn}}]{Botter2009Quantum}%
  \BibitemOpen
  \bibfield  {author} {\bibinfo {author} {\bibnamefont {{B}otter},
  \bibfnamefont {T.}}, \bibinfo {author} {\bibfnamefont {D.}~\bibnamefont
  {{B}rooks}}, \bibinfo {author} {\bibfnamefont {S.}~\bibnamefont {{G}upta}},
  \bibinfo {author} {\bibfnamefont {M.~A.}\ \bibnamefont {{Zhao}-{Yuan}}},
  \bibinfo {author} {\bibfnamefont {K.~L.}\ \bibnamefont {{M}oore}}, \bibinfo
  {author} {\bibfnamefont {K.~W.}\ \bibnamefont {{M}urch}}, \bibinfo {author}
  {\bibfnamefont {T.}~\bibnamefont {{Purdy}}}, \ and\ \bibinfo {author}
  {\bibfnamefont {D.~M.}\ \bibnamefont {{S}tamper {K}urn}}} (\bibinfo {year}
  {2009}),\ in\ \href
  {http://books.google.com/books?hl=en\&amp;lr=\&amp;id=6pMYIP6mWEkC\&amp;oi=fnd\&amp;pg=PA117\&amp;dq=Quantum+micro-mechanics+with+ultracold+atoms\&amp;ots=zFZWnYyb9c\&amp;sig=rQwr6WZD9sDCyTV6yr9SFOdzbLc}
  {\emph {\bibinfo {booktitle} {21st International Conference on Atomic
  Physics}}}\ (\bibinfo  {publisher} {World Scientific Pub Co Inc})\ p.\
  \bibinfo {pages} {117},\ \Eprint {http://arxiv.org/abs/arXiv:0810.3841v1}
  {arXiv:arXiv:0810.3841v1} \BibitemShut {NoStop}%
\bibitem [{\citenamefont {Brahms}\ \emph {et~al.}(2012)\citenamefont {Brahms},
  \citenamefont {Botter}, \citenamefont {Schreppler}, \citenamefont {Brooks},\
  and\ \citenamefont {Stamper~Kurn}}]{Brahms2012Optical}%
  \BibitemOpen
  \bibfield  {author} {\bibinfo {author} {\bibnamefont {Brahms}, \bibfnamefont
  {N.}}, \bibinfo {author} {\bibfnamefont {T.}~\bibnamefont {Botter}}, \bibinfo
  {author} {\bibfnamefont {S.}~\bibnamefont {Schreppler}}, \bibinfo {author}
  {\bibfnamefont {D.~W.~C.}\ \bibnamefont {Brooks}}, \ and\ \bibinfo {author}
  {\bibfnamefont {D.~M.}\ \bibnamefont {Stamper~Kurn}}} (\bibinfo {year}
  {2012}),\ \href {\doibase 10.1103/PhysRevLett.108.133601} {\bibfield
  {journal} {\bibinfo  {journal} {Physical Review Letters}\ }\textbf {\bibinfo
  {volume} {108}},\ \bibinfo {pages} {133601+}}\BibitemShut {NoStop}%
\bibitem [{\citenamefont {Brahms}\ and\ \citenamefont
  {Kurn}(2010)}]{Brahms2010Spin}%
  \BibitemOpen
  \bibfield  {author} {\bibinfo {author} {\bibnamefont {Brahms}, \bibfnamefont
  {N.}}, \ and\ \bibinfo {author} {\bibfnamefont {D.~M.~S.}\ \bibnamefont
  {Kurn}}} (\bibinfo {year} {2010}),\ \href {\doibase
  10.1103/PhysRevA.82.041804} {\bibfield  {journal} {\bibinfo  {journal}
  {Physical Review A}\ }\textbf {\bibinfo {volume} {82}}~(\bibinfo {number}
  {4}),\ \bibinfo {pages} {041804+}}\BibitemShut {NoStop}%
\bibitem [{\citenamefont {Brazovskii}(1975)}]{Brazovskii1975Phase}%
  \BibitemOpen
  \bibfield  {author} {\bibinfo {author} {\bibnamefont {Brazovskii},
  \bibfnamefont {S.}}} (\bibinfo {year} {1975}),\ \href@noop {} {\bibfield
  {journal} {\bibinfo  {journal} {Sov Phys JETP}\ }\textbf {\bibinfo {volume}
  {41}}~(\bibinfo {number} {1}),\ \bibinfo {pages} {85}}\BibitemShut {NoStop}%
\bibitem [{\citenamefont {Brennecke}\ \emph {et~al.}(2007)\citenamefont
  {Brennecke}, \citenamefont {Donner}, \citenamefont {Ritter}, \citenamefont
  {Bourdel}, \citenamefont {K\"{o}hl},\ and\ \citenamefont
  {Esslinger}}]{Brennecke2007Cavity}%
  \BibitemOpen
  \bibfield  {author} {\bibinfo {author} {\bibnamefont {Brennecke},
  \bibfnamefont {F.}}, \bibinfo {author} {\bibfnamefont {T.}~\bibnamefont
  {Donner}}, \bibinfo {author} {\bibfnamefont {S.}~\bibnamefont {Ritter}},
  \bibinfo {author} {\bibfnamefont {T.}~\bibnamefont {Bourdel}}, \bibinfo
  {author} {\bibfnamefont {M.}~\bibnamefont {K\"{o}hl}}, \ and\ \bibinfo
  {author} {\bibfnamefont {T.}~\bibnamefont {Esslinger}}} (\bibinfo {year}
  {2007}),\ \href {\doibase 10.1038/nature06120} {\bibfield  {journal}
  {\bibinfo  {journal} {Nature}\ }\textbf {\bibinfo {volume} {450}}~(\bibinfo
  {number} {7167}),\ \bibinfo {pages} {268}}\BibitemShut {NoStop}%
\bibitem [{\citenamefont {Brennecke}\ \emph {et~al.}(2008)\citenamefont
  {Brennecke}, \citenamefont {Ritter}, \citenamefont {Donner},\ and\
  \citenamefont {Esslinger}}]{Brennecke2008Cavity}%
  \BibitemOpen
  \bibfield  {author} {\bibinfo {author} {\bibnamefont {Brennecke},
  \bibfnamefont {F.}}, \bibinfo {author} {\bibfnamefont {S.}~\bibnamefont
  {Ritter}}, \bibinfo {author} {\bibfnamefont {T.}~\bibnamefont {Donner}}, \
  and\ \bibinfo {author} {\bibfnamefont {T.}~\bibnamefont {Esslinger}}}
  (\bibinfo {year} {2008}),\ \href {\doibase 10.1126/science.1163218}
  {\bibfield  {journal} {\bibinfo  {journal} {Science}\ }\textbf {\bibinfo
  {volume} {322}}~(\bibinfo {number} {5899}),\ \bibinfo {pages}
  {235}}\BibitemShut {NoStop}%
\bibitem [{\citenamefont {Brooks}\ \emph {et~al.}(2012)\citenamefont {Brooks},
  \citenamefont {Botter}, \citenamefont {Schreppler}, \citenamefont {Purdy},
  \citenamefont {Brahms},\ and\ \citenamefont
  {Stamper-Kurn}}]{Brooks2012Nonclassical}%
  \BibitemOpen
  \bibfield  {author} {\bibinfo {author} {\bibnamefont {Brooks}, \bibfnamefont
  {D.~W.~C.}}, \bibinfo {author} {\bibfnamefont {T.}~\bibnamefont {Botter}},
  \bibinfo {author} {\bibfnamefont {S.}~\bibnamefont {Schreppler}}, \bibinfo
  {author} {\bibfnamefont {T.~P.}\ \bibnamefont {Purdy}}, \bibinfo {author}
  {\bibfnamefont {N.}~\bibnamefont {Brahms}}, \ and\ \bibinfo {author}
  {\bibfnamefont {D.~M.}\ \bibnamefont {Stamper-Kurn}}} (\bibinfo {year}
  {2012}),\ \href {\doibase 10.1038/nature11325} {\bibfield  {journal}
  {\bibinfo  {journal} {Nature}\ }\textbf {\bibinfo {volume} {488}}~(\bibinfo
  {number} {7412}),\ \bibinfo {pages} {476}}\BibitemShut {NoStop}%
\bibitem [{\citenamefont {Brune}\ \emph {et~al.}(1990)\citenamefont {Brune},
  \citenamefont {Haroche}, \citenamefont {Lefevre}, \citenamefont {Raimond},\
  and\ \citenamefont {Zagury}}]{Brune1990Quantum}%
  \BibitemOpen
  \bibfield  {author} {\bibinfo {author} {\bibnamefont {Brune}, \bibfnamefont
  {M.}}, \bibinfo {author} {\bibfnamefont {S.}~\bibnamefont {Haroche}},
  \bibinfo {author} {\bibfnamefont {V.}~\bibnamefont {Lefevre}}, \bibinfo
  {author} {\bibfnamefont {J.~M.}\ \bibnamefont {Raimond}}, \ and\ \bibinfo
  {author} {\bibfnamefont {N.}~\bibnamefont {Zagury}}} (\bibinfo {year}
  {1990}),\ \href {\doibase 10.1103/PhysRevLett.65.976} {\bibfield  {journal}
  {\bibinfo  {journal} {Physical Review Letters}\ }\textbf {\bibinfo {volume}
  {65}},\ \bibinfo {pages} {976}}\BibitemShut {NoStop}%
\bibitem [{\citenamefont {Brzozowska}\ \emph {et~al.}(2006)\citenamefont
  {Brzozowska}, \citenamefont {Brzozowski}, \citenamefont {Zachorowski},\ and\
  \citenamefont {Gawlik}}]{Brzozowska2006Bound}%
  \BibitemOpen
  \bibfield  {author} {\bibinfo {author} {\bibnamefont {Brzozowska},
  \bibfnamefont {M.}}, \bibinfo {author} {\bibfnamefont {T.~M.}\ \bibnamefont
  {Brzozowski}}, \bibinfo {author} {\bibfnamefont {J.}~\bibnamefont
  {Zachorowski}}, \ and\ \bibinfo {author} {\bibfnamefont {W.}~\bibnamefont
  {Gawlik}}} (\bibinfo {year} {2006}),\ \href {\doibase
  10.1103/PhysRevA.73.063414} {\bibfield  {journal} {\bibinfo  {journal}
  {Physical Review A}\ }\textbf {\bibinfo {volume} {73}},\ \bibinfo {pages}
  {063414+}}\BibitemShut {NoStop}%
\bibitem [{\citenamefont {B\"{u}chler}\ and\ \citenamefont
  {Blatter}(2003)}]{Buchler2003Supersolid}%
  \BibitemOpen
  \bibfield  {author} {\bibinfo {author} {\bibnamefont {B\"{u}chler},
  \bibfnamefont {H.~P.}}, \ and\ \bibinfo {author} {\bibfnamefont
  {G.}~\bibnamefont {Blatter}}} (\bibinfo {year} {2003}),\ \href {\doibase
  10.1103/PhysRevLett.91.130404} {\bibfield  {journal} {\bibinfo  {journal}
  {Physical Review Letters}\ }\textbf {\bibinfo {volume} {91}},\ \bibinfo
  {pages} {130404+}}\BibitemShut {NoStop}%
\bibitem [{\citenamefont {Bux}\ \emph {et~al.}(2011)\citenamefont {Bux},
  \citenamefont {Gnahm}, \citenamefont {Maier}, \citenamefont {Zimmermann},\
  and\ \citenamefont {Courteille}}]{Bux2011CavityControlled}%
  \BibitemOpen
  \bibfield  {author} {\bibinfo {author} {\bibnamefont {Bux}, \bibfnamefont
  {S.}}, \bibinfo {author} {\bibfnamefont {C.}~\bibnamefont {Gnahm}}, \bibinfo
  {author} {\bibfnamefont {R.~A.~W.}\ \bibnamefont {Maier}}, \bibinfo {author}
  {\bibfnamefont {C.}~\bibnamefont {Zimmermann}}, \ and\ \bibinfo {author}
  {\bibfnamefont {P.~W.}\ \bibnamefont {Courteille}}} (\bibinfo {year}
  {2011}),\ \href {\doibase 10.1103/PhysRevLett.106.203601} {\bibfield
  {journal} {\bibinfo  {journal} {Physical Review Letters}\ }\textbf {\bibinfo
  {volume} {106}},\ \bibinfo {pages} {203601+}}\BibitemShut {NoStop}%
\bibitem [{\citenamefont {Carmichael}(2003)}]{Carmichael2003Statistical}%
  \BibitemOpen
  \bibfield  {author} {\bibinfo {author} {\bibnamefont {Carmichael},
  \bibfnamefont {H.~J.}}} (\bibinfo {year} {2003}),\ \href@noop {} {\emph
  {\bibinfo {title} {Statistical Methods in Quantum Optics 1: Master Equations
  and Fokker-Planck Equations}}}\ (\bibinfo  {publisher}
  {Springer})\BibitemShut {NoStop}%
\bibitem [{\citenamefont {Carmichael}\ \emph {et~al.}(1973)\citenamefont
  {Carmichael}, \citenamefont {Gardiner},\ and\ \citenamefont
  {Walls}}]{Carmichael1973Higher}%
  \BibitemOpen
  \bibfield  {author} {\bibinfo {author} {\bibnamefont {Carmichael},
  \bibfnamefont {H.~J.}}, \bibinfo {author} {\bibfnamefont {C.~W.}\
  \bibnamefont {Gardiner}}, \ and\ \bibinfo {author} {\bibfnamefont {D.~F.}\
  \bibnamefont {Walls}}} (\bibinfo {year} {1973}),\ \href {\doibase
  10.1016/0375-9601(73)90679-8} {\bibfield  {journal} {\bibinfo  {journal}
  {Physics Letters A}\ }\textbf {\bibinfo {volume} {46}}~(\bibinfo {number}
  {1}),\ \bibinfo {pages} {47}}\BibitemShut {NoStop}%
\bibitem [{\citenamefont {Caves}(1980)}]{Caves1980QuantumMechanical}%
  \BibitemOpen
  \bibfield  {author} {\bibinfo {author} {\bibnamefont {Caves}, \bibfnamefont
  {C.~M.}}} (\bibinfo {year} {1980}),\ \href {\doibase
  10.1103/PhysRevLett.45.75} {\bibfield  {journal} {\bibinfo  {journal}
  {Physical Review Letters}\ }\textbf {\bibinfo {volume} {45}}~(\bibinfo
  {number} {2}),\ \bibinfo {pages} {75}}\BibitemShut {NoStop}%
\bibitem [{\citenamefont {Chan}\ \emph {et~al.}(2003)\citenamefont {Chan},
  \citenamefont {Black},\ and\ \citenamefont
  {Vuleti\'{c}}}]{Chan2003Observation}%
  \BibitemOpen
  \bibfield  {author} {\bibinfo {author} {\bibnamefont {Chan}, \bibfnamefont
  {H.~W.}}, \bibinfo {author} {\bibfnamefont {A.~T.}\ \bibnamefont {Black}}, \
  and\ \bibinfo {author} {\bibfnamefont {V.}~\bibnamefont {Vuleti\'{c}}}}
  (\bibinfo {year} {2003}),\ \href {\doibase 10.1103/PhysRevLett.90.063003}
  {\bibfield  {journal} {\bibinfo  {journal} {Physical Review Letters}\
  }\textbf {\bibinfo {volume} {90}}~(\bibinfo {number} {6}),\ \bibinfo {pages}
  {063003+}}\BibitemShut {NoStop}%
\bibitem [{\citenamefont {Chang}\ \emph {et~al.}(2009)\citenamefont {Chang},
  \citenamefont {Regal}, \citenamefont {Papp}, \citenamefont {Wilson},
  \citenamefont {Ye}, \citenamefont {Painter}, \citenamefont {Kimble},\ and\
  \citenamefont {Zoller}}]{Chang2009Cavity}%
  \BibitemOpen
  \bibfield  {author} {\bibinfo {author} {\bibnamefont {Chang}, \bibfnamefont
  {D.~E.}}, \bibinfo {author} {\bibfnamefont {C.~A.}\ \bibnamefont {Regal}},
  \bibinfo {author} {\bibfnamefont {S.~B.}\ \bibnamefont {Papp}}, \bibinfo
  {author} {\bibfnamefont {D.~J.}\ \bibnamefont {Wilson}}, \bibinfo {author}
  {\bibfnamefont {J.}~\bibnamefont {Ye}}, \bibinfo {author} {\bibfnamefont
  {O.}~\bibnamefont {Painter}}, \bibinfo {author} {\bibfnamefont {H.~J.}\
  \bibnamefont {Kimble}}, \ and\ \bibinfo {author} {\bibfnamefont
  {P.}~\bibnamefont {Zoller}}} (\bibinfo {year} {2009}),\ \href {\doibase
  10.1073/pnas.0912969107} {\bibfield  {journal} {\bibinfo  {journal}
  {Proceedings of the National Academy of Sciences}\ }\textbf {\bibinfo
  {volume} {107}}~(\bibinfo {number} {3}),\ \bibinfo {pages}
  {1005}}\BibitemShut {NoStop}%
\bibitem [{\citenamefont {Chavanis}(2007)}]{Chavanis2007Kinetic}%
  \BibitemOpen
  \bibfield  {author} {\bibinfo {author} {\bibnamefont {Chavanis},
  \bibfnamefont {P.}}} (\bibinfo {year} {2007}),\ \href {\doibase
  10.1016/j.physa.2006.11.078} {\bibfield  {journal} {\bibinfo  {journal}
  {Physica A: Statistical Mechanics and its Applications}\ }\textbf {\bibinfo
  {volume} {377}}~(\bibinfo {number} {2}),\ \bibinfo {pages} {469}}\BibitemShut
  {NoStop}%
\bibitem [{\citenamefont {Chen}\ \emph {et~al.}(2011)\citenamefont {Chen},
  \citenamefont {Jiang}, \citenamefont {Li},\ and\ \citenamefont
  {Di~Zhu}}]{Chen2011Alloptical}%
  \BibitemOpen
  \bibfield  {author} {\bibinfo {author} {\bibnamefont {Chen}, \bibfnamefont
  {B.}}, \bibinfo {author} {\bibfnamefont {C.}~\bibnamefont {Jiang}}, \bibinfo
  {author} {\bibfnamefont {J.~J.}\ \bibnamefont {Li}}, \ and\ \bibinfo {author}
  {\bibfnamefont {K.}~\bibnamefont {Di~Zhu}}} (\bibinfo {year} {2011}),\ \href
  {\doibase 10.1103/PhysRevA.84.055802} {\bibfield  {journal} {\bibinfo
  {journal} {Physical Review A}\ }\textbf {\bibinfo {volume} {84}},\ \bibinfo
  {pages} {055802+}}\BibitemShut {NoStop}%
\bibitem [{\citenamefont {Chen}\ \emph {et~al.}(2010)\citenamefont {Chen},
  \citenamefont {Goldbaum}, \citenamefont {Bhattacharya},\ and\ \citenamefont
  {Meystre}}]{Chen2010Classical}%
  \BibitemOpen
  \bibfield  {author} {\bibinfo {author} {\bibnamefont {Chen}, \bibfnamefont
  {W.}}, \bibinfo {author} {\bibfnamefont {D.~S.}\ \bibnamefont {Goldbaum}},
  \bibinfo {author} {\bibfnamefont {M.}~\bibnamefont {Bhattacharya}}, \ and\
  \bibinfo {author} {\bibfnamefont {P.}~\bibnamefont {Meystre}}} (\bibinfo
  {year} {2010}),\ \href {\doibase 10.1103/PhysRevA.81.053833} {\bibfield
  {journal} {\bibinfo  {journal} {Physical Review A}\ }\textbf {\bibinfo
  {volume} {81}}~(\bibinfo {number} {5}),\ \bibinfo {pages}
  {053833+}}\BibitemShut {NoStop}%
\bibitem [{\citenamefont {Chen}\ \emph {et~al.}(2007)\citenamefont {Chen},
  \citenamefont {Meiser},\ and\ \citenamefont {Meystre}}]{Chen2007Cavity}%
  \BibitemOpen
  \bibfield  {author} {\bibinfo {author} {\bibnamefont {Chen}, \bibfnamefont
  {W.}}, \bibinfo {author} {\bibfnamefont {D.}~\bibnamefont {Meiser}}, \ and\
  \bibinfo {author} {\bibfnamefont {P.}~\bibnamefont {Meystre}}} (\bibinfo
  {year} {2007}),\ \href {\doibase 10.1103/PhysRevA.75.023812} {\bibfield
  {journal} {\bibinfo  {journal} {Physical Review A}\ }\textbf {\bibinfo
  {volume} {75}},\ \bibinfo {pages} {023812+}}\BibitemShut {NoStop}%
\bibitem [{\citenamefont {Chen}\ and\ \citenamefont
  {Meystre}(2009)}]{Chen2009Cavity}%
  \BibitemOpen
  \bibfield  {author} {\bibinfo {author} {\bibnamefont {Chen}, \bibfnamefont
  {W.}}, \ and\ \bibinfo {author} {\bibfnamefont {P.}~\bibnamefont {Meystre}}}
  (\bibinfo {year} {2009}),\ \href {\doibase 10.1103/PhysRevA.79.043801}
  {\bibfield  {journal} {\bibinfo  {journal} {Physical Review A}\ }\textbf
  {\bibinfo {volume} {79}}~(\bibinfo {number} {4}),\ \bibinfo {pages}
  {043801+}}\BibitemShut {NoStop}%
\bibitem [{\citenamefont {Chen}\ \emph {et~al.}(2009)\citenamefont {Chen},
  \citenamefont {Zhang}, \citenamefont {Goldbaum}, \citenamefont
  {Bhattacharya},\ and\ \citenamefont {Meystre}}]{Chen2009Bistable}%
  \BibitemOpen
  \bibfield  {author} {\bibinfo {author} {\bibnamefont {Chen}, \bibfnamefont
  {W.}}, \bibinfo {author} {\bibfnamefont {K.}~\bibnamefont {Zhang}}, \bibinfo
  {author} {\bibfnamefont {D.~S.}\ \bibnamefont {Goldbaum}}, \bibinfo {author}
  {\bibfnamefont {M.}~\bibnamefont {Bhattacharya}}, \ and\ \bibinfo {author}
  {\bibfnamefont {P.}~\bibnamefont {Meystre}}} (\bibinfo {year} {2009}),\ \href
  {\doibase 10.1103/PhysRevA.80.011801} {\bibfield  {journal} {\bibinfo
  {journal} {Physical Review A}\ }\textbf {\bibinfo {volume} {80}}~(\bibinfo
  {number} {1}),\ \bibinfo {pages} {011801+}}\BibitemShut {NoStop}%
\bibitem [{\citenamefont {Chu}(1998)}]{Chu1998Nobel}%
  \BibitemOpen
  \bibfield  {author} {\bibinfo {author} {\bibnamefont {Chu}, \bibfnamefont
  {S.}}} (\bibinfo {year} {1998}),\ \href {\doibase 10.1103/RevModPhys.70.685}
  {\bibfield  {journal} {\bibinfo  {journal} {Reviews of Modern Physics}\
  }\textbf {\bibinfo {volume} {70}}~(\bibinfo {number} {3}),\ \bibinfo {pages}
  {685}}\BibitemShut {NoStop}%
\bibitem [{\citenamefont {Cirac}\ \emph {et~al.}(1995)\citenamefont {Cirac},
  \citenamefont {Lewenstein},\ and\ \citenamefont {Zoller}}]{Cirac1995Laser}%
  \BibitemOpen
  \bibfield  {author} {\bibinfo {author} {\bibnamefont {Cirac}, \bibfnamefont
  {J.~I.}}, \bibinfo {author} {\bibfnamefont {M.}~\bibnamefont {Lewenstein}}, \
  and\ \bibinfo {author} {\bibfnamefont {P.}~\bibnamefont {Zoller}}} (\bibinfo
  {year} {1995}),\ \href {\doibase 10.1103/PhysRevA.51.1650} {\bibfield
  {journal} {\bibinfo  {journal} {Physical Review A}\ }\textbf {\bibinfo
  {volume} {51}}~(\bibinfo {number} {2}),\ \bibinfo {pages} {1650}}\BibitemShut
  {NoStop}%
\bibitem [{\citenamefont {Cirac}\ \emph {et~al.}(1997)\citenamefont {Cirac},
  \citenamefont {Zoller}, \citenamefont {Kimble},\ and\ \citenamefont
  {Mabuchi}}]{Cirac1997Quantum}%
  \BibitemOpen
  \bibfield  {author} {\bibinfo {author} {\bibnamefont {Cirac}, \bibfnamefont
  {J.~I.}}, \bibinfo {author} {\bibfnamefont {P.}~\bibnamefont {Zoller}},
  \bibinfo {author} {\bibfnamefont {H.~J.}\ \bibnamefont {Kimble}}, \ and\
  \bibinfo {author} {\bibfnamefont {H.}~\bibnamefont {Mabuchi}}} (\bibinfo
  {year} {1997}),\ \href {\doibase 10.1103/PhysRevLett.78.3221} {\bibfield
  {journal} {\bibinfo  {journal} {Physical Review Letters}\ }\textbf {\bibinfo
  {volume} {78}}~(\bibinfo {number} {16}),\ \bibinfo {pages}
  {3221}}\BibitemShut {NoStop}%
\bibitem [{\citenamefont {Cohen-Tannoudji}(1992)}]{cct}%
  \BibitemOpen
  \bibfield  {author} {\bibinfo {author} {\bibnamefont {Cohen-Tannoudji},
  \bibfnamefont {C.}}} (\bibinfo {year} {1992}),\ in\ \href@noop {} {\emph
  {\bibinfo {booktitle} {Fundamental Systems in Quantum Optics, Proceedings of
  the Les Houches Summer School, Session LIII}}},\ \bibinfo {editor} {edited
  by\ \bibinfo {editor} {\bibfnamefont {J.}~\bibnamefont {Dalibard}}, \bibinfo
  {editor} {\bibfnamefont {J.-M.}\ \bibnamefont {Raimond}}, \ and\ \bibinfo
  {editor} {\bibfnamefont {J.}~\bibnamefont {Zinn-Justin}}}\ (\bibinfo
  {publisher} {North-Holland, Amsterdam})\ pp.\ \bibinfo {pages}
  {1--164}\BibitemShut {NoStop}%
\bibitem [{\citenamefont {Cohen~Tannoudji}(1998)}]{CohenTannoudji1998Nobel}%
  \BibitemOpen
  \bibfield  {author} {\bibinfo {author} {\bibnamefont {Cohen~Tannoudji},
  \bibfnamefont {C.~N.}}} (\bibinfo {year} {1998}),\ \href {\doibase
  10.1103/RevModPhys.70.707} {\bibfield  {journal} {\bibinfo  {journal}
  {Reviews of Modern Physics}\ }\textbf {\bibinfo {volume} {70}}~(\bibinfo
  {number} {3}),\ \bibinfo {pages} {707}}\BibitemShut {NoStop}%
\bibitem [{\citenamefont {Colombe}\ \emph {et~al.}(2007)\citenamefont
  {Colombe}, \citenamefont {Steinmetz}, \citenamefont {Dubois}, \citenamefont
  {Linke}, \citenamefont {Hunger},\ and\ \citenamefont
  {Reichel}}]{Colombe2007Strong}%
  \BibitemOpen
  \bibfield  {author} {\bibinfo {author} {\bibnamefont {Colombe}, \bibfnamefont
  {Y.}}, \bibinfo {author} {\bibfnamefont {T.}~\bibnamefont {Steinmetz}},
  \bibinfo {author} {\bibfnamefont {G.}~\bibnamefont {Dubois}}, \bibinfo
  {author} {\bibfnamefont {F.}~\bibnamefont {Linke}}, \bibinfo {author}
  {\bibfnamefont {D.}~\bibnamefont {Hunger}}, \ and\ \bibinfo {author}
  {\bibfnamefont {J.}~\bibnamefont {Reichel}}} (\bibinfo {year} {2007}),\
  \bibfield  {booktitle} {\emph {\bibinfo {booktitle} {Nature}},\ }\href
  {\doibase 10.1038/nature06331} {\bibfield  {journal} {\bibinfo  {journal}
  {Nature}\ }\textbf {\bibinfo {volume} {450}}~(\bibinfo {number} {7167}),\
  \bibinfo {pages} {272}}\BibitemShut {NoStop}%
\bibitem [{\citenamefont {Cornell}\ and\ \citenamefont
  {Wieman}(2002)}]{Cornell2002Nobel}%
  \BibitemOpen
  \bibfield  {author} {\bibinfo {author} {\bibnamefont {Cornell}, \bibfnamefont
  {E.~A.}}, \ and\ \bibinfo {author} {\bibfnamefont {C.~E.}\ \bibnamefont
  {Wieman}}} (\bibinfo {year} {2002}),\ \href {\doibase
  10.1103/RevModPhys.74.875} {\bibfield  {journal} {\bibinfo  {journal}
  {Reviews of Modern Physics}\ }\textbf {\bibinfo {volume} {74}}~(\bibinfo
  {number} {3}),\ \bibinfo {pages} {875}}\BibitemShut {NoStop}%
\bibitem [{\citenamefont {Courtois}\ \emph {et~al.}(1994)\citenamefont
  {Courtois}, \citenamefont {Grynberg}, \citenamefont {Lounis},\ and\
  \citenamefont {Verkerk}}]{Courtois1994Recoil}%
  \BibitemOpen
  \bibfield  {author} {\bibinfo {author} {\bibnamefont {Courtois},
  \bibfnamefont {J.-Y.}}, \bibinfo {author} {\bibfnamefont {G.}~\bibnamefont
  {Grynberg}}, \bibinfo {author} {\bibfnamefont {B.}~\bibnamefont {Lounis}}, \
  and\ \bibinfo {author} {\bibfnamefont {P.}~\bibnamefont {Verkerk}}} (\bibinfo
  {year} {1994}),\ \href {\doibase 10.1103/PhysRevLett.72.3017} {\bibfield
  {journal} {\bibinfo  {journal} {Phys. Rev. Lett.}\ }\textbf {\bibinfo
  {volume} {72}},\ \bibinfo {pages} {3017}}\BibitemShut {NoStop}%
\bibitem [{\citenamefont {Cube}\ \emph {et~al.}(2004)\citenamefont {Cube},
  \citenamefont {Slama}, \citenamefont {Kruse}, \citenamefont {Zimmermann},
  \citenamefont {Courteille}, \citenamefont {Robb}, \citenamefont {Piovella},\
  and\ \citenamefont {Bonifacio}}]{Cube2004SelfSynchronization}%
  \BibitemOpen
  \bibfield  {author} {\bibinfo {author} {\bibnamefont {Cube}, \bibfnamefont
  {C.}}, \bibinfo {author} {\bibfnamefont {S.}~\bibnamefont {Slama}}, \bibinfo
  {author} {\bibfnamefont {D.}~\bibnamefont {Kruse}}, \bibinfo {author}
  {\bibfnamefont {C.}~\bibnamefont {Zimmermann}}, \bibinfo {author}
  {\bibfnamefont {P.~W.}\ \bibnamefont {Courteille}}, \bibinfo {author}
  {\bibfnamefont {G.~R.~M.}\ \bibnamefont {Robb}}, \bibinfo {author}
  {\bibfnamefont {N.}~\bibnamefont {Piovella}}, \ and\ \bibinfo {author}
  {\bibfnamefont {R.}~\bibnamefont {Bonifacio}}} (\bibinfo {year} {2004}),\
  \href@noop {} {\bibfield  {journal} {\bibinfo  {journal} {Phys. Rev. Lett.}\
  }\textbf {\bibinfo {volume} {93}},\ \bibinfo {pages} {083601+}}\BibitemShut
  {NoStop}%
\bibitem [{\citenamefont {Dalibard}\ and\ \citenamefont
  {Cohen-Tannoudji}(1985)}]{Dalibard1985Atomic}%
  \BibitemOpen
  \bibfield  {author} {\bibinfo {author} {\bibnamefont {Dalibard},
  \bibfnamefont {J.}}, \ and\ \bibinfo {author} {\bibfnamefont
  {C.}~\bibnamefont {Cohen-Tannoudji}}} (\bibinfo {year} {1985}),\ \href
  {\doibase 10.1088/0022-3700/18/8/019} {\bibfield  {journal} {\bibinfo
  {journal} {Journal of Physics B: Atomic and Molecular Physics}\ }\textbf
  {\bibinfo {volume} {18}}~(\bibinfo {number} {8}),\ \bibinfo {pages}
  {1661+}}\BibitemShut {NoStop}%
\bibitem [{\citenamefont {De~Chiara}\ \emph {et~al.}(2011)\citenamefont
  {De~Chiara}, \citenamefont {Paternostro},\ and\ \citenamefont
  {Palma}}]{DeChiara2011Entanglement}%
  \BibitemOpen
  \bibfield  {author} {\bibinfo {author} {\bibnamefont {De~Chiara},
  \bibfnamefont {G.}}, \bibinfo {author} {\bibfnamefont {M.}~\bibnamefont
  {Paternostro}}, \ and\ \bibinfo {author} {\bibfnamefont {G.~M.}\ \bibnamefont
  {Palma}}} (\bibinfo {year} {2011}),\ \href {\doibase
  10.1103/PhysRevA.83.052324} {\bibfield  {journal} {\bibinfo  {journal}
  {Physical Review A}\ }\textbf {\bibinfo {volume} {83}}~(\bibinfo {number}
  {5}),\ \bibinfo {pages} {052324+}}\BibitemShut {NoStop}%
\bibitem [{\citenamefont {Deachapunya}\ \emph {et~al.}(2008)\citenamefont
  {Deachapunya}, \citenamefont {Fagan}, \citenamefont {Major}, \citenamefont
  {Reiger}, \citenamefont {Ritsch}, \citenamefont {Stefanov}, \citenamefont
  {Ulbricht},\ and\ \citenamefont {Arndt}}]{Deachapunya2008Slow}%
  \BibitemOpen
  \bibfield  {author} {\bibinfo {author} {\bibnamefont {Deachapunya},
  \bibfnamefont {S.}}, \bibinfo {author} {\bibfnamefont {P.~J.}\ \bibnamefont
  {Fagan}}, \bibinfo {author} {\bibfnamefont {A.~G.}\ \bibnamefont {Major}},
  \bibinfo {author} {\bibfnamefont {E.}~\bibnamefont {Reiger}}, \bibinfo
  {author} {\bibfnamefont {H.}~\bibnamefont {Ritsch}}, \bibinfo {author}
  {\bibfnamefont {A.}~\bibnamefont {Stefanov}}, \bibinfo {author}
  {\bibfnamefont {H.}~\bibnamefont {Ulbricht}}, \ and\ \bibinfo {author}
  {\bibfnamefont {M.}~\bibnamefont {Arndt}}} (\bibinfo {year} {2008}),\ \href
  {\doibase 10.1140/epjd/e2007-00301-8} {\bibfield  {journal} {\bibinfo
  {journal} {The European Physical Journal D - Atomic, Molecular, Optical and
  Plasma Physics}\ }\textbf {\bibinfo {volume} {46}}~(\bibinfo {number} {2}),\
  \bibinfo {pages} {307}}\BibitemShut {NoStop}%
\bibitem [{\citenamefont {DeVoe}\ and\ \citenamefont
  {Brewer}(1996)}]{DeVoe1996Observation}%
  \BibitemOpen
  \bibfield  {author} {\bibinfo {author} {\bibnamefont {DeVoe}, \bibfnamefont
  {R.~G.}}, \ and\ \bibinfo {author} {\bibfnamefont {R.~G.}\ \bibnamefont
  {Brewer}}} (\bibinfo {year} {1996}),\ \href {\doibase
  10.1103/PhysRevLett.76.2049} {\bibfield  {journal} {\bibinfo  {journal}
  {Physical Review Letters}\ }\textbf {\bibinfo {volume} {76}},\ \bibinfo
  {pages} {2049}}\BibitemShut {NoStop}%
\bibitem [{\citenamefont {Dicke}(1954)}]{Dicke1954Coherence}%
  \BibitemOpen
  \bibfield  {author} {\bibinfo {author} {\bibnamefont {Dicke}, \bibfnamefont
  {R.~H.}}} (\bibinfo {year} {1954}),\ \href {\doibase 10.1103/PhysRev.93.99}
  {\bibfield  {journal} {\bibinfo  {journal} {Physical Review Online Archive
  (Prola)}\ }\textbf {\bibinfo {volume} {93}}~(\bibinfo {number} {1}),\
  \bibinfo {pages} {99}}\BibitemShut {NoStop}%
\bibitem [{\citenamefont {Dimer}\ \emph {et~al.}(2007)\citenamefont {Dimer},
  \citenamefont {Estienne}, \citenamefont {Parkins},\ and\ \citenamefont
  {Carmichael}}]{Dimer2007Proposed}%
  \BibitemOpen
  \bibfield  {author} {\bibinfo {author} {\bibnamefont {Dimer}, \bibfnamefont
  {F.}}, \bibinfo {author} {\bibfnamefont {B.}~\bibnamefont {Estienne}},
  \bibinfo {author} {\bibfnamefont {A.~S.}\ \bibnamefont {Parkins}}, \ and\
  \bibinfo {author} {\bibfnamefont {H.~J.}\ \bibnamefont {Carmichael}}}
  (\bibinfo {year} {2007}),\ \href {\doibase 10.1103/PhysRevA.75.013804}
  {\bibfield  {journal} {\bibinfo  {journal} {Physical Review A}\ }\textbf
  {\bibinfo {volume} {75}}~(\bibinfo {number} {1}),\ \bibinfo {pages}
  {013804+}}\BibitemShut {NoStop}%
\bibitem [{\citenamefont {Doherty}\ \emph {et~al.}(2000)\citenamefont
  {Doherty}, \citenamefont {Lynn}, \citenamefont {Hood},\ and\ \citenamefont
  {Kimble}}]{Doherty2000Trapping}%
  \BibitemOpen
  \bibfield  {author} {\bibinfo {author} {\bibnamefont {Doherty}, \bibfnamefont
  {A.~C.}}, \bibinfo {author} {\bibfnamefont {T.~W.}\ \bibnamefont {Lynn}},
  \bibinfo {author} {\bibfnamefont {C.~J.}\ \bibnamefont {Hood}}, \ and\
  \bibinfo {author} {\bibfnamefont {H.~J.}\ \bibnamefont {Kimble}}} (\bibinfo
  {year} {2000}),\ \href {\doibase 10.1103/PhysRevA.63.013401} {\bibfield
  {journal} {\bibinfo  {journal} {Physical Review A}\ }\textbf {\bibinfo
  {volume} {63}}~(\bibinfo {number} {1}),\ \bibinfo {pages}
  {013401+}}\BibitemShut {NoStop}%
\bibitem [{\citenamefont {Domokos}\ \emph {et~al.}(2001)\citenamefont
  {Domokos}, \citenamefont {Horak},\ and\ \citenamefont
  {Ritsch}}]{Domokos2001Semiclassical}%
  \BibitemOpen
  \bibfield  {author} {\bibinfo {author} {\bibnamefont {Domokos}, \bibfnamefont
  {P.}}, \bibinfo {author} {\bibfnamefont {P.}~\bibnamefont {Horak}}, \ and\
  \bibinfo {author} {\bibfnamefont {H.}~\bibnamefont {Ritsch}}} (\bibinfo
  {year} {2001}),\ \href {http://iopscience.iop.org/0953-4075/34/2/306}
  {\bibfield  {journal} {\bibinfo  {journal} {J. Phys. B}\ }\textbf {\bibinfo
  {volume} {34}},\ \bibinfo {pages} {187+}}\BibitemShut {NoStop}%
\bibitem [{\citenamefont {Domokos}\ and\ \citenamefont
  {Ritsch}(2002)}]{Domokos2002Collective}%
  \BibitemOpen
  \bibfield  {author} {\bibinfo {author} {\bibnamefont {Domokos}, \bibfnamefont
  {P.}}, \ and\ \bibinfo {author} {\bibfnamefont {H.}~\bibnamefont {Ritsch}}}
  (\bibinfo {year} {2002}),\ \href {\doibase 10.1103/PhysRevLett.89.253003}
  {\bibfield  {journal} {\bibinfo  {journal} {Physical Review Letters}\
  }\textbf {\bibinfo {volume} {89}}~(\bibinfo {number} {25}),\ \bibinfo {pages}
  {253003+}}\BibitemShut {NoStop}%
\bibitem [{\citenamefont {Domokos}\ and\ \citenamefont
  {Ritsch}(2003)}]{Domokos2003Mechanical}%
  \BibitemOpen
  \bibfield  {author} {\bibinfo {author} {\bibnamefont {Domokos}, \bibfnamefont
  {P.}}, \ and\ \bibinfo {author} {\bibfnamefont {H.}~\bibnamefont {Ritsch}}}
  (\bibinfo {year} {2003}),\ \href {\doibase 10.1364/JOSAB.20.001098}
  {\bibfield  {journal} {\bibinfo  {journal} {J. Opt. Soc. Am. B}\ }\textbf
  {\bibinfo {volume} {20}}~(\bibinfo {number} {5}),\ \bibinfo {pages}
  {1098}}\BibitemShut {NoStop}%
\bibitem [{\citenamefont {Domokos}\ \emph {et~al.}(2002)\citenamefont
  {Domokos}, \citenamefont {Salzburger},\ and\ \citenamefont
  {Ritsch}}]{Domokos2002Dissipative}%
  \BibitemOpen
  \bibfield  {author} {\bibinfo {author} {\bibnamefont {Domokos}, \bibfnamefont
  {P.}}, \bibinfo {author} {\bibfnamefont {T.}~\bibnamefont {Salzburger}}, \
  and\ \bibinfo {author} {\bibfnamefont {H.}~\bibnamefont {Ritsch}}} (\bibinfo
  {year} {2002}),\ \href {\doibase 10.1103/PhysRevA.66.043406} {\bibfield
  {journal} {\bibinfo  {journal} {Physical Review A}\ }\textbf {\bibinfo
  {volume} {66}}~(\bibinfo {number} {4}),\ \bibinfo {pages}
  {043406+}}\BibitemShut {NoStop}%
\bibitem [{\citenamefont {Domokos}\ \emph {et~al.}(2004)\citenamefont
  {Domokos}, \citenamefont {Vukics},\ and\ \citenamefont
  {Ritsch}}]{Domokos2004Anomalous}%
  \BibitemOpen
  \bibfield  {author} {\bibinfo {author} {\bibnamefont {Domokos}, \bibfnamefont
  {P.}}, \bibinfo {author} {\bibfnamefont {A.}~\bibnamefont {Vukics}}, \ and\
  \bibinfo {author} {\bibfnamefont {H.}~\bibnamefont {Ritsch}}} (\bibinfo
  {year} {2004}),\ \href {\doibase 10.1103/PhysRevLett.92.103601} {\bibfield
  {journal} {\bibinfo  {journal} {Physical Review Letters}\ }\textbf {\bibinfo
  {volume} {92}}~(\bibinfo {number} {10}),\ \bibinfo {pages}
  {103601+}}\BibitemShut {NoStop}%
\bibitem [{\citenamefont {Dotsenko}\ \emph {et~al.}(2005)\citenamefont
  {Dotsenko}, \citenamefont {Alt}, \citenamefont {Khudaverdyan}, \citenamefont
  {Kuhr}, \citenamefont {Meschede}, \citenamefont {Miroshnychenko},
  \citenamefont {Schrader},\ and\ \citenamefont
  {Rauschenbeutel}}]{Dotsenko2005Submicrometer}%
  \BibitemOpen
  \bibfield  {author} {\bibinfo {author} {\bibnamefont {Dotsenko},
  \bibfnamefont {I.}}, \bibinfo {author} {\bibfnamefont {W.}~\bibnamefont
  {Alt}}, \bibinfo {author} {\bibfnamefont {M.}~\bibnamefont {Khudaverdyan}},
  \bibinfo {author} {\bibfnamefont {S.}~\bibnamefont {Kuhr}}, \bibinfo {author}
  {\bibfnamefont {D.}~\bibnamefont {Meschede}}, \bibinfo {author}
  {\bibfnamefont {Y.}~\bibnamefont {Miroshnychenko}}, \bibinfo {author}
  {\bibfnamefont {D.}~\bibnamefont {Schrader}}, \ and\ \bibinfo {author}
  {\bibfnamefont {A.}~\bibnamefont {Rauschenbeutel}}} (\bibinfo {year}
  {2005}),\ \href {\doibase 10.1103/PhysRevLett.95.033002} {\bibfield
  {journal} {\bibinfo  {journal} {Physical Review Letters}\ }\textbf {\bibinfo
  {volume} {95}}~(\bibinfo {number} {3}),\ \bibinfo {pages}
  {033002+}}\BibitemShut {NoStop}%
\bibitem [{\citenamefont {Elsasser}\ \emph {et~al.}(2003)\citenamefont
  {Elsasser}, \citenamefont {Nagorny},\ and\ \citenamefont
  {Hemmerich}}]{Elsasser2003Collective}%
  \BibitemOpen
  \bibfield  {author} {\bibinfo {author} {\bibnamefont {Elsasser},
  \bibfnamefont {T.}}, \bibinfo {author} {\bibfnamefont {B.}~\bibnamefont
  {Nagorny}}, \ and\ \bibinfo {author} {\bibfnamefont {A.}~\bibnamefont
  {Hemmerich}}} (\bibinfo {year} {2003}),\ \href {\doibase
  10.1103/PhysRevA.67.051401} {\bibfield  {journal} {\bibinfo  {journal}
  {Physical Review A}\ }\textbf {\bibinfo {volume} {67}}~(\bibinfo {number}
  {5}),\ \bibinfo {pages} {051401+}}\BibitemShut {NoStop}%
\bibitem [{\citenamefont {Elsasser}\ \emph {et~al.}(2004)\citenamefont
  {Elsasser}, \citenamefont {Nagorny},\ and\ \citenamefont
  {Hemmerich}}]{Elsasser2004Optical}%
  \BibitemOpen
  \bibfield  {author} {\bibinfo {author} {\bibnamefont {Elsasser},
  \bibfnamefont {T.}}, \bibinfo {author} {\bibfnamefont {B.}~\bibnamefont
  {Nagorny}}, \ and\ \bibinfo {author} {\bibfnamefont {A.}~\bibnamefont
  {Hemmerich}}} (\bibinfo {year} {2004}),\ \href {\doibase
  10.1103/PhysRevA.69.033403} {\bibfield  {journal} {\bibinfo  {journal}
  {Physical Review A}\ }\textbf {\bibinfo {volume} {69}}~(\bibinfo {number}
  {3}),\ \bibinfo {pages} {033403+}}\BibitemShut {NoStop}%
\bibitem [{\citenamefont {van Enk}\ \emph {et~al.}(2001)\citenamefont {van
  Enk}, \citenamefont {McKeever}, \citenamefont {Kimble},\ and\ \citenamefont
  {Ye}}]{vanEnk2001Cooling}%
  \BibitemOpen
  \bibfield  {author} {\bibinfo {author} {\bibnamefont {van Enk}, \bibfnamefont
  {S.~J.}}, \bibinfo {author} {\bibfnamefont {J.}~\bibnamefont {McKeever}},
  \bibinfo {author} {\bibfnamefont {H.~J.}\ \bibnamefont {Kimble}}, \ and\
  \bibinfo {author} {\bibfnamefont {J.}~\bibnamefont {Ye}}} (\bibinfo {year}
  {2001}),\ \href {\doibase 10.1103/PhysRevA.64.013407} {\bibfield  {journal}
  {\bibinfo  {journal} {Physical Review A}\ }\textbf {\bibinfo {volume}
  {64}}~(\bibinfo {number} {1}),\ \bibinfo {pages} {013407+}}\BibitemShut
  {NoStop}%
\bibitem [{\citenamefont {Fabre}\ \emph {et~al.}(1994)\citenamefont {Fabre},
  \citenamefont {Pinard}, \citenamefont {Bourzeix}, \citenamefont {Heidmann},
  \citenamefont {Giacobino},\ and\ \citenamefont
  {Reynaud}}]{Fabre1994Quantumnoise}%
  \BibitemOpen
  \bibfield  {author} {\bibinfo {author} {\bibnamefont {Fabre}, \bibfnamefont
  {C.}}, \bibinfo {author} {\bibfnamefont {M.}~\bibnamefont {Pinard}}, \bibinfo
  {author} {\bibfnamefont {S.}~\bibnamefont {Bourzeix}}, \bibinfo {author}
  {\bibfnamefont {A.}~\bibnamefont {Heidmann}}, \bibinfo {author}
  {\bibfnamefont {E.}~\bibnamefont {Giacobino}}, \ and\ \bibinfo {author}
  {\bibfnamefont {S.}~\bibnamefont {Reynaud}}} (\bibinfo {year} {1994}),\ \href
  {\doibase 10.1103/PhysRevA.49.1337} {\bibfield  {journal} {\bibinfo
  {journal} {Physical Review A}\ }\textbf {\bibinfo {volume} {49}}~(\bibinfo
  {number} {2}),\ \bibinfo {pages} {1337}}\BibitemShut {NoStop}%
\bibitem [{\citenamefont {Figueroa}\ \emph {et~al.}(2011)\citenamefont
  {Figueroa}, \citenamefont {M{\"u}cke}, \citenamefont {Bochmann},
  \citenamefont {Hahn}, \citenamefont {Murr}, \citenamefont {Ritter},
  \citenamefont {Villas-Boas},\ and\ \citenamefont
  {Rempe}}]{Figueroa2011Electromagnetically}%
  \BibitemOpen
  \bibfield  {author} {\bibinfo {author} {\bibnamefont {Figueroa},
  \bibfnamefont {E.}}, \bibinfo {author} {\bibfnamefont {M.}~\bibnamefont
  {M{\"u}cke}}, \bibinfo {author} {\bibfnamefont {J.}~\bibnamefont {Bochmann}},
  \bibinfo {author} {\bibfnamefont {C.}~\bibnamefont {Hahn}}, \bibinfo {author}
  {\bibfnamefont {K.}~\bibnamefont {Murr}}, \bibinfo {author} {\bibfnamefont
  {S.}~\bibnamefont {Ritter}}, \bibinfo {author} {\bibfnamefont
  {C.}~\bibnamefont {Villas-Boas}}, \ and\ \bibinfo {author} {\bibfnamefont
  {G.}~\bibnamefont {Rempe}}} (\bibinfo {year} {2011}),\ in\ \href@noop {}
  {\emph {\bibinfo {booktitle} {AIP Conference Proceedings}}},\ Vol.\ \bibinfo
  {volume} {1363},\ p.\ \bibinfo {pages} {389}\BibitemShut {NoStop}%
\bibitem [{\citenamefont {Fischer}\ \emph {et~al.}(2002)\citenamefont
  {Fischer}, \citenamefont {Maunz}, \citenamefont {Pinkse}, \citenamefont
  {Puppe},\ and\ \citenamefont {Rempe}}]{Fischer2002Feedback}%
  \BibitemOpen
  \bibfield  {author} {\bibinfo {author} {\bibnamefont {Fischer}, \bibfnamefont
  {T.}}, \bibinfo {author} {\bibfnamefont {P.}~\bibnamefont {Maunz}}, \bibinfo
  {author} {\bibfnamefont {P.~W.~H.}\ \bibnamefont {Pinkse}}, \bibinfo {author}
  {\bibfnamefont {T.}~\bibnamefont {Puppe}}, \ and\ \bibinfo {author}
  {\bibfnamefont {G.}~\bibnamefont {Rempe}}} (\bibinfo {year} {2002}),\ \href
  {\doibase 10.1103/PhysRevLett.88.163002} {\bibfield  {journal} {\bibinfo
  {journal} {Physical Review Letters}\ }\textbf {\bibinfo {volume}
  {88}}~(\bibinfo {number} {16}),\ \bibinfo {pages} {163002+}}\BibitemShut
  {NoStop}%
\bibitem [{\citenamefont {Fischer}\ \emph {et~al.}(2001)\citenamefont
  {Fischer}, \citenamefont {Maunz}, \citenamefont {Puppe}, \citenamefont
  {Pinkse},\ and\ \citenamefont {Rempe}}]{Fischer2001Collective}%
  \BibitemOpen
  \bibfield  {author} {\bibinfo {author} {\bibnamefont {Fischer}, \bibfnamefont
  {T.}}, \bibinfo {author} {\bibfnamefont {P.}~\bibnamefont {Maunz}}, \bibinfo
  {author} {\bibfnamefont {T.}~\bibnamefont {Puppe}}, \bibinfo {author}
  {\bibfnamefont {P.~W.~H.}\ \bibnamefont {Pinkse}}, \ and\ \bibinfo {author}
  {\bibfnamefont {G.}~\bibnamefont {Rempe}}} (\bibinfo {year} {2001}),\ \href
  {\doibase 10.1088/1367-2630/3/1/311} {\bibfield  {journal} {\bibinfo
  {journal} {New Journal of Physics}\ }\textbf {\bibinfo {volume}
  {3}}~(\bibinfo {number} {1}),\ \bibinfo {pages} {11}}\BibitemShut {NoStop}%
\bibitem [{\citenamefont {Fisher}\ \emph {et~al.}(1989)\citenamefont {Fisher},
  \citenamefont {Weichman}, \citenamefont {Grinstein},\ and\ \citenamefont
  {Fisher}}]{Fisher1989Boson}%
  \BibitemOpen
  \bibfield  {author} {\bibinfo {author} {\bibnamefont {Fisher}, \bibfnamefont
  {M.~P.~A.}}, \bibinfo {author} {\bibfnamefont {P.~B.}\ \bibnamefont
  {Weichman}}, \bibinfo {author} {\bibfnamefont {G.}~\bibnamefont {Grinstein}},
  \ and\ \bibinfo {author} {\bibfnamefont {D.~S.}\ \bibnamefont {Fisher}}}
  (\bibinfo {year} {1989}),\ \href {\doibase 10.1103/PhysRevB.40.546}
  {\bibfield  {journal} {\bibinfo  {journal} {Physical Review B}\ }\textbf
  {\bibinfo {volume} {40}}~(\bibinfo {number} {1}),\ \bibinfo {pages}
  {546}}\BibitemShut {NoStop}%
\bibitem [{\citenamefont {Gangl}\ \emph {et~al.}(2000)\citenamefont {Gangl},
  \citenamefont {Horak},\ and\ \citenamefont {Ritsch}}]{Gangl2000Cooling}%
  \BibitemOpen
  \bibfield  {author} {\bibinfo {author} {\bibnamefont {Gangl}, \bibfnamefont
  {M.}}, \bibinfo {author} {\bibfnamefont {P.}~\bibnamefont {Horak}}, \ and\
  \bibinfo {author} {\bibfnamefont {H.}~\bibnamefont {Ritsch}}} (\bibinfo
  {year} {2000}),\ \href {\doibase 10.1080/09500340008232194} {\bibfield
  {journal} {\bibinfo  {journal} {Journal of Modern Optics}\ }\textbf {\bibinfo
  {volume} {47}}~(\bibinfo {number} {14-15}),\ \bibinfo {pages}
  {2741}}\BibitemShut {NoStop}%
\bibitem [{\citenamefont {Gangl}\ and\ \citenamefont
  {Ritsch}(1999)}]{Gangl1999Collective}%
  \BibitemOpen
  \bibfield  {author} {\bibinfo {author} {\bibnamefont {Gangl}, \bibfnamefont
  {M.}}, \ and\ \bibinfo {author} {\bibfnamefont {H.}~\bibnamefont {Ritsch}}}
  (\bibinfo {year} {1999}),\ \href {\doibase 10.1103/PhysRevA.61.011402}
  {\bibfield  {journal} {\bibinfo  {journal} {Physical Review A}\ }\textbf
  {\bibinfo {volume} {61}}~(\bibinfo {number} {1}),\ \bibinfo {pages}
  {011402+}}\BibitemShut {NoStop}%
\bibitem [{\citenamefont {Gangl}\ and\ \citenamefont
  {Ritsch}(2000)}]{Gangl2000Cold}%
  \BibitemOpen
  \bibfield  {author} {\bibinfo {author} {\bibnamefont {Gangl}, \bibfnamefont
  {M.}}, \ and\ \bibinfo {author} {\bibfnamefont {H.}~\bibnamefont {Ritsch}}}
  (\bibinfo {year} {2000}),\ \href {\doibase 10.1103/PhysRevA.61.043405}
  {\bibfield  {journal} {\bibinfo  {journal} {Physical Review A}\ }\textbf
  {\bibinfo {volume} {61}}~(\bibinfo {number} {4}),\ \bibinfo {pages}
  {043405+}}\BibitemShut {NoStop}%
\bibitem [{\citenamefont {Gangl}\ and\ \citenamefont
  {Ritsch}(2001)}]{Gangl2001Cavitymediated}%
  \BibitemOpen
  \bibfield  {author} {\bibinfo {author} {\bibnamefont {Gangl}, \bibfnamefont
  {M.}}, \ and\ \bibinfo {author} {\bibfnamefont {H.}~\bibnamefont {Ritsch}}}
  (\bibinfo {year} {2001}),\ \href {\doibase 10.1103/PhysRevA.64.063414}
  {\bibfield  {journal} {\bibinfo  {journal} {Physical Review A}\ }\textbf
  {\bibinfo {volume} {64}}~(\bibinfo {number} {6}),\ \bibinfo {pages}
  {063414+}}\BibitemShut {NoStop}%
\bibitem [{\citenamefont {Gardiner}\ and\ \citenamefont
  {Zoller}(2004)}]{Gardiner1991Quantum}%
  \BibitemOpen
  \bibfield  {author} {\bibinfo {author} {\bibnamefont {Gardiner},
  \bibfnamefont {C.~W.}}, \ and\ \bibinfo {author} {\bibfnamefont
  {P.}~\bibnamefont {Zoller}}} (\bibinfo {year} {2004}),\ \href@noop {} {\emph
  {\bibinfo {title} {Quantum Noise: A Handbook Of Markovian And Non-markovian
  Quantum Stochastic Methods with Applications to Quantum Optics}}},\ \bibinfo
  {edition} {third edition}\ ed.\ (\bibinfo  {publisher} {Springer Berlin /
  Heidelberg})\BibitemShut {NoStop}%
\bibitem [{\citenamefont {Gardiner}\ \emph {et~al.}(2001)\citenamefont
  {Gardiner}, \citenamefont {Gheri},\ and\ \citenamefont
  {Zoller}}]{Gardiner2001Cavityassisted}%
  \BibitemOpen
  \bibfield  {author} {\bibinfo {author} {\bibnamefont {Gardiner},
  \bibfnamefont {S.~A.}}, \bibinfo {author} {\bibfnamefont {K.~M.}\
  \bibnamefont {Gheri}}, \ and\ \bibinfo {author} {\bibfnamefont
  {P.}~\bibnamefont {Zoller}}} (\bibinfo {year} {2001}),\ \href {\doibase
  10.1103/PhysRevA.63.051603} {\bibfield  {journal} {\bibinfo  {journal}
  {Physical Review A}\ }\textbf {\bibinfo {volume} {63}}~(\bibinfo {number}
  {5}),\ \bibinfo {pages} {051603+}}\BibitemShut {NoStop}%
\bibitem [{\citenamefont {Genes}\ \emph {et~al.}(2009)\citenamefont {Genes},
  \citenamefont {Ritsch},\ and\ \citenamefont
  {Vitali}}]{Genes2009Micromechanical}%
  \BibitemOpen
  \bibfield  {author} {\bibinfo {author} {\bibnamefont {Genes}, \bibfnamefont
  {C.}}, \bibinfo {author} {\bibfnamefont {H.}~\bibnamefont {Ritsch}}, \ and\
  \bibinfo {author} {\bibfnamefont {D.}~\bibnamefont {Vitali}}} (\bibinfo
  {year} {2009}),\ \href {\doibase 10.1103/PhysRevA.80.061803} {\bibfield
  {journal} {\bibinfo  {journal} {Physical Review A}\ }\textbf {\bibinfo
  {volume} {80}}~(\bibinfo {number} {6}),\ \bibinfo {pages}
  {061803+}}\BibitemShut {NoStop}%
\bibitem [{\citenamefont {Genes}\ \emph {et~al.}(2008)\citenamefont {Genes},
  \citenamefont {Vitali}, \citenamefont {Tombesi}, \citenamefont {Gigan},\ and\
  \citenamefont {Aspelmeyer}}]{Genes2008Groundstate}%
  \BibitemOpen
  \bibfield  {author} {\bibinfo {author} {\bibnamefont {Genes}, \bibfnamefont
  {C.}}, \bibinfo {author} {\bibfnamefont {D.}~\bibnamefont {Vitali}}, \bibinfo
  {author} {\bibfnamefont {P.}~\bibnamefont {Tombesi}}, \bibinfo {author}
  {\bibfnamefont {S.}~\bibnamefont {Gigan}}, \ and\ \bibinfo {author}
  {\bibfnamefont {M.}~\bibnamefont {Aspelmeyer}}} (\bibinfo {year} {2008}),\
  \href
  {http://scitation.aip.org/getabs/servlet/GetabsServlet?prog=normal\&id=PLRAAN000077000003033804000001\&idtype=cvips\&gifs=yes}
  {\bibfield  {journal} {\bibinfo  {journal} {Physical Review A (Atomic,
  Molecular, and Optical Physics)}\ }\textbf {\bibinfo {volume} {77}}~(\bibinfo
  {number} {3}),\ \bibinfo {pages} {033804+}}\BibitemShut {NoStop}%
\bibitem [{\citenamefont {Gigan}\ \emph {et~al.}(2006)\citenamefont {Gigan},
  \citenamefont {Bohm}, \citenamefont {Paternostro}, \citenamefont {Blaser},
  \citenamefont {Langer}, \citenamefont {Hertzberg}, \citenamefont {Schwab},
  \citenamefont {Bauerle}, \citenamefont {Aspelmeyer},\ and\ \citenamefont
  {Zeilinger}}]{Gigan2006}%
  \BibitemOpen
  \bibfield  {author} {\bibinfo {author} {\bibnamefont {Gigan}, \bibfnamefont
  {S.}}, \bibinfo {author} {\bibfnamefont {H.~R.}\ \bibnamefont {Bohm}},
  \bibinfo {author} {\bibfnamefont {M.}~\bibnamefont {Paternostro}}, \bibinfo
  {author} {\bibfnamefont {F.}~\bibnamefont {Blaser}}, \bibinfo {author}
  {\bibfnamefont {G.}~\bibnamefont {Langer}}, \bibinfo {author} {\bibfnamefont
  {J.~B.}\ \bibnamefont {Hertzberg}}, \bibinfo {author} {\bibfnamefont {K.~C.}\
  \bibnamefont {Schwab}}, \bibinfo {author} {\bibfnamefont {D.}~\bibnamefont
  {Bauerle}}, \bibinfo {author} {\bibfnamefont {M.}~\bibnamefont {Aspelmeyer}},
  \ and\ \bibinfo {author} {\bibfnamefont {A.}~\bibnamefont {Zeilinger}}}
  (\bibinfo {year} {2006}),\ \href {\doibase doi:10.1038/nature05273}
  {\bibfield  {journal} {\bibinfo  {journal} {Nature}\ }\textbf {\bibinfo
  {volume} {444}}~(\bibinfo {number} {7115}),\ \bibinfo {pages}
  {67}}\BibitemShut {NoStop}%
\bibitem [{\citenamefont {Giorgini}\ \emph {et~al.}(2008)\citenamefont
  {Giorgini}, \citenamefont {Pitaevskii},\ and\ \citenamefont
  {Stringari}}]{Giorgini2008Theory}%
  \BibitemOpen
  \bibfield  {author} {\bibinfo {author} {\bibnamefont {Giorgini},
  \bibfnamefont {S.}}, \bibinfo {author} {\bibfnamefont {L.~P.}\ \bibnamefont
  {Pitaevskii}}, \ and\ \bibinfo {author} {\bibfnamefont {S.}~\bibnamefont
  {Stringari}}} (\bibinfo {year} {2008}),\ \href {\doibase
  10.1103/RevModPhys.80.1215} {\bibfield  {journal} {\bibinfo  {journal}
  {Reviews of Modern Physics}\ }\textbf {\bibinfo {volume} {80}}~(\bibinfo
  {number} {4}),\ \bibinfo {pages} {1215}}\BibitemShut {NoStop}%
\bibitem [{\citenamefont {Gopalakrishnan}\ \emph {et~al.}(2009)\citenamefont
  {Gopalakrishnan}, \citenamefont {Lev},\ and\ \citenamefont
  {Goldbart}}]{Gopalakrishnan2009Emergent}%
  \BibitemOpen
  \bibfield  {author} {\bibinfo {author} {\bibnamefont {Gopalakrishnan},
  \bibfnamefont {S.}}, \bibinfo {author} {\bibfnamefont {B.~L.}\ \bibnamefont
  {Lev}}, \ and\ \bibinfo {author} {\bibfnamefont {P.~M.}\ \bibnamefont
  {Goldbart}}} (\bibinfo {year} {2009}),\ \href {\doibase 10.1038/nphys1403}
  {\bibfield  {journal} {\bibinfo  {journal} {Nature Physics}\ }\textbf
  {\bibinfo {volume} {5}}~(\bibinfo {number} {11}),\ \bibinfo {pages}
  {845}}\BibitemShut {NoStop}%
\bibitem [{\citenamefont {Gopalakrishnan}\ \emph {et~al.}(2010)\citenamefont
  {Gopalakrishnan}, \citenamefont {Lev},\ and\ \citenamefont
  {Goldbart}}]{Gopalakrishnan2010Atomlight}%
  \BibitemOpen
  \bibfield  {author} {\bibinfo {author} {\bibnamefont {Gopalakrishnan},
  \bibfnamefont {S.}}, \bibinfo {author} {\bibfnamefont {B.~L.}\ \bibnamefont
  {Lev}}, \ and\ \bibinfo {author} {\bibfnamefont {P.~M.}\ \bibnamefont
  {Goldbart}}} (\bibinfo {year} {2010}),\ \href {\doibase
  10.1103/PhysRevA.82.043612} {\bibfield  {journal} {\bibinfo  {journal}
  {Physical Review A}\ }\textbf {\bibinfo {volume} {82}}~(\bibinfo {number}
  {4}),\ \bibinfo {pages} {043612+}}\BibitemShut {NoStop}%
\bibitem [{\citenamefont {Gopalakrishnan}\ \emph
  {et~al.}(2011{\natexlab{a}})\citenamefont {Gopalakrishnan}, \citenamefont
  {Lev},\ and\ \citenamefont {Goldbart}}]{Gopalakrishnan2011Exploring}%
  \BibitemOpen
  \bibfield  {author} {\bibinfo {author} {\bibnamefont {Gopalakrishnan},
  \bibfnamefont {S.}}, \bibinfo {author} {\bibfnamefont {B.~L.}\ \bibnamefont
  {Lev}}, \ and\ \bibinfo {author} {\bibfnamefont {P.~M.}\ \bibnamefont
  {Goldbart}}} (\bibinfo {year} {2011}{\natexlab{a}}),\ \href {\doibase
  10.1080/14786435.2011.637980} {\bibfield  {journal} {\bibinfo  {journal}
  {Philosophical Magazine}\ }\textbf {\bibinfo {volume} {92}}~(\bibinfo
  {number} {1-3}),\ \bibinfo {pages} {353}}\BibitemShut {NoStop}%
\bibitem [{\citenamefont {Gopalakrishnan}\ \emph
  {et~al.}(2011{\natexlab{b}})\citenamefont {Gopalakrishnan}, \citenamefont
  {Lev},\ and\ \citenamefont {Goldbart}}]{Gopalakrishnan2011Frustration}%
  \BibitemOpen
  \bibfield  {author} {\bibinfo {author} {\bibnamefont {Gopalakrishnan},
  \bibfnamefont {S.}}, \bibinfo {author} {\bibfnamefont {B.~L.}\ \bibnamefont
  {Lev}}, \ and\ \bibinfo {author} {\bibfnamefont {P.~M.}\ \bibnamefont
  {Goldbart}}} (\bibinfo {year} {2011}{\natexlab{b}}),\ \href {\doibase
  10.1103/PhysRevLett.107.277201} {\bibfield  {journal} {\bibinfo  {journal}
  {Physical Review Letters}\ }\textbf {\bibinfo {volume} {107}},\ \bibinfo
  {pages} {277201+}}\BibitemShut {NoStop}%
\bibitem [{\citenamefont {Gordon}\ and\ \citenamefont
  {Ashkin}(1980)}]{Gordon1980Motion}%
  \BibitemOpen
  \bibfield  {author} {\bibinfo {author} {\bibnamefont {Gordon}, \bibfnamefont
  {J.~P.}}, \ and\ \bibinfo {author} {\bibfnamefont {A.}~\bibnamefont
  {Ashkin}}} (\bibinfo {year} {1980}),\ \href {\doibase
  10.1103/PhysRevA.21.1606} {\bibfield  {journal} {\bibinfo  {journal}
  {Physical Review A}\ }\textbf {\bibinfo {volume} {21}}~(\bibinfo {number}
  {5}),\ \bibinfo {pages} {1606}}\BibitemShut {NoStop}%
\bibitem [{\citenamefont {Grangier}\ and\ \citenamefont
  {Poizat}(1998)}]{Grangier1998Simple}%
  \BibitemOpen
  \bibfield  {author} {\bibinfo {author} {\bibnamefont {Grangier},
  \bibfnamefont {P.}}, \ and\ \bibinfo {author} {\bibnamefont {Poizat}}}
  (\bibinfo {year} {1998}),\ \href {\doibase 10.1007/s100530050069} {\bibfield
  {journal} {\bibinfo  {journal} {The European Physical Journal D - Atomic,
  Molecular, Optical and Plasma Physics}\ }\textbf {\bibinfo {volume}
  {1}}~(\bibinfo {number} {1}),\ \bibinfo {pages} {97}}\BibitemShut {NoStop}%
\bibitem [{\citenamefont {Grie{\ss}er}\ \emph {et~al.}(2011)\citenamefont
  {Grie{\ss}er}, \citenamefont {Niedenzu},\ and\ \citenamefont
  {Ritsch}}]{Griesser2011Selforganisation}%
  \BibitemOpen
  \bibfield  {author} {\bibinfo {author} {\bibnamefont {Grie{\ss}er},
  \bibfnamefont {T.}}, \bibinfo {author} {\bibfnamefont {W.}~\bibnamefont
  {Niedenzu}}, \ and\ \bibinfo {author} {\bibfnamefont {H.}~\bibnamefont
  {Ritsch}}} (\bibinfo {year} {2011}),\ \href {http://arxiv.org/abs/1106.2340}
  {\ }\Eprint {http://arxiv.org/abs/1106.2340} {arXiv:1106.2340} \BibitemShut
  {NoStop}%
\bibitem [{\citenamefont {Grie{\ss}er}\ and\ \citenamefont
  {Ritsch}(2011)}]{Griesser2011Nonlinear}%
  \BibitemOpen
  \bibfield  {author} {\bibinfo {author} {\bibnamefont {Grie{\ss}er},
  \bibfnamefont {T.}}, \ and\ \bibinfo {author} {\bibfnamefont
  {H.}~\bibnamefont {Ritsch}}} (\bibinfo {year} {2011}),\ \href {\doibase
  10.1364/OE.19.011242} {\bibfield  {journal} {\bibinfo  {journal} {Opt.
  Express}\ }\textbf {\bibinfo {volume} {19}}~(\bibinfo {number} {12}),\
  \bibinfo {pages} {11242}}\BibitemShut {NoStop}%
\bibitem [{\citenamefont {Grie{\ss}er}\ \emph {et~al.}(2010)\citenamefont
  {Grie{\ss}er}, \citenamefont {Ritsch}, \citenamefont {Hemmerling},\ and\
  \citenamefont {Robb}}]{Griesser2010Vlasov}%
  \BibitemOpen
  \bibfield  {author} {\bibinfo {author} {\bibnamefont {Grie{\ss}er},
  \bibfnamefont {T.}}, \bibinfo {author} {\bibfnamefont {H.}~\bibnamefont
  {Ritsch}}, \bibinfo {author} {\bibfnamefont {M.}~\bibnamefont {Hemmerling}},
  \ and\ \bibinfo {author} {\bibnamefont {Robb}}} (\bibinfo {year} {2010}),\
  \href {\doibase 10.1140/epjd/e2010-00113-9} {\bibfield  {journal} {\bibinfo
  {journal} {The European Physical Journal D - Atomic, Molecular, Optical and
  Plasma Physics}\ }\textbf {\bibinfo {volume} {58}}~(\bibinfo {number} {3}),\
  \bibinfo {pages} {349}}\BibitemShut {NoStop}%
\bibitem [{\citenamefont {Grimm}\ \emph {et~al.}(2000)\citenamefont {Grimm},
  \citenamefont {Weidem\"{u}ller},\ and\ \citenamefont
  {Ovchinnikov}}]{Grimm2000Optical}%
  \BibitemOpen
  \bibfield  {author} {\bibinfo {author} {\bibnamefont {Grimm}, \bibfnamefont
  {R.}}, \bibinfo {author} {\bibfnamefont {M.}~\bibnamefont {Weidem\"{u}ller}},
  \ and\ \bibinfo {author} {\bibfnamefont {Y.~B.}\ \bibnamefont {Ovchinnikov}}}
  (\bibinfo {year} {2000}),\ \enquote {\bibinfo {title} {{Optical Dipole Traps
  for Neutral Atoms}},}\ \ (\bibinfo  {publisher} {Elsevier})\ pp.\ \bibinfo
  {pages} {95--170}\BibitemShut {NoStop}%
\bibitem [{\citenamefont {Gross}\ and\ \citenamefont
  {Haroche}(1982)}]{Gross1982Superradiance}%
  \BibitemOpen
  \bibfield  {author} {\bibinfo {author} {\bibnamefont {Gross}, \bibfnamefont
  {M.}}, \ and\ \bibinfo {author} {\bibfnamefont {S.}~\bibnamefont {Haroche}}}
  (\bibinfo {year} {1982}),\ \href {\doibase 10.1016/0370-1573(82)90102-8}
  {\bibfield  {journal} {\bibinfo  {journal} {Physics Reports}\ }\textbf
  {\bibinfo {volume} {93}}~(\bibinfo {number} {5}),\ \bibinfo {pages}
  {301}}\BibitemShut {NoStop}%
\bibitem [{\citenamefont {Guerlin}\ \emph {et~al.}(2007)\citenamefont
  {Guerlin}, \citenamefont {Bernu}, \citenamefont {Deleglise}, \citenamefont
  {Sayrin}, \citenamefont {Gleyzes}, \citenamefont {Kuhr}, \citenamefont
  {Brune}, \citenamefont {Raimond},\ and\ \citenamefont
  {Haroche}}]{Guerlin2007Progressive}%
  \BibitemOpen
  \bibfield  {author} {\bibinfo {author} {\bibnamefont {Guerlin}, \bibfnamefont
  {C.}}, \bibinfo {author} {\bibfnamefont {J.}~\bibnamefont {Bernu}}, \bibinfo
  {author} {\bibfnamefont {S.}~\bibnamefont {Deleglise}}, \bibinfo {author}
  {\bibfnamefont {C.}~\bibnamefont {Sayrin}}, \bibinfo {author} {\bibfnamefont
  {S.}~\bibnamefont {Gleyzes}}, \bibinfo {author} {\bibfnamefont
  {S.}~\bibnamefont {Kuhr}}, \bibinfo {author} {\bibfnamefont {M.}~\bibnamefont
  {Brune}}, \bibinfo {author} {\bibfnamefont {J.-M.}\ \bibnamefont {Raimond}},
  \ and\ \bibinfo {author} {\bibfnamefont {S.}~\bibnamefont {Haroche}}}
  (\bibinfo {year} {2007}),\ \href {\doibase 10.1038/nature06057} {\bibfield
  {journal} {\bibinfo  {journal} {Nature}\ }\textbf {\bibinfo {volume}
  {448}}~(\bibinfo {number} {7156}),\ \bibinfo {pages} {889}}\BibitemShut
  {NoStop}%
\bibitem [{\citenamefont {Gupta}\ \emph {et~al.}(2007)\citenamefont {Gupta},
  \citenamefont {Moore}, \citenamefont {Murch},\ and\ \citenamefont
  {Stamper-Kurn}}]{Gupta2007Cavity}%
  \BibitemOpen
  \bibfield  {author} {\bibinfo {author} {\bibnamefont {Gupta}, \bibfnamefont
  {S.}}, \bibinfo {author} {\bibfnamefont {K.~L.}\ \bibnamefont {Moore}},
  \bibinfo {author} {\bibfnamefont {K.~W.}\ \bibnamefont {Murch}}, \ and\
  \bibinfo {author} {\bibfnamefont {D.~M.}\ \bibnamefont {Stamper-Kurn}}}
  (\bibinfo {year} {2007}),\ \href {\doibase 10.1103/PhysRevLett.99.213601}
  {\bibfield  {journal} {\bibinfo  {journal} {Physical Review Letters}\
  }\textbf {\bibinfo {volume} {99}}~(\bibinfo {number} {21}),\ \bibinfo {pages}
  {213601+}}\BibitemShut {NoStop}%
\bibitem [{\citenamefont {Happer}(1972)}]{Happer1972Optical}%
  \BibitemOpen
  \bibfield  {author} {\bibinfo {author} {\bibnamefont {Happer}, \bibfnamefont
  {W.}}} (\bibinfo {year} {1972}),\ \href {\doibase 10.1103/RevModPhys.44.169}
  {\bibfield  {journal} {\bibinfo  {journal} {Reviews of Modern Physics}\
  }\textbf {\bibinfo {volume} {44}}~(\bibinfo {number} {2}),\ \bibinfo {pages}
  {169}}\BibitemShut {NoStop}%
\bibitem [{\citenamefont {Haroche}(1992)}]{Haroche1992Cqed}%
  \BibitemOpen
  \bibfield  {author} {\bibinfo {author} {\bibnamefont {Haroche}, \bibfnamefont
  {S.}}} (\bibinfo {year} {1992}),\ in\ \href@noop {} {\emph {\bibinfo
  {booktitle} {Fundamental Systems in Quantum Optics, Proceedings of the Les
  Houches Summer School, Session LIII}}},\ \bibinfo {editor} {edited by\
  \bibinfo {editor} {\bibfnamefont {J.}~\bibnamefont {Dalibard}}, \bibinfo
  {editor} {\bibfnamefont {J.-M.}\ \bibnamefont {Raimond}}, \ and\ \bibinfo
  {editor} {\bibfnamefont {J.}~\bibnamefont {Zinn-Justin}}}\ (\bibinfo
  {publisher} {North-Holland, Amsterdam})\ p.\ \bibinfo {pages}
  {165}\BibitemShut {NoStop}%
\bibitem [{\citenamefont {Hechenblaikner}\ \emph {et~al.}(1998)\citenamefont
  {Hechenblaikner}, \citenamefont {Gangl}, \citenamefont {Horak},\ and\
  \citenamefont {Ritsch}}]{Hechenblaikner1998Cooling}%
  \BibitemOpen
  \bibfield  {author} {\bibinfo {author} {\bibnamefont {Hechenblaikner},
  \bibfnamefont {G.}}, \bibinfo {author} {\bibfnamefont {M.}~\bibnamefont
  {Gangl}}, \bibinfo {author} {\bibfnamefont {P.}~\bibnamefont {Horak}}, \ and\
  \bibinfo {author} {\bibfnamefont {H.}~\bibnamefont {Ritsch}}} (\bibinfo
  {year} {1998}),\ \href {\doibase 10.1103/PhysRevA.58.3030} {\bibfield
  {journal} {\bibinfo  {journal} {Physical Review A}\ }\textbf {\bibinfo
  {volume} {58}}~(\bibinfo {number} {4}),\ \bibinfo {pages} {3030}}\BibitemShut
  {NoStop}%
\bibitem [{\citenamefont {Henschel}\ \emph {et~al.}(2010)\citenamefont
  {Henschel}, \citenamefont {Majer}, \citenamefont {Schmiedmayer},\ and\
  \citenamefont {Ritsch}}]{Henschel2010Cavity}%
  \BibitemOpen
  \bibfield  {author} {\bibinfo {author} {\bibnamefont {Henschel},
  \bibfnamefont {K.}}, \bibinfo {author} {\bibfnamefont {J.}~\bibnamefont
  {Majer}}, \bibinfo {author} {\bibfnamefont {J.}~\bibnamefont {Schmiedmayer}},
  \ and\ \bibinfo {author} {\bibfnamefont {H.}~\bibnamefont {Ritsch}}}
  (\bibinfo {year} {2010}),\ \href {\doibase 10.1103/PhysRevA.82.033810}
  {\bibfield  {journal} {\bibinfo  {journal} {Physical Review A}\ }\textbf
  {\bibinfo {volume} {82}},\ \bibinfo {pages} {033810+}}\BibitemShut {NoStop}%
\bibitem [{\citenamefont {Hepp}\ and\ \citenamefont
  {Lieb}(1973)}]{Hepp1973Superradiant}%
  \BibitemOpen
  \bibfield  {author} {\bibinfo {author} {\bibnamefont {Hepp}, \bibfnamefont
  {K.}}, \ and\ \bibinfo {author} {\bibfnamefont {E.~H.}\ \bibnamefont {Lieb}}}
  (\bibinfo {year} {1973}),\ \href {\doibase 10.1016/0003-4916(73)90039-0}
  {\bibfield  {journal} {\bibinfo  {journal} {Annals of Physics}\ }\textbf
  {\bibinfo {volume} {76}}~(\bibinfo {number} {2}),\ \bibinfo {pages}
  {360}}\BibitemShut {NoStop}%
\bibitem [{\citenamefont {Herskind}\ \emph {et~al.}(2009)\citenamefont
  {Herskind}, \citenamefont {Dantan}, \citenamefont {Marler}, \citenamefont
  {Albert},\ and\ \citenamefont {Drewsen}}]{Herskind2009Realization}%
  \BibitemOpen
  \bibfield  {author} {\bibinfo {author} {\bibnamefont {Herskind},
  \bibfnamefont {P.~F.}}, \bibinfo {author} {\bibfnamefont {A.}~\bibnamefont
  {Dantan}}, \bibinfo {author} {\bibfnamefont {J.~P.}\ \bibnamefont {Marler}},
  \bibinfo {author} {\bibfnamefont {M.}~\bibnamefont {Albert}}, \ and\ \bibinfo
  {author} {\bibfnamefont {M.}~\bibnamefont {Drewsen}}} (\bibinfo {year}
  {2009}),\ \href {\doibase 10.1038/nphys1302} {\bibfield  {journal} {\bibinfo
  {journal} {Nat Phys}\ }\textbf {\bibinfo {volume} {5}}~(\bibinfo {number}
  {7}),\ \bibinfo {pages} {494}}\BibitemShut {NoStop}%
\bibitem [{\citenamefont {Hood}\ \emph {et~al.}(1998)\citenamefont {Hood},
  \citenamefont {Chapman}, \citenamefont {Lynn},\ and\ \citenamefont
  {Kimble}}]{Hood1998RealTime}%
  \BibitemOpen
  \bibfield  {author} {\bibinfo {author} {\bibnamefont {Hood}, \bibfnamefont
  {C.~J.}}, \bibinfo {author} {\bibfnamefont {M.~S.}\ \bibnamefont {Chapman}},
  \bibinfo {author} {\bibfnamefont {T.~W.}\ \bibnamefont {Lynn}}, \ and\
  \bibinfo {author} {\bibfnamefont {H.~J.}\ \bibnamefont {Kimble}}} (\bibinfo
  {year} {1998}),\ \href {\doibase 10.1103/PhysRevLett.80.4157} {\bibfield
  {journal} {\bibinfo  {journal} {Physical Review Letters}\ }\textbf {\bibinfo
  {volume} {80}}~(\bibinfo {number} {19}),\ \bibinfo {pages}
  {4157}}\BibitemShut {NoStop}%
\bibitem [{\citenamefont {Hood}\ \emph {et~al.}(2000)\citenamefont {Hood},
  \citenamefont {Lynn}, \citenamefont {Doherty}, \citenamefont {Parkins},\ and\
  \citenamefont {Kimble}}]{Hood2000AtomCavity}%
  \BibitemOpen
  \bibfield  {author} {\bibinfo {author} {\bibnamefont {Hood}, \bibfnamefont
  {C.~J.}}, \bibinfo {author} {\bibfnamefont {T.~W.}\ \bibnamefont {Lynn}},
  \bibinfo {author} {\bibfnamefont {A.~C.}\ \bibnamefont {Doherty}}, \bibinfo
  {author} {\bibfnamefont {A.~S.}\ \bibnamefont {Parkins}}, \ and\ \bibinfo
  {author} {\bibfnamefont {H.~J.}\ \bibnamefont {Kimble}}} (\bibinfo {year}
  {2000}),\ \href {\doibase 10.1126/science.287.5457.1447} {\bibfield
  {journal} {\bibinfo  {journal} {Science}\ }\textbf {\bibinfo {volume}
  {287}}~(\bibinfo {number} {5457}),\ \bibinfo {pages} {1447}}\BibitemShut
  {NoStop}%
\bibitem [{\citenamefont {Horak}\ \emph {et~al.}(2000)\citenamefont {Horak},
  \citenamefont {Barnett},\ and\ \citenamefont {Ritsch}}]{Horak2000Coherent}%
  \BibitemOpen
  \bibfield  {author} {\bibinfo {author} {\bibnamefont {Horak}, \bibfnamefont
  {P.}}, \bibinfo {author} {\bibfnamefont {S.~M.}\ \bibnamefont {Barnett}}, \
  and\ \bibinfo {author} {\bibfnamefont {H.}~\bibnamefont {Ritsch}}} (\bibinfo
  {year} {2000}),\ \href {\doibase 10.1103/PhysRevA.61.033609} {\bibfield
  {journal} {\bibinfo  {journal} {Physical Review A}\ }\textbf {\bibinfo
  {volume} {61}}~(\bibinfo {number} {3}),\ \bibinfo {pages}
  {033609+}}\BibitemShut {NoStop}%
\bibitem [{\citenamefont {Horak}\ \emph {et~al.}(1997)\citenamefont {Horak},
  \citenamefont {Hechenblaikner}, \citenamefont {Gheri}, \citenamefont
  {Stecher},\ and\ \citenamefont {Ritsch}}]{Horak1997CavityInduced}%
  \BibitemOpen
  \bibfield  {author} {\bibinfo {author} {\bibnamefont {Horak}, \bibfnamefont
  {P.}}, \bibinfo {author} {\bibfnamefont {G.}~\bibnamefont {Hechenblaikner}},
  \bibinfo {author} {\bibfnamefont {K.~M.}\ \bibnamefont {Gheri}}, \bibinfo
  {author} {\bibfnamefont {H.}~\bibnamefont {Stecher}}, \ and\ \bibinfo
  {author} {\bibfnamefont {H.}~\bibnamefont {Ritsch}}} (\bibinfo {year}
  {1997}),\ \href {\doibase 10.1103/PhysRevLett.79.4974} {\bibfield  {journal}
  {\bibinfo  {journal} {Physical Review Letters}\ }\textbf {\bibinfo {volume}
  {79}}~(\bibinfo {number} {25}),\ \bibinfo {pages} {4974}}\BibitemShut
  {NoStop}%
\bibitem [{\citenamefont {Horak}\ and\ \citenamefont
  {Ritsch}(2001{\natexlab{a}})}]{Horak2001Dissipative}%
  \BibitemOpen
  \bibfield  {author} {\bibinfo {author} {\bibnamefont {Horak}, \bibfnamefont
  {P.}}, \ and\ \bibinfo {author} {\bibfnamefont {H.}~\bibnamefont {Ritsch}}}
  (\bibinfo {year} {2001}{\natexlab{a}}),\ \href {\doibase
  10.1103/PhysRevA.63.023603} {\bibfield  {journal} {\bibinfo  {journal}
  {Physical Review A}\ }\textbf {\bibinfo {volume} {63}}~(\bibinfo {number}
  {2}),\ \bibinfo {pages} {023603+}}\BibitemShut {NoStop}%
\bibitem [{\citenamefont {Horak}\ and\ \citenamefont
  {Ritsch}(2001{\natexlab{b}})}]{Horak2001Manipulating}%
  \BibitemOpen
  \bibfield  {author} {\bibinfo {author} {\bibnamefont {Horak}, \bibfnamefont
  {P.}}, \ and\ \bibinfo {author} {\bibfnamefont {H.}~\bibnamefont {Ritsch}}}
  (\bibinfo {year} {2001}{\natexlab{b}}),\ \href {\doibase
  10.1007/s100530170277} {\bibfield  {journal} {\bibinfo  {journal} {The
  European Physical Journal D - Atomic, Molecular, Optical and Plasma Physics}\
  }\textbf {\bibinfo {volume} {13}}~(\bibinfo {number} {2}),\ \bibinfo {pages}
  {279}}\BibitemShut {NoStop}%
\bibitem [{\citenamefont {Horak}\ and\ \citenamefont
  {Ritsch}(2001{\natexlab{c}})}]{Horak2001Scaling}%
  \BibitemOpen
  \bibfield  {author} {\bibinfo {author} {\bibnamefont {Horak}, \bibfnamefont
  {P.}}, \ and\ \bibinfo {author} {\bibfnamefont {H.}~\bibnamefont {Ritsch}}}
  (\bibinfo {year} {2001}{\natexlab{c}}),\ \href {\doibase
  10.1103/PhysRevA.64.033422} {\bibfield  {journal} {\bibinfo  {journal}
  {Physical Review A}\ }\textbf {\bibinfo {volume} {64}}~(\bibinfo {number}
  {3}),\ \bibinfo {pages} {033422+}}\BibitemShut {NoStop}%
\bibitem [{\citenamefont {Horak}\ \emph {et~al.}(2002)\citenamefont {Horak},
  \citenamefont {Ritsch}, \citenamefont {Fischer}, \citenamefont {Maunz},
  \citenamefont {Puppe}, \citenamefont {Pinkse},\ and\ \citenamefont
  {Rempe}}]{Horak2002Optical}%
  \BibitemOpen
  \bibfield  {author} {\bibinfo {author} {\bibnamefont {Horak}, \bibfnamefont
  {P.}}, \bibinfo {author} {\bibfnamefont {H.}~\bibnamefont {Ritsch}}, \bibinfo
  {author} {\bibfnamefont {T.}~\bibnamefont {Fischer}}, \bibinfo {author}
  {\bibfnamefont {P.}~\bibnamefont {Maunz}}, \bibinfo {author} {\bibfnamefont
  {T.}~\bibnamefont {Puppe}}, \bibinfo {author} {\bibfnamefont {P.~W.~H.}\
  \bibnamefont {Pinkse}}, \ and\ \bibinfo {author} {\bibfnamefont
  {G.}~\bibnamefont {Rempe}}} (\bibinfo {year} {2002}),\ \href {\doibase
  10.1103/PhysRevLett.88.043601} {\bibfield  {journal} {\bibinfo  {journal}
  {Physical Review Letters}\ }\textbf {\bibinfo {volume} {88}}~(\bibinfo
  {number} {4}),\ \bibinfo {pages} {043601+}}\BibitemShut {NoStop}%
\bibitem [{\citenamefont {Imamo{g}lu}\ \emph {et~al.}(1997)\citenamefont
  {Imamo{g}lu}, \citenamefont {Schmidt}, \citenamefont {Woods},\ and\
  \citenamefont {Deutsch}}]{Imamoglu1997Strongly}%
  \BibitemOpen
  \bibfield  {author} {\bibinfo {author} {\bibnamefont {Imamo{g}lu},
  \bibfnamefont {A.}}, \bibinfo {author} {\bibfnamefont {H.}~\bibnamefont
  {Schmidt}}, \bibinfo {author} {\bibfnamefont {G.}~\bibnamefont {Woods}}, \
  and\ \bibinfo {author} {\bibfnamefont {M.}~\bibnamefont {Deutsch}}} (\bibinfo
  {year} {1997}),\ \href {\doibase 10.1103/PhysRevLett.79.1467} {\bibfield
  {journal} {\bibinfo  {journal} {Physical Review Letters}\ }\textbf {\bibinfo
  {volume} {79}}~(\bibinfo {number} {8}),\ \bibinfo {pages} {1467}}\BibitemShut
  {NoStop}%
\bibitem [{\citenamefont {Inouye}\ \emph {et~al.}(1999)\citenamefont {Inouye},
  \citenamefont {Chikkatur}, \citenamefont {Stamper-Kurn}, \citenamefont
  {Stenger}, \citenamefont {Pritchard},\ and\ \citenamefont
  {Ketterle}}]{Inouye1999Superradiant}%
  \BibitemOpen
  \bibfield  {author} {\bibinfo {author} {\bibnamefont {Inouye}, \bibfnamefont
  {S.}}, \bibinfo {author} {\bibfnamefont {A.~P.}\ \bibnamefont {Chikkatur}},
  \bibinfo {author} {\bibfnamefont {D.~M.}\ \bibnamefont {Stamper-Kurn}},
  \bibinfo {author} {\bibfnamefont {J.}~\bibnamefont {Stenger}}, \bibinfo
  {author} {\bibfnamefont {D.~E.}\ \bibnamefont {Pritchard}}, \ and\ \bibinfo
  {author} {\bibfnamefont {W.}~\bibnamefont {Ketterle}}} (\bibinfo {year}
  {1999}),\ \href {\doibase 10.1126/science.285.5427.571} {\bibfield  {journal}
  {\bibinfo  {journal} {Science}\ }\textbf {\bibinfo {volume} {285}}~(\bibinfo
  {number} {5427}),\ \bibinfo {pages} {571}}\BibitemShut {NoStop}%
\bibitem [{\citenamefont {Jayich}\ \emph {et~al.}(2011)\citenamefont {Jayich},
  \citenamefont {Sankey}, \citenamefont {Petrenko},\ and\ \citenamefont
  {Harris}}]{Jayich2011Resolved}%
  \BibitemOpen
  \bibfield  {author} {\bibinfo {author} {\bibnamefont {Jayich}, \bibfnamefont
  {A.}}, \bibinfo {author} {\bibfnamefont {J.}~\bibnamefont {Sankey}}, \bibinfo
  {author} {\bibfnamefont {A.}~\bibnamefont {Petrenko}}, \ and\ \bibinfo
  {author} {\bibfnamefont {J.}~\bibnamefont {Harris}}} (\bibinfo {year}
  {2011}),\ in\ \href
  {http://www.opticsinfobase.org/abstract.cfm?URI=QELS-2011-QThM3} {\emph
  {\bibinfo {booktitle} {Quantum Electronics and Laser Science Conference}}}\
  (\bibinfo  {publisher} {Optical Society of America})\ p.\ \bibinfo {pages}
  {QThM3}\BibitemShut {NoStop}%
\bibitem [{\citenamefont {Jayich}\ \emph {et~al.}(2008)\citenamefont {Jayich},
  \citenamefont {Sankey}, \citenamefont {Zwickl}, \citenamefont {Yang},
  \citenamefont {Thompson}, \citenamefont {Girvin}, \citenamefont {Clerk},
  \citenamefont {Marquardt},\ and\ \citenamefont
  {Harris}}]{Jayich2008Dispersive}%
  \BibitemOpen
  \bibfield  {author} {\bibinfo {author} {\bibnamefont {Jayich}, \bibfnamefont
  {A.~M.}}, \bibinfo {author} {\bibfnamefont {J.~C.}\ \bibnamefont {Sankey}},
  \bibinfo {author} {\bibfnamefont {B.~M.}\ \bibnamefont {Zwickl}}, \bibinfo
  {author} {\bibfnamefont {C.}~\bibnamefont {Yang}}, \bibinfo {author}
  {\bibfnamefont {J.~D.}\ \bibnamefont {Thompson}}, \bibinfo {author}
  {\bibfnamefont {S.~M.}\ \bibnamefont {Girvin}}, \bibinfo {author}
  {\bibfnamefont {A.~A.}\ \bibnamefont {Clerk}}, \bibinfo {author}
  {\bibfnamefont {F.}~\bibnamefont {Marquardt}}, \ and\ \bibinfo {author}
  {\bibfnamefont {J.~G.~E.}\ \bibnamefont {Harris}}} (\bibinfo {year} {2008}),\
  \href {\doibase 10.1088/1367-2630/10/9/095008} {\bibfield  {journal}
  {\bibinfo  {journal} {New Journal of Physics}\ }\textbf {\bibinfo {volume}
  {10}}~(\bibinfo {number} {9}),\ \bibinfo {pages} {095008+}}\BibitemShut
  {NoStop}%
\bibitem [{\citenamefont {Jaynes}\ and\ \citenamefont
  {Cummings}(1963)}]{Jaynes1963Comparison}%
  \BibitemOpen
  \bibfield  {author} {\bibinfo {author} {\bibnamefont {Jaynes}, \bibfnamefont
  {E.~T.}}, \ and\ \bibinfo {author} {\bibfnamefont {F.~W.}\ \bibnamefont
  {Cummings}}} (\bibinfo {year} {1963}),\ \href
  {http://ieeexplore.ieee.org/xpl/freeabs\_all.jsp?arnumber=1443594\&\#38;abstractAccess=no\&\#38;userType=inst}
  {\bibfield  {journal} {\bibinfo  {journal} {Proceedings of the IEEE}\
  }\textbf {\bibinfo {volume} {51}}~(\bibinfo {number} {1}),\ \bibinfo {pages}
  {89+}}\BibitemShut {NoStop}%
\bibitem [{\citenamefont {Jing}\ \emph {et~al.}(2011)\citenamefont {Jing},
  \citenamefont {Goldbaum}, \citenamefont {Buchmann},\ and\ \citenamefont
  {Meystre}}]{Jing2011Quantum}%
  \BibitemOpen
  \bibfield  {author} {\bibinfo {author} {\bibnamefont {Jing}, \bibfnamefont
  {H.}}, \bibinfo {author} {\bibfnamefont {D.~S.}\ \bibnamefont {Goldbaum}},
  \bibinfo {author} {\bibfnamefont {L.}~\bibnamefont {Buchmann}}, \ and\
  \bibinfo {author} {\bibfnamefont {P.}~\bibnamefont {Meystre}}} (\bibinfo
  {year} {2011}),\ \href {\doibase 10.1103/PhysRevLett.106.223601} {\bibfield
  {journal} {\bibinfo  {journal} {Physical Review Letters}\ }\textbf {\bibinfo
  {volume} {106}}~(\bibinfo {number} {22}),\ \bibinfo {pages}
  {223601+}}\BibitemShut {NoStop}%
\bibitem [{\citenamefont {Kampschulte}\ \emph {et~al.}(2010)\citenamefont
  {Kampschulte}, \citenamefont {Alt}, \citenamefont {Brakhane}, \citenamefont
  {Eckstein}, \citenamefont {Reimann}, \citenamefont {Widera},\ and\
  \citenamefont {Meschede}}]{Kampschulte2010Optical}%
  \BibitemOpen
  \bibfield  {author} {\bibinfo {author} {\bibnamefont {Kampschulte},
  \bibfnamefont {T.}}, \bibinfo {author} {\bibfnamefont {W.}~\bibnamefont
  {Alt}}, \bibinfo {author} {\bibfnamefont {S.}~\bibnamefont {Brakhane}},
  \bibinfo {author} {\bibfnamefont {M.}~\bibnamefont {Eckstein}}, \bibinfo
  {author} {\bibfnamefont {R.}~\bibnamefont {Reimann}}, \bibinfo {author}
  {\bibfnamefont {A.}~\bibnamefont {Widera}}, \ and\ \bibinfo {author}
  {\bibfnamefont {D.}~\bibnamefont {Meschede}}} (\bibinfo {year} {2010}),\
  \href {\doibase 10.1103/PhysRevLett.105.153603} {\bibfield  {journal}
  {\bibinfo  {journal} {Physical Review Letters}\ }\textbf {\bibinfo {volume}
  {105}}~(\bibinfo {number} {15}),\ \bibinfo {pages} {153603+}}\BibitemShut
  {NoStop}%
\bibitem [{\citenamefont {Kanamoto}\ and\ \citenamefont
  {Meystre}(2010)}]{Kanamoto2010Optomechanics}%
  \BibitemOpen
  \bibfield  {author} {\bibinfo {author} {\bibnamefont {Kanamoto},
  \bibfnamefont {R.}}, \ and\ \bibinfo {author} {\bibfnamefont
  {P.}~\bibnamefont {Meystre}}} (\bibinfo {year} {2010}),\ \href {\doibase
  10.1103/PhysRevLett.104.063601} {\bibfield  {journal} {\bibinfo  {journal}
  {Physical Review Letters}\ }\textbf {\bibinfo {volume} {104}}~(\bibinfo
  {number} {6}),\ \bibinfo {pages} {063601+}}\BibitemShut {NoStop}%
\bibitem [{\citenamefont {Keeling}\ \emph
  {et~al.}(2010{\natexlab{a}})\citenamefont {Keeling}, \citenamefont
  {Bhaseen},\ and\ \citenamefont {Simons}}]{Keeling2010Liquid}%
  \BibitemOpen
  \bibfield  {author} {\bibinfo {author} {\bibnamefont {Keeling}, \bibfnamefont
  {J.}}, \bibinfo {author} {\bibfnamefont {J.}~\bibnamefont {Bhaseen}}, \ and\
  \bibinfo {author} {\bibfnamefont {B.}~\bibnamefont {Simons}}} (\bibinfo
  {year} {2010}{\natexlab{a}}),\ \href {\doibase 10.1103/Physics.3.88}
  {\bibfield  {journal} {\bibinfo  {journal} {Physics}\ }\textbf {\bibinfo
  {volume} {3}},\ \bibinfo {pages} {88}}\BibitemShut {NoStop}%
\bibitem [{\citenamefont {Keeling}\ \emph
  {et~al.}(2010{\natexlab{b}})\citenamefont {Keeling}, \citenamefont
  {Bhaseen},\ and\ \citenamefont {Simons}}]{Keeling2010Collective}%
  \BibitemOpen
  \bibfield  {author} {\bibinfo {author} {\bibnamefont {Keeling}, \bibfnamefont
  {J.}}, \bibinfo {author} {\bibfnamefont {M.~J.}\ \bibnamefont {Bhaseen}}, \
  and\ \bibinfo {author} {\bibfnamefont {B.~D.}\ \bibnamefont {Simons}}}
  (\bibinfo {year} {2010}{\natexlab{b}}),\ \href {\doibase
  10.1103/PhysRevLett.105.043001} {\bibfield  {journal} {\bibinfo  {journal}
  {Physical Review Letters}\ }\textbf {\bibinfo {volume} {105}}~(\bibinfo
  {number} {4}),\ \bibinfo {pages} {043001+}}\BibitemShut {NoStop}%
\bibitem [{\citenamefont {Keller}\ \emph {et~al.}(2004)\citenamefont {Keller},
  \citenamefont {Lange}, \citenamefont {Hayasaka}, \citenamefont {Lange},\ and\
  \citenamefont {Walther}}]{Keller2004Continuous}%
  \BibitemOpen
  \bibfield  {author} {\bibinfo {author} {\bibnamefont {Keller}, \bibfnamefont
  {M.}}, \bibinfo {author} {\bibfnamefont {B.}~\bibnamefont {Lange}}, \bibinfo
  {author} {\bibfnamefont {K.}~\bibnamefont {Hayasaka}}, \bibinfo {author}
  {\bibfnamefont {W.}~\bibnamefont {Lange}}, \ and\ \bibinfo {author}
  {\bibfnamefont {H.}~\bibnamefont {Walther}}} (\bibinfo {year} {2004}),\
  \href@noop {} {\bibfield  {journal} {\bibinfo  {journal} {Nature}\ }\textbf
  {\bibinfo {volume} {431}},\ \bibinfo {pages} {1075}}\BibitemShut {NoStop}%
\bibitem [{\citenamefont {Ketterle}(2002)}]{Ketterle2002Nobel}%
  \BibitemOpen
  \bibfield  {author} {\bibinfo {author} {\bibnamefont {Ketterle},
  \bibfnamefont {W.}}} (\bibinfo {year} {2002}),\ \href {\doibase
  10.1103/RevModPhys.74.1131} {\bibfield  {journal} {\bibinfo  {journal}
  {Reviews of Modern Physics}\ }\textbf {\bibinfo {volume} {74}}~(\bibinfo
  {number} {4}),\ \bibinfo {pages} {1131}}\BibitemShut {NoStop}%
\bibitem [{\citenamefont {Khudaverdyan}\ \emph {et~al.}(2008)\citenamefont
  {Khudaverdyan}, \citenamefont {Alt}, \citenamefont {Dotsenko}, \citenamefont
  {Kampschulte}, \citenamefont {Lenhard}, \citenamefont {Rauschenbeutel},
  \citenamefont {Reick}, \citenamefont {Sch\"{o}rner}, \citenamefont {Widera},\
  and\ \citenamefont {Meschede}}]{Khudaverdyan2008Controlled}%
  \BibitemOpen
  \bibfield  {author} {\bibinfo {author} {\bibnamefont {Khudaverdyan},
  \bibfnamefont {M.}}, \bibinfo {author} {\bibfnamefont {W.}~\bibnamefont
  {Alt}}, \bibinfo {author} {\bibfnamefont {I.}~\bibnamefont {Dotsenko}},
  \bibinfo {author} {\bibfnamefont {T.}~\bibnamefont {Kampschulte}}, \bibinfo
  {author} {\bibfnamefont {K.}~\bibnamefont {Lenhard}}, \bibinfo {author}
  {\bibfnamefont {A.}~\bibnamefont {Rauschenbeutel}}, \bibinfo {author}
  {\bibfnamefont {S.}~\bibnamefont {Reick}}, \bibinfo {author} {\bibfnamefont
  {K.}~\bibnamefont {Sch\"{o}rner}}, \bibinfo {author} {\bibfnamefont
  {A.}~\bibnamefont {Widera}}, \ and\ \bibinfo {author} {\bibfnamefont
  {D.}~\bibnamefont {Meschede}}} (\bibinfo {year} {2008}),\ \href {\doibase
  10.1088/1367-2630/10/7/073023} {\bibfield  {journal} {\bibinfo  {journal}
  {New Journal of Physics}\ }\textbf {\bibinfo {volume} {10}}~(\bibinfo
  {number} {7}),\ \bibinfo {pages} {073023+}}\BibitemShut {NoStop}%
\bibitem [{\citenamefont {{K}imble}(1998)}]{Kimble1998Strong}%
  \BibitemOpen
  \bibfield  {author} {\bibinfo {author} {\bibnamefont {{K}imble},
  \bibfnamefont {H.~J.}}} (\bibinfo {year} {1998}),\ \href@noop {} {\bibfield
  {journal} {\bibinfo  {journal} {{P}hysica {S}cripta}\ }\textbf {\bibinfo
  {volume} {T76}},\ \bibinfo {pages} {127}}\BibitemShut {NoStop}%
\bibitem [{\citenamefont {{K}ippenberg}\ and\ \citenamefont
  {{V}ahala}(2008)}]{Kippenberg2008Cavity}%
  \BibitemOpen
  \bibfield  {author} {\bibinfo {author} {\bibnamefont {{K}ippenberg},
  \bibfnamefont {T.~J.}}, \ and\ \bibinfo {author} {\bibfnamefont {K.~J.}\
  \bibnamefont {{V}ahala}}} (\bibinfo {year} {2008}),\ \href {\doibase
  10.1126/science.1156032} {\bibfield  {journal} {\bibinfo  {journal}
  {{S}cience ({N}ew {Y}ork, {N}.{Y}.)}\ }\textbf {\bibinfo {volume}
  {321}}~(\bibinfo {number} {5893}),\ \bibinfo {pages} {1172}}\BibitemShut
  {NoStop}%
\bibitem [{\citenamefont {Kleckner}\ and\ \citenamefont
  {Bouwmeester}(2006)}]{Kleckner2006}%
  \BibitemOpen
  \bibfield  {author} {\bibinfo {author} {\bibnamefont {Kleckner},
  \bibfnamefont {D.}}, \ and\ \bibinfo {author} {\bibfnamefont
  {D.}~\bibnamefont {Bouwmeester}}} (\bibinfo {year} {2006}),\ \href {\doibase
  10.1038/nature05231} {\bibfield  {journal} {\bibinfo  {journal} {Nature}\
  }\textbf {\bibinfo {volume} {444}}~(\bibinfo {number} {7115}),\ \bibinfo
  {pages} {75}}\BibitemShut {NoStop}%
\bibitem [{\citenamefont {Klinner}\ \emph {et~al.}(2006)\citenamefont
  {Klinner}, \citenamefont {Lindholdt}, \citenamefont {Nagorny},\ and\
  \citenamefont {Hemmerich}}]{Klinner2006Normal}%
  \BibitemOpen
  \bibfield  {author} {\bibinfo {author} {\bibnamefont {Klinner}, \bibfnamefont
  {J.}}, \bibinfo {author} {\bibfnamefont {M.}~\bibnamefont {Lindholdt}},
  \bibinfo {author} {\bibfnamefont {B.}~\bibnamefont {Nagorny}}, \ and\
  \bibinfo {author} {\bibfnamefont {A.}~\bibnamefont {Hemmerich}}} (\bibinfo
  {year} {2006}),\ \href {\doibase 10.1103/PhysRevLett.96.023002} {\bibfield
  {journal} {\bibinfo  {journal} {Physical Review Letters}\ }\textbf {\bibinfo
  {volume} {96}}~(\bibinfo {number} {2}),\ \bibinfo {pages}
  {023002+}}\BibitemShut {NoStop}%
\bibitem [{\citenamefont {K\'{o}nya}\ \emph {et~al.}(2011)\citenamefont
  {K\'{o}nya}, \citenamefont {Szirmai},\ and\ \citenamefont
  {Domokos}}]{Konya2011Multimode}%
  \BibitemOpen
  \bibfield  {author} {\bibinfo {author} {\bibnamefont {K\'{o}nya},
  \bibfnamefont {G.}}, \bibinfo {author} {\bibfnamefont {G.}~\bibnamefont
  {Szirmai}}, \ and\ \bibinfo {author} {\bibfnamefont {P.}~\bibnamefont
  {Domokos}}} (\bibinfo {year} {2011}),\ \href {\doibase
  10.1140/epjd/e2011-20050-3} {\bibfield  {journal} {\bibinfo  {journal} {The
  European Physical Journal D - Atomic, Molecular, Optical and Plasma Physics}\
  }\textbf {\bibinfo {volume} {65}}~(\bibinfo {number} {1}),\ \bibinfo {pages}
  {33}}\BibitemShut {NoStop}%
\bibitem [{\citenamefont {Kowalewski}\ \emph {et~al.}(2007)\citenamefont
  {Kowalewski}, \citenamefont {Morigi}, \citenamefont {Pinkse},\ and\
  \citenamefont {de~Vivie-Riedle}}]{Kowalewski2007Cavity}%
  \BibitemOpen
  \bibfield  {author} {\bibinfo {author} {\bibnamefont {Kowalewski},
  \bibfnamefont {M.}}, \bibinfo {author} {\bibfnamefont {G.}~\bibnamefont
  {Morigi}}, \bibinfo {author} {\bibfnamefont {P.~W.~H.}\ \bibnamefont
  {Pinkse}}, \ and\ \bibinfo {author} {\bibfnamefont {R.}~\bibnamefont
  {de~Vivie-Riedle}}} (\bibinfo {year} {2007}),\ \href {\doibase
  10.1007/s00340-007-2860-y} {\bibfield  {journal} {\bibinfo  {journal}
  {Applied Physics B}\ }\textbf {\bibinfo {volume} {89}}~(\bibinfo {number}
  {4}),\ \bibinfo {pages} {459}}\BibitemShut {NoStop}%
\bibitem [{\citenamefont {Kowalewski}\ \emph {et~al.}(2011)\citenamefont
  {Kowalewski}, \citenamefont {Morigi}, \citenamefont {Pinkse},\ and\
  \citenamefont {de~Vivie~Riedle}}]{Kowalewski2011Cavity}%
  \BibitemOpen
  \bibfield  {author} {\bibinfo {author} {\bibnamefont {Kowalewski},
  \bibfnamefont {M.}}, \bibinfo {author} {\bibfnamefont {G.}~\bibnamefont
  {Morigi}}, \bibinfo {author} {\bibfnamefont {P.~W.~H.}\ \bibnamefont
  {Pinkse}}, \ and\ \bibinfo {author} {\bibfnamefont {R.}~\bibnamefont
  {de~Vivie~Riedle}}} (\bibinfo {year} {2011}),\ \href {\doibase
  10.1103/PhysRevA.84.033408} {\bibfield  {journal} {\bibinfo  {journal}
  {Physical Review A}\ }\textbf {\bibinfo {volume} {84}},\ \bibinfo {pages}
  {033408+}}\BibitemShut {NoStop}%
\bibitem [{\citenamefont {Kruse}\ \emph
  {et~al.}(2003{\natexlab{a}})\citenamefont {Kruse}, \citenamefont {von Cube},
  \citenamefont {Zimmermann},\ and\ \citenamefont
  {Courteille}}]{Kruse2003Observation}%
  \BibitemOpen
  \bibfield  {author} {\bibinfo {author} {\bibnamefont {Kruse}, \bibfnamefont
  {D.}}, \bibinfo {author} {\bibfnamefont {C.}~\bibnamefont {von Cube}},
  \bibinfo {author} {\bibfnamefont {C.}~\bibnamefont {Zimmermann}}, \ and\
  \bibinfo {author} {\bibfnamefont {P.~W.}\ \bibnamefont {Courteille}}}
  (\bibinfo {year} {2003}{\natexlab{a}}),\ \href {\doibase
  10.1103/PhysRevLett.91.183601} {\bibfield  {journal} {\bibinfo  {journal}
  {Phys. Rev. Lett.}\ }\textbf {\bibinfo {volume} {91}},\ \bibinfo {pages}
  {183601}}\BibitemShut {NoStop}%
\bibitem [{\citenamefont {Kruse}\ \emph
  {et~al.}(2003{\natexlab{b}})\citenamefont {Kruse}, \citenamefont {Ruder},
  \citenamefont {Benhelm}, \citenamefont {von Cube}, \citenamefont
  {Zimmermann}, \citenamefont {Ph}, \citenamefont {Th}, \citenamefont
  {Nagorny},\ and\ \citenamefont {Hemmerich}}]{Kruse2003Cold}%
  \BibitemOpen
  \bibfield  {author} {\bibinfo {author} {\bibnamefont {Kruse}, \bibfnamefont
  {D.}}, \bibinfo {author} {\bibfnamefont {M.}~\bibnamefont {Ruder}}, \bibinfo
  {author} {\bibfnamefont {J.}~\bibnamefont {Benhelm}}, \bibinfo {author}
  {\bibfnamefont {C.}~\bibnamefont {von Cube}}, \bibinfo {author}
  {\bibfnamefont {C.}~\bibnamefont {Zimmermann}}, \bibinfo {author}
  {\bibnamefont {Ph}}, \bibinfo {author} {\bibnamefont {Th}}, \bibinfo {author}
  {\bibfnamefont {B.}~\bibnamefont {Nagorny}}, \ and\ \bibinfo {author}
  {\bibfnamefont {A.}~\bibnamefont {Hemmerich}}} (\bibinfo {year}
  {2003}{\natexlab{b}}),\ \href {\doibase 10.1103/PhysRevA.67.051802}
  {\bibfield  {journal} {\bibinfo  {journal} {Physical Review A}\ }\textbf
  {\bibinfo {volume} {67}}~(\bibinfo {number} {5}),\ \bibinfo {pages}
  {051802+}}\BibitemShut {NoStop}%
\bibitem [{\citenamefont {Kubanek}\ \emph {et~al.}(2009)\citenamefont
  {Kubanek}, \citenamefont {Koch}, \citenamefont {Sames}, \citenamefont
  {Ourjoumtsev}, \citenamefont {Pinkse}, \citenamefont {Murr},\ and\
  \citenamefont {Rempe}}]{Kubanek2009Photonbyphoton}%
  \BibitemOpen
  \bibfield  {author} {\bibinfo {author} {\bibnamefont {Kubanek}, \bibfnamefont
  {A.}}, \bibinfo {author} {\bibfnamefont {M.}~\bibnamefont {Koch}}, \bibinfo
  {author} {\bibfnamefont {C.}~\bibnamefont {Sames}}, \bibinfo {author}
  {\bibfnamefont {A.}~\bibnamefont {Ourjoumtsev}}, \bibinfo {author}
  {\bibfnamefont {P.~W.~H.}\ \bibnamefont {Pinkse}}, \bibinfo {author}
  {\bibfnamefont {K.}~\bibnamefont {Murr}}, \ and\ \bibinfo {author}
  {\bibfnamefont {G.}~\bibnamefont {Rempe}}} (\bibinfo {year} {2009}),\ \href
  {\doibase 10.1038/nature08563} {\bibfield  {journal} {\bibinfo  {journal}
  {Nature}\ }\textbf {\bibinfo {volume} {462}}~(\bibinfo {number} {7275}),\
  \bibinfo {pages} {898}}\BibitemShut {NoStop}%
\bibitem [{\citenamefont {Kubanek}\ \emph {et~al.}(2011)\citenamefont
  {Kubanek}, \citenamefont {Koch}, \citenamefont {Sames}, \citenamefont
  {Ourjoumtsev}, \citenamefont {Wilk}, \citenamefont {Pinkse}, \citenamefont
  {Rempe}, \citenamefont {Kubanek}, \citenamefont {Koch}, \citenamefont
  {Sames}, \citenamefont {Ourjoumtsev}, \citenamefont {Wilk}, \citenamefont
  {Pinkse},\ and\ \citenamefont {Rempe}}]{Kubanek2011Feedback}%
  \BibitemOpen
  \bibfield  {author} {\bibinfo {author} {\bibnamefont {Kubanek}, \bibfnamefont
  {A.}}, \bibinfo {author} {\bibfnamefont {M.}~\bibnamefont {Koch}}, \bibinfo
  {author} {\bibfnamefont {C.}~\bibnamefont {Sames}}, \bibinfo {author}
  {\bibfnamefont {A.}~\bibnamefont {Ourjoumtsev}}, \bibinfo {author}
  {\bibfnamefont {T.}~\bibnamefont {Wilk}}, \bibinfo {author} {\bibfnamefont
  {P.}~\bibnamefont {Pinkse}}, \bibinfo {author} {\bibfnamefont
  {G.}~\bibnamefont {Rempe}}, \bibinfo {author} {\bibfnamefont
  {A.}~\bibnamefont {Kubanek}}, \bibinfo {author} {\bibfnamefont
  {M.}~\bibnamefont {Koch}}, \bibinfo {author} {\bibfnamefont {C.}~\bibnamefont
  {Sames}}, \bibinfo {author} {\bibfnamefont {A.}~\bibnamefont {Ourjoumtsev}},
  \bibinfo {author} {\bibfnamefont {T.}~\bibnamefont {Wilk}}, \bibinfo {author}
  {\bibfnamefont {P.~W.~H.}\ \bibnamefont {Pinkse}}, \ and\ \bibinfo {author}
  {\bibfnamefont {G.}~\bibnamefont {Rempe}}} (\bibinfo {year} {2011}),\ \href
  {\doibase 10.1007/s00340-011-4410-x} {\bibfield  {journal} {\bibinfo
  {journal} {Applied Physics B: Lasers and Optics}\ }\textbf {\bibinfo {volume}
  {102}}~(\bibinfo {number} {3}),\ \bibinfo {pages} {433}}\BibitemShut
  {NoStop}%
\bibitem [{\citenamefont {Kubanek}\ \emph {et~al.}(2008)\citenamefont
  {Kubanek}, \citenamefont {Ourjoumtsev}, \citenamefont {Schuster},
  \citenamefont {Koch}, \citenamefont {Pinkse}, \citenamefont {Murr},\ and\
  \citenamefont {Rempe}}]{Kubanek2008Twophoton}%
  \BibitemOpen
  \bibfield  {author} {\bibinfo {author} {\bibnamefont {Kubanek}, \bibfnamefont
  {A.}}, \bibinfo {author} {\bibfnamefont {A.}~\bibnamefont {Ourjoumtsev}},
  \bibinfo {author} {\bibfnamefont {I.}~\bibnamefont {Schuster}}, \bibinfo
  {author} {\bibfnamefont {M.}~\bibnamefont {Koch}}, \bibinfo {author}
  {\bibfnamefont {P.~W.~H.}\ \bibnamefont {Pinkse}}, \bibinfo {author}
  {\bibfnamefont {K.}~\bibnamefont {Murr}}, \ and\ \bibinfo {author}
  {\bibfnamefont {G.}~\bibnamefont {Rempe}}} (\bibinfo {year} {2008}),\ \href
  {\doibase 10.1103/PhysRevLett.101.203602} {\bibfield  {journal} {\bibinfo
  {journal} {Physical Review Letters}\ }\textbf {\bibinfo {volume}
  {101}}~(\bibinfo {number} {20}),\ \bibinfo {pages} {203602+}}\BibitemShut
  {NoStop}%
\bibitem [{\citenamefont {Kuhn}\ \emph {et~al.}(2002)\citenamefont {Kuhn},
  \citenamefont {Hennrich},\ and\ \citenamefont
  {Rempe}}]{Kuhn2002Deterministic}%
  \BibitemOpen
  \bibfield  {author} {\bibinfo {author} {\bibnamefont {Kuhn}, \bibfnamefont
  {A.}}, \bibinfo {author} {\bibfnamefont {M.}~\bibnamefont {Hennrich}}, \ and\
  \bibinfo {author} {\bibfnamefont {G.}~\bibnamefont {Rempe}}} (\bibinfo {year}
  {2002}),\ \href {\doibase 10.1103/PhysRevLett.89.067901} {\bibfield
  {journal} {\bibinfo  {journal} {Physical Review Letters}\ }\textbf {\bibinfo
  {volume} {89}}~(\bibinfo {number} {6}),\ \bibinfo {pages}
  {067901+}}\BibitemShut {NoStop}%
\bibitem [{\citenamefont {Kuhr}\ \emph {et~al.}(2003)\citenamefont {Kuhr},
  \citenamefont {Alt}, \citenamefont {Schrader}, \citenamefont {Dotsenko},
  \citenamefont {Miroshnychenko}, \citenamefont {Rosenfeld}, \citenamefont
  {Khudaverdyan}, \citenamefont {Gomer}, \citenamefont {Rauschenbeutel},\ and\
  \citenamefont {Meschede}}]{Kuhr2003Coherence}%
  \BibitemOpen
  \bibfield  {author} {\bibinfo {author} {\bibnamefont {Kuhr}, \bibfnamefont
  {S.}}, \bibinfo {author} {\bibfnamefont {W.}~\bibnamefont {Alt}}, \bibinfo
  {author} {\bibfnamefont {D.}~\bibnamefont {Schrader}}, \bibinfo {author}
  {\bibfnamefont {I.}~\bibnamefont {Dotsenko}}, \bibinfo {author}
  {\bibfnamefont {Y.}~\bibnamefont {Miroshnychenko}}, \bibinfo {author}
  {\bibfnamefont {W.}~\bibnamefont {Rosenfeld}}, \bibinfo {author}
  {\bibfnamefont {M.}~\bibnamefont {Khudaverdyan}}, \bibinfo {author}
  {\bibfnamefont {V.}~\bibnamefont {Gomer}}, \bibinfo {author} {\bibfnamefont
  {A.}~\bibnamefont {Rauschenbeutel}}, \ and\ \bibinfo {author} {\bibfnamefont
  {D.}~\bibnamefont {Meschede}}} (\bibinfo {year} {2003}),\ \href {\doibase
  10.1103/PhysRevLett.91.213002} {\bibfield  {journal} {\bibinfo  {journal}
  {Physical Review Letters}\ }\textbf {\bibinfo {volume} {91}}~(\bibinfo
  {number} {21}),\ \bibinfo {pages} {213002+}}\BibitemShut {NoStop}%
\bibitem [{\citenamefont {Kuramoto}(1975)}]{Kuramoto1975Selfentrainment}%
  \BibitemOpen
  \bibfield  {author} {\bibinfo {author} {\bibnamefont {Kuramoto},
  \bibfnamefont {Y.}}} (\bibinfo {year} {1975}),\ in\ \href
  {http://dx.doi.org/10.1007/BFb0013365} {\emph {\bibinfo {booktitle}
  {International Symposium on Mathematical Problems in Theoretical Physics}}},\
  \bibinfo {series} {Lecture Notes in Physics}, Vol.~\bibinfo {volume} {39},\
  \bibinfo {editor} {edited by\ \bibinfo {editor} {\bibfnamefont
  {H.}~\bibnamefont {Araki}}}\ (\bibinfo  {publisher} {Springer Berlin /
  Heidelberg})\ pp.\ \bibinfo {pages} {420--422},\ \bibinfo {note}
  {10.1007/BFb0013365}\BibitemShut {NoStop}%
\bibitem [{\citenamefont {Lambert}\ \emph {et~al.}(2004)\citenamefont
  {Lambert}, \citenamefont {Emary},\ and\ \citenamefont
  {Brandes}}]{Lambert2004Entanglement}%
  \BibitemOpen
  \bibfield  {author} {\bibinfo {author} {\bibnamefont {Lambert}, \bibfnamefont
  {N.}}, \bibinfo {author} {\bibfnamefont {C.}~\bibnamefont {Emary}}, \ and\
  \bibinfo {author} {\bibfnamefont {T.}~\bibnamefont {Brandes}}} (\bibinfo
  {year} {2004}),\ \href {\doibase 10.1103/PhysRevLett.92.073602} {\bibfield
  {journal} {\bibinfo  {journal} {Physical Review Letters}\ }\textbf {\bibinfo
  {volume} {92}}~(\bibinfo {number} {7}),\ \bibinfo {pages}
  {073602+}}\BibitemShut {NoStop}%
\bibitem [{\citenamefont {Larson}\ \emph
  {et~al.}(2008{\natexlab{a}})\citenamefont {Larson}, \citenamefont {Damski},
  \citenamefont {Morigi},\ and\ \citenamefont
  {Lewenstein}}]{Larson2008MottInsulator}%
  \BibitemOpen
  \bibfield  {author} {\bibinfo {author} {\bibnamefont {Larson}, \bibfnamefont
  {J.}}, \bibinfo {author} {\bibfnamefont {B.}~\bibnamefont {Damski}}, \bibinfo
  {author} {\bibfnamefont {G.}~\bibnamefont {Morigi}}, \ and\ \bibinfo {author}
  {\bibfnamefont {M.}~\bibnamefont {Lewenstein}}} (\bibinfo {year}
  {2008}{\natexlab{a}}),\ \href {\doibase 10.1103/PhysRevLett.100.050401}
  {\bibfield  {journal} {\bibinfo  {journal} {Physical Review Letters}\
  }\textbf {\bibinfo {volume} {100}}~(\bibinfo {number} {5}),\ \bibinfo {pages}
  {050401+}}\BibitemShut {NoStop}%
\bibitem [{\citenamefont {Larson}\ \emph
  {et~al.}(2008{\natexlab{b}})\citenamefont {Larson}, \citenamefont
  {Fern\'{a}ndez-Vidal}, \citenamefont {Morigi},\ and\ \citenamefont
  {Lewenstein}}]{Larson2008Quantum}%
  \BibitemOpen
  \bibfield  {author} {\bibinfo {author} {\bibnamefont {Larson}, \bibfnamefont
  {J.}}, \bibinfo {author} {\bibfnamefont {S.}~\bibnamefont
  {Fern\'{a}ndez-Vidal}}, \bibinfo {author} {\bibfnamefont {G.}~\bibnamefont
  {Morigi}}, \ and\ \bibinfo {author} {\bibfnamefont {M.}~\bibnamefont
  {Lewenstein}}} (\bibinfo {year} {2008}{\natexlab{b}}),\ \href {\doibase
  10.1088/1367-2630/10/4/045002} {\bibfield  {journal} {\bibinfo  {journal}
  {New Journal of Physics}\ }\textbf {\bibinfo {volume} {10}}~(\bibinfo
  {number} {4}),\ \bibinfo {pages} {045002+}}\BibitemShut {NoStop}%
\bibitem [{\citenamefont {Larson}\ and\ \citenamefont
  {Lewenstein}(2009)}]{Larson2009Dilute}%
  \BibitemOpen
  \bibfield  {author} {\bibinfo {author} {\bibnamefont {Larson}, \bibfnamefont
  {J.}}, \ and\ \bibinfo {author} {\bibfnamefont {M.}~\bibnamefont
  {Lewenstein}}} (\bibinfo {year} {2009}),\ \href {\doibase
  10.1088/1367-2630/11/6/063027} {\bibfield  {journal} {\bibinfo  {journal}
  {New Journal of Physics}\ }\textbf {\bibinfo {volume} {11}}~(\bibinfo
  {number} {6}),\ \bibinfo {pages} {063027+}}\BibitemShut {NoStop}%
\bibitem [{\citenamefont {Larson}\ \emph
  {et~al.}(2008{\natexlab{c}})\citenamefont {Larson}, \citenamefont {Morigi},\
  and\ \citenamefont {Lewenstein}}]{Larson2008Cold}%
  \BibitemOpen
  \bibfield  {author} {\bibinfo {author} {\bibnamefont {Larson}, \bibfnamefont
  {J.}}, \bibinfo {author} {\bibfnamefont {G.}~\bibnamefont {Morigi}}, \ and\
  \bibinfo {author} {\bibfnamefont {M.}~\bibnamefont {Lewenstein}}} (\bibinfo
  {year} {2008}{\natexlab{c}}),\ \href@noop {} {\bibfield  {journal} {\bibinfo
  {journal} {Physical Review A}\ }\textbf {\bibinfo {volume} {78}}~(\bibinfo
  {number} {2}),\ \bibinfo {pages} {023815}}\BibitemShut {NoStop}%
\bibitem [{\citenamefont {Leggett}(1970)}]{Leggett1970Can}%
  \BibitemOpen
  \bibfield  {author} {\bibinfo {author} {\bibnamefont {Leggett}, \bibfnamefont
  {A.~J.}}} (\bibinfo {year} {1970}),\ \href {\doibase
  10.1103/PhysRevLett.25.1543} {\bibfield  {journal} {\bibinfo  {journal}
  {Physical Review Letters}\ }\textbf {\bibinfo {volume} {25}}~(\bibinfo
  {number} {22}),\ \bibinfo {pages} {1543}}\BibitemShut {NoStop}%
\bibitem [{\citenamefont {Leibrandt}\ \emph {et~al.}(2009)\citenamefont
  {Leibrandt}, \citenamefont {Labaziewicz}, \citenamefont {Vuleti\'{c}},\ and\
  \citenamefont {Chuang}}]{Leibrandt2009Cavity}%
  \BibitemOpen
  \bibfield  {author} {\bibinfo {author} {\bibnamefont {Leibrandt},
  \bibfnamefont {D.~R.}}, \bibinfo {author} {\bibfnamefont {J.}~\bibnamefont
  {Labaziewicz}}, \bibinfo {author} {\bibfnamefont {V.}~\bibnamefont
  {Vuleti\'{c}}}, \ and\ \bibinfo {author} {\bibfnamefont {I.~L.}\ \bibnamefont
  {Chuang}}} (\bibinfo {year} {2009}),\ \href {\doibase
  10.1103/PhysRevLett.103.103001} {\bibfield  {journal} {\bibinfo  {journal}
  {Physical Review Letters}\ }\textbf {\bibinfo {volume} {103}}~(\bibinfo
  {number} {10}),\ \bibinfo {pages} {103001+}}\BibitemShut {NoStop}%
\bibitem [{\citenamefont {Lev}\ \emph {et~al.}(2008)\citenamefont {Lev},
  \citenamefont {Vukics}, \citenamefont {Hudson}, \citenamefont {Sawyer},
  \citenamefont {Domokos}, \citenamefont {Ritsch},\ and\ \citenamefont
  {Ye}}]{Lev2008Prospects}%
  \BibitemOpen
  \bibfield  {author} {\bibinfo {author} {\bibnamefont {Lev}, \bibfnamefont
  {B.~L.}}, \bibinfo {author} {\bibfnamefont {A.}~\bibnamefont {Vukics}},
  \bibinfo {author} {\bibfnamefont {E.~R.}\ \bibnamefont {Hudson}}, \bibinfo
  {author} {\bibfnamefont {B.~C.}\ \bibnamefont {Sawyer}}, \bibinfo {author}
  {\bibfnamefont {P.}~\bibnamefont {Domokos}}, \bibinfo {author} {\bibfnamefont
  {H.}~\bibnamefont {Ritsch}}, \ and\ \bibinfo {author} {\bibfnamefont
  {J.}~\bibnamefont {Ye}}} (\bibinfo {year} {2008}),\ \href {\doibase
  10.1103/PhysRevA.77.023402} {\bibfield  {journal} {\bibinfo  {journal}
  {Physical Review A (Atomic, Molecular, and Optical Physics)}\ }\textbf
  {\bibinfo {volume} {77}}~(\bibinfo {number} {2}),\ \bibinfo {pages}
  {023402+}}\BibitemShut {NoStop}%
\bibitem [{\citenamefont {Lewenstein}\ and\ \citenamefont
  {Roso}(1993)}]{Lewenstein1993Cooling}%
  \BibitemOpen
  \bibfield  {author} {\bibinfo {author} {\bibnamefont {Lewenstein},
  \bibfnamefont {M.}}, \ and\ \bibinfo {author} {\bibfnamefont
  {L.}~\bibnamefont {Roso}}} (\bibinfo {year} {1993}),\ \href {\doibase
  10.1103/PhysRevA.47.3385} {\bibfield  {journal} {\bibinfo  {journal}
  {Physical Review A}\ }\textbf {\bibinfo {volume} {47}}~(\bibinfo {number}
  {4}),\ \bibinfo {pages} {3385}}\BibitemShut {NoStop}%
\bibitem [{\citenamefont {Lewenstein}\ \emph {et~al.}(2007)\citenamefont
  {Lewenstein}, \citenamefont {Sanpera}, \citenamefont {Ahufinger},
  \citenamefont {Damski}, \citenamefont {SenDe},\ and\ \citenamefont
  {Sen}}]{Lewenstein2007Ultracold}%
  \BibitemOpen
  \bibfield  {author} {\bibinfo {author} {\bibnamefont {Lewenstein},
  \bibfnamefont {M.}}, \bibinfo {author} {\bibfnamefont {A.}~\bibnamefont
  {Sanpera}}, \bibinfo {author} {\bibfnamefont {V.}~\bibnamefont {Ahufinger}},
  \bibinfo {author} {\bibfnamefont {B.}~\bibnamefont {Damski}}, \bibinfo
  {author} {\bibfnamefont {A.}~\bibnamefont {SenDe}}, \ and\ \bibinfo {author}
  {\bibfnamefont {U.}~\bibnamefont {Sen}}} (\bibinfo {year} {2007}),\ \href
  {\doibase 10.1080/00018730701223200} {\bibfield  {journal} {\bibinfo
  {journal} {Advances in Physics}\ }\textbf {\bibinfo {volume} {56}}~(\bibinfo
  {number} {2}),\ \bibinfo {pages} {243}}\BibitemShut {NoStop}%
\bibitem [{\citenamefont {Liu}\ \emph {et~al.}(2011)\citenamefont {Liu},
  \citenamefont {Lian}, \citenamefont {Ma}, \citenamefont {Xiao}, \citenamefont
  {Chen}, \citenamefont {Liang},\ and\ \citenamefont
  {Jia}}]{Liu2011Lightshiftinduced}%
  \BibitemOpen
  \bibfield  {author} {\bibinfo {author} {\bibnamefont {Liu}, \bibfnamefont
  {N.}}, \bibinfo {author} {\bibfnamefont {J.}~\bibnamefont {Lian}}, \bibinfo
  {author} {\bibfnamefont {J.}~\bibnamefont {Ma}}, \bibinfo {author}
  {\bibfnamefont {L.}~\bibnamefont {Xiao}}, \bibinfo {author} {\bibfnamefont
  {G.}~\bibnamefont {Chen}}, \bibinfo {author} {\bibfnamefont {J.~Q.}\
  \bibnamefont {Liang}}, \ and\ \bibinfo {author} {\bibfnamefont
  {S.}~\bibnamefont {Jia}}} (\bibinfo {year} {2011}),\ \href {\doibase
  10.1103/PhysRevA.83.033601} {\bibfield  {journal} {\bibinfo  {journal}
  {Physical Review A}\ }\textbf {\bibinfo {volume} {83}}~(\bibinfo {number}
  {3}),\ \bibinfo {pages} {033601+}}\BibitemShut {NoStop}%
\bibitem [{\citenamefont {Lu}\ \emph {et~al.}(2007)\citenamefont {Lu},
  \citenamefont {Zhao},\ and\ \citenamefont {Barker}}]{Lu2007Cooling}%
  \BibitemOpen
  \bibfield  {author} {\bibinfo {author} {\bibnamefont {Lu}, \bibfnamefont
  {W.}}, \bibinfo {author} {\bibfnamefont {Y.}~\bibnamefont {Zhao}}, \ and\
  \bibinfo {author} {\bibfnamefont {P.~F.}\ \bibnamefont {Barker}}} (\bibinfo
  {year} {2007}),\ \href {\doibase 10.1103/PhysRevA.76.013417} {\bibfield
  {journal} {\bibinfo  {journal} {Physical Review A}\ }\textbf {\bibinfo
  {volume} {76}}~(\bibinfo {number} {1}),\ \bibinfo {pages}
  {013417+}}\BibitemShut {NoStop}%
\bibitem [{\citenamefont {Luciani}\ and\ \citenamefont
  {Pellat}(1987)}]{Luciani1987Kinetic}%
  \BibitemOpen
  \bibfield  {author} {\bibinfo {author} {\bibnamefont {Luciani}, \bibfnamefont
  {J.}}, \ and\ \bibinfo {author} {\bibfnamefont {R.}~\bibnamefont {Pellat}}}
  (\bibinfo {year} {1987}),\ \href@noop {} {\bibfield  {journal} {\bibinfo
  {journal} {Journal de Physique}\ }\textbf {\bibinfo {volume} {48}}~(\bibinfo
  {number} {4}),\ \bibinfo {pages} {591}}\BibitemShut {NoStop}%
\bibitem [{\citenamefont {Ludwig}\ \emph {et~al.}(2008)\citenamefont {Ludwig},
  \citenamefont {Kubala},\ and\ \citenamefont
  {Marquardt}}]{Ludwig2008optomechanical}%
  \BibitemOpen
  \bibfield  {author} {\bibinfo {author} {\bibnamefont {Ludwig}, \bibfnamefont
  {M.}}, \bibinfo {author} {\bibfnamefont {B.}~\bibnamefont {Kubala}}, \ and\
  \bibinfo {author} {\bibfnamefont {F.}~\bibnamefont {Marquardt}}} (\bibinfo
  {year} {2008}),\ \href {\doibase 10.1088/1367-2630/10/9/095013} {\bibfield
  {journal} {\bibinfo  {journal} {New Journal of Physics}\ }\textbf {\bibinfo
  {volume} {10}}~(\bibinfo {number} {9}),\ \bibinfo {pages}
  {095013+}}\BibitemShut {NoStop}%
\bibitem [{\citenamefont {Lugiato}(1984)}]{Lugiato1984II}%
  \BibitemOpen
  \bibfield  {author} {\bibinfo {author} {\bibnamefont {Lugiato}, \bibfnamefont
  {L.~A.}}} (\bibinfo {year} {1984}),\ \enquote {\bibinfo {title} {{II Theory
  of Optical Bistability}},}\ \ (\bibinfo  {publisher} {Elsevier})\ pp.\
  \bibinfo {pages} {69--216}\BibitemShut {NoStop}%
\bibitem [{\citenamefont {Mabuchi}\ and\ \citenamefont
  {Doherty}(2002)}]{Mabuchi2002Cavity}%
  \BibitemOpen
  \bibfield  {author} {\bibinfo {author} {\bibnamefont {Mabuchi}, \bibfnamefont
  {H.}}, \ and\ \bibinfo {author} {\bibfnamefont {A.~C.}\ \bibnamefont
  {Doherty}}} (\bibinfo {year} {2002}),\ \href {\doibase
  10.1126/science.1078446} {\bibfield  {journal} {\bibinfo  {journal}
  {Science}\ }\textbf {\bibinfo {volume} {298}}~(\bibinfo {number} {5597}),\
  \bibinfo {pages} {1372}}\BibitemShut {NoStop}%
\bibitem [{\citenamefont {Mabuchi}\ \emph {et~al.}(1996)\citenamefont
  {Mabuchi}, \citenamefont {Turchette}, \citenamefont {Chapman},\ and\
  \citenamefont {Kimble}}]{Mabuchi1996Realtime}%
  \BibitemOpen
  \bibfield  {author} {\bibinfo {author} {\bibnamefont {Mabuchi}, \bibfnamefont
  {H.}}, \bibinfo {author} {\bibfnamefont {Q.~A.}\ \bibnamefont {Turchette}},
  \bibinfo {author} {\bibfnamefont {M.~S.}\ \bibnamefont {Chapman}}, \ and\
  \bibinfo {author} {\bibfnamefont {H.~J.}\ \bibnamefont {Kimble}}} (\bibinfo
  {year} {1996}),\ \href {\doibase 10.1364/OL.21.001393} {\bibfield  {journal}
  {\bibinfo  {journal} {Opt. Lett.}\ }\textbf {\bibinfo {volume}
  {21}}~(\bibinfo {number} {17}),\ \bibinfo {pages} {1393}}\BibitemShut
  {NoStop}%
\bibitem [{\citenamefont {Mancini}\ and\ \citenamefont
  {Tombesi}(1994)}]{Mancini1994Quantum}%
  \BibitemOpen
  \bibfield  {author} {\bibinfo {author} {\bibnamefont {Mancini}, \bibfnamefont
  {S.}}, \ and\ \bibinfo {author} {\bibfnamefont {P.}~\bibnamefont {Tombesi}}}
  (\bibinfo {year} {1994}),\ \href {\doibase 10.1103/PhysRevA.49.4055}
  {\bibfield  {journal} {\bibinfo  {journal} {Physical Review A}\ }\textbf
  {\bibinfo {volume} {49}},\ \bibinfo {pages} {4055}}\BibitemShut {NoStop}%
\bibitem [{\citenamefont {Marquardt}\ \emph {et~al.}(2007)\citenamefont
  {Marquardt}, \citenamefont {Chen}, \citenamefont {Clerk},\ and\ \citenamefont
  {Girvin}}]{Marquardt2007Quantum}%
  \BibitemOpen
  \bibfield  {author} {\bibinfo {author} {\bibnamefont {Marquardt},
  \bibfnamefont {F.}}, \bibinfo {author} {\bibfnamefont {J.~P.}\ \bibnamefont
  {Chen}}, \bibinfo {author} {\bibfnamefont {A.~A.}\ \bibnamefont {Clerk}}, \
  and\ \bibinfo {author} {\bibfnamefont {S.~M.}\ \bibnamefont {Girvin}}}
  (\bibinfo {year} {2007}),\ \href {\doibase 10.1103/PhysRevLett.99.093902}
  {\bibfield  {journal} {\bibinfo  {journal} {Physical Review Letters}\
  }\textbf {\bibinfo {volume} {99}}~(\bibinfo {number} {9}),\ \bibinfo {pages}
  {093902+}}\BibitemShut {NoStop}%
\bibitem [{\citenamefont {Maschler}\ \emph {et~al.}(2008)\citenamefont
  {Maschler}, \citenamefont {Mekhov},\ and\ \citenamefont
  {Ritsch}}]{Maschler2008Ultracold}%
  \BibitemOpen
  \bibfield  {author} {\bibinfo {author} {\bibnamefont {Maschler},
  \bibfnamefont {C.}}, \bibinfo {author} {\bibfnamefont {I.~B.}\ \bibnamefont
  {Mekhov}}, \ and\ \bibinfo {author} {\bibfnamefont {H.}~\bibnamefont
  {Ritsch}}} (\bibinfo {year} {2008}),\ \href {\doibase
  10.1140/epjd/e2008-00016-4} {\bibfield  {journal} {\bibinfo  {journal} {The
  European Physical Journal D - Atomic, Molecular, Optical and Plasma Physics}\
  }\textbf {\bibinfo {volume} {46}}~(\bibinfo {number} {3}),\ \bibinfo {pages}
  {545}}\BibitemShut {NoStop}%
\bibitem [{\citenamefont {Maschler}\ and\ \citenamefont
  {Ritsch}(2005)}]{Maschler2005Cold}%
  \BibitemOpen
  \bibfield  {author} {\bibinfo {author} {\bibnamefont {Maschler},
  \bibfnamefont {C.}}, \ and\ \bibinfo {author} {\bibfnamefont
  {H.}~\bibnamefont {Ritsch}}} (\bibinfo {year} {2005}),\ \href {\doibase
  10.1103/PhysRevLett.95.260401} {\bibfield  {journal} {\bibinfo  {journal}
  {Physical Review Letters}\ }\textbf {\bibinfo {volume} {95}}~(\bibinfo
  {number} {26}),\ \bibinfo {pages} {260401+}}\BibitemShut {NoStop}%
\bibitem [{\citenamefont {Maschler}\ \emph {et~al.}(2007)\citenamefont
  {Maschler}, \citenamefont {Ritsch}, \citenamefont {Vukics},\ and\
  \citenamefont {Domokos}}]{Maschler2007Entanglement}%
  \BibitemOpen
  \bibfield  {author} {\bibinfo {author} {\bibnamefont {Maschler},
  \bibfnamefont {C.}}, \bibinfo {author} {\bibfnamefont {H.}~\bibnamefont
  {Ritsch}}, \bibinfo {author} {\bibfnamefont {A.}~\bibnamefont {Vukics}}, \
  and\ \bibinfo {author} {\bibfnamefont {P.}~\bibnamefont {Domokos}}} (\bibinfo
  {year} {2007}),\ \href {\doibase 10.1016/j.optcom.2007.01.069} {\bibfield
  {journal} {\bibinfo  {journal} {Optics Communications}\ }\textbf {\bibinfo
  {volume} {273}}~(\bibinfo {number} {2}),\ \bibinfo {pages} {446}}\BibitemShut
  {NoStop}%
\bibitem [{\citenamefont {Maunz}\ \emph {et~al.}(2003)\citenamefont {Maunz},
  \citenamefont {Puppe}, \citenamefont {Fischer}, \citenamefont {Pinkse},\ and\
  \citenamefont {Rempe}}]{Maunz2003Emission}%
  \BibitemOpen
  \bibfield  {author} {\bibinfo {author} {\bibnamefont {Maunz}, \bibfnamefont
  {P.}}, \bibinfo {author} {\bibfnamefont {T.}~\bibnamefont {Puppe}}, \bibinfo
  {author} {\bibfnamefont {T.}~\bibnamefont {Fischer}}, \bibinfo {author}
  {\bibfnamefont {P.~W.~H.}\ \bibnamefont {Pinkse}}, \ and\ \bibinfo {author}
  {\bibfnamefont {G.}~\bibnamefont {Rempe}}} (\bibinfo {year} {2003}),\ \href
  {\doibase 10.1364/OL.28.000046} {\bibfield  {journal} {\bibinfo  {journal}
  {Opt. Lett.}\ }\textbf {\bibinfo {volume} {28}}~(\bibinfo {number} {1}),\
  \bibinfo {pages} {46}}\BibitemShut {NoStop}%
\bibitem [{\citenamefont {Maunz}\ \emph {et~al.}(2004)\citenamefont {Maunz},
  \citenamefont {Puppe}, \citenamefont {Schuster}, \citenamefont {Syassen},
  \citenamefont {Pinkse},\ and\ \citenamefont {Rempe}}]{Maunz2004Cavity}%
  \BibitemOpen
  \bibfield  {author} {\bibinfo {author} {\bibnamefont {Maunz}, \bibfnamefont
  {P.}}, \bibinfo {author} {\bibfnamefont {T.}~\bibnamefont {Puppe}}, \bibinfo
  {author} {\bibfnamefont {I.}~\bibnamefont {Schuster}}, \bibinfo {author}
  {\bibfnamefont {N.}~\bibnamefont {Syassen}}, \bibinfo {author} {\bibfnamefont
  {P.~W.~H.}\ \bibnamefont {Pinkse}}, \ and\ \bibinfo {author} {\bibfnamefont
  {G.}~\bibnamefont {Rempe}}} (\bibinfo {year} {2004}),\ \href {\doibase
  10.1038/nature02387} {\bibfield  {journal} {\bibinfo  {journal} {Nature}\
  }\textbf {\bibinfo {volume} {428}}~(\bibinfo {number} {6978}),\ \bibinfo
  {pages} {50}}\BibitemShut {NoStop}%
\bibitem [{\citenamefont {Maunz}\ \emph {et~al.}(2005)\citenamefont {Maunz},
  \citenamefont {Puppe}, \citenamefont {Schuster}, \citenamefont {Syassen},
  \citenamefont {Pinkse},\ and\ \citenamefont {Rempe}}]{Maunz2005NormalMode}%
  \BibitemOpen
  \bibfield  {author} {\bibinfo {author} {\bibnamefont {Maunz}, \bibfnamefont
  {P.}}, \bibinfo {author} {\bibfnamefont {T.}~\bibnamefont {Puppe}}, \bibinfo
  {author} {\bibfnamefont {I.}~\bibnamefont {Schuster}}, \bibinfo {author}
  {\bibfnamefont {N.}~\bibnamefont {Syassen}}, \bibinfo {author} {\bibfnamefont
  {P.~W.~H.}\ \bibnamefont {Pinkse}}, \ and\ \bibinfo {author} {\bibfnamefont
  {G.}~\bibnamefont {Rempe}}} (\bibinfo {year} {2005}),\ \href {\doibase
  10.1103/PhysRevLett.94.033002} {\bibfield  {journal} {\bibinfo  {journal}
  {Physical Review Letters}\ }\textbf {\bibinfo {volume} {94}}~(\bibinfo
  {number} {3}),\ \bibinfo {pages} {033002+}}\BibitemShut {NoStop}%
\bibitem [{\citenamefont {McKeever}\ \emph {et~al.}(2004)\citenamefont
  {McKeever}, \citenamefont {Boca}, \citenamefont {Boozer}, \citenamefont
  {Miller}, \citenamefont {Buck}, \citenamefont {Kuzmich},\ and\ \citenamefont
  {Kimble}}]{McKeever2004Deterministic}%
  \BibitemOpen
  \bibfield  {author} {\bibinfo {author} {\bibnamefont {McKeever},
  \bibfnamefont {J.}}, \bibinfo {author} {\bibfnamefont {A.}~\bibnamefont
  {Boca}}, \bibinfo {author} {\bibfnamefont {A.~D.}\ \bibnamefont {Boozer}},
  \bibinfo {author} {\bibfnamefont {R.}~\bibnamefont {Miller}}, \bibinfo
  {author} {\bibfnamefont {J.~R.}\ \bibnamefont {Buck}}, \bibinfo {author}
  {\bibfnamefont {A.}~\bibnamefont {Kuzmich}}, \ and\ \bibinfo {author}
  {\bibfnamefont {H.~J.}\ \bibnamefont {Kimble}}} (\bibinfo {year} {2004}),\
  \href {\doibase 10.1126/science.1095232} {\bibfield  {journal} {\bibinfo
  {journal} {Science}\ }\textbf {\bibinfo {volume} {303}}~(\bibinfo {number}
  {5666}),\ \bibinfo {pages} {1992}}\BibitemShut {NoStop}%
\bibitem [{\citenamefont {McKeever}\ \emph {et~al.}(2003)\citenamefont
  {McKeever}, \citenamefont {Buck}, \citenamefont {Boozer}, \citenamefont
  {Kuzmich}, \citenamefont {N\"{a}gerl}, \citenamefont {Kurn},\ and\
  \citenamefont {Kimble}}]{McKeever2003StateInsensitive}%
  \BibitemOpen
  \bibfield  {author} {\bibinfo {author} {\bibnamefont {McKeever},
  \bibfnamefont {J.}}, \bibinfo {author} {\bibfnamefont {J.~R.}\ \bibnamefont
  {Buck}}, \bibinfo {author} {\bibfnamefont {A.~D.}\ \bibnamefont {Boozer}},
  \bibinfo {author} {\bibfnamefont {A.}~\bibnamefont {Kuzmich}}, \bibinfo
  {author} {\bibfnamefont {H.~C.}\ \bibnamefont {N\"{a}gerl}}, \bibinfo
  {author} {\bibfnamefont {D.~M.~S.}\ \bibnamefont {Kurn}}, \ and\ \bibinfo
  {author} {\bibfnamefont {H.~J.}\ \bibnamefont {Kimble}}} (\bibinfo {year}
  {2003}),\ \href {\doibase 10.1103/PhysRevLett.90.133602} {\bibfield
  {journal} {\bibinfo  {journal} {Physical Review Letters}\ }\textbf {\bibinfo
  {volume} {90}}~(\bibinfo {number} {13}),\ \bibinfo {pages}
  {133602+}}\BibitemShut {NoStop}%
\bibitem [{\citenamefont {Mekhov}\ and\ \citenamefont
  {Ritsch}(2012{\natexlab{a}})}]{Mekhov2012Review}%
  \BibitemOpen
  \bibfield  {author} {\bibinfo {author} {\bibnamefont {Mekhov}, \bibfnamefont
  {I.}}, \ and\ \bibinfo {author} {\bibfnamefont {H.}~\bibnamefont {Ritsch}}}
  (\bibinfo {year} {2012}{\natexlab{a}}),\ \href@noop {} {\bibinfo  {journal}
  {Arxiv preprint arXiv:1203.0552}\ }\BibitemShut {NoStop}%
\bibitem [{\citenamefont {Mekhov}\ \emph
  {et~al.}(2007{\natexlab{a}})\citenamefont {Mekhov}, \citenamefont
  {Maschler},\ and\ \citenamefont {Ritsch}}]{Mekhov2007CavityEnhanced}%
  \BibitemOpen
\bibfield  {journal} {  }\bibfield  {author} {\bibinfo {author} {\bibnamefont
  {Mekhov}, \bibfnamefont {I.~B.}}, \bibinfo {author} {\bibfnamefont
  {C.}~\bibnamefont {Maschler}}, \ and\ \bibinfo {author} {\bibfnamefont
  {H.}~\bibnamefont {Ritsch}}} (\bibinfo {year} {2007}{\natexlab{a}}),\ \href
  {\doibase 10.1103/PhysRevLett.98.100402} {\bibfield  {journal} {\bibinfo
  {journal} {Physical Review Letters}\ }\textbf {\bibinfo {volume}
  {98}}~(\bibinfo {number} {10}),\ \bibinfo {pages} {100402+}}\BibitemShut
  {NoStop}%
\bibitem [{\citenamefont {Mekhov}\ \emph
  {et~al.}(2007{\natexlab{b}})\citenamefont {Mekhov}, \citenamefont
  {Maschler},\ and\ \citenamefont {Ritsch}}]{Mekhov2007Light}%
  \BibitemOpen
  \bibfield  {author} {\bibinfo {author} {\bibnamefont {Mekhov}, \bibfnamefont
  {I.~B.}}, \bibinfo {author} {\bibfnamefont {C.}~\bibnamefont {Maschler}}, \
  and\ \bibinfo {author} {\bibfnamefont {H.}~\bibnamefont {Ritsch}}} (\bibinfo
  {year} {2007}{\natexlab{b}}),\ \href {\doibase 10.1103/PhysRevA.76.053618}
  {\bibfield  {journal} {\bibinfo  {journal} {Physical Review A}\ }\textbf
  {\bibinfo {volume} {76}}~(\bibinfo {number} {5}),\ \bibinfo {pages}
  {053618+}}\BibitemShut {NoStop}%
\bibitem [{\citenamefont {Mekhov}\ \emph
  {et~al.}(2007{\natexlab{c}})\citenamefont {Mekhov}, \citenamefont
  {Maschler},\ and\ \citenamefont {Ritsch}}]{Mekhov2007Probing}%
  \BibitemOpen
  \bibfield  {author} {\bibinfo {author} {\bibnamefont {Mekhov}, \bibfnamefont
  {I.~B.}}, \bibinfo {author} {\bibfnamefont {C.}~\bibnamefont {Maschler}}, \
  and\ \bibinfo {author} {\bibfnamefont {H.}~\bibnamefont {Ritsch}}} (\bibinfo
  {year} {2007}{\natexlab{c}}),\ \href {\doibase 10.1038/nphys571} {\bibfield
  {journal} {\bibinfo  {journal} {Nat Phys}\ }\textbf {\bibinfo {volume}
  {3}}~(\bibinfo {number} {5}),\ \bibinfo {pages} {319}}\BibitemShut {NoStop}%
\bibitem [{\citenamefont {Mekhov}\ and\ \citenamefont
  {Ritsch}(2009)}]{Mekhov2009Quantum}%
  \BibitemOpen
  \bibfield  {author} {\bibinfo {author} {\bibnamefont {Mekhov}, \bibfnamefont
  {I.~B.}}, \ and\ \bibinfo {author} {\bibfnamefont {H.}~\bibnamefont
  {Ritsch}}} (\bibinfo {year} {2009}),\ \href {\doibase
  10.1103/PhysRevA.80.013604} {\bibfield  {journal} {\bibinfo  {journal}
  {Physical Review A}\ }\textbf {\bibinfo {volume} {80}}~(\bibinfo {number}
  {1}),\ \bibinfo {pages} {013604+}}\BibitemShut {NoStop}%
\bibitem [{\citenamefont {Mekhov}\ and\ \citenamefont
  {Ritsch}(2012{\natexlab{b}})}]{Mekhov2012Quantum}%
  \BibitemOpen
  \bibfield  {author} {\bibinfo {author} {\bibnamefont {Mekhov}, \bibfnamefont
  {I.~B.}}, \ and\ \bibinfo {author} {\bibfnamefont {H.}~\bibnamefont
  {Ritsch}}} (\bibinfo {year} {2012}{\natexlab{b}}),\ \href {\doibase
  10.1088/0953-4075/45/10/102001} {\bibfield  {journal} {\bibinfo  {journal}
  {Journal of Physics B: Atomic, Molecular and Optical Physics}\ }\textbf
  {\bibinfo {volume} {45}}~(\bibinfo {number} {10}),\ \bibinfo {pages}
  {102001+}}\BibitemShut {NoStop}%
\bibitem [{\citenamefont {Metzger}\ and\ \citenamefont
  {Karrai}(2004)}]{Metzger2004}%
  \BibitemOpen
  \bibfield  {author} {\bibinfo {author} {\bibnamefont {Metzger}, \bibfnamefont
  {C.~H.}}, \ and\ \bibinfo {author} {\bibfnamefont {K.}~\bibnamefont
  {Karrai}}} (\bibinfo {year} {2004}),\ \href {\doibase 10.1038/nature03118}
  {\bibfield  {journal} {\bibinfo  {journal} {Nature}\ }\textbf {\bibinfo
  {volume} {432}}~(\bibinfo {number} {7020}),\ \bibinfo {pages}
  {1002}}\BibitemShut {NoStop}%
\bibitem [{\citenamefont {Miyake}\ \emph {et~al.}(2011)\citenamefont {Miyake},
  \citenamefont {Siviloglou}, \citenamefont {Puentes}, \citenamefont
  {Pritchard}, \citenamefont {Ketterle},\ and\ \citenamefont
  {Weld}}]{Miyake2011Bragg}%
  \BibitemOpen
  \bibfield  {author} {\bibinfo {author} {\bibnamefont {Miyake}, \bibfnamefont
  {H.}}, \bibinfo {author} {\bibfnamefont {G.~A.}\ \bibnamefont {Siviloglou}},
  \bibinfo {author} {\bibfnamefont {G.}~\bibnamefont {Puentes}}, \bibinfo
  {author} {\bibfnamefont {D.~E.}\ \bibnamefont {Pritchard}}, \bibinfo {author}
  {\bibfnamefont {W.}~\bibnamefont {Ketterle}}, \ and\ \bibinfo {author}
  {\bibfnamefont {D.~M.}\ \bibnamefont {Weld}}} (\bibinfo {year} {2011}),\
  \href {\doibase 10.1103/PhysRevLett.107.175302} {\bibfield  {journal}
  {\bibinfo  {journal} {Physical Review Letters}\ }\textbf {\bibinfo {volume}
  {107}},\ \bibinfo {pages} {175302+}}\BibitemShut {NoStop}%
\bibitem [{\citenamefont {Montgomery}(1971)}]{Montgomery1971Theory}%
  \BibitemOpen
  \bibfield  {author} {\bibinfo {author} {\bibnamefont {Montgomery},
  \bibfnamefont {D.~C.}}} (\bibinfo {year} {1971}),\ \href@noop {} {\emph
  {\bibinfo {title} {Theory of the Unmagnetized Plasma}}}\ (\bibinfo
  {publisher} {Gordon and Breach})\BibitemShut {NoStop}%
\bibitem [{\citenamefont {Moore}\ \emph {et~al.}(1999)\citenamefont {Moore},
  \citenamefont {Zobay},\ and\ \citenamefont {Meystre}}]{Moore1999Quantum}%
  \BibitemOpen
  \bibfield  {author} {\bibinfo {author} {\bibnamefont {Moore}, \bibfnamefont
  {M.~G.}}, \bibinfo {author} {\bibfnamefont {O.}~\bibnamefont {Zobay}}, \ and\
  \bibinfo {author} {\bibfnamefont {P.}~\bibnamefont {Meystre}}} (\bibinfo
  {year} {1999}),\ \href@noop {} {\bibfield  {journal} {\bibinfo  {journal}
  {Physical Review A}\ }\textbf {\bibinfo {volume} {60}}~(\bibinfo {number}
  {2}),\ \bibinfo {pages} {1491}}\BibitemShut {NoStop}%
\bibitem [{\citenamefont {Morigi}\ \emph {et~al.}(2000)\citenamefont {Morigi},
  \citenamefont {Eschner},\ and\ \citenamefont {Keitel}}]{Morigi2000Ground}%
  \BibitemOpen
  \bibfield  {author} {\bibinfo {author} {\bibnamefont {Morigi}, \bibfnamefont
  {G.}}, \bibinfo {author} {\bibfnamefont {J.}~\bibnamefont {Eschner}}, \ and\
  \bibinfo {author} {\bibfnamefont {C.~H.}\ \bibnamefont {Keitel}}} (\bibinfo
  {year} {2000}),\ \href {\doibase 10.1103/PhysRevLett.85.4458} {\bibfield
  {journal} {\bibinfo  {journal} {Physical Review Letters}\ }\textbf {\bibinfo
  {volume} {85}}~(\bibinfo {number} {21}),\ \bibinfo {pages}
  {4458}}\BibitemShut {NoStop}%
\bibitem [{\citenamefont {Morigi}\ \emph {et~al.}(2007)\citenamefont {Morigi},
  \citenamefont {Pinkse}, \citenamefont {Kowalewski},\ and\ \citenamefont
  {de~Vivie~Riedle}}]{Morigi2007Cavity}%
  \BibitemOpen
  \bibfield  {author} {\bibinfo {author} {\bibnamefont {Morigi}, \bibfnamefont
  {G.}}, \bibinfo {author} {\bibfnamefont {P.~W.~H.}\ \bibnamefont {Pinkse}},
  \bibinfo {author} {\bibfnamefont {M.}~\bibnamefont {Kowalewski}}, \ and\
  \bibinfo {author} {\bibfnamefont {R.}~\bibnamefont {de~Vivie~Riedle}}}
  (\bibinfo {year} {2007}),\ \href {\doibase 10.1103/PhysRevLett.99.073001}
  {\bibfield  {journal} {\bibinfo  {journal} {Physical Review Letters}\
  }\textbf {\bibinfo {volume} {99}}~(\bibinfo {number} {7}),\ \bibinfo {pages}
  {073001+}}\BibitemShut {NoStop}%
\bibitem [{\citenamefont {Mossberg}\ \emph {et~al.}(1991)\citenamefont
  {Mossberg}, \citenamefont {Lewenstein},\ and\ \citenamefont
  {Gauthier}}]{Mossberg1991Trapping}%
  \BibitemOpen
  \bibfield  {author} {\bibinfo {author} {\bibnamefont {Mossberg},
  \bibfnamefont {T.~W.}}, \bibinfo {author} {\bibfnamefont {M.}~\bibnamefont
  {Lewenstein}}, \ and\ \bibinfo {author} {\bibfnamefont {D.~J.}\ \bibnamefont
  {Gauthier}}} (\bibinfo {year} {1991}),\ \href {\doibase
  10.1103/PhysRevLett.67.1723} {\bibfield  {journal} {\bibinfo  {journal}
  {Physical Review Letters}\ }\textbf {\bibinfo {volume} {67}}~(\bibinfo
  {number} {13}),\ \bibinfo {pages} {1723}}\BibitemShut {NoStop}%
\bibitem [{\citenamefont {Mottl}\ \emph {et~al.}(2012)\citenamefont {Mottl},
  \citenamefont {Brennecke}, \citenamefont {Baumann}, \citenamefont {Landig},
  \citenamefont {Donner},\ and\ \citenamefont
  {Esslinger}}]{Mottl2012RotonType}%
  \BibitemOpen
  \bibfield  {author} {\bibinfo {author} {\bibnamefont {Mottl}, \bibfnamefont
  {R.}}, \bibinfo {author} {\bibfnamefont {F.}~\bibnamefont {Brennecke}},
  \bibinfo {author} {\bibfnamefont {K.}~\bibnamefont {Baumann}}, \bibinfo
  {author} {\bibfnamefont {R.}~\bibnamefont {Landig}}, \bibinfo {author}
  {\bibfnamefont {T.}~\bibnamefont {Donner}}, \ and\ \bibinfo {author}
  {\bibfnamefont {T.}~\bibnamefont {Esslinger}}} (\bibinfo {year} {2012}),\
  \href {\doibase 10.1126/science.1220314} {\bibfield  {journal} {\bibinfo
  {journal} {Science}\ }\textbf {\bibinfo {volume} {336}}~(\bibinfo {number}
  {6088}),\ \bibinfo {pages} {1570}}\BibitemShut {NoStop}%
\bibitem [{\citenamefont {M\"{u}ller}\ \emph {et~al.}(2012)\citenamefont
  {M\"{u}ller}, \citenamefont {Strack},\ and\ \citenamefont
  {Sachdev}}]{Muller2012Quantum}%
  \BibitemOpen
  \bibfield  {author} {\bibinfo {author} {\bibnamefont {M\"{u}ller},
  \bibfnamefont {M.}}, \bibinfo {author} {\bibfnamefont {P.}~\bibnamefont
  {Strack}}, \ and\ \bibinfo {author} {\bibfnamefont {S.}~\bibnamefont
  {Sachdev}}} (\bibinfo {year} {2012}),\ \href {\doibase
  10.1103/PhysRevA.86.023604} {\bibfield  {journal} {\bibinfo  {journal}
  {Physical Review A}\ }\textbf {\bibinfo {volume} {86}},\ \bibinfo {pages}
  {023604+}}\BibitemShut {NoStop}%
\bibitem [{\citenamefont {M\"{u}nstermann}\ \emph {et~al.}(1999)\citenamefont
  {M\"{u}nstermann}, \citenamefont {Fischer}, \citenamefont {Maunz},
  \citenamefont {Pinkse},\ and\ \citenamefont
  {Rempe}}]{Munstermann1999Dynamics}%
  \BibitemOpen
  \bibfield  {author} {\bibinfo {author} {\bibnamefont {M\"{u}nstermann},
  \bibfnamefont {P.}}, \bibinfo {author} {\bibfnamefont {T.}~\bibnamefont
  {Fischer}}, \bibinfo {author} {\bibfnamefont {P.}~\bibnamefont {Maunz}},
  \bibinfo {author} {\bibfnamefont {P.~W.~H.}\ \bibnamefont {Pinkse}}, \ and\
  \bibinfo {author} {\bibfnamefont {G.}~\bibnamefont {Rempe}}} (\bibinfo {year}
  {1999}),\ \href {\doibase 10.1103/PhysRevLett.82.3791} {\bibfield  {journal}
  {\bibinfo  {journal} {Physical Review Letters}\ }\textbf {\bibinfo {volume}
  {82}}~(\bibinfo {number} {19}),\ \bibinfo {pages} {3791}}\BibitemShut
  {NoStop}%
\bibitem [{\citenamefont {M\"{u}nstermann}\ \emph {et~al.}(2000)\citenamefont
  {M\"{u}nstermann}, \citenamefont {Fischer}, \citenamefont {Maunz},
  \citenamefont {Pinkse},\ and\ \citenamefont
  {Rempe}}]{Munstermann2000Observation}%
  \BibitemOpen
  \bibfield  {author} {\bibinfo {author} {\bibnamefont {M\"{u}nstermann},
  \bibfnamefont {P.}}, \bibinfo {author} {\bibfnamefont {T.}~\bibnamefont
  {Fischer}}, \bibinfo {author} {\bibfnamefont {P.}~\bibnamefont {Maunz}},
  \bibinfo {author} {\bibfnamefont {P.~W.~H.}\ \bibnamefont {Pinkse}}, \ and\
  \bibinfo {author} {\bibfnamefont {G.}~\bibnamefont {Rempe}}} (\bibinfo {year}
  {2000}),\ \href {\doibase 10.1103/PhysRevLett.84.4068} {\bibfield  {journal}
  {\bibinfo  {journal} {Physical Review Letters}\ }\textbf {\bibinfo {volume}
  {84}}~(\bibinfo {number} {18}),\ \bibinfo {pages} {4068}}\BibitemShut
  {NoStop}%
\bibitem [{\citenamefont {Murch}\ \emph {et~al.}(2008)\citenamefont {Murch},
  \citenamefont {Moore}, \citenamefont {Gupta},\ and\ \citenamefont
  {Stamper-Kurn}}]{Murch2008Observation}%
  \BibitemOpen
  \bibfield  {author} {\bibinfo {author} {\bibnamefont {Murch}, \bibfnamefont
  {K.~W.}}, \bibinfo {author} {\bibfnamefont {K.~L.}\ \bibnamefont {Moore}},
  \bibinfo {author} {\bibfnamefont {S.}~\bibnamefont {Gupta}}, \ and\ \bibinfo
  {author} {\bibfnamefont {D.~M.}\ \bibnamefont {Stamper-Kurn}}} (\bibinfo
  {year} {2008}),\ \href {\doibase 10.1038/nphys965} {\bibfield  {journal}
  {\bibinfo  {journal} {Nature Physics}\ }\textbf {\bibinfo {volume}
  {4}}~(\bibinfo {number} {7}),\ \bibinfo {pages} {561}}\BibitemShut {NoStop}%
\bibitem [{\citenamefont {Murr}(2003)}]{Murr2003Suppression}%
  \BibitemOpen
  \bibfield  {author} {\bibinfo {author} {\bibnamefont {Murr}, \bibfnamefont
  {K.}}} (\bibinfo {year} {2003}),\ \href {\doibase
  10.1088/0953-4075/36/12/311} {\bibfield  {journal} {\bibinfo  {journal}
  {Journal of Physics B: Atomic, Molecular and Optical Physics}\ }\textbf
  {\bibinfo {volume} {36}}~(\bibinfo {number} {12}),\ \bibinfo {pages}
  {2515}}\BibitemShut {NoStop}%
\bibitem [{\citenamefont {Murr}(2006)}]{Murr2006Large}%
  \BibitemOpen
  \bibfield  {author} {\bibinfo {author} {\bibnamefont {Murr}, \bibfnamefont
  {K.}}} (\bibinfo {year} {2006}),\ \href {\doibase
  10.1103/PhysRevLett.96.253001} {\bibfield  {journal} {\bibinfo  {journal}
  {Physical Review Letters}\ }\textbf {\bibinfo {volume} {96}}~(\bibinfo
  {number} {25}),\ \bibinfo {pages} {253001+}}\BibitemShut {NoStop}%
\bibitem [{\citenamefont {Murr}\ \emph
  {et~al.}(2006{\natexlab{a}})\citenamefont {Murr}, \citenamefont {Maunz},
  \citenamefont {Pinkse}, \citenamefont {Puppe}, \citenamefont {Schuster},
  \citenamefont {Vitali},\ and\ \citenamefont {Rempe}}]{Murr2006Momentum}%
  \BibitemOpen
  \bibfield  {author} {\bibinfo {author} {\bibnamefont {Murr}, \bibfnamefont
  {K.}}, \bibinfo {author} {\bibfnamefont {P.}~\bibnamefont {Maunz}}, \bibinfo
  {author} {\bibfnamefont {P.~W.~H.}\ \bibnamefont {Pinkse}}, \bibinfo {author}
  {\bibfnamefont {T.}~\bibnamefont {Puppe}}, \bibinfo {author} {\bibfnamefont
  {I.}~\bibnamefont {Schuster}}, \bibinfo {author} {\bibfnamefont
  {D.}~\bibnamefont {Vitali}}, \ and\ \bibinfo {author} {\bibfnamefont
  {G.}~\bibnamefont {Rempe}}} (\bibinfo {year} {2006}{\natexlab{a}}),\ \href
  {\doibase 10.1103/PhysRevA.74.043412} {\bibfield  {journal} {\bibinfo
  {journal} {Physical Review A}\ }\textbf {\bibinfo {volume} {74}}~(\bibinfo
  {number} {4}),\ \bibinfo {pages} {043412+}}\BibitemShut {NoStop}%
\bibitem [{\citenamefont {Murr}\ \emph
  {et~al.}(2006{\natexlab{b}})\citenamefont {Murr}, \citenamefont {Nuss{}mann},
  \citenamefont {Puppe}, \citenamefont {Hijlkema}, \citenamefont {Weber},
  \citenamefont {Webster}, \citenamefont {Kuhn},\ and\ \citenamefont
  {Rempe}}]{Murr2006Threedimensional}%
  \BibitemOpen
  \bibfield  {author} {\bibinfo {author} {\bibnamefont {Murr}, \bibfnamefont
  {K.}}, \bibinfo {author} {\bibfnamefont {S.}~\bibnamefont {Nuss{}mann}},
  \bibinfo {author} {\bibfnamefont {T.}~\bibnamefont {Puppe}}, \bibinfo
  {author} {\bibfnamefont {M.}~\bibnamefont {Hijlkema}}, \bibinfo {author}
  {\bibfnamefont {B.}~\bibnamefont {Weber}}, \bibinfo {author} {\bibfnamefont
  {S.~C.}\ \bibnamefont {Webster}}, \bibinfo {author} {\bibfnamefont
  {A.}~\bibnamefont {Kuhn}}, \ and\ \bibinfo {author} {\bibfnamefont
  {G.}~\bibnamefont {Rempe}}} (\bibinfo {year} {2006}{\natexlab{b}}),\ \href
  {\doibase 10.1103/PhysRevA.73.063415} {\bibfield  {journal} {\bibinfo
  {journal} {Physical Review A}\ }\textbf {\bibinfo {volume} {73}}~(\bibinfo
  {number} {6}),\ \bibinfo {pages} {063415+}}\BibitemShut {NoStop}%
\bibitem [{\citenamefont {Nagorny}\ \emph {et~al.}(2003)\citenamefont
  {Nagorny}, \citenamefont {Elsasser},\ and\ \citenamefont
  {Hemmerich}}]{Nagorny2003Collective}%
  \BibitemOpen
  \bibfield  {author} {\bibinfo {author} {\bibnamefont {Nagorny}, \bibfnamefont
  {B.}}, \bibinfo {author} {\bibfnamefont {T.}~\bibnamefont {Elsasser}}, \ and\
  \bibinfo {author} {\bibfnamefont {A.}~\bibnamefont {Hemmerich}}} (\bibinfo
  {year} {2003}),\ \href {\doibase 10.1103/PhysRevLett.91.153003} {\bibfield
  {journal} {\bibinfo  {journal} {Physical Review Letters}\ }\textbf {\bibinfo
  {volume} {91}}~(\bibinfo {number} {15}),\ \bibinfo {pages} {153003+}},\
  \Eprint {http://arxiv.org/abs/quant-ph/0305164} {quant-ph/0305164}
  \BibitemShut {NoStop}%
\bibitem [{\citenamefont {Nagy}\ \emph
  {et~al.}(2006{\natexlab{a}})\citenamefont {Nagy}, \citenamefont
  {Asb{\'o}th},\ and\ \citenamefont {Domokos}}]{Nagy2006Collective}%
  \BibitemOpen
  \bibfield  {author} {\bibinfo {author} {\bibnamefont {Nagy}, \bibfnamefont
  {D.}}, \bibinfo {author} {\bibfnamefont {J.}~\bibnamefont {Asb{\'o}th}}, \
  and\ \bibinfo {author} {\bibfnamefont {P.}~\bibnamefont {Domokos}}} (\bibinfo
  {year} {2006}{\natexlab{a}}),\ \href
  {http://dx.doi.org/10.1556/APH.26.2006.1-2.16} {\bibfield  {journal}
  {\bibinfo  {journal} {Acta Physica Hungarica B) Quantum Electronics}\
  }\textbf {\bibinfo {volume} {26}},\ \bibinfo {pages} {141}},\ \bibinfo {note}
  {10.1556/APH.26.2006.1-2.16}\BibitemShut {NoStop}%
\bibitem [{\citenamefont {Nagy}\ \emph
  {et~al.}(2006{\natexlab{b}})\citenamefont {Nagy}, \citenamefont {Asboth},
  \citenamefont {Domokos},\ and\ \citenamefont
  {Ritsch}}]{Nagy2006Selforganization}%
  \BibitemOpen
  \bibfield  {author} {\bibinfo {author} {\bibnamefont {Nagy}, \bibfnamefont
  {D.}}, \bibinfo {author} {\bibfnamefont {J.~K.}\ \bibnamefont {Asboth}},
  \bibinfo {author} {\bibfnamefont {P.}~\bibnamefont {Domokos}}, \ and\
  \bibinfo {author} {\bibfnamefont {H.}~\bibnamefont {Ritsch}}} (\bibinfo
  {year} {2006}{\natexlab{b}}),\ \href {\doibase 10.1209/epl/i2005-10521-4}
  {\bibfield  {journal} {\bibinfo  {journal} {EPL (Europhysics Letters)}\
  }\textbf {\bibinfo {volume} {74}}~(\bibinfo {number} {2}),\ \bibinfo {pages}
  {254}}\BibitemShut {NoStop}%
\bibitem [{\citenamefont {Nagy}\ \emph {et~al.}(2009)\citenamefont {Nagy},
  \citenamefont {Domokos}, \citenamefont {Vukics},\ and\ \citenamefont
  {Ritsch}}]{Nagy2009Nonlinear}%
  \BibitemOpen
  \bibfield  {author} {\bibinfo {author} {\bibnamefont {Nagy}, \bibfnamefont
  {D.}}, \bibinfo {author} {\bibfnamefont {P.}~\bibnamefont {Domokos}},
  \bibinfo {author} {\bibfnamefont {A.}~\bibnamefont {Vukics}}, \ and\ \bibinfo
  {author} {\bibfnamefont {H.}~\bibnamefont {Ritsch}}} (\bibinfo {year}
  {2009}),\ \href {\doibase 10.1140/epjd/e2009-00265-7} {\bibfield  {journal}
  {\bibinfo  {journal} {The European Physical Journal D - Atomic, Molecular,
  Optical and Plasma Physics}\ }\textbf {\bibinfo {volume} {55}}~(\bibinfo
  {number} {3}),\ \bibinfo {pages} {659}}\BibitemShut {NoStop}%
\bibitem [{\citenamefont {Nagy}\ \emph {et~al.}(2010)\citenamefont {Nagy},
  \citenamefont {K\'{o}nya}, \citenamefont {Szirmai},\ and\ \citenamefont
  {Domokos}}]{Nagy2010DickeModel}%
  \BibitemOpen
  \bibfield  {author} {\bibinfo {author} {\bibnamefont {Nagy}, \bibfnamefont
  {D.}}, \bibinfo {author} {\bibfnamefont {G.}~\bibnamefont {K\'{o}nya}},
  \bibinfo {author} {\bibfnamefont {G.}~\bibnamefont {Szirmai}}, \ and\
  \bibinfo {author} {\bibfnamefont {P.}~\bibnamefont {Domokos}}} (\bibinfo
  {year} {2010}),\ \href {\doibase 10.1103/PhysRevLett.104.130401} {\bibfield
  {journal} {\bibinfo  {journal} {Physical Review Letters}\ }\textbf {\bibinfo
  {volume} {104}}~(\bibinfo {number} {13}),\ \bibinfo {pages}
  {130401+}}\BibitemShut {NoStop}%
\bibitem [{\citenamefont {Nagy}\ \emph {et~al.}(2008)\citenamefont {Nagy},
  \citenamefont {Szirmai},\ and\ \citenamefont
  {Domokos}}]{Nagy2008Selforganization}%
  \BibitemOpen
  \bibfield  {author} {\bibinfo {author} {\bibnamefont {Nagy}, \bibfnamefont
  {D.}}, \bibinfo {author} {\bibfnamefont {G.}~\bibnamefont {Szirmai}}, \ and\
  \bibinfo {author} {\bibfnamefont {P.}~\bibnamefont {Domokos}}} (\bibinfo
  {year} {2008}),\ \href {\doibase 10.1140/epjd/e2008-00074-6} {\bibfield
  {journal} {\bibinfo  {journal} {The European Physical Journal D - Atomic,
  Molecular, Optical and Plasma Physics}\ }\textbf {\bibinfo {volume}
  {48}}~(\bibinfo {number} {1}),\ \bibinfo {pages} {127}}\BibitemShut {NoStop}%
\bibitem [{\citenamefont {Nagy}\ \emph {et~al.}(2011)\citenamefont {Nagy},
  \citenamefont {Szirmai},\ and\ \citenamefont {Domokos}}]{Nagy2011Critical}%
  \BibitemOpen
  \bibfield  {author} {\bibinfo {author} {\bibnamefont {Nagy}, \bibfnamefont
  {D.}}, \bibinfo {author} {\bibfnamefont {G.}~\bibnamefont {Szirmai}}, \ and\
  \bibinfo {author} {\bibfnamefont {P.}~\bibnamefont {Domokos}}} (\bibinfo
  {year} {2011}),\ \href {\doibase 10.1103/PhysRevA.84.043637} {\bibfield
  {journal} {\bibinfo  {journal} {Physical Review A}\ }\textbf {\bibinfo
  {volume} {84}},\ \bibinfo {pages} {043637+}}\BibitemShut {NoStop}%
\bibitem [{\citenamefont {Nicholson}\ \emph {et~al.}(2012)\citenamefont
  {Nicholson}, \citenamefont {Williams}, \citenamefont {Bloom}, \citenamefont
  {Campbell}, \citenamefont {Martin}, \citenamefont {Swallows}, \citenamefont
  {Bishof},\ and\ \citenamefont {Ye}}]{Nicholson2012Cavity}%
  \BibitemOpen
  \bibfield  {author} {\bibinfo {author} {\bibnamefont {Nicholson},
  \bibfnamefont {T.}}, \bibinfo {author} {\bibfnamefont {J.}~\bibnamefont
  {Williams}}, \bibinfo {author} {\bibfnamefont {B.}~\bibnamefont {Bloom}},
  \bibinfo {author} {\bibfnamefont {S.}~\bibnamefont {Campbell}}, \bibinfo
  {author} {\bibfnamefont {M.}~\bibnamefont {Martin}}, \bibinfo {author}
  {\bibfnamefont {M.}~\bibnamefont {Swallows}}, \bibinfo {author}
  {\bibfnamefont {M.}~\bibnamefont {Bishof}}, \ and\ \bibinfo {author}
  {\bibfnamefont {J.}~\bibnamefont {Ye}}} (\bibinfo {year} {2012}),\ \href@noop
  {} {\bibfield  {journal} {\bibinfo  {journal} {Bulletin of the American
  Physical Society}\ }\textbf {\bibinfo {volume} {57}}}\BibitemShut {NoStop}%
\bibitem [{\citenamefont {Niedenzu}\ \emph {et~al.}(2011)\citenamefont
  {Niedenzu}, \citenamefont {Grie{\ss}er},\ and\ \citenamefont
  {Ritsch}}]{Niedenzu2011Kinetic}%
  \BibitemOpen
  \bibfield  {author} {\bibinfo {author} {\bibnamefont {Niedenzu},
  \bibfnamefont {W.}}, \bibinfo {author} {\bibfnamefont {T.}~\bibnamefont
  {Grie{\ss}er}}, \ and\ \bibinfo {author} {\bibfnamefont {H.}~\bibnamefont
  {Ritsch}}} (\bibinfo {year} {2011}),\ \href {\doibase
  10.1209/0295-5075/96/43001} {\bibinfo  {journal} {EPL (Europhysics Letters)}\
  ,\ \bibinfo {pages} {43001+}}\BibitemShut {NoStop}%
\bibitem [{\citenamefont {Niedenzu}\ \emph {et~al.}(2010)\citenamefont
  {Niedenzu}, \citenamefont {Schulze}, \citenamefont {Vukics},\ and\
  \citenamefont {Ritsch}}]{Niedenzu2010Microscopic}%
  \BibitemOpen
\bibfield  {journal} {  }\bibfield  {author} {\bibinfo {author} {\bibnamefont
  {Niedenzu}, \bibfnamefont {W.}}, \bibinfo {author} {\bibfnamefont
  {R.}~\bibnamefont {Schulze}}, \bibinfo {author} {\bibfnamefont
  {A.}~\bibnamefont {Vukics}}, \ and\ \bibinfo {author} {\bibfnamefont
  {H.}~\bibnamefont {Ritsch}}} (\bibinfo {year} {2010}),\ \href {\doibase
  10.1103/PhysRevA.82.043605} {\bibfield  {journal} {\bibinfo  {journal}
  {Physical Review A}\ }\textbf {\bibinfo {volume} {82}}~(\bibinfo {number}
  {4}),\ \bibinfo {pages} {043605+}}\BibitemShut {NoStop}%
\bibitem [{\citenamefont {Nimmrichter}\ \emph {et~al.}(2010)\citenamefont
  {Nimmrichter}, \citenamefont {Hammerer}, \citenamefont {Asenbaum},
  \citenamefont {Ritsch},\ and\ \citenamefont {Arndt}}]{Nimmrichter2010Master}%
  \BibitemOpen
  \bibfield  {author} {\bibinfo {author} {\bibnamefont {Nimmrichter},
  \bibfnamefont {S.}}, \bibinfo {author} {\bibfnamefont {K.}~\bibnamefont
  {Hammerer}}, \bibinfo {author} {\bibfnamefont {P.}~\bibnamefont {Asenbaum}},
  \bibinfo {author} {\bibfnamefont {H.}~\bibnamefont {Ritsch}}, \ and\ \bibinfo
  {author} {\bibfnamefont {M.}~\bibnamefont {Arndt}}} (\bibinfo {year}
  {2010}),\ \href {\doibase 10.1088/1367-2630/12/8/083003} {\bibfield
  {journal} {\bibinfo  {journal} {New Journal of Physics}\ }\textbf {\bibinfo
  {volume} {12}}~(\bibinfo {number} {8}),\ \bibinfo {pages}
  {083003+}}\BibitemShut {NoStop}%
\bibitem [{\citenamefont {Nuss{}mann}\ \emph {et~al.}(2005)\citenamefont
  {Nuss{}mann}, \citenamefont {Hijlkema}, \citenamefont {Weber}, \citenamefont
  {Rohde}, \citenamefont {Rempe},\ and\ \citenamefont
  {Kuhn}}]{Nussmann2005Submicron}%
  \BibitemOpen
  \bibfield  {author} {\bibinfo {author} {\bibnamefont {Nuss{}mann},
  \bibfnamefont {S.}}, \bibinfo {author} {\bibfnamefont {M.}~\bibnamefont
  {Hijlkema}}, \bibinfo {author} {\bibfnamefont {B.}~\bibnamefont {Weber}},
  \bibinfo {author} {\bibfnamefont {F.}~\bibnamefont {Rohde}}, \bibinfo
  {author} {\bibfnamefont {G.}~\bibnamefont {Rempe}}, \ and\ \bibinfo {author}
  {\bibfnamefont {A.}~\bibnamefont {Kuhn}}} (\bibinfo {year} {2005}),\ \href
  {\doibase 10.1103/PhysRevLett.95.173602} {\bibfield  {journal} {\bibinfo
  {journal} {Physical Review Letters}\ }\textbf {\bibinfo {volume}
  {95}}~(\bibinfo {number} {17}),\ \bibinfo {pages} {173602+}}\BibitemShut
  {NoStop}%
\bibitem [{\citenamefont {Nussmann}\ \emph {et~al.}(2005)\citenamefont
  {Nussmann}, \citenamefont {Murr}, \citenamefont {Hijlkema}, \citenamefont
  {Weber}, \citenamefont {Kuhn},\ and\ \citenamefont
  {Rempe}}]{Nussmann2005Vacuumstimulated}%
  \BibitemOpen
  \bibfield  {author} {\bibinfo {author} {\bibnamefont {Nussmann},
  \bibfnamefont {S.}}, \bibinfo {author} {\bibfnamefont {K.}~\bibnamefont
  {Murr}}, \bibinfo {author} {\bibfnamefont {M.}~\bibnamefont {Hijlkema}},
  \bibinfo {author} {\bibfnamefont {B.}~\bibnamefont {Weber}}, \bibinfo
  {author} {\bibfnamefont {A.}~\bibnamefont {Kuhn}}, \ and\ \bibinfo {author}
  {\bibfnamefont {G.}~\bibnamefont {Rempe}}} (\bibinfo {year} {2005}),\
  \bibfield  {booktitle} {\emph {\bibinfo {booktitle} {Nat Phys}},\ }\href
  {\doibase 10.1038/nphys120} {\bibfield  {journal} {\bibinfo  {journal}
  {Nature Physics}\ }\textbf {\bibinfo {volume} {1}}~(\bibinfo {number} {2}),\
  \bibinfo {pages} {122}}\BibitemShut {NoStop}%
\bibitem [{\citenamefont {Ourjoumtsev}\ \emph {et~al.}(2011)\citenamefont
  {Ourjoumtsev}, \citenamefont {Kubanek}, \citenamefont {Koch}, \citenamefont
  {Sames}, \citenamefont {Pinkse}, \citenamefont {Rempe},\ and\ \citenamefont
  {Murr}}]{Ourjoumtsev2011Observation}%
  \BibitemOpen
  \bibfield  {author} {\bibinfo {author} {\bibnamefont {Ourjoumtsev},
  \bibfnamefont {A.}}, \bibinfo {author} {\bibfnamefont {A.}~\bibnamefont
  {Kubanek}}, \bibinfo {author} {\bibfnamefont {M.}~\bibnamefont {Koch}},
  \bibinfo {author} {\bibfnamefont {C.}~\bibnamefont {Sames}}, \bibinfo
  {author} {\bibfnamefont {P.~W.~H.}\ \bibnamefont {Pinkse}}, \bibinfo {author}
  {\bibfnamefont {G.}~\bibnamefont {Rempe}}, \ and\ \bibinfo {author}
  {\bibfnamefont {K.}~\bibnamefont {Murr}}} (\bibinfo {year} {2011}),\ \href
  {\doibase 10.1038/nature10170} {\bibfield  {journal} {\bibinfo  {journal}
  {Nature}\ }\textbf {\bibinfo {volume} {474}}~(\bibinfo {number} {7353}),\
  \bibinfo {pages} {623}}\BibitemShut {NoStop}%
\bibitem [{\citenamefont {\"{O}ztop}\ \emph {et~al.}(2012)\citenamefont
  {\"{O}ztop}, \citenamefont {Bordyuh}, \citenamefont
  {M\"{u}stecapl{\i}o\u{g}lu},\ and\ \citenamefont
  {T\"{u}reci}}]{Oztop2012Excitations}%
  \BibitemOpen
  \bibfield  {author} {\bibinfo {author} {\bibnamefont {\"{O}ztop},
  \bibfnamefont {B.}}, \bibinfo {author} {\bibfnamefont {M.}~\bibnamefont
  {Bordyuh}}, \bibinfo {author} {\bibfnamefont {O.~E.}\ \bibnamefont
  {M\"{u}stecapl{\i}o\u{g}lu}}, \ and\ \bibinfo {author} {\bibfnamefont
  {H.~E.}\ \bibnamefont {T\"{u}reci}}} (\bibinfo {year} {2012}),\ \href
  {\doibase 10.1088/1367-2630/14/8/085011} {\bibfield  {journal} {\bibinfo
  {journal} {New Journal of Physics}\ }\textbf {\bibinfo {volume}
  {14}}~(\bibinfo {number} {8}),\ \bibinfo {pages} {085011+}}\BibitemShut
  {NoStop}%
\bibitem [{\citenamefont {Perrin}\ \emph {et~al.}(2001)\citenamefont {Perrin},
  \citenamefont {Lippi},\ and\ \citenamefont {Politi}}]{Perrin2001Phase}%
  \BibitemOpen
  \bibfield  {author} {\bibinfo {author} {\bibnamefont {Perrin}, \bibfnamefont
  {M.}}, \bibinfo {author} {\bibfnamefont {G.~L.}\ \bibnamefont {Lippi}}, \
  and\ \bibinfo {author} {\bibfnamefont {A.}~\bibnamefont {Politi}}} (\bibinfo
  {year} {2001}),\ \href {\doibase 10.1103/PhysRevLett.86.4520} {\bibfield
  {journal} {\bibinfo  {journal} {Physical Review Letters}\ }\textbf {\bibinfo
  {volume} {86}},\ \bibinfo {pages} {4520}}\BibitemShut {NoStop}%
\bibitem [{\citenamefont {Perrin}\ \emph {et~al.}(2002)\citenamefont {Perrin},
  \citenamefont {Ye},\ and\ \citenamefont {Narducci}}]{Perrin2002Microscopic}%
  \BibitemOpen
  \bibfield  {author} {\bibinfo {author} {\bibnamefont {Perrin}, \bibfnamefont
  {M.}}, \bibinfo {author} {\bibfnamefont {Z.}~\bibnamefont {Ye}}, \ and\
  \bibinfo {author} {\bibfnamefont {L.~M.}\ \bibnamefont {Narducci}}} (\bibinfo
  {year} {2002}),\ \href {\doibase 10.1103/PhysRevA.66.043809} {\bibfield
  {journal} {\bibinfo  {journal} {Physical Review A}\ }\textbf {\bibinfo
  {volume} {66}},\ \bibinfo {pages} {043809+}}\BibitemShut {NoStop}%
\bibitem [{\citenamefont {Phillips}(1998)}]{Phillips1998Nobel}%
  \BibitemOpen
  \bibfield  {author} {\bibinfo {author} {\bibnamefont {Phillips},
  \bibfnamefont {W.~D.}}} (\bibinfo {year} {1998}),\ \href {\doibase
  10.1103/RevModPhys.70.721} {\bibfield  {journal} {\bibinfo  {journal}
  {Reviews of Modern Physics}\ }\textbf {\bibinfo {volume} {70}}~(\bibinfo
  {number} {3}),\ \bibinfo {pages} {721}}\BibitemShut {NoStop}%
\bibitem [{\citenamefont {Pinkse}\ and\ \citenamefont
  {Rempe}(2002)}]{Pinkse2002Single}%
  \BibitemOpen
  \bibfield  {author} {\bibinfo {author} {\bibnamefont {Pinkse}, \bibfnamefont
  {P.}}, \ and\ \bibinfo {author} {\bibfnamefont {G.}~\bibnamefont {Rempe}}}
  (\bibinfo {year} {2002}),\ \enquote {\bibinfo {title} {{Single Atoms Moving
  in a High-Finesse Cavity}},}\ in\ \href@noop {} {\emph {\bibinfo {booktitle}
  {Cavity Enhanced Spectroscopies}}},\ \bibinfo {editor} {edited by\ \bibinfo
  {editor} {\bibfnamefont {R.~D.}\ \bibnamefont {van Zee}}\ and\ \bibinfo
  {editor} {\bibfnamefont {J.~P.}\ \bibnamefont {Looney}}},\ Chap.~\bibinfo
  {chapter} {13}\ (\bibinfo  {publisher} {Academic Press})\ pp.\ \bibinfo
  {pages} {255--295}\BibitemShut {NoStop}%
\bibitem [{\citenamefont {Pinkse}\ \emph {et~al.}(2000)\citenamefont {Pinkse},
  \citenamefont {Fischer}, \citenamefont {Maunz},\ and\ \citenamefont
  {Rempe}}]{Pinkse2000Trapping}%
  \BibitemOpen
  \bibfield  {author} {\bibinfo {author} {\bibnamefont {Pinkse}, \bibfnamefont
  {P.~W.~H.}}, \bibinfo {author} {\bibfnamefont {T.}~\bibnamefont {Fischer}},
  \bibinfo {author} {\bibfnamefont {P.}~\bibnamefont {Maunz}}, \ and\ \bibinfo
  {author} {\bibfnamefont {G.}~\bibnamefont {Rempe}}} (\bibinfo {year}
  {2000}),\ \href {\doibase 10.1038/35006006} {\bibfield  {journal} {\bibinfo
  {journal} {Nature}\ }\textbf {\bibinfo {volume} {404}}~(\bibinfo {number}
  {6776}),\ \bibinfo {pages} {365}}\BibitemShut {NoStop}%
\bibitem [{\citenamefont {{P}itaevskii}\ and\ \citenamefont
  {{S}tringari}(2003)}]{Pitaevskii2003Bose-Einstein}%
  \BibitemOpen
  \bibfield  {author} {\bibinfo {author} {\bibnamefont {{P}itaevskii},
  \bibfnamefont {L.}}, \ and\ \bibinfo {author} {\bibfnamefont
  {S.}~\bibnamefont {{S}tringari}}} (\bibinfo {year} {2003}),\ \href@noop {}
  {\emph {\bibinfo {title} {\{B\}ose-\{E\}instein \{C\}ondensation}}},\ edited
  by\ \bibinfo {editor} {\bibfnamefont {J.}~\bibnamefont {Birman}}, \bibinfo
  {editor} {\bibfnamefont {S.~F.}\ \bibnamefont {Friend}}, \bibinfo {editor}
  {\bibfnamefont {C.~H.}\ \bibnamefont {Llewellyn-Smith}}, \bibinfo {editor}
  {\bibfnamefont {M.}~\bibnamefont {Rees}}, \bibinfo {editor} {\bibfnamefont
  {D.}~\bibnamefont {Sherrington}}, \ and\ \bibinfo {editor} {\bibfnamefont
  {G.}~\bibnamefont {Veneziano}}\ (\bibinfo  {publisher} {Oxford University
  Press},\ \bibinfo {address} {Oxford})\BibitemShut {NoStop}%
\bibitem [{\citenamefont {Posch}\ and\ \citenamefont
  {Thirring}(2005)}]{Posch2005Stellar}%
  \BibitemOpen
  \bibfield  {author} {\bibinfo {author} {\bibnamefont {Posch}, \bibfnamefont
  {H.}}, \ and\ \bibinfo {author} {\bibfnamefont {W.}~\bibnamefont {Thirring}}}
  (\bibinfo {year} {2005}),\ \href@noop {} {\bibfield  {journal} {\bibinfo
  {journal} {Physical Review Letters}\ }\textbf {\bibinfo {volume}
  {95}}~(\bibinfo {number} {25}),\ \bibinfo {pages} {251101}}\BibitemShut
  {NoStop}%
\bibitem [{\citenamefont {Puppe}\ \emph
  {et~al.}(2007{\natexlab{a}})\citenamefont {Puppe}, \citenamefont {Schuster},
  \citenamefont {Grothe}, \citenamefont {Kubanek}, \citenamefont {Murr},
  \citenamefont {Pinkse},\ and\ \citenamefont {Rempe}}]{Puppe2007Trapping}%
  \BibitemOpen
  \bibfield  {author} {\bibinfo {author} {\bibnamefont {Puppe}, \bibfnamefont
  {T.}}, \bibinfo {author} {\bibfnamefont {I.}~\bibnamefont {Schuster}},
  \bibinfo {author} {\bibfnamefont {A.}~\bibnamefont {Grothe}}, \bibinfo
  {author} {\bibfnamefont {A.}~\bibnamefont {Kubanek}}, \bibinfo {author}
  {\bibfnamefont {K.}~\bibnamefont {Murr}}, \bibinfo {author} {\bibfnamefont
  {P.~W.~H.}\ \bibnamefont {Pinkse}}, \ and\ \bibinfo {author} {\bibfnamefont
  {G.}~\bibnamefont {Rempe}}} (\bibinfo {year} {2007}{\natexlab{a}}),\ \href
  {\doibase 10.1103/PhysRevLett.99.013002} {\bibfield  {journal} {\bibinfo
  {journal} {Physical Review Letters}\ }\textbf {\bibinfo {volume}
  {99}}~(\bibinfo {number} {1}),\ \bibinfo {pages} {013002+}}\BibitemShut
  {NoStop}%
\bibitem [{\citenamefont {Puppe}\ \emph
  {et~al.}(2007{\natexlab{b}})\citenamefont {Puppe}, \citenamefont {Schuster},
  \citenamefont {Maunz}, \citenamefont {Murr}, \citenamefont {Pinkse},\ and\
  \citenamefont {Rempe}}]{Puppe2007Light}%
  \BibitemOpen
  \bibfield  {author} {\bibinfo {author} {\bibnamefont {Puppe}, \bibfnamefont
  {T.}}, \bibinfo {author} {\bibfnamefont {I.}~\bibnamefont {Schuster}},
  \bibinfo {author} {\bibfnamefont {P.}~\bibnamefont {Maunz}}, \bibinfo
  {author} {\bibfnamefont {K.}~\bibnamefont {Murr}}, \bibinfo {author}
  {\bibfnamefont {P.~W.~H.}\ \bibnamefont {Pinkse}}, \ and\ \bibinfo {author}
  {\bibfnamefont {G.}~\bibnamefont {Rempe}}} (\bibinfo {year}
  {2007}{\natexlab{b}}),\ \href {\doibase 10.1080/09500340701415578} {\bibfield
   {journal} {\bibinfo  {journal} {Journal of Modern Optics}\ }\textbf
  {\bibinfo {volume} {54}}~(\bibinfo {number} {13-15}),\ \bibinfo {pages}
  {1927}}\BibitemShut {NoStop}%
\bibitem [{\citenamefont {Purcell}(1946)}]{Purcell1946Spontaneous}%
  \BibitemOpen
  \bibfield  {author} {\bibinfo {author} {\bibnamefont {Purcell}, \bibfnamefont
  {E.~M.}}} (\bibinfo {year} {1946}),\ \href@noop {} {\bibfield  {journal}
  {\bibinfo  {journal} {Phys. Rev.}\ }\textbf {\bibinfo {volume} {69}},\
  \bibinfo {pages} {681+}}\BibitemShut {NoStop}%
\bibitem [{\citenamefont {Purdy}\ \emph {et~al.}(2010)\citenamefont {Purdy},
  \citenamefont {Brooks}, \citenamefont {Botter}, \citenamefont {Brahms},
  \citenamefont {Ma},\ and\ \citenamefont {Kurn}}]{Purdy2010Tunable}%
  \BibitemOpen
  \bibfield  {author} {\bibinfo {author} {\bibnamefont {Purdy}, \bibfnamefont
  {T.~P.}}, \bibinfo {author} {\bibfnamefont {D.~W.~C.}\ \bibnamefont
  {Brooks}}, \bibinfo {author} {\bibfnamefont {T.}~\bibnamefont {Botter}},
  \bibinfo {author} {\bibfnamefont {N.}~\bibnamefont {Brahms}}, \bibinfo
  {author} {\bibfnamefont {Z.~Y.}\ \bibnamefont {Ma}}, \ and\ \bibinfo {author}
  {\bibfnamefont {D.~M.~S.}\ \bibnamefont {Kurn}}} (\bibinfo {year} {2010}),\
  \href {\doibase 10.1103/PhysRevLett.105.133602} {\bibfield  {journal}
  {\bibinfo  {journal} {Physical Review Letters}\ }\textbf {\bibinfo {volume}
  {105}}~(\bibinfo {number} {13}),\ \bibinfo {pages} {133602+}}\BibitemShut
  {NoStop}%
\bibitem [{\citenamefont {Raimond}\ \emph {et~al.}(2001)\citenamefont
  {Raimond}, \citenamefont {Brune},\ and\ \citenamefont
  {Haroche}}]{Raimond2001Manipulating}%
  \BibitemOpen
  \bibfield  {author} {\bibinfo {author} {\bibnamefont {Raimond}, \bibfnamefont
  {J.~M.}}, \bibinfo {author} {\bibfnamefont {M.}~\bibnamefont {Brune}}, \ and\
  \bibinfo {author} {\bibfnamefont {S.}~\bibnamefont {Haroche}}} (\bibinfo
  {year} {2001}),\ \href {\doibase 10.1103/RevModPhys.73.565} {\bibfield
  {journal} {\bibinfo  {journal} {Reviews of Modern Physics}\ }\textbf
  {\bibinfo {volume} {73}}~(\bibinfo {number} {3}),\ \bibinfo {pages}
  {565}}\BibitemShut {NoStop}%
\bibitem [{\citenamefont {Raizen}\ \emph {et~al.}(1989)\citenamefont {Raizen},
  \citenamefont {Thompson}, \citenamefont {Brecha}, \citenamefont {Kimble},\
  and\ \citenamefont {Carmichael}}]{Raizen1989Normalmode}%
  \BibitemOpen
  \bibfield  {author} {\bibinfo {author} {\bibnamefont {Raizen}, \bibfnamefont
  {M.~G.}}, \bibinfo {author} {\bibfnamefont {R.~J.}\ \bibnamefont {Thompson}},
  \bibinfo {author} {\bibfnamefont {R.~J.}\ \bibnamefont {Brecha}}, \bibinfo
  {author} {\bibfnamefont {H.~J.}\ \bibnamefont {Kimble}}, \ and\ \bibinfo
  {author} {\bibfnamefont {H.~J.}\ \bibnamefont {Carmichael}}} (\bibinfo {year}
  {1989}),\ \href {\doibase 10.1103/PhysRevLett.63.240} {\bibfield  {journal}
  {\bibinfo  {journal} {Phys. Rev. Lett.}\ }\textbf {\bibinfo {volume} {63}},\
  \bibinfo {pages} {240}}\BibitemShut {NoStop}%
\bibitem [{\citenamefont {Rangwala}\ \emph {et~al.}(2003)\citenamefont
  {Rangwala}, \citenamefont {Junglen}, \citenamefont {Rieger}, \citenamefont
  {Pinkse},\ and\ \citenamefont {Rempe}}]{Rangwala2003continuous}%
  \BibitemOpen
  \bibfield  {author} {\bibinfo {author} {\bibnamefont {Rangwala},
  \bibfnamefont {S.~A.}}, \bibinfo {author} {\bibfnamefont {T.}~\bibnamefont
  {Junglen}}, \bibinfo {author} {\bibfnamefont {T.}~\bibnamefont {Rieger}},
  \bibinfo {author} {\bibfnamefont {P.~W.~H.}\ \bibnamefont {Pinkse}}, \ and\
  \bibinfo {author} {\bibfnamefont {G.}~\bibnamefont {Rempe}}} (\bibinfo {year}
  {2003}),\ \href {\doibase 10.1103/PhysRevA.67.043406} {\bibfield  {journal}
  {\bibinfo  {journal} {Physical Review A}\ }\textbf {\bibinfo {volume}
  {67}}~(\bibinfo {number} {4}),\ \bibinfo {pages} {043406+}}\BibitemShut
  {NoStop}%
\bibitem [{\citenamefont {Rist}\ \emph {et~al.}(2010)\citenamefont {Rist},
  \citenamefont {Menotti},\ and\ \citenamefont {Morigi}}]{Rist2010Light}%
  \BibitemOpen
  \bibfield  {author} {\bibinfo {author} {\bibnamefont {Rist}, \bibfnamefont
  {S.}}, \bibinfo {author} {\bibfnamefont {C.}~\bibnamefont {Menotti}}, \ and\
  \bibinfo {author} {\bibfnamefont {G.}~\bibnamefont {Morigi}}} (\bibinfo
  {year} {2010}),\ \href {\doibase 10.1103/PhysRevA.81.013404} {\bibfield
  {journal} {\bibinfo  {journal} {Physical Review A}\ }\textbf {\bibinfo
  {volume} {81}}~(\bibinfo {number} {1}),\ \bibinfo {pages}
  {013404+}}\BibitemShut {NoStop}%
\bibitem [{\citenamefont {Ritsch}(2009)}]{Ritsch2009Crystals}%
  \BibitemOpen
  \bibfield  {author} {\bibinfo {author} {\bibnamefont {Ritsch}, \bibfnamefont
  {H.}}} (\bibinfo {year} {2009}),\ \href {\doibase 10.1038/nphys1435}
  {\bibfield  {journal} {\bibinfo  {journal} {Nat Phys}\ }\textbf {\bibinfo
  {volume} {5}}~(\bibinfo {number} {11}),\ \bibinfo {pages} {781}}\BibitemShut
  {NoStop}%
\bibitem [{\citenamefont {Ritter}\ \emph {et~al.}(2009)\citenamefont {Ritter},
  \citenamefont {Brennecke}, \citenamefont {Baumann}, \citenamefont {Donner},
  \citenamefont {Guerlin},\ and\ \citenamefont
  {Esslinger}}]{Ritter2009Dynamical}%
  \BibitemOpen
  \bibfield  {author} {\bibinfo {author} {\bibnamefont {Ritter}, \bibfnamefont
  {S.}}, \bibinfo {author} {\bibfnamefont {F.}~\bibnamefont {Brennecke}},
  \bibinfo {author} {\bibfnamefont {K.}~\bibnamefont {Baumann}}, \bibinfo
  {author} {\bibfnamefont {T.}~\bibnamefont {Donner}}, \bibinfo {author}
  {\bibfnamefont {C.}~\bibnamefont {Guerlin}}, \ and\ \bibinfo {author}
  {\bibfnamefont {T.}~\bibnamefont {Esslinger}}} (\bibinfo {year} {2009}),\
  \href {\doibase 10.1007/s00340-009-3436-9} {\bibfield  {journal} {\bibinfo
  {journal} {Applied Physics B: Lasers and Optics}\ }\textbf {\bibinfo {volume}
  {95}}~(\bibinfo {number} {2}),\ \bibinfo {pages} {213}}\BibitemShut {NoStop}%
\bibitem [{\citenamefont {Robb}\ \emph {et~al.}(2004)\citenamefont {Robb},
  \citenamefont {Piovella}, \citenamefont {Ferraro}, \citenamefont {Bonifacio},
  \citenamefont {Courteille},\ and\ \citenamefont
  {Zimmermann}}]{Robb2004Collective}%
  \BibitemOpen
  \bibfield  {author} {\bibinfo {author} {\bibnamefont {Robb}, \bibfnamefont
  {G.~R.~M.}}, \bibinfo {author} {\bibfnamefont {N.}~\bibnamefont {Piovella}},
  \bibinfo {author} {\bibfnamefont {A.}~\bibnamefont {Ferraro}}, \bibinfo
  {author} {\bibfnamefont {R.}~\bibnamefont {Bonifacio}}, \bibinfo {author}
  {\bibfnamefont {P.~W.}\ \bibnamefont {Courteille}}, \ and\ \bibinfo {author}
  {\bibfnamefont {C.}~\bibnamefont {Zimmermann}}} (\bibinfo {year} {2004}),\
  \href {\doibase 10.1103/PhysRevA.69.041403} {\bibfield  {journal} {\bibinfo
  {journal} {Physical Review A}\ }\textbf {\bibinfo {volume} {69}}~(\bibinfo
  {number} {4}),\ \bibinfo {pages} {041403+}}\BibitemShut {NoStop}%
\bibitem [{\citenamefont {{Rolston}}\ and\ \citenamefont
  {{Phillips}}(2002)}]{Rolston2002Nonlinear}%
  \BibitemOpen
  \bibfield  {author} {\bibinfo {author} {\bibnamefont {{Rolston}},
  \bibfnamefont {S.~L.}}, \ and\ \bibinfo {author} {\bibfnamefont {W.~D.}\
  \bibnamefont {{Phillips}}}} (\bibinfo {year} {2002}),\ \href {\doibase
  10.1038/416219a} {\bibfield  {journal} {\bibinfo  {journal} {\nat}\ }\textbf
  {\bibinfo {volume} {416}},\ \bibinfo {pages} {219}}\BibitemShut {NoStop}%
\bibitem [{\citenamefont {Romero-Isart}\ \emph {et~al.}(2010)\citenamefont
  {Romero-Isart}, \citenamefont {Juan}, \citenamefont {Quidant},\ and\
  \citenamefont {Cirac}}]{Romero2010Toward}%
  \BibitemOpen
  \bibfield  {author} {\bibinfo {author} {\bibnamefont {Romero-Isart},
  \bibfnamefont {O.}}, \bibinfo {author} {\bibfnamefont {M.}~\bibnamefont
  {Juan}}, \bibinfo {author} {\bibfnamefont {R.}~\bibnamefont {Quidant}}, \
  and\ \bibinfo {author} {\bibfnamefont {J.}~\bibnamefont {Cirac}}} (\bibinfo
  {year} {2010}),\ \href@noop {} {\bibfield  {journal} {\bibinfo  {journal}
  {New Journal of Physics}\ }\textbf {\bibinfo {volume} {12}},\ \bibinfo
  {pages} {033015}}\BibitemShut {NoStop}%
\bibitem [{\citenamefont {Rza\.{z}ewski}\ \emph {et~al.}(1975)\citenamefont
  {Rza\.{z}ewski}, \citenamefont {W\'{o}dkiewicz},\ and\ \citenamefont
  {\.{Z}akowicz}}]{Rzazewski1975Phase}%
  \BibitemOpen
  \bibfield  {author} {\bibinfo {author} {\bibnamefont {Rza\.{z}ewski},
  \bibfnamefont {K.}}, \bibinfo {author} {\bibfnamefont {K.}~\bibnamefont
  {W\'{o}dkiewicz}}, \ and\ \bibinfo {author} {\bibfnamefont {W.}~\bibnamefont
  {\.{Z}akowicz}}} (\bibinfo {year} {1975}),\ \href {\doibase
  10.1103/PhysRevLett.35.432} {\bibfield  {journal} {\bibinfo  {journal}
  {Physical Review Letters}\ }\textbf {\bibinfo {volume} {35}}~(\bibinfo
  {number} {7}),\ \bibinfo {pages} {432}}\BibitemShut {NoStop}%
\bibitem [{\citenamefont {Salzburger}\ \emph {et~al.}(2002)\citenamefont
  {Salzburger}, \citenamefont {Domokos},\ and\ \citenamefont
  {Ritsch}}]{Salzburger2002Enhanced}%
  \BibitemOpen
  \bibfield  {author} {\bibinfo {author} {\bibnamefont {Salzburger},
  \bibfnamefont {T.}}, \bibinfo {author} {\bibfnamefont {P.}~\bibnamefont
  {Domokos}}, \ and\ \bibinfo {author} {\bibfnamefont {H.}~\bibnamefont
  {Ritsch}}} (\bibinfo {year} {2002}),\ \href
  {http://www.opticsinfobase.org/abstract.cfm?id=70295} {\bibfield  {journal}
  {\bibinfo  {journal} {Opt. Express}\ }\textbf {\bibinfo {volume}
  {10}}~(\bibinfo {number} {21}),\ \bibinfo {pages} {1204}}\BibitemShut
  {NoStop}%
\bibitem [{\citenamefont {Salzburger}\ \emph {et~al.}(2005)\citenamefont
  {Salzburger}, \citenamefont {Domokos},\ and\ \citenamefont
  {Ritsch}}]{Salzburger2005Theory}%
  \BibitemOpen
  \bibfield  {author} {\bibinfo {author} {\bibnamefont {Salzburger},
  \bibfnamefont {T.}}, \bibinfo {author} {\bibfnamefont {P.}~\bibnamefont
  {Domokos}}, \ and\ \bibinfo {author} {\bibfnamefont {H.}~\bibnamefont
  {Ritsch}}} (\bibinfo {year} {2005}),\ \href {\doibase
  10.1103/PhysRevA.72.033805} {\bibfield  {journal} {\bibinfo  {journal}
  {Physical Review A}\ }\textbf {\bibinfo {volume} {72}}~(\bibinfo {number}
  {3}),\ \bibinfo {pages} {033805+}}\BibitemShut {NoStop}%
\bibitem [{\citenamefont {Salzburger}\ and\ \citenamefont
  {Ritsch}(2004)}]{Salzburger2004Atomic}%
  \BibitemOpen
  \bibfield  {author} {\bibinfo {author} {\bibnamefont {Salzburger},
  \bibfnamefont {T.}}, \ and\ \bibinfo {author} {\bibfnamefont
  {H.}~\bibnamefont {Ritsch}}} (\bibinfo {year} {2004}),\ \href {\doibase
  10.1103/PhysRevLett.93.063002} {\bibfield  {journal} {\bibinfo  {journal}
  {Physical Review Letters}\ }\textbf {\bibinfo {volume} {93}}~(\bibinfo
  {number} {6}),\ \bibinfo {pages} {063002+}}\BibitemShut {NoStop}%
\bibitem [{\citenamefont {Salzburger}\ and\ \citenamefont
  {Ritsch}(2006)}]{Salzburger2006Lasing}%
  \BibitemOpen
  \bibfield  {author} {\bibinfo {author} {\bibnamefont {Salzburger},
  \bibfnamefont {T.}}, \ and\ \bibinfo {author} {\bibfnamefont
  {H.}~\bibnamefont {Ritsch}}} (\bibinfo {year} {2006}),\ \href {\doibase
  10.1103/PhysRevA.74.033806} {\bibfield  {journal} {\bibinfo  {journal}
  {Physical Review A}\ }\textbf {\bibinfo {volume} {74}}~(\bibinfo {number}
  {3}),\ \bibinfo {pages} {033806+}}\BibitemShut {NoStop}%
\bibitem [{\citenamefont {Salzburger}\ and\ \citenamefont
  {Ritsch}(2007)}]{Salzburger2007Atomphoton}%
  \BibitemOpen
  \bibfield  {author} {\bibinfo {author} {\bibnamefont {Salzburger},
  \bibfnamefont {T.}}, \ and\ \bibinfo {author} {\bibfnamefont
  {H.}~\bibnamefont {Ritsch}}} (\bibinfo {year} {2007}),\ \href {\doibase
  10.1103/PhysRevA.75.061601} {\bibfield  {journal} {\bibinfo  {journal}
  {Physical Review A}\ }\textbf {\bibinfo {volume} {75}}~(\bibinfo {number}
  {6}),\ \bibinfo {pages} {061601+}}\BibitemShut {NoStop}%
\bibitem [{\citenamefont {Salzburger}\ and\ \citenamefont
  {Ritsch}(2008)}]{Salzburger2008Twin}%
  \BibitemOpen
  \bibfield  {author} {\bibinfo {author} {\bibnamefont {Salzburger},
  \bibfnamefont {T.}}, \ and\ \bibinfo {author} {\bibfnamefont
  {H.}~\bibnamefont {Ritsch}}} (\bibinfo {year} {2008}),\ \href {\doibase
  10.1103/PhysRevA.77.063620} {\bibfield  {journal} {\bibinfo  {journal}
  {Physical Review A}\ }\textbf {\bibinfo {volume} {77}},\ \bibinfo {pages}
  {063620+}}\BibitemShut {NoStop}%
\bibitem [{\citenamefont {Salzburger}\ and\ \citenamefont
  {Ritsch}(2009)}]{Salzburger2009Collective}%
  \BibitemOpen
  \bibfield  {author} {\bibinfo {author} {\bibnamefont {Salzburger},
  \bibfnamefont {T.}}, \ and\ \bibinfo {author} {\bibfnamefont
  {H.}~\bibnamefont {Ritsch}}} (\bibinfo {year} {2009}),\ \href {\doibase
  10.1088/1367-2630/11/5/055025} {\bibfield  {journal} {\bibinfo  {journal}
  {New Journal of Physics}\ }\textbf {\bibinfo {volume} {11}}~(\bibinfo
  {number} {5}),\ \bibinfo {pages} {055025+}}\BibitemShut {NoStop}%
\bibitem [{\citenamefont {Sauer}\ \emph {et~al.}(2004)\citenamefont {Sauer},
  \citenamefont {Fortier}, \citenamefont {Chang}, \citenamefont {Hamley},\ and\
  \citenamefont {Chapman}}]{Sauer2004Cavity}%
  \BibitemOpen
  \bibfield  {author} {\bibinfo {author} {\bibnamefont {Sauer}, \bibfnamefont
  {J.~A.}}, \bibinfo {author} {\bibfnamefont {K.~M.}\ \bibnamefont {Fortier}},
  \bibinfo {author} {\bibfnamefont {M.~S.}\ \bibnamefont {Chang}}, \bibinfo
  {author} {\bibfnamefont {C.~D.}\ \bibnamefont {Hamley}}, \ and\ \bibinfo
  {author} {\bibfnamefont {M.~S.}\ \bibnamefont {Chapman}}} (\bibinfo {year}
  {2004}),\ \href {\doibase 10.1103/PhysRevA.69.051804} {\bibfield  {journal}
  {\bibinfo  {journal} {Physical Review A}\ }\textbf {\bibinfo {volume}
  {69}}~(\bibinfo {number} {5}),\ \bibinfo {pages} {051804+}}\BibitemShut
  {NoStop}%
\bibitem [{\citenamefont {Schleier-Smith}\ \emph {et~al.}(2011)\citenamefont
  {Schleier-Smith}, \citenamefont {Leroux}, \citenamefont {Zhang},
  \citenamefont {Van~Camp},\ and\ \citenamefont {Vuleti\ifmmode~\acute{c}\else
  \'{c}\fi{}}}]{Schleier2011Optomechanical}%
  \BibitemOpen
  \bibfield  {author} {\bibinfo {author} {\bibnamefont {Schleier-Smith},
  \bibfnamefont {M.~H.}}, \bibinfo {author} {\bibfnamefont {I.~D.}\
  \bibnamefont {Leroux}}, \bibinfo {author} {\bibfnamefont {H.}~\bibnamefont
  {Zhang}}, \bibinfo {author} {\bibfnamefont {M.~A.}\ \bibnamefont {Van~Camp}},
  \ and\ \bibinfo {author} {\bibfnamefont {V.}~\bibnamefont
  {Vuleti\ifmmode~\acute{c}\else \'{c}\fi{}}}} (\bibinfo {year} {2011}),\ \href
  {\doibase 10.1103/PhysRevLett.107.143005} {\bibfield  {journal} {\bibinfo
  {journal} {Phys. Rev. Lett.}\ }\textbf {\bibinfo {volume} {107}},\ \bibinfo
  {pages} {143005}}\BibitemShut {NoStop}%
\bibitem [{\citenamefont {Schliesser}\ \emph {et~al.}(2009)\citenamefont
  {Schliesser}, \citenamefont {Arcizet}, \citenamefont {Rivi{\`{e}}re},
  \citenamefont {Anetsberger},\ and\ \citenamefont
  {Kippenberg}}]{Schliesser2009}%
  \BibitemOpen
  \bibfield  {author} {\bibinfo {author} {\bibnamefont {Schliesser},
  \bibfnamefont {A.}}, \bibinfo {author} {\bibfnamefont {O.}~\bibnamefont
  {Arcizet}}, \bibinfo {author} {\bibfnamefont {R.}~\bibnamefont
  {Rivi{\`{e}}re}}, \bibinfo {author} {\bibfnamefont {G.}~\bibnamefont
  {Anetsberger}}, \ and\ \bibinfo {author} {\bibfnamefont {T.~J.}\ \bibnamefont
  {Kippenberg}}} (\bibinfo {year} {2009}),\ \href@noop {} {\bibfield  {journal}
  {\bibinfo  {journal} {Nat. Phys.}\ }\textbf {\bibinfo {volume} {5}},\
  \bibinfo {pages} {509}}\BibitemShut {NoStop}%
\bibitem [{\citenamefont {Schulze}\ \emph {et~al.}(2010)\citenamefont
  {Schulze}, \citenamefont {Genes},\ and\ \citenamefont
  {Ritsch}}]{Schulze2010Optomechanical}%
  \BibitemOpen
  \bibfield  {author} {\bibinfo {author} {\bibnamefont {Schulze}, \bibfnamefont
  {R.~J.}}, \bibinfo {author} {\bibfnamefont {C.}~\bibnamefont {Genes}}, \ and\
  \bibinfo {author} {\bibfnamefont {H.}~\bibnamefont {Ritsch}}} (\bibinfo
  {year} {2010}),\ \href {\doibase 10.1103/PhysRevA.81.063820} {\bibfield
  {journal} {\bibinfo  {journal} {Physical Review A}\ }\textbf {\bibinfo
  {volume} {81}}~(\bibinfo {number} {6}),\ \bibinfo {pages}
  {063820+}}\BibitemShut {NoStop}%
\bibitem [{\citenamefont {Schuster}\ \emph {et~al.}(2008)\citenamefont
  {Schuster}, \citenamefont {Kubanek}, \citenamefont {Fuhrmanek}, \citenamefont
  {Puppe}, \citenamefont {Pinkse}, \citenamefont {Murr},\ and\ \citenamefont
  {Rempe}}]{Schuster2008Nonlinear}%
  \BibitemOpen
  \bibfield  {author} {\bibinfo {author} {\bibnamefont {Schuster},
  \bibfnamefont {I.}}, \bibinfo {author} {\bibfnamefont {A.}~\bibnamefont
  {Kubanek}}, \bibinfo {author} {\bibfnamefont {A.}~\bibnamefont {Fuhrmanek}},
  \bibinfo {author} {\bibfnamefont {T.}~\bibnamefont {Puppe}}, \bibinfo
  {author} {\bibfnamefont {P.~W.~H.}\ \bibnamefont {Pinkse}}, \bibinfo {author}
  {\bibfnamefont {K.}~\bibnamefont {Murr}}, \ and\ \bibinfo {author}
  {\bibfnamefont {G.}~\bibnamefont {Rempe}}} (\bibinfo {year} {2008}),\ \href
  {\doibase 10.1038/nphys940} {\bibfield  {journal} {\bibinfo  {journal}
  {Nature Physics}\ }\textbf {\bibinfo {volume} {4}}~(\bibinfo {number} {5}),\
  \bibinfo {pages} {382}}\BibitemShut {NoStop}%
\bibitem [{\citenamefont {Shore}\ and\ \citenamefont
  {Knight}(1993)}]{Shore1993JaynesCummings}%
  \BibitemOpen
  \bibfield  {author} {\bibinfo {author} {\bibnamefont {Shore}, \bibfnamefont
  {B.~W.}}, \ and\ \bibinfo {author} {\bibfnamefont {P.~L.}\ \bibnamefont
  {Knight}}} (\bibinfo {year} {1993}),\ \href {\doibase
  10.1080/09500349314551321} {\bibfield  {journal} {\bibinfo  {journal}
  {Journal of Modern Optics}\ }\textbf {\bibinfo {volume} {40}}~(\bibinfo
  {number} {7}),\ \bibinfo {pages} {1195}}\BibitemShut {NoStop}%
\bibitem [{\citenamefont {Shuman}\ \emph {et~al.}(2010)\citenamefont {Shuman},
  \citenamefont {Barry},\ and\ \citenamefont {DeMille}}]{Shuman2010Laser}%
  \BibitemOpen
  \bibfield  {author} {\bibinfo {author} {\bibnamefont {Shuman}, \bibfnamefont
  {E.~S.}}, \bibinfo {author} {\bibfnamefont {J.~F.}\ \bibnamefont {Barry}}, \
  and\ \bibinfo {author} {\bibfnamefont {D.}~\bibnamefont {DeMille}}} (\bibinfo
  {year} {2010}),\ \href {\doibase 10.1038/nature09443} {\bibfield  {journal}
  {\bibinfo  {journal} {Nature}\ }\textbf {\bibinfo {volume} {467}}~(\bibinfo
  {number} {7317}),\ \bibinfo {pages} {820}}\BibitemShut {NoStop}%
\bibitem [{\citenamefont {Silver}\ \emph {et~al.}(2010)\citenamefont {Silver},
  \citenamefont {Hohenadler}, \citenamefont {Bhaseen},\ and\ \citenamefont
  {Simons}}]{Silver2010BoseHubbard}%
  \BibitemOpen
  \bibfield  {author} {\bibinfo {author} {\bibnamefont {Silver}, \bibfnamefont
  {A.~O.}}, \bibinfo {author} {\bibfnamefont {M.}~\bibnamefont {Hohenadler}},
  \bibinfo {author} {\bibfnamefont {M.~J.}\ \bibnamefont {Bhaseen}}, \ and\
  \bibinfo {author} {\bibfnamefont {B.~D.}\ \bibnamefont {Simons}}} (\bibinfo
  {year} {2010}),\ \href {\doibase 10.1103/PhysRevA.81.023617} {\bibfield
  {journal} {\bibinfo  {journal} {Physical Review A}\ }\textbf {\bibinfo
  {volume} {81}},\ \bibinfo {pages} {023617+}}\BibitemShut {NoStop}%
\bibitem [{\citenamefont {Slama}\ \emph
  {et~al.}(2007{\natexlab{a}})\citenamefont {Slama}, \citenamefont {Bux},
  \citenamefont {Krenz}, \citenamefont {Zimmermann},\ and\ \citenamefont
  {Courteille}}]{Slama2007Superradiant}%
  \BibitemOpen
  \bibfield  {author} {\bibinfo {author} {\bibnamefont {Slama}, \bibfnamefont
  {S.}}, \bibinfo {author} {\bibfnamefont {S.}~\bibnamefont {Bux}}, \bibinfo
  {author} {\bibfnamefont {G.}~\bibnamefont {Krenz}}, \bibinfo {author}
  {\bibfnamefont {C.}~\bibnamefont {Zimmermann}}, \ and\ \bibinfo {author}
  {\bibfnamefont {P.~W.}\ \bibnamefont {Courteille}}} (\bibinfo {year}
  {2007}{\natexlab{a}}),\ \href {\doibase 10.1103/PhysRevLett.98.053603}
  {\bibfield  {journal} {\bibinfo  {journal} {Physical Review Letters}\
  }\textbf {\bibinfo {volume} {98}}~(\bibinfo {number} {5}),\ \bibinfo {pages}
  {053603+}}\BibitemShut {NoStop}%
\bibitem [{\citenamefont {Slama}\ \emph
  {et~al.}(2007{\natexlab{b}})\citenamefont {Slama}, \citenamefont {Krenz},
  \citenamefont {Bux}, \citenamefont {Zimmermann},\ and\ \citenamefont
  {Courteille}}]{Slama2007Cavityenhanced}%
  \BibitemOpen
  \bibfield  {author} {\bibinfo {author} {\bibnamefont {Slama}, \bibfnamefont
  {S.}}, \bibinfo {author} {\bibfnamefont {G.}~\bibnamefont {Krenz}}, \bibinfo
  {author} {\bibfnamefont {S.}~\bibnamefont {Bux}}, \bibinfo {author}
  {\bibfnamefont {C.}~\bibnamefont {Zimmermann}}, \ and\ \bibinfo {author}
  {\bibfnamefont {P.~W.}\ \bibnamefont {Courteille}}} (\bibinfo {year}
  {2007}{\natexlab{b}}),\ \href {\doibase 10.1103/PhysRevA.75.063620}
  {\bibfield  {journal} {\bibinfo  {journal} {Physical Review A}\ }\textbf
  {\bibinfo {volume} {75}}~(\bibinfo {number} {6}),\ \bibinfo {pages}
  {063620+}}\BibitemShut {NoStop}%
\bibitem [{\citenamefont {de~Souza}\ and\ \citenamefont
  {Tsallis}(1997)}]{deSouza1997Students}%
  \BibitemOpen
  \bibfield  {author} {\bibinfo {author} {\bibnamefont {de~Souza},
  \bibfnamefont {A.}}, \ and\ \bibinfo {author} {\bibfnamefont
  {C.}~\bibnamefont {Tsallis}}} (\bibinfo {year} {1997}),\ \href {\doibase
  10.1016/S0378-4371(96)00395-0} {\bibfield  {journal} {\bibinfo  {journal}
  {Physica A: Statistical Mechanics and its Applications}\ }\textbf {\bibinfo
  {volume} {236}}~(\bibinfo {number} {1-2}),\ \bibinfo {pages}
  {52}}\BibitemShut {NoStop}%
\bibitem [{\citenamefont {Specht}\ \emph {et~al.}(2011)\citenamefont {Specht},
  \citenamefont {Nolleke}, \citenamefont {Reiserer}, \citenamefont {Uphoff},
  \citenamefont {Figueroa}, \citenamefont {Ritter},\ and\ \citenamefont
  {Rempe}}]{Specht2011Singleatom}%
  \BibitemOpen
  \bibfield  {author} {\bibinfo {author} {\bibnamefont {Specht}, \bibfnamefont
  {H.~P.}}, \bibinfo {author} {\bibfnamefont {C.}~\bibnamefont {Nolleke}},
  \bibinfo {author} {\bibfnamefont {A.}~\bibnamefont {Reiserer}}, \bibinfo
  {author} {\bibfnamefont {M.}~\bibnamefont {Uphoff}}, \bibinfo {author}
  {\bibfnamefont {E.}~\bibnamefont {Figueroa}}, \bibinfo {author}
  {\bibfnamefont {S.}~\bibnamefont {Ritter}}, \ and\ \bibinfo {author}
  {\bibfnamefont {G.}~\bibnamefont {Rempe}}} (\bibinfo {year} {2011}),\ \href
  {\doibase 10.1038/nature09997} {\bibfield  {journal} {\bibinfo  {journal}
  {Nature}\ }\textbf {\bibinfo {volume} {473}}~(\bibinfo {number} {7346}),\
  \bibinfo {pages} {190}}\BibitemShut {NoStop}%
\bibitem [{\citenamefont {Stamper-Kurn}(2012)}]{StamperKurn2012Optomechanics}%
  \BibitemOpen
  \bibfield  {author} {\bibinfo {author} {\bibnamefont {Stamper-Kurn},
  \bibfnamefont {D.~M.}}} (\bibinfo {year} {2012}),\ \enquote {\bibinfo {title}
  {Cavity optomechanics with cold atoms},}\ in\ \href@noop {} {\emph {\bibinfo
  {booktitle} {Cavity optomechanics}}},\ \bibinfo {editor} {edited by\ \bibinfo
  {editor} {\bibfnamefont {M.}~\bibnamefont {Aspelmeyer}}, \bibinfo {editor}
  {\bibfnamefont {T.}~\bibnamefont {Kippenberg}}, \ and\ \bibinfo {editor}
  {\bibfnamefont {F.}~\bibnamefont {Marquardt}}}\ (\bibinfo  {publisher}
  {Springer})\ p.\ \bibinfo {pages} {(arXiv:1204.4351)}\BibitemShut {NoStop}%
\bibitem [{\citenamefont {Stenger}\ \emph {et~al.}(1999)\citenamefont
  {Stenger}, \citenamefont {Inouye}, \citenamefont {Chikkatur}, \citenamefont
  {Kurn}, \citenamefont {Pritchard},\ and\ \citenamefont
  {Ketterle}}]{Stenger1999Bragg}%
  \BibitemOpen
  \bibfield  {author} {\bibinfo {author} {\bibnamefont {Stenger}, \bibfnamefont
  {J.}}, \bibinfo {author} {\bibfnamefont {S.}~\bibnamefont {Inouye}}, \bibinfo
  {author} {\bibfnamefont {A.~P.}\ \bibnamefont {Chikkatur}}, \bibinfo {author}
  {\bibfnamefont {D.~M.~S.}\ \bibnamefont {Kurn}}, \bibinfo {author}
  {\bibfnamefont {D.~E.}\ \bibnamefont {Pritchard}}, \ and\ \bibinfo {author}
  {\bibfnamefont {W.}~\bibnamefont {Ketterle}}} (\bibinfo {year} {1999}),\
  \href {\doibase 10.1103/PhysRevLett.82.4569} {\bibfield  {journal} {\bibinfo
  {journal} {Physical Review Letters}\ }\textbf {\bibinfo {volume}
  {82}}~(\bibinfo {number} {23}),\ \bibinfo {pages} {4569}}\BibitemShut
  {NoStop}%
\bibitem [{\citenamefont {Strack}\ and\ \citenamefont
  {Sachdev}(2011)}]{Strack2011Dicke}%
  \BibitemOpen
  \bibfield  {author} {\bibinfo {author} {\bibnamefont {Strack}, \bibfnamefont
  {P.}}, \ and\ \bibinfo {author} {\bibfnamefont {S.}~\bibnamefont {Sachdev}}}
  (\bibinfo {year} {2011}),\ \href {\doibase 10.1103/PhysRevLett.107.277202}
  {\bibfield  {journal} {\bibinfo  {journal} {Physical Review Letters}\
  }\textbf {\bibinfo {volume} {107}},\ \bibinfo {pages} {277202+}}\BibitemShut
  {NoStop}%
\bibitem [{\citenamefont {Strogatz}(2000)}]{Strogatz2000From}%
  \BibitemOpen
  \bibfield  {author} {\bibinfo {author} {\bibnamefont {Strogatz},
  \bibfnamefont {S.}}} (\bibinfo {year} {2000}),\ \href {\doibase
  10.1016/S0167-2789(00)00094-4} {\bibfield  {journal} {\bibinfo  {journal}
  {Physica D: Nonlinear Phenomena}\ }\textbf {\bibinfo {volume}
  {143}}~(\bibinfo {number} {1-4}),\ \bibinfo {pages} {1}}\BibitemShut
  {NoStop}%
\bibitem [{\citenamefont {Sun}\ \emph {et~al.}(2011)\citenamefont {Sun},
  \citenamefont {Hu}, \citenamefont {Ji},\ and\ \citenamefont
  {Liu}}]{Sun2011Dynamics}%
  \BibitemOpen
  \bibfield  {author} {\bibinfo {author} {\bibnamefont {Sun}, \bibfnamefont
  {Q.}}, \bibinfo {author} {\bibfnamefont {X.~H.}\ \bibnamefont {Hu}}, \bibinfo
  {author} {\bibfnamefont {A.~C.}\ \bibnamefont {Ji}}, \ and\ \bibinfo {author}
  {\bibfnamefont {W.~M.}\ \bibnamefont {Liu}}} (\bibinfo {year} {2011}),\ \href
  {\doibase 10.1103/PhysRevA.83.043606} {\bibfield  {journal} {\bibinfo
  {journal} {Physical Review A}\ }\textbf {\bibinfo {volume} {83}}~(\bibinfo
  {number} {4}),\ \bibinfo {pages} {043606+}}\BibitemShut {NoStop}%
\bibitem [{\citenamefont {Szirmai}\ \emph {et~al.}(2009)\citenamefont
  {Szirmai}, \citenamefont {Nagy},\ and\ \citenamefont
  {Domokos}}]{Szirmai2009Excess}%
  \BibitemOpen
  \bibfield  {author} {\bibinfo {author} {\bibnamefont {Szirmai}, \bibfnamefont
  {G.}}, \bibinfo {author} {\bibfnamefont {D.}~\bibnamefont {Nagy}}, \ and\
  \bibinfo {author} {\bibfnamefont {P.}~\bibnamefont {Domokos}}} (\bibinfo
  {year} {2009}),\ \href {\doibase 10.1103/PhysRevLett.102.080401} {\bibfield
  {journal} {\bibinfo  {journal} {Physical Review Letters}\ }\textbf {\bibinfo
  {volume} {102}}~(\bibinfo {number} {8}),\ \bibinfo {pages}
  {080401+}}\BibitemShut {NoStop}%
\bibitem [{\citenamefont {Szirmai}\ \emph {et~al.}(2010)\citenamefont
  {Szirmai}, \citenamefont {Nagy},\ and\ \citenamefont
  {Domokos}}]{Szirmai2010Quantum}%
  \BibitemOpen
  \bibfield  {author} {\bibinfo {author} {\bibnamefont {Szirmai}, \bibfnamefont
  {G.}}, \bibinfo {author} {\bibfnamefont {D.}~\bibnamefont {Nagy}}, \ and\
  \bibinfo {author} {\bibfnamefont {P.}~\bibnamefont {Domokos}}} (\bibinfo
  {year} {2010}),\ \href {\doibase 10.1103/PhysRevA.81.043639} {\bibfield
  {journal} {\bibinfo  {journal} {Physical Review A}\ }\textbf {\bibinfo
  {volume} {81}}~(\bibinfo {number} {4}),\ \bibinfo {pages}
  {043639+}}\BibitemShut {NoStop}%
\bibitem [{\citenamefont {Tavis}\ and\ \citenamefont
  {Cummings}(1968)}]{Tavis1968Exact}%
  \BibitemOpen
  \bibfield  {author} {\bibinfo {author} {\bibnamefont {Tavis}, \bibfnamefont
  {M.}}, \ and\ \bibinfo {author} {\bibfnamefont {F.~W.}\ \bibnamefont
  {Cummings}}} (\bibinfo {year} {1968}),\ \href {\doibase
  10.1103/PhysRev.170.379} {\bibfield  {journal} {\bibinfo  {journal} {Physical
  Review Online Archive (Prola)}\ }\textbf {\bibinfo {volume} {170}}~(\bibinfo
  {number} {2}),\ \bibinfo {pages} {379}}\BibitemShut {NoStop}%
\bibitem [{\citenamefont {Teper}\ \emph {et~al.}(2006)\citenamefont {Teper},
  \citenamefont {Lin},\ and\ \citenamefont
  {Vuleti\'{c}}}]{Teper2006ResonatorAided}%
  \BibitemOpen
  \bibfield  {author} {\bibinfo {author} {\bibnamefont {Teper}, \bibfnamefont
  {I.}}, \bibinfo {author} {\bibfnamefont {Y.~J.}\ \bibnamefont {Lin}}, \ and\
  \bibinfo {author} {\bibfnamefont {V.}~\bibnamefont {Vuleti\'{c}}}} (\bibinfo
  {year} {2006}),\ \href {\doibase 10.1103/PhysRevLett.97.023002} {\bibfield
  {journal} {\bibinfo  {journal} {Physical Review Letters}\ }\textbf {\bibinfo
  {volume} {97}}~(\bibinfo {number} {2}),\ \bibinfo {pages}
  {023002+}}\BibitemShut {NoStop}%
\bibitem [{\citenamefont {{T}hompson}\ \emph {et~al.}(2008)\citenamefont
  {{T}hompson}, \citenamefont {{Z}wickl}, \citenamefont {{J}ayich},
  \citenamefont {{M}arquardt}, \citenamefont {{G}irvin},\ and\ \citenamefont
  {{H}arris}}]{Thompson2008Strong}%
  \BibitemOpen
  \bibfield  {author} {\bibinfo {author} {\bibnamefont {{T}hompson},
  \bibfnamefont {J.~D.}}, \bibinfo {author} {\bibfnamefont {B.~M.}\
  \bibnamefont {{Z}wickl}}, \bibinfo {author} {\bibfnamefont {A.~M.}\
  \bibnamefont {{J}ayich}}, \bibinfo {author} {\bibfnamefont {F.}~\bibnamefont
  {{M}arquardt}}, \bibinfo {author} {\bibfnamefont {S.~M.}\ \bibnamefont
  {{G}irvin}}, \ and\ \bibinfo {author} {\bibfnamefont {J.~G.~E.}\ \bibnamefont
  {{H}arris}}} (\bibinfo {year} {2008}),\ \href
  {http://dx.doi.org/10.1038/nature06715} {\bibfield  {journal} {\bibinfo
  {journal} {\{N\}ature}\ }\textbf {\bibinfo {volume} {452}}~(\bibinfo {number}
  {7183}),\ \bibinfo {pages} {72}}\BibitemShut {NoStop}%
\bibitem [{\citenamefont {Tuchman}\ \emph {et~al.}(2006)\citenamefont
  {Tuchman}, \citenamefont {Long}, \citenamefont {Vrijsen}, \citenamefont
  {Boudet}, \citenamefont {Lee},\ and\ \citenamefont
  {Kasevich}}]{Tuchman2006Normalmode}%
  \BibitemOpen
  \bibfield  {author} {\bibinfo {author} {\bibnamefont {Tuchman}, \bibfnamefont
  {A.~K.}}, \bibinfo {author} {\bibfnamefont {R.}~\bibnamefont {Long}},
  \bibinfo {author} {\bibfnamefont {G.}~\bibnamefont {Vrijsen}}, \bibinfo
  {author} {\bibfnamefont {J.}~\bibnamefont {Boudet}}, \bibinfo {author}
  {\bibfnamefont {J.}~\bibnamefont {Lee}}, \ and\ \bibinfo {author}
  {\bibfnamefont {M.~A.}\ \bibnamefont {Kasevich}}} (\bibinfo {year}
  {{2006}}),\ \href {\doibase \%7B10.1103/PhysRevA.74.053821\%7D} {\bibfield
  {journal} {\bibinfo  {journal} {{PHYSICAL REVIEW A}}\ }\textbf {\bibinfo
  {volume} {{74}}}~(\bibinfo {number} {{5}}),\
  \%7B10.1103/PhysRevA.74.053821\%7D}\BibitemShut {NoStop}%
\bibitem [{\citenamefont {Vidal}\ \emph {et~al.}(2010)\citenamefont {Vidal},
  \citenamefont {De~Chiara}, \citenamefont {Larson},\ and\ \citenamefont
  {Morigi}}]{Vidal2010Quantum}%
  \BibitemOpen
  \bibfield  {author} {\bibinfo {author} {\bibnamefont {Vidal}, \bibfnamefont
  {S.~F.}}, \bibinfo {author} {\bibfnamefont {G.}~\bibnamefont {De~Chiara}},
  \bibinfo {author} {\bibfnamefont {J.}~\bibnamefont {Larson}}, \ and\ \bibinfo
  {author} {\bibfnamefont {G.}~\bibnamefont {Morigi}}} (\bibinfo {year}
  {2010}),\ \href {\doibase 10.1103/PhysRevA.81.043407} {\bibfield  {journal}
  {\bibinfo  {journal} {Physical Review A}\ }\textbf {\bibinfo {volume}
  {81}}~(\bibinfo {number} {4}),\ \bibinfo {pages} {043407+}}\BibitemShut
  {NoStop}%
\bibitem [{\citenamefont {Vidal}\ \emph {et~al.}(2007)\citenamefont {Vidal},
  \citenamefont {Zippilli},\ and\ \citenamefont {Morigi}}]{Vidal2007Nonlinear}%
  \BibitemOpen
  \bibfield  {author} {\bibinfo {author} {\bibnamefont {Vidal}, \bibfnamefont
  {S.~F.}}, \bibinfo {author} {\bibfnamefont {S.}~\bibnamefont {Zippilli}}, \
  and\ \bibinfo {author} {\bibfnamefont {G.}~\bibnamefont {Morigi}}} (\bibinfo
  {year} {2007}),\ \href {\doibase 10.1103/PhysRevA.76.053829} {\bibfield
  {journal} {\bibinfo  {journal} {Physical Review A}\ }\textbf {\bibinfo
  {volume} {76}},\ \bibinfo {pages} {053829+}}\BibitemShut {NoStop}%
\bibitem [{\citenamefont {Vilensky}\ \emph {et~al.}(2007)\citenamefont
  {Vilensky}, \citenamefont {Prior},\ and\ \citenamefont
  {Sh}}]{Vilensky2007Cooling}%
  \BibitemOpen
  \bibfield  {author} {\bibinfo {author} {\bibnamefont {Vilensky},
  \bibfnamefont {M.~Y.}}, \bibinfo {author} {\bibfnamefont {Y.}~\bibnamefont
  {Prior}}, \ and\ \bibinfo {author} {\bibfnamefont {I.}~\bibnamefont {Sh}}}
  (\bibinfo {year} {2007}),\ \href {\doibase 10.1103/PhysRevLett.99.103002}
  {\bibfield  {journal} {\bibinfo  {journal} {Physical Review Letters}\
  }\textbf {\bibinfo {volume} {99}}~(\bibinfo {number} {10}),\ \bibinfo {pages}
  {103002+}}\BibitemShut {NoStop}%
\bibitem [{\citenamefont {Vukics}(2012)}]{Vukics2012CQEDv2}%
  \BibitemOpen
  \bibfield  {author} {\bibinfo {author} {\bibnamefont {Vukics}, \bibfnamefont
  {A.}}} (\bibinfo {year} {2012}),\ \href {\doibase 10.1016/j.cpc.2012.02.004}
  {\bibfield  {journal} {\bibinfo  {journal} {Computer Physics Communications}\
  }\textbf {\bibinfo {volume} {183}}~(\bibinfo {number} {6}),\ \bibinfo {pages}
  {1381}}\BibitemShut {NoStop}%
\bibitem [{\citenamefont {Vukics}\ and\ \citenamefont
  {Domokos}(2005)}]{Vukics2005Simultaneous}%
  \BibitemOpen
  \bibfield  {author} {\bibinfo {author} {\bibnamefont {Vukics}, \bibfnamefont
  {A.}}, \ and\ \bibinfo {author} {\bibfnamefont {P.}~\bibnamefont {Domokos}}}
  (\bibinfo {year} {2005}),\ \href {\doibase 10.1103/PhysRevA.72.031401}
  {\bibfield  {journal} {\bibinfo  {journal} {Physical Review A}\ }\textbf
  {\bibinfo {volume} {72}}~(\bibinfo {number} {3}),\ \bibinfo {pages}
  {031401+}}\BibitemShut {NoStop}%
\bibitem [{\citenamefont {Vukics}\ \emph {et~al.}(2004)\citenamefont {Vukics},
  \citenamefont {Domokos},\ and\ \citenamefont
  {Ritsch}}]{Vukics2004Multidimensional}%
  \BibitemOpen
  \bibfield  {author} {\bibinfo {author} {\bibnamefont {Vukics}, \bibfnamefont
  {A.}}, \bibinfo {author} {\bibfnamefont {P.}~\bibnamefont {Domokos}}, \ and\
  \bibinfo {author} {\bibfnamefont {H.}~\bibnamefont {Ritsch}}} (\bibinfo
  {year} {2004}),\ \href@noop {} {\bibfield  {journal} {\bibinfo  {journal} {J.
  Opt. B: Quantum Semiclass. Opt.}\ }\textbf {\bibinfo {volume} {6}},\ \bibinfo
  {pages} {143}}\BibitemShut {NoStop}%
\bibitem [{\citenamefont {Vukics}\ \emph {et~al.}(2005)\citenamefont {Vukics},
  \citenamefont {Janszky},\ and\ \citenamefont {Domokos}}]{Vukics2005Cavity}%
  \BibitemOpen
  \bibfield  {author} {\bibinfo {author} {\bibnamefont {Vukics}, \bibfnamefont
  {A.}}, \bibinfo {author} {\bibfnamefont {J.}~\bibnamefont {Janszky}}, \ and\
  \bibinfo {author} {\bibfnamefont {P.}~\bibnamefont {Domokos}}} (\bibinfo
  {year} {2005}),\ \href {\doibase 10.1088/0953-4075/38/10/005} {\bibfield
  {journal} {\bibinfo  {journal} {Journal of Physics B: Atomic, Molecular and
  Optical Physics}\ }\textbf {\bibinfo {volume} {38}}~(\bibinfo {number}
  {10}),\ \bibinfo {pages} {1453}}\BibitemShut {NoStop}%
\bibitem [{\citenamefont {Vukics}\ \emph {et~al.}(2007)\citenamefont {Vukics},
  \citenamefont {Maschler},\ and\ \citenamefont
  {Ritsch}}]{Vukics2007Microscopic}%
  \BibitemOpen
  \bibfield  {author} {\bibinfo {author} {\bibnamefont {Vukics}, \bibfnamefont
  {A.}}, \bibinfo {author} {\bibfnamefont {C.}~\bibnamefont {Maschler}}, \ and\
  \bibinfo {author} {\bibfnamefont {H.}~\bibnamefont {Ritsch}}} (\bibinfo
  {year} {2007}),\ \href {\doibase 10.1088/1367-2630/9/8/255} {\bibfield
  {journal} {\bibinfo  {journal} {New J. Phys.}\ }\textbf {\bibinfo {volume}
  {9}}~(\bibinfo {number} {8}),\ \bibinfo {pages} {255+}}\BibitemShut {NoStop}%
\bibitem [{\citenamefont {Vukics}\ \emph {et~al.}(2009)\citenamefont {Vukics},
  \citenamefont {Niedenzu},\ and\ \citenamefont {Ritsch}}]{Vukics2009Cavity}%
  \BibitemOpen
  \bibfield  {author} {\bibinfo {author} {\bibnamefont {Vukics}, \bibfnamefont
  {A.}}, \bibinfo {author} {\bibfnamefont {W.}~\bibnamefont {Niedenzu}}, \ and\
  \bibinfo {author} {\bibfnamefont {H.}~\bibnamefont {Ritsch}}} (\bibinfo
  {year} {2009}),\ \href {\doibase 10.1103/PhysRevA.79.013828} {\bibfield
  {journal} {\bibinfo  {journal} {Physical Review A}\ }\textbf {\bibinfo
  {volume} {79}},\ \bibinfo {pages} {013828+}}\BibitemShut {NoStop}%
\bibitem [{\citenamefont {Vukics}\ and\ \citenamefont
  {Ritsch}(2007)}]{Vukics2007CQED}%
  \BibitemOpen
  \bibfield  {author} {\bibinfo {author} {\bibnamefont {Vukics}, \bibfnamefont
  {A.}}, \ and\ \bibinfo {author} {\bibfnamefont {H.}~\bibnamefont {Ritsch}}}
  (\bibinfo {year} {2007}),\ \href {\doibase 10.1140/epjd/e2007-00210-x}
  {\bibfield  {journal} {\bibinfo  {journal} {The European Physical Journal D -
  Atomic, Molecular, Optical and Plasma Physics}\ }\textbf {\bibinfo {volume}
  {44}}~(\bibinfo {number} {3}),\ \bibinfo {pages} {585}}\BibitemShut {NoStop}%
\bibitem [{\citenamefont {Vuletic}(2001)}]{Vuletic2001Cavity}%
  \BibitemOpen
  \bibfield  {author} {\bibinfo {author} {\bibnamefont {Vuletic}, \bibfnamefont
  {V.}}} (\bibinfo {year} {2001}),\ in\ \href@noop {} {\emph {\bibinfo
  {booktitle} {Laser physics at the limits}}},\ \bibinfo {editor} {edited by\
  \bibinfo {editor} {\bibfnamefont {H.}~\bibnamefont {Figger}}, \bibinfo
  {editor} {\bibfnamefont {D.}~\bibnamefont {Meschede}}, \ and\ \bibinfo
  {editor} {\bibfnamefont {C.}~\bibnamefont {Zimmermann}}}\ (\bibinfo
  {publisher} {Springer, New York})\ pp.\ \bibinfo {pages} {67--74}\BibitemShut
  {NoStop}%
\bibitem [{\citenamefont {Vuleti\'{c}}\ \emph {et~al.}(2001)\citenamefont
  {Vuleti\'{c}}, \citenamefont {Chan},\ and\ \citenamefont
  {Black}}]{Vuletic2001Threedimensional}%
  \BibitemOpen
  \bibfield  {author} {\bibinfo {author} {\bibnamefont {Vuleti\'{c}},
  \bibfnamefont {V.}}, \bibinfo {author} {\bibfnamefont {H.~W.}\ \bibnamefont
  {Chan}}, \ and\ \bibinfo {author} {\bibfnamefont {A.~T.}\ \bibnamefont
  {Black}}} (\bibinfo {year} {2001}),\ \href {\doibase
  10.1103/PhysRevA.64.033405} {\bibfield  {journal} {\bibinfo  {journal}
  {Physical Review A}\ }\textbf {\bibinfo {volume} {64}}~(\bibinfo {number}
  {3}),\ \bibinfo {pages} {033405+}}\BibitemShut {NoStop}%
\bibitem [{\citenamefont {Vuleti\'{c}}\ and\ \citenamefont
  {Chu}(2000)}]{Vuletic2000Laser}%
  \BibitemOpen
  \bibfield  {author} {\bibinfo {author} {\bibnamefont {Vuleti\'{c}},
  \bibfnamefont {V.}}, \ and\ \bibinfo {author} {\bibfnamefont
  {S.}~\bibnamefont {Chu}}} (\bibinfo {year} {2000}),\ \href {\doibase
  10.1103/PhysRevLett.84.3787} {\bibfield  {journal} {\bibinfo  {journal}
  {Physical Review Letters}\ }\textbf {\bibinfo {volume} {84}}~(\bibinfo
  {number} {17}),\ \bibinfo {pages} {3787}}\BibitemShut {NoStop}%
\bibitem [{\citenamefont {{W}alther}(2002)}]{Walther2002Quantum}%
  \BibitemOpen
  \bibfield  {author} {\bibinfo {author} {\bibnamefont {{W}alther},
  \bibfnamefont {H.}}} (\bibinfo {year} {2002}),\ \href@noop {} {\bibfield
  {journal} {\bibinfo  {journal} {{A}dv. {C}hem. {P}hys.}\ }\textbf {\bibinfo
  {volume} {122}},\ \bibinfo {pages} {167}}\BibitemShut {NoStop}%
\bibitem [{\citenamefont {Wang}\ and\ \citenamefont
  {Hioe}(1973)}]{Wang1973Phase}%
  \BibitemOpen
  \bibfield  {author} {\bibinfo {author} {\bibnamefont {Wang}, \bibfnamefont
  {Y.~K.}}, \ and\ \bibinfo {author} {\bibfnamefont {F.~T.}\ \bibnamefont
  {Hioe}}} (\bibinfo {year} {1973}),\ \href {\doibase 10.1103/PhysRevA.7.831}
  {\bibfield  {journal} {\bibinfo  {journal} {Physical Review A}\ }\textbf
  {\bibinfo {volume} {7}}~(\bibinfo {number} {3}),\ \bibinfo {pages}
  {831}}\BibitemShut {NoStop}%
\bibitem [{\citenamefont {Wilk}\ \emph
  {et~al.}(2007{\natexlab{a}})\citenamefont {Wilk}, \citenamefont {Webster},
  \citenamefont {Kuhn},\ and\ \citenamefont {Rempe}}]{Wilk2007SingleAtom}%
  \BibitemOpen
  \bibfield  {author} {\bibinfo {author} {\bibnamefont {Wilk}, \bibfnamefont
  {T.}}, \bibinfo {author} {\bibfnamefont {S.~C.}\ \bibnamefont {Webster}},
  \bibinfo {author} {\bibfnamefont {A.}~\bibnamefont {Kuhn}}, \ and\ \bibinfo
  {author} {\bibfnamefont {G.}~\bibnamefont {Rempe}}} (\bibinfo {year}
  {2007}{\natexlab{a}}),\ \href {\doibase 10.1126/science.1143835} {\bibfield
  {journal} {\bibinfo  {journal} {Science}\ }\textbf {\bibinfo {volume}
  {317}}~(\bibinfo {number} {5837}),\ \bibinfo {pages} {488}}\BibitemShut
  {NoStop}%
\bibitem [{\citenamefont {Wilk}\ \emph
  {et~al.}(2007{\natexlab{b}})\citenamefont {Wilk}, \citenamefont {Webster},
  \citenamefont {Specht}, \citenamefont {Rempe},\ and\ \citenamefont
  {Kuhn}}]{Wilk2007PolarizationControlled}%
  \BibitemOpen
  \bibfield  {author} {\bibinfo {author} {\bibnamefont {Wilk}, \bibfnamefont
  {T.}}, \bibinfo {author} {\bibfnamefont {S.~C.}\ \bibnamefont {Webster}},
  \bibinfo {author} {\bibfnamefont {H.~P.}\ \bibnamefont {Specht}}, \bibinfo
  {author} {\bibfnamefont {G.}~\bibnamefont {Rempe}}, \ and\ \bibinfo {author}
  {\bibfnamefont {A.}~\bibnamefont {Kuhn}}} (\bibinfo {year}
  {2007}{\natexlab{b}}),\ \href {\doibase 10.1103/PhysRevLett.98.063601}
  {\bibfield  {journal} {\bibinfo  {journal} {Physical Review Letters}\
  }\textbf {\bibinfo {volume} {98}}~(\bibinfo {number} {6}),\ \bibinfo {pages}
  {063601+}}\BibitemShut {NoStop}%
\bibitem [{\citenamefont {Wolke}\ \emph {et~al.}(2012)\citenamefont {Wolke},
  \citenamefont {Klinner}, \citenamefont {Ke{\ss}ler},\ and\ \citenamefont
  {Hemmerich}}]{Wolke2012Cavity}%
  \BibitemOpen
  \bibfield  {author} {\bibinfo {author} {\bibnamefont {Wolke}, \bibfnamefont
  {M.}}, \bibinfo {author} {\bibfnamefont {J.}~\bibnamefont {Klinner}},
  \bibinfo {author} {\bibfnamefont {H.}~\bibnamefont {Ke{\ss}ler}}, \ and\
  \bibinfo {author} {\bibfnamefont {A.}~\bibnamefont {Hemmerich}}} (\bibinfo
  {year} {2012}),\ \href {\doibase 10.1126/science.1219166} {\bibfield
  {journal} {\bibinfo  {journal} {Science}\ }\textbf {\bibinfo {volume}
  {337}}~(\bibinfo {number} {6090}),\ \bibinfo {pages} {75}}\BibitemShut
  {NoStop}%
\bibitem [{\citenamefont {Xuereb}\ \emph {et~al.}(2009)\citenamefont {Xuereb},
  \citenamefont {Domokos}, \citenamefont {Asb\'{o}th}, \citenamefont {Horak},\
  and\ \citenamefont {Freegarde}}]{Xuereb2009Scattering}%
  \BibitemOpen
  \bibfield  {author} {\bibinfo {author} {\bibnamefont {Xuereb}, \bibfnamefont
  {A.}}, \bibinfo {author} {\bibfnamefont {P.}~\bibnamefont {Domokos}},
  \bibinfo {author} {\bibfnamefont {J.}~\bibnamefont {Asb\'{o}th}}, \bibinfo
  {author} {\bibfnamefont {P.}~\bibnamefont {Horak}}, \ and\ \bibinfo {author}
  {\bibfnamefont {T.}~\bibnamefont {Freegarde}}} (\bibinfo {year} {2009}),\
  \href {\doibase 10.1103/PhysRevA.79.053810} {\bibfield  {journal} {\bibinfo
  {journal} {Physical Review A}\ }\textbf {\bibinfo {volume} {79}}~(\bibinfo
  {number} {5}),\ \bibinfo {pages} {053810+}}\BibitemShut {NoStop}%
\bibitem [{\citenamefont {Xuereb}\ \emph {et~al.}(2010)\citenamefont {Xuereb},
  \citenamefont {Freegarde}, \citenamefont {Horak},\ and\ \citenamefont
  {Domokos}}]{Xuereb2010Optomechanical}%
  \BibitemOpen
  \bibfield  {author} {\bibinfo {author} {\bibnamefont {Xuereb}, \bibfnamefont
  {A.}}, \bibinfo {author} {\bibfnamefont {T.}~\bibnamefont {Freegarde}},
  \bibinfo {author} {\bibfnamefont {P.}~\bibnamefont {Horak}}, \ and\ \bibinfo
  {author} {\bibfnamefont {P.}~\bibnamefont {Domokos}}} (\bibinfo {year}
  {2010}),\ \href {\doibase 10.1103/PhysRevLett.105.013602} {\bibfield
  {journal} {\bibinfo  {journal} {Physical Review Letters}\ }\textbf {\bibinfo
  {volume} {105}}~(\bibinfo {number} {1}),\ \bibinfo {pages}
  {013602+}}\BibitemShut {NoStop}%
\bibitem [{\citenamefont {Xuereb}\ \emph {et~al.}(2012)\citenamefont {Xuereb},
  \citenamefont {Genes},\ and\ \citenamefont {Dantan}}]{Xuereb2012Strong}%
  \BibitemOpen
  \bibfield  {author} {\bibinfo {author} {\bibnamefont {Xuereb}, \bibfnamefont
  {A.}}, \bibinfo {author} {\bibfnamefont {C.}~\bibnamefont {Genes}}, \ and\
  \bibinfo {author} {\bibfnamefont {A.}~\bibnamefont {Dantan}}} (\bibinfo
  {year} {2012}),\ \href@noop {} {\bibinfo  {journal} {Arxiv preprint
  arXiv:1202.6210}\ }\BibitemShut {NoStop}%
\bibitem [{\citenamefont {Zeppenfeld}\ \emph {et~al.}(2009)\citenamefont
  {Zeppenfeld}, \citenamefont {Motsch}, \citenamefont {Pinkse},\ and\
  \citenamefont {Rempe}}]{Zeppenfeld2009Optoelectrical}%
  \BibitemOpen
\bibfield  {journal} {  }\bibfield  {author} {\bibinfo {author} {\bibnamefont
  {Zeppenfeld}, \bibfnamefont {M.}}, \bibinfo {author} {\bibfnamefont
  {M.}~\bibnamefont {Motsch}}, \bibinfo {author} {\bibfnamefont {P.~W.~H.}\
  \bibnamefont {Pinkse}}, \ and\ \bibinfo {author} {\bibfnamefont
  {G.}~\bibnamefont {Rempe}}} (\bibinfo {year} {2009}),\ \href {\doibase
  10.1103/PhysRevA.80.041401} {\bibfield  {journal} {\bibinfo  {journal}
  {Physical Review A}\ }\textbf {\bibinfo {volume} {80}}~(\bibinfo {number}
  {4}),\ \bibinfo {pages} {041401+}}\BibitemShut {NoStop}%
\bibitem [{\citenamefont {Zhang}\ \emph {et~al.}(2009)\citenamefont {Zhang},
  \citenamefont {Cui}, \citenamefont {Zhou},\ and\ \citenamefont
  {Liu}}]{Zhang2009Nonlinear}%
  \BibitemOpen
  \bibfield  {author} {\bibinfo {author} {\bibnamefont {Zhang}, \bibfnamefont
  {J.~M.}}, \bibinfo {author} {\bibfnamefont {F.~C.}\ \bibnamefont {Cui}},
  \bibinfo {author} {\bibfnamefont {D.~L.}\ \bibnamefont {Zhou}}, \ and\
  \bibinfo {author} {\bibfnamefont {W.~M.}\ \bibnamefont {Liu}}} (\bibinfo
  {year} {2009}),\ \href
  {http://scitation.aip.org/getabs/servlet/GetabsServlet?prog=normal\&id=PLRAAN000079000003033401000001\&idtype=cvips\&gifs=yes}
  {\bibfield  {journal} {\bibinfo  {journal} {Physical Review A (Atomic,
  Molecular, and Optical Physics)}\ }\textbf {\bibinfo {volume} {79}}~(\bibinfo
  {number} {3}),\ \bibinfo {pages} {033401+}}\BibitemShut {NoStop}%
\bibitem [{\citenamefont {Zippilli}\ \emph
  {et~al.}(2004{\natexlab{a}})\citenamefont {Zippilli}, \citenamefont {Asboth},
  \citenamefont {Morigi},\ and\ \citenamefont {Ritsch}}]{Zippilli2004Forces}%
  \BibitemOpen
  \bibfield  {author} {\bibinfo {author} {\bibnamefont {Zippilli},
  \bibfnamefont {S.}}, \bibinfo {author} {\bibfnamefont {J.}~\bibnamefont
  {Asboth}}, \bibinfo {author} {\bibfnamefont {G.}~\bibnamefont {Morigi}}, \
  and\ \bibinfo {author} {\bibfnamefont {H.}~\bibnamefont {Ritsch}}} (\bibinfo
  {year} {2004}{\natexlab{a}}),\ \href {\doibase 10.1007/s00340-004-1660-x}
  {\bibfield  {journal} {\bibinfo  {journal} {Applied Physics B: Lasers and
  Optics}\ }\textbf {\bibinfo {volume} {79}}~(\bibinfo {number} {8}),\ \bibinfo
  {pages} {969}}\BibitemShut {NoStop}%
\bibitem [{\citenamefont {Zippilli}\ and\ \citenamefont
  {Morigi}(2005)}]{Zippilli2005Cooling}%
  \BibitemOpen
  \bibfield  {author} {\bibinfo {author} {\bibnamefont {Zippilli},
  \bibfnamefont {S.}}, \ and\ \bibinfo {author} {\bibfnamefont
  {G.}~\bibnamefont {Morigi}}} (\bibinfo {year} {2005}),\ \href {\doibase
  10.1103/PhysRevLett.95.143001} {\bibfield  {journal} {\bibinfo  {journal}
  {Physical Review Letters}\ }\textbf {\bibinfo {volume} {95}}~(\bibinfo
  {number} {14}),\ \bibinfo {pages} {143001+}}\BibitemShut {NoStop}%
\bibitem [{\citenamefont {Zippilli}\ and\ \citenamefont
  {Morigi}(2007)}]{Zippilli2007Mechanical}%
  \BibitemOpen
  \bibfield  {author} {\bibinfo {author} {\bibnamefont {Zippilli},
  \bibfnamefont {S.}}, \ and\ \bibinfo {author} {\bibfnamefont
  {G.}~\bibnamefont {Morigi}}} (\bibinfo {year} {2007}),\ \href {\doibase
  10.1103/PhysRevA.72.053408} {\bibfield  {journal} {\bibinfo  {journal}
  {Physical Review A}\ }\textbf {\bibinfo {volume} {72}}~(\bibinfo {number}
  {5}),\ \bibinfo {pages} {053408+}},\ \Eprint
  {http://arxiv.org/abs/quant-ph/0508075} {quant-ph/0508075} \BibitemShut
  {NoStop}%
\bibitem [{\citenamefont {Zippilli}\ \emph
  {et~al.}(2004{\natexlab{b}})\citenamefont {Zippilli}, \citenamefont
  {Morigi},\ and\ \citenamefont {Ritsch}}]{Zippilli2004Collective}%
  \BibitemOpen
  \bibfield  {author} {\bibinfo {author} {\bibnamefont {Zippilli},
  \bibfnamefont {S.}}, \bibinfo {author} {\bibfnamefont {G.}~\bibnamefont
  {Morigi}}, \ and\ \bibinfo {author} {\bibfnamefont {H.}~\bibnamefont
  {Ritsch}}} (\bibinfo {year} {2004}{\natexlab{b}}),\ \href {\doibase
  10.1140/epjd/e2004-00137-8} {\bibfield  {journal} {\bibinfo  {journal} {The
  European Physical Journal D - Atomic, Molecular, Optical and Plasma Physics}\
  }\textbf {\bibinfo {volume} {31}}~(\bibinfo {number} {3}),\ \bibinfo {pages}
  {507}}\BibitemShut {NoStop}%
\end{thebibliography}%

\end{document}